\documentclass[final,twoside,12pt]{article}

\usepackage{latexsym}
\usepackage{graphicx}
\usepackage{amsfonts}
\usepackage{amssymb,amscd}
\usepackage[dvips]{epsfig}                   
\usepackage{array}
\usepackage{wrapfig}
\usepackage{graphics}
\usepackage{amsmath}
\usepackage{enumitem}
\usepackage[numbers,sort&compress]{natbib}

\newcommand*{\La}{{\cal{L}}}
\newcommand*{\no}{\noindent}
\newcommand*{\bea}{\begin{eqnarray}}
\newcommand*{\eea}{\end{eqnarray}}
\newcommand*{\be}{\begin{equation}}
\newcommand*{\ee}{\end{equation}}
\newcommand*{\pd}{\partial}
\newcommand*{\pdm}{\pd_{\mu}}
\newcommand*{\pdn}{\pd_{\nu}}

\newcommand*{\pref}[1]{(\ref{#1})}

\newcommand*{\mn}{{\mu\nu}}

\newcommand*{\prefr}[2]{(\ref{#1}-\ref{#2})} 
\newcommand*{\nn}{\nonumber}

\newcommand*{\tr}{\mathrm{tr}}

\newcommand*{\diag}{\mathrm{diag}}

\newcommand*{\tl}{\mathrm{tl}}

\newcommand*{\la}{\left\langle}
\newcommand*{\ra}{\right\rangle}
\newcommand*{\bma}{\begin{matrix}}
\newcommand*{\ema}{\end{matrix}}
\newcommand*{\bpm}{\left(\bma}
\newcommand*{\epm}{\ema\right)}
\newcommand*{\rn}{\mathbb{R}}
\newcommand*{\cn}{\mathbb{C}}
\newcommand*{\cd}{{\cal D}}
\newcommand*{\pint}{\int\cd}
\newcommand*{\op}{{\cal O}}
\newcommand*{\ym}{{\text{YM}}}

\begin{document}
 
\title{Brout-Englert-Higgs physics: From foundations to phenomenology}

\author{Axel Maas\\
Institute of Physics, NAWI Graz,\\ University of Graz, Universit\"atsplatz 5, A-8010 Graz, Austria}

\maketitle

\begin{abstract}

Brout-Englert Higgs physics is one of the most central and successful parts of the standard model. It is also part of a multitude of beyond-the-standard-model scenarios.

The aim of this review is to describe the field-theoretical foundations of Brout-Englert-Higgs physics, and to show how the usual phenomenology arises from it. This requires to give a precise and gauge-invariant meaning to the underlying physics. This is complicated by the fact that concepts like the Higgs vacuum expectation value or the separation between confinement and the Brout-Englert-Higgs effect loose their meaning beyond perturbation theory. This is addressed by carefully constructing the corresponding theory space and the quantum phase diagram of theories with elementary Higgs fields and gauge interactions.

The physical spectrum needs then to be also given in terms of gauge-invariant, i.\ e.\ composite, states. Using gauge-invariant perturbation theory, as developed by Fr\"ohlich, Morchio, and Strocchi, it is possible to rederive conventional perturbation theory in this framework. This derivation explicitly shows why the description of the standard model in terms of the unphysical, gauge-dependent, elementary states of the Higgs and $W$-bosons and $Z$-boson, but also of the elementary fermions, is adequate and successful.

These are unavoidable consequences of the field theory underlying the standard model, from which the usual picture emerges. The validity of this emergence can only be tested non-perturbatively. Such tests, in particular using lattice gauge theory, will be reviewed as well. They fully confirm the underlying mechanisms.

In this course it will be seen that the structure of the standard model is very special, and qualitative changes occur beyond it. The extension beyond the standard model will therefore also be reviewed. Particular attention will be given to structural differences arising for phenomenology. Again, non-perturbative tests of these results will be reviewed.

Finally, to make this review self-contained a brief discussion of issues like the triviality and hierarchy problem, and how they fit into a fundamental field-theoretical formulation, is included.

\end{abstract}


\pagenumbering{roman}
\setcounter{secnumdepth}{4}
\setcounter{tocdepth}{4}
\tableofcontents
\newpage
\pagenumbering{arabic}

\section{Introduction}\label{sintro}

\subsection{Context and topic}

The primary subject of this review are the foundations of the Brout-Englert-Higgs (BEH) effect \cite{Englert:1964et,Higgs:1964pj,Higgs:1964ia,Higgs:1966ev,Guralnik:1964eu,Kibble:1967sv,Englert:1966uea} in the standard model and beyond, and how from these foundations phenomenological predictions can be made.

However, it will become clear that it is actually already very hard to unambiguously and in physical terms define what is meant by this expression. Nonetheless, with the discovery \cite{Aad:2012tfa,Chatrchyan:2012ufa} and establishing \cite{pdg,Khachatryan:2016vau} of the Higgs particle experimentally, it is necessary to fully understand what is actually meant by Brout-Englert-Higgs physics, and what its implications are for more general theories.\\[0.5mm]

{\em The aim of this review is to collect what we know on this question and its answer. Both the question and the answer are much more subtle than commonly assumed.}\\[0.5mm]

In particular, this review is rather a status report than a final answer. In this, it summarizes almost half a century of research. This research has to a large extent proceeded in parallel to standard phenomenology using perturbation theory, though there had been occasional connections. This research has deep links to the lattice formulations of quantum field theories. But it had also little connection to the mainstream lattice literature which concentrated on strong interactions, though much of which is to be discussed applies irrespective of the strength of interactions. As a consequence, many of the results appear unfamiliar, or seem to even contradict the common lore found in many textbooks. One purpose of this review is therefore to show how these apparent contradictions are resolved, and how the insights of all of these fields interconnect, and represent different facets of the same physics. It will also become clear that the standard model, as beautifully as it is, is deceptive from the vantage point presented here: It hides many of the subtleties of quantum field theory, because of its very special structure. In fact, the ultimate quest of the decades of research presented here is not to change the standard model. Much to the contrary, the goal is to provide a single, consistent description of the standard model, and any theory beyond, as long as it is a quantum gauge field theory. 

But it is not the aim to have a ultraviolet-complete, non-perturbative definition of such theories. Rather, what will be discussed will apply also to any low-energy effective theory, whether it is renormalizable or not, provided the details of the ultraviolet completion do not interfere with the low-energy physics. This is a modest condition, and a requirement quite generally put on models of particle physics. It is also not the aim to discuss the epistemological and ontological relevance of the theories and degrees of freedom  to be described here, which is a subject of the philosophy of physics \cite{Lyre:2008af,Francois:2017aa}. Although the technical description presented here in fact realizes such views.

Because superficially the research to be discussed in this review seems to be in conflict with what is written in many textbooks, there is appendix \ref{s:faq} with frequently asked questions. Besides a very brief answer to the most often asked questions it contains pointers to both the sections of this review as well as the most pertinent original literature where they are answered in detail.

\subsection{Setting the scene}

After these preliminary remarks on the context of this review, it is now time to start describing what are the underlying questions. To approach the difficulties and subtleties involved, it is probably best to define first the class of theories which will be addressed. The central ingredient is a non-Abelian gauge theory of Yang-Mills type, with some gauge group $G$ and gauge bosons $W_\mu$. In addition, there are one or more scalar bosons $\phi$, called Higgs bosons, which belong to some representation $R$ of the gauge group. They can exhibit some global symmetry group ${\cal G}$, especially if there are more than one of them in the same representation. This global symmetry will be called custodial symmetry in the following.

These Higgs bosons can interact with each other in some way described by the Higgs potential $V$, which involves only gauge-invariant combinations of the Higgs bosons, and which may or may not break the global symmetries explicitly in part or fully. There may also be further particles, especially fermions, in the theory, but their structure does not matter right now, and these will be absorbed in a Lagrangian ${\La}_r$. The theory therefore has the structure
\bea
{\La}&=&-\frac{1}{4}W_\mn^aW^\mn_a+(D_\mu\phi_\alpha)^\dagger D^\mu\phi_\alpha-V(\phi)+{\La}_r\label{la:gen}\\
W_\mn^a&=&\pdm W_\nu^a-\pdn W_\mu^a-g f^{abc} W_\mu^b W_\nu^c\nn\\
D_\mu^{ij}&=&\pd_\mu\delta^{ij}-igW_\mu^aT^{R_\alpha ij}_a\label{covderiv},
\eea
\no where the index $\alpha$ counts the Higgs particles and representations, $f^{abc}$ are the structure constants of the gauge group, and $g$ is the gauge coupling. If the gauge group is not simple, multiple gauge couplings can arise. The $T^{R_\alpha}$ are the generators of the gauge group $G$ in the respective representation.

The simplest case is obtained when reducing this theory to the gauge group SU(2) with a complex doublet of Higgs fields in the fundamental representation. These are in total four real degrees of freedom for the Higgs, which exhibit a custodial symmetry group SU(2). Furthermore, a polynomial potential renormalizable by power-counting is chosen, and no other particles are involved. The Lagrangian reduces then to
\be
\La=-\frac{1}{4}W_\mn^aW^\mn_a+(D_\mu\phi)^\dagger D^\mu\phi-\lambda\left((\phi\phi^\dagger)^2-f^2\right)^2\label{la:beh}.
\ee
\no The parameters $f$ and $\lambda$ are couplings of the potential. This is the prototype theory for BEH physics. Most of the features and problems to be discussed in this review can already be made explicit for this theory, and it will therefore serve as the workhorse throughout this review. Still, some features, especially those discussed in sections \ref{ss:morse}, \ref{ss:bsm}, and \ref{ss:qscattering}, will require to return to a more extended theory.

The standard approach to the theory \pref{la:beh} \cite{Bohm:2001yx,Weinberg:1996kr} is to note that there is a classical minimum at $\phi^\dagger\phi=f^2$. In the next step the Higgs field is split as 
\be
\phi(x)=v+\eta(x)\label{split},
\ee
\no with $\eta$ a new fluctuation field and setting the constant field $v$ such that $|v|=f$ and thus $\la\eta\ra=0$. The value $v$ is also called the vacuum expectation value of the Higgs field. Inserting this split into the Lagrangian \pref{la:beh} results at tree-level in the usual phenomenology of massive gauge bosons and a single massive Higgs particle, as well as a number of would-be Goldstone bosons. Performing perturbation theory leads to the quite successful phenomenology of the standard model today \cite{pdg}. This will be discussed in detail in section \ref{ss:qulev}.

\subsection{The central issue}

The fundamental field-theoretical problem arising is that $v$, in contrast to the parameter $f$ of the Lagrangian \pref{la:beh}, is dependent on the gauge. In fact, it is possible to construct gauges, as will be described in section \ref{s:local}, in which $\la\phi\ra=0$ always holds. Thus no vacuum expectation value exists. This has been noticed already very early on \cite{Lee:1974zg}. However, in such gauges the gauge bosons remain massless to all orders in perturbation theory, as can be shown using standard methods \cite{Bohm:2001yx}. This immediately implies that perturbative results on the masses of the $W$ and $Z$ gauge bosons would be gauge-dependent, and thus potentially unphysical. This, of course, leaves immediately the question of why it then works so well \cite{pdg}.

The answer to this question has two parts. The first is how to gauge-invariantly formulate the phenomenology of this theory. The second is why the standard approach yields the same result. The first question will be addressed in the first part of this review. The second question will not be addressed until section \ref{ss:gipt}. In the course of answering the second question, it will be seen that the standard model is quite special, and it was therefore historically not really necessary to fully answer the first question yet\footnote{As noted, this question has been discussed in the philosophy of science literature, see e.\ g.\ \cite{Lyre:2008af,Francois:2017aa}. There, the problem of the gauge-dependence has also been recognized, and led to a quite similar conclusion as the arguments leading up to here. A general historical overview of BEH physics without focus on these questions can be found e.\ g.\ in \cite{Borrelli:2015goa}.}.

The whole problem starts when considering what a split like \pref{split} actually means on a field-theoretical level. In classical physics, this is very well defined. In a quantum theory, this is much less so, and requires already in the case of global symmetries a well-defined definition of what symmetry breaking actually means. This will be discussed in section \ref{s:global}. When then taking the transition to gauge symmetries in section \ref{s:local}, it will become clear that there is no gauge-invariant way of how to distinguish the existence or absence of a vacuum expectation value of the Higgs.

If there is no gauge-invariant way of defining the vacuum expectation value of the Higgs, the whole standard procedure \cite{Bohm:2001yx} of doing BEH physics seems to collapse on itself at first. This seems to question the whole basis of the current theoretical description of experiments, despite its huge success \cite{pdg}. But it does not. Rather, it can be shown that a treatment taking the gauge-dependence into account will lead to, essentially, the same results for the standard model.

How this works out is exactly the core of this review: The construction of a fully gauge-invariant description of this physics, and how the standard phenomenology follows from it. This starts in section \ref{s:beh}. There, a gauge-invariant description of the theory \pref{la:beh} will be set up. This is the first step of developing a comprehensive, field-theoretical consistent view on BEH physics. This view is continued by establishing the link to phenomenology, and experiment, in section \ref{s:spectrum}. This link was originally devised in \cite{Frohlich:1980gj,Frohlich:1981yi} by Fr\"ohlich, Morchio, and Strocchi (FMS). This approach will be developed in section \ref{s:spectrum}, and was dubbed gauge-invariant perturbation theory in \cite{Seiler:2015rwa}. It will be used in section \ref{ss:gipt} to determine the static properties of the spectrum, i.\ e.\ particle types, multiplicities and masses, in a manifestly gauge-invariant way.

Of course, also the dynamics, and especially scattering processes, have been investigated experimentally exceedingly well, and need to be recovered. While this has not yet been developed as far as the static properties, it is possible to do so, as will be shown in section \ref{s:scattering}. These results will finally resolve the paradox, as it will show that quantitatively everything reduces (almost) to the results of ordinary perturbation theory, thereby sanctifying what has been done so far. The agreement of this approach with experiment follows therefore immediately. Still, doing consciously phenomenology in this framework is a rather recent development, and thus much still needs to be done. However, many of the lattice investigations of BEH physics reviewed here did took this implicitly into account, without explicitly spelling the relation out. The reason is that performing lattice simulations forces you to decide implicitly how you identify observable particles, as will be discussed in section \ref{s:spectrum} in detail.

However, the agreement between standard phenomenology based on perturbation theory and a manifest gauge-invariant formulation is a special feature of the standard model. In more general theories like \pref{la:gen} the results of both approaches may or may not differ qualitatively. This has far reaching implications for BSM phenomenology, and there is not yet a systematic way to decide in which kind of theories agreement occurs, and in which not. This will be discussed, together with the lattice support for such qualitative differences, in section \ref{ss:bsm}.

While gauge-invariant perturbation theory actually describes physics quite accurately, it is not obvious that it should work. To test the described framework theoretically requires non-perturbative methods, which can deal, within their systematic limitations, with this type of physics. Most of the support stems from lattice simulations, which will be briefly introduced in section \ref{ss:lattice}. The most relevant support for the developed picture of BEH physics will be presented in sections \ref{ss:pdlat}, \ref{ss:latgipt}, and \ref{sss:gut}. Unfortunately, for reasons to be discussed in section \ref{ss:lattice}, lattice simulations cannot yet address the standard model as a whole, much less any extensions of it. Therefore only subsectors have been investigated with these methods.

As the gauge symmetry is the key to the concepts presented here, only such investigations will be covered which include the gauge nature of the weak interactions. There is a substantial literature on the ungauged Yukawa sector, including extensions, see e.\ g.\ \cite{Fodor:2007fn,Gerhold:2007gx,Gerhold:2010bh,Gerhold:2011mx,Chu:2015nha,Bulava:2012rb,Gies:2014xha,Capitani:2019syo}, which will not be covered. Also not reviewed will be many beyond-the-standard model extensions without BEH effect, which have been investigated on the lattice, see e.\ g.\ \cite{DeGrand:2015zxa} for a review. However, the present subject has also bearings on such scenarios, as will be discussed in section \ref{ss:tc}.

The core of this review is a physical picture of BEH physics. From this somewhat separate stand issues like triviality or the vacuum stability problem. However, they are not entirely disconnected, and this will be discussed in section \ref{s:further}. This includes the questions of the finite-temperature electroweak phase transition and its implications for cosmology. This subject will only be briefly touched upon in section \ref{ss:ft}, and forms a vast enough topic to require an own review. A summary and a list of open problems and future directions will be provided in section \ref{s:sum}.

\section{Global symmetries}\label{s:global}

Before embarking to gauge symmetries, it is quite helpful to reconsider global symmetries. Also here a full field-theoretical treatment \cite{Strocchi:2015uaa} leads to insights which seem at first counter-intuitive. But in the end, they give a much clearer notion of what symmetry (breaking) at the quantum level really is. At the same time, the standard results and notions will be recovered, embedded in a bigger picture. Thus, the purpose of this section is not to alter anything about global symmetry breaking. Rather, the aim is to make explicit various steps which in textbook treatments are often left implicit.

It should be noted that most of the following depends strongly on the space-time dimension. In particular, most mechanisms do not operate at the quantum level in one and two dimensions \cite{Frohlich:1978px,Strocchi:2015uaa}, due to infrared fluctuations. This the celebrated Coleman-Mermin-Wagner theorem \cite{Strocchi:2015uaa,Mermin:1966fe,Coleman:1973ci}. In the following it will always be implicitly assumed that the dimensionality is sufficient, i.\ e.\ usually three or more, to allow for the described phenomena to take place. This is of course with hindsight to the application to particle physics.

\subsection{Setup}

For simplicity, consider a set of real or complex\footnote{It is always possible to map a complex representation to a, not necessarily linear, real representation \cite{Bohm:2001yx,O'Raifeartaigh:1978kv,Sartori:1992ib}, although, as the example of the groups SU(2) and SU(2)/Z$_2\approx$SO(3) shows, the global structure of the group may induce subtleties \cite{O'Raifeartaigh:1978kv,Heissenberg:2015tji}.} fields $\phi_i$, where $i$ labels the fields. The theory should be described by a Lagrangian
\be
\La=\pdm\phi^\dagger_i\pd^\mu\phi_i-V(\phi)\label{linsig},
\ee
\no where $V$ is a potential term. Now, let the kinetic term be invariant under a symmetry transformation $\phi_i\to\phi_i+\delta\phi_i$. These transformations form a group ${\cal G}$, usually a Lie group, called the symmetry group. The fields are then in a representation ${\cal R}$ of this group.

Assuming that the transformation is linear, it acts as $\phi\to g\phi$, where the matrix $g$ is a representation of some group element $\bar{g}\in{\cal G}$. If the representation ${\cal R}$ is not faithful, the related group element is not unique. To avoid this complication, the following will focus on the representations $g(\bar{g})$ of the group elements $\bar{g}$, and identify it with the corresponding group element(s).

For global symmetries the group elements are space-time-independent. Any value of $\phi$, which can be related by some $g$ to a fixed field configuration $\Phi$, is said to be in the same orbit as $\Phi$ \cite{O'Raifeartaigh:1978kv}. I.\ e.\ a (group) orbit is defined as $O_\Phi=\{\phi=g(\bar{g})\Phi,\bar{g}\in{\cal G}\}$, similar to a ray in linear algebra. $\Phi$ plays the role of a representative field configuration of the orbit. Of course, $\Phi$ can be exchanged for any other element of the orbit, its choice is arbitrary.

For simplicity, it will be assumed in the following that the group is a simple Lie group and that the representation is irreducible, but not necessarily faithful. The latter is necessary to include, e.\ g., explicitly the adjoint representation of many groups, especially the special unitary groups.

Probably the best known example of a theory like \pref{linsig} is the linear $\sigma$-model \cite{Bohm:2001yx}, where the fields are real and the symmetry group is the orthogonal group O($N$), with $N$ the number of real fields. In principle the fields may also be in higher tensor representations of the group, but this does add little conceptually new here, and will therefore not be considered explicitly. Especially as this raises the technical complexity considerably \cite{O'Raifeartaigh:1978kv}. 

\subsection{Explicit breaking}\label{ss:gsym}

It is now up to the potential $V$, whether the whole Lagrangian is invariant under the symmetry \cite{O'Raifeartaigh:1978kv,Sartori:1992ib}. For the sake of simplicity, let the potential be a polynomial in the fields. If $V$ can be written as a sum of terms where each term is an invariant of the representation, the potential is invariant. Invariants are (for real fields) expressions $a_{i_1,...,i_k}\phi_{i_1}...\phi_{i_k}$ such that the expression is invariant under the symmetry transformation, which implies
\be
a_{i_1...i_k}g_{i_1 j_1}...g_{i_kj_k}\phi_{j_1}...\phi_{j_k}=a_{i_1...i_k}\phi_{i_1}...\phi_{i_k}\nn.
\ee
\no E.\ g.\ for the linear-$\sigma$ model $a\phi^2=a\delta_{ij}\phi_i\phi_j=a_{ij}\phi_i\phi_j$ is of this type, but $a_i\phi_i$, is not, as it singles out a direction. If the potential is not build from invariants, the Lagrangian is no longer invariant under the symmetry group, and the symmetry is called explicitly broken. If the potential is invariant under some true subgroup ${\cal H}\subset{\cal G}$ of the symmetry group only, it is said that the symmetry is explicitly broken down to this subgroup.

\subsection{Classical spontaneous breaking}\label{ss:cssb}

Consider now classically the potential
\be
V(\phi)=-\frac{\mu^2}{2}\phi^2+\lambda\phi^4\label{esb:pot},
\ee
\no with $\mu^2\ge0$, $\lambda>0$, and with the fields in the fundamental representation of ${\cal G}=$O($N$). Then the potential is invariant under ${\cal G}$.

But orbits can still be distinguished in subclasses. One subclass $O^{\cal G}$ are those orbits, which are invariant under the whole group, i.\ e.\
\be
O_\Phi=O_\Phi^{\cal G}\implies g(\bar{g})\Phi=\Phi\quad\forall \bar{g}\in{\cal G}\nn.
\ee
\no In the present example, this would only be the orbit characterized by $\Phi=0$. Then there are other orbits $O^{{\cal H}_i}$, which are invariant under some true subgroup ${\cal H}_i\subset{\cal G}$, where $i$ labels the subgroups of the group ${\cal G}$, which may be both continuous and discrete. I.\ e., these orbits satisfy
\be
O_\Phi=O_\Phi^{{\cal H}_i}\implies h(\bar{h})\Phi=\Phi\quad\forall \bar{h}\in{\cal H}_i\nn.
\ee
\no  It is important to note that if another representative $\Phi$ will be chosen, this will lead to a conjugate subgroup ${\cal H}_i'$, and thus a similarity transformation on the level of the $h$. It is therefore a statement about whole groups, rather than a particular set of generators. Elements from $O^{{\cal H}_i}$ will not be invariant under any group element of ${\cal G}$ outside the subgroup ${\cal H}_i$, but trivially so under any subgroups of ${\cal H}_i$.

For the present example, this would be any field configuration of arbitrary, but fixed, length $\phi^2=a^2$, and the (single) subgroup is ${\cal H}=$O$(N-1)$. Thus, for every orbit the corresponding subgroup may differ, as it will be the one leaving a fixed $\Phi$ invariant. However, while the particular rotation axis differs for each orbit, the rotation group O($N-1$) does not. In this example the orbits are also characterized by the length $|\Phi|$, as elements with different lengths belong to different orbits.

Note that there can be one or multiple orbits fulfilling the same condition. The collections of orbits invariant under the same subgroup is called stratum, ${\cal S}^{{\cal H}_i}$, which includes the case ${\cal H}_i={\cal G}$. The groups ${\cal H}_i\neq{\cal G}$ are in this context called little groups. A complete classification of possible little groups and strata is not entirely trivial, especially for non-simple Lie groups \cite{O'Raifeartaigh:1978kv}, but for any explicit case possible.

Return to the example \pref{esb:pot}. If $\mu=0$ the stratum ${\cal S}^{\text{O}(N)}$ is special, as it minimizes the action. If $\mu\neq 0$, this changes. Now, those orbits are (classically) preferred which satisfy $\phi^2=v^2=\mu^2/(2\lambda)$, and thus have a particular length. However, they are still invariant under the little group associated with the stratum the singled out orbit belongs to, in this case ${\cal S}^{\text{O}(N-1)}$. As a consequence, everything remains still symmetric under the corresponding little group. This residual symmetry is thus determined by the stratum membership of the classically preferred orbit. Especially, a change of $\mu^2$, as long as is remains non-zero, will only select a different orbit from this stratum, but not a different stratum. In this sense, the physics of classical symmetry (breaking) is actually a physics of stratum and orbit selection.

But in classical physics initial conditions are required. The initial conditions will then ultimately not only select an orbit, but actually a member of the orbit. Thus, the existence of the initial conditions can finally select a particular direction and therefore break the little group explicitly. Thus, the necessary initial conditions of classical physics turns spontaneous symmetry breaking into explicit symmetry breaking. Classically, spontaneous symmetry breaking can therefore be considered to be restricting the choice of possible initial conditions to a particular stratum and orbit. But such a final solution will never respect the symmetry, even of the associated little group. It is, of course, possible to select symmetric initial conditions, e.\ g.\ $\phi=0$ for \pref{esb:pot}. But since all directions are now equally preferred, the system again only moves along trajectories which keep this feature. In this example, it will actually remain at $\phi=0$, as without additional perturbation of the initial conditions or the potential this is the suitable solution to the equations of motion. This is a metastable situation.

The important insight is that symmetry breaking is classically due to either of two reasons. Either the symmetry is intrinsically explicitly broken by a term in the Lagrangian. Or it is after spontaneous breaking explicitly broken by the initial conditions. Since the initial conditions are not predictable from the Lagrangian or equation of motions, they will be considered an external influence.

However, it should be noted that it is possible to demand for a mechanical system that trajectories have to be stable under time evolution \cite{Strocchi:2005yk,Strocchi:2012ir}. If this additional constraint is imposed, only solutions can be obtained which are not metastable, and therefore correspond to absolute minima of the potential. This demand effectively reduces the space of possible initial conditions but guarantees a stable time evolution under perturbations. In the following this restriction will not be imposed.

While this is a straightforward example, this is mainly due to the choice of an irreducible representation where there is only the trivial stratum ${\cal S}^{\cal G}$ and a single non-trivial stratum. The stratum structure becomes quickly more involved in more general cases \cite{O'Raifeartaigh:1978kv}, as does the physics. For now, only the case with a single non-trivial stratum will be considered. The situation with multiple non-trivial strata will be considered again in section \ref{ss:strata}.

\subsection{At the quantum level}\label{s:hidssb}

Before discussing the quantum version of the classical situation, it should be noted that there is the additional possibility that the path integral\footnote{It should be understood that a rigorous definition of the path integral requires infrared and ultraviolet regulation \cite{Seiler:1982pw,Rivers:1987hi}. This can be achieved, e.\ g., using a lattice as is done in section \ref{ss:lattice}. This will be assumed tacitly in the following.} measure can also be non-invariant under a symmetry \cite{Bohm:2001yx,Bertlmann:1996xk}. This is a so-called anomaly. In the following it will always be assumed that such anomalies do not occur. This is mainly for the purpose of simplicity. Otherwise, they act in the present context of a quantum theory essentially in the same way as an explicit breaking by the Lagrangian, as the symmetry is not realized at the quantum level.

\subsubsection{Global symmetries and the path integral}\label{sss:metastable}

Consider now the quantum case \cite{O'Raifeartaigh:1978kv,Sartori:1992ib} without explicit symmetry breaking. Adding explicit symmetry breaking will not change the following, except for reducing the full symmetry group to its unbroken subgroup. It is best to study an explicit example, and then generalize the concepts. Take, e.\ g., the same situation as before and thus still with
\be
V(\phi)=-\frac{\mu^2}{2}\phi^2+\lambda\phi^4\label{hvsb:pot},
\ee
\no being the potential of the (interacting) linear $\sigma$-model \cite{Bohm:2001yx}. Classically, as discussed in section \ref{ss:cssb}, the potential \pref{hvsb:pot} is minimized by the orbit with a fixed length\footnote{Note that it is often customary to rewrite the parameters of the Lagrangian such that, e.\ g., $\lambda$ is written as a ratio of $\mu$ and $v$. This will not be done here, as this intermingles two independent quantities, the length of an orbit and a parameter of the Lagrangian, which need to be carefully distinguished in the following.} $\phi^2=v^2=\mu^2/(2\lambda)$. However, rather than an individual field configuration, the whole orbit minimizes the potential. As noted before, in classical physics the initial conditions determine which member of an orbit would be chosen, yielding the classical field configuration. In quantum physics, this is different. This is best understood in the context of a path-integral treatment, and an even more simplified model.

Consider the zero-dimensional, i.\ e.\ space-time-independent, case of the linear-$\sigma$ model. Then the kinetic part is dropped. The remainder is equivalent to the situation of a constant field, having still the same minimum. The path integral can then be explicitly evaluated, as can expectation values involving an odd or even number of fields, yielding
\bea
Z&=&\int d^N\phi e^{-i\left(\frac{\mu^2}{2}\phi^2+\lambda\phi^4\right)}=\Omega(N)\int r^{N-1}dr e^{-i\left(\frac{\mu^2}{2}r^2+\lambda r^4\right)}=\Omega(N)f(\mu,\lambda)\nn\\
\la\phi_j g(\phi^2)\ra&=&\int d^N\phi \phi_j g(\phi^2) e^{-i\left(\frac{\mu^2}{2}\phi^2+\lambda\phi^4\right)}=0\label{hvsb:vvev}\\
\la g(\phi^2)\ra&=&\int d^N\phi g(\phi^2) e^{-i\left(\frac{\mu^2}{2}\phi^2+\lambda\phi^4\right)}=\Omega(N)h(\mu,\lambda)\nn
\eea
\no where $\Omega(N)$ is the $N$-dimensional surface of the unit sphere, $g$ is some arbitrary function, and the functions $f$ and $h$ can be calculated exactly. These results hold always, no matter the values of the parameters. In particular, the vacuum expectation value $\la\phi_j\ra=0$ for all values, even in cases where a fixed-length field is classically preferred.

It should be noted that also the length $\la\phi^2\ra$ of the field does not distinguish the cases where the minima of \pref{hvsb:pot} are at length zero or at a finite length. In both cases, the explicit evaluation of the integral gives a non-zero value.

Understanding the result \pref{hvsb:vvev} is straightforward from a path-integral perspective. Because the minimization of \pref{hvsb:pot} fixed only the length, but not the direction, of the minimizing field configuration, the path integral (or, in the present case, the integral) still averages over all directions. Since all directions are equal, any quantity which has no preferred direction has to vanish. Thus at the level of expectation values the symmetry is still maintained: All expectation values are invariant under the symmetry.

This is not true for an individual measurement, however. The statement \pref{hvsb:vvev} does only say that when performing many measurements, the orientation vanishes on average. A single measurement may, or may not, have an orientation. The probability for this depends on the parameters of the theory, and \pref{hvsb:vvev} only states that if it has an orientation, every orientation is equally likely.

As an aside, note that in a path integral treatment it is not necessary to define a vacuum state. If one insists on doing so \cite{Weinberg:1996kr}, the vacuum state is a superposition of fields with all directions. An actual measurement of the vacuum state will then also yield a state with a particular condensate and orientation. But this condensate will again be zero on the average.

This seems at first at odds with the usual picture of spontaneous symmetry breaking in quantum theories, where it is usually assured that a particular direction is 'chosen', leading to a non-zero vacuum expectation value \cite{Bohm:2001yx}.

\subsubsection{Sources}

However, both views can be made compatible, provided they are carefully enough defined. To do so, introduce an external source,
\be
Z[j]=\int d^N\phi_i e^{-i\left(\frac{\mu^2}{2}\phi^2+\lambda\phi^4+j\phi\right)}\label{d0model}.
\ee
\no Then
\be
\la\phi_i g(\phi^2)\ra_{j\neq 0}\neq 0\nn,
\ee
\no because now a direction is preferred. However, this is because the source breaks the symmetry explicitly, and this leads back to the case of section \ref{ss:gsym}. 

The analogy carries only this far, because the expression \pref{d0model} is analytic in $j$. Returning therefore back to the field theory case \pref{hvsb:pot}, some statements carry over \cite{Perez:2008fv}. The first is that when classically no spontaneous symmetry breaking occurs, the same argument applies at the quantum level: There is no preferred direction, and thus any expectation value which has a direction, like \pref{hvsb:vvev}, is necessarily zero.

But in the quantum theory, the situation changes compared to the classical case if spontaneous symmetry breaking could occur. The argument above still holds true in the quantum case, for any value of the parameters of the potential. There is no explicitly preferred direction, and the path integral averages over all possible directions equally, with equal measure, as the action and measure is invariant. To put it differently, the path integral represents here the quantum phenomena of averaging incoherently over all possible breaking patterns. Non-invariant expectation values therefore still vanish, as if the symmetry would be intact.

The reason is really the presence of the path integral's measure, and that this measure is invariant under the symmetry and therefore includes all possible field configurations, and thus in particular all possible (global) directions of the field. As this is just the quantum mechanical superposition principle, it can be said that quantum effects restore the classically spontaneously broken symmetry at the level of expectation values. However, this does not imply that there are no consequences of the classical situation.

To understand how they arise, proceed as in \pref{d0model}. Introduce a source to explicitly break the symmetry, say in the 1-direction, i.\ e. a source $j\delta_{i1}$. Then this direction is preferred, yielding
\be
\la\phi_i\ra_{j\neq 0}=\delta_{i1}f(j)\label{vev}.
\ee
\no The big difference is now that there is a non-analyticity of the function $f$ in $j$,
\be
\lim_{j\to 0}f(j)=v\neq 0\label{vevlimit}.
\ee
\no This feature is what can be properly called spontaneous symmetry breaking: This limit does not vanish. Especially, this implies
\be
\lim_{j\to 0}\la\phi_i\ra_{j}=v\delta_{i1}\label{limssb},
\ee
\no and there is a conventional vacuum expectation value\footnote{Note that this does not need to be in one-to-one-correspondence with classical spontaneous symmetry breaking. Quantum effects can lift a classical symmetry breaking, as will be seen in explicit examples in section \ref{ss:pd}, or lead to additional non-analyticities \cite{Maas:2013sca}. Potentially, also the reverse is possible. This stems from quantum corrections to the potential, but the details play no role for the conceptual considerations of this section.}. The non-analyticity becomes manifest by noting that the value of this limit depends on the source. If, e.\ g., the source would prefer direction 2, the value would be $v\delta_{i2}$ instead of $v\delta_{i1}$. The same effect can also be achieved by introducing boundary conditions in space and time which break the symmetry explicitly \cite{Strocchi:2005yk,Frohlich:1976it}.

Nonetheless, since the integration argument above is still valid, this implies
\be
0=f(0)\neq\lim_{j\to 0}f(j)\label{limnssb}.
\ee
\no In addition, because not only the value \pref{limnssb} at the point $j=0$ differs from the limit \pref{limssb}, but also the value of the limit \pref{limssb} differs for different sources, $\la\phi\ra_{j\to 0}$ is necessarily non-analytic in $j$ at $j=0$. Since the vacuum expectation value \pref{vev} will be non-zero for even an infinitesimally small value of $j$, this implies that the system is metastable with respect to external sources in terms of $j$. Thus, without an external source, the system should be regarded as metastable, rather than spontaneously broken. Whether there is then a vacuum expectation value in an actual theory depends on whether there is an external influence like a source.

\subsubsection{Detecting symmetry breaking without a source}

In fact, the metastability can actually also be observed at $j=0$ without taking an explicit limit. A possible, non-local, operator to detect it is \cite{Langfeld:2002ic,Caudy:2007sf}
\be
m_r=\left\langle\left(\int d^dx\phi_i(x)\right)^2\right\rangle\label{vevd}.
\ee
\no The important difference between the operator \pref{vev} and \pref{vevd} is that \pref{vev} measures the average absolute orientation. Without explicit symmetry breaking it therefore needs to be zero, as all directions are treated equally in the path integral. The expression \pref{vevd} measures a relative orientation for each field configuration, and therefore tests, whether there is a preference in such a single configuration. The path-integral is then a weighted average only over the yes-or-no statement that there is or is not an orientation in any given configuration. Therefore, it can detect even in the case of all directions equal a potential for a direction, and thus the metastability.

Conversely, a quantity like $\la\phi^2\ra$ cannot be used to detect either the presence or absence of metastability. This quantity merely measures the average size of fluctuations. It will therefore only be zero if the field itself is zero except for some measure-zero region of space-time. However, such configurations are only a measure-zero part of the path integral, and therefore this quantity will never vanish.

An important issue to note is that this discussion is about expectation values. I.\ e.\ averages over many measurements. A statement like \pref{limnssb} needs therefore to be read as that without source there is {\bf on the average} no preferred direction. Performing {\bf an individual measurement} without source will still yield a result with preferred, but random, direction, if the system is metastable, i.\ e.\ if a quantity like \pref{vevd} is non-zero. This is the same as in the limit of non-vanishing source, except for the fact that the measured orientation is random, while with source it will be aligned with the source\footnote{Actually, not every measurement will yield something which is perfectly aligned with the source, only on the average.}. This distinction between expectation values and measurements is crucial for the distinction between metastable and spontaneously broken systems. Especially, in the metastable system there is no vacuum expectation value on average, while there is one on average in the spontaneously broken case. In the metastable case, expectation values are (trivially by being zero) invariant under symmetry transformations, while they are not invariant in the spontaneously broken case. But measurements will in both cases yield an explicit direction.

In a sense, measurements take thus over the role of initial conditions of the classical case. After all, the measurement process is not part of the path integral description, and thus not described by it. It would be quite interesting to understand how these effects would play out in the context of quantum probability theory \cite{Frohlich:2015qpt}.

\subsubsection{Sources as parameters of external effects}

Whether to use sources is now a question of context, and mainly depends on the embedding of a theory. If the theory is considered as part of a larger system, either by having an environment or if it is considered to be a low-energy effective theory, the environment or the overarching theory can provide an infinitesimal external influence. Then a description using a source is surely adequate. If there is no surrounding, or the surrounding itself should not break the symmetry explicitly, then the theory should also not be treated using the limiting procedure\footnote{Though this may be still a useful technical trick.}. In particular, in particle physics, where there is usually not considered any outside, it is often hard to argue where such an explicit breaking should come from. Subsectors of particle physics may be different. E.\ g.\ the quark masses in pure QCD can be taken as an example of such an external source giving the chiral symmetry breaking a preferred direction. Otherwise, chiral symmetry would only be metastable rather than spontaneously broken.

This is a view driven by phenomenology which aims at considering the system as closed as possible. Alternatively, it is possible to put a mathematical definition of the theory first, which allows for the appearance of boundary conditions\footnote{Boundary conditions are here additional conditions imposed, e.\ g., on the space-time manifold or the integration manifold of the path integral. They are not necessary to define the partition function, as the path integral can be extended over the whole integration manifold, and no conditions need to be imposed on (flat) space-time. They are therefore different from the classical initial conditions.} and/or limits of external sources as defining properties of a theory \cite{Strocchi:2005yk,Frohlich:1976it}. This demands the theory to be stable under unitary time evolution in the presence of external perturbations. In such a case, the theory is defined in terms of a (vanishing) source or boundary conditions. That is the equivalence of restricting to trajectories with stable time evolution in the classical case. However, in the following the phenomenological view will be taken.

To stay in line with standard textbooks \cite{Bohm:2001yx,Weinberg:1996kr}, the situation with an explicit source will be referred to as 'spontaneous symmetry breaking'. The situation without source will be referred to as 'metastable'. That it is even possible to define the difference is due to the global nature of the symmetry. This will change very much for local symmetries in section \ref{s:beh}.

This leaves the question, if the usual phenomena associated with spontaneous symmetry breaking, most prominently Goldstone bosons, still occur in the metastable case. The answer to this is that they will not arise in the usual way, as will be discussed in section \ref{ss:goldstone}. 

\subsubsection{Finite volume}

Since in the following lattice methods will be used, it is necessary to make a few observations on how the above discussed issues will change if there are only a finite number of degrees of freedom, in particular a finite volume $V$ \cite{Landau:2005mc,Perez:2008fv,Birman:2013gaa}. In such systems there is no non-analyticity possible. Therefore, the limit of $f(j)$ for $j$ to zero will always vanish. Physically, this is because tunneling processes allow to again sample all possible orientations. Therefore, the order of limits is important. To have a preferred direction requires
\be
\lim_{j\to 0}\lim_{V\to\infty}f(j)\nn.
\ee
\no as the appropriate order of limits. Of course, this plays no role when working in the metastable case, i.\ e.\ when evaluating $f(0)$. It is important to note that an infinite volume will make tunneling between different orientations impossible. However, the path integral integration range is still over all orientations. Thus the qualitative features of the metastable case are volume-independent.

\subsection{Hidden symmetry}\label{sss:hidden}

In addition to any of the above, there is an additional effect due to the usually employed shift \pref{split}. Irrespective of the status of the symmetry, this hides the symmetry, i.\ e.\ makes it not longer manifest \cite{O'Raifeartaigh:1978kv}. This has profound implications of its own, as the presence or absence of a symmetry makes itself only indirectly felt.

Consider for concreteness an O(2) model with potential\footnote{It should be noted, as an aside, that it is always possible to shift the potential such that the lowest energy state has energy zero. In this case, the potential takes the form
\be
V=\frac{\mu^2}{2f^2}\left(\phi\phi^\dagger-\frac{f^2}{\sqrt{2}}\right)^2.\label{o2model}
\ee
\no This is the same as the potential \pref{complexpot}, up to a constant term of size $\mu^2f^2/8$, which is irrelevant.}
\be
V(\phi,\phi^\dagger)=-\frac{1}{2}\mu^2\phi^\dagger\phi+\frac{1}{2}\frac{\mu^2}{f^2}(\phi^\dagger\phi)^2\label{complexpot}.
\ee
\no The potential is invariant under the transformation
\be
\phi\to e^{-i\theta}\phi\approx\phi-i\theta\phi.
\ee
\no Hiding proceeds now by rewriting the complex field in terms of its real and imaginary part, $\phi=\sigma+i\chi\nn$. The complete Lagrangian then takes the form
\be
\La=\frac{1}{2}(\pdm\sigma\pd^\mu\sigma+\pdm\chi\pd^\mu\chi)+\frac{\mu^2}{2}(\sigma^2+\chi^2)-\frac{1}{2}\frac{\mu^2}{f^2}(\sigma^2+\chi^2)^2\nn,
\ee
\no and therefore describes two real scalar fields, which interact with each other and having the same (tachyonic) tree-level mass $\mu$. The symmetry is still manifest, as only the combination $\sigma^2+\chi^2$ enters the Lagrangian. The corresponding transformations take the (infinitesimal) form
\bea
\sigma&\to&\sigma+\theta\chi\label{sxtrafo1}\\
\chi&\to&\chi-\theta\sigma\label{sxtrafo2},
\eea
\no and therefore mix the two degrees of freedom.

The extrema of the potential occur at $\sigma=\chi=0$ and at 
\be
\sigma^2+\chi^2=\frac{f^2}{\sqrt{2}}=\phi^\dagger\phi\label{mincond}.
\ee
\no Whether these extrema are maxima or minima is obtained from the second derivatives of the potential. The second derivatives are smaller or equal to zero at zero field, and thus this  is a local maximum. Fields satisfying \pref{mincond} yield minima. The situation at the minima is symmetric, so it is possible to make any choice to split the $f^2/\sqrt{2}$ between $\sigma$ and $\chi$.

It is possible to perform an arbitrary linear field redefinition. E.\ g.\ shift
\bea
\sigma&\to&\sigma+\frac{v}{\sqrt{2}}\label{sigmashift}\\
\chi&\to&\chi\nn.
\eea
\no This yields a new (and equally well-defined) Lagrangian
\bea
\La&=&\frac{1}{2}(\pdm\sigma\pd^\mu\sigma+\pdm\chi\pd^\mu\chi)+\frac{\mu^2v}{\sqrt{2}}\left(\frac{v^2}{f^2}-1\right)\sigma+\frac{\mu^2v^2}{4}\left(\frac{v^2}{2f^2}-1\right)\nn\\
&&-\mu^2\left(\frac{3}{2}\frac{v^2}{f^2}-\frac{1}{2}\right)\sigma^2-\mu^2\left(\frac{1}{2}\frac{v^2}{f^2}-\frac{1}{2}\right)\chi^2+\frac{\sqrt{2}\mu^2}{f}\frac{v}{f}\sigma(\sigma^2+\chi^2)+\frac{\mu^2}{2f^2}(\sigma^2+\chi^2)^2\label{beforeid2}\\
&\stackrel{v=f}{=}&\frac{1}{2}(\pdm\sigma\pd^\mu\sigma+\pdm\chi\pd^\mu\chi)-\mu^2\sigma^2+\frac{\sqrt{2}\mu^2}{f}\sigma(\sigma^2+\chi^2)+\frac{1}{2}\frac{\mu^2}{f^2}(\sigma^2+\chi^2)^2-\frac{\mu^2f^2}{8}\label{lag:global}.
\eea
\no In the last line, \pref{lag:global}, the usual \cite{Bohm:2001yx} identification $v=f$ has been done.

The Lagrangian \pref{lag:global} describes the fluctuation field $\sigma$ and the tree-level-massless field $\chi$. These interact with cubic and quartic interactions. There is no obvious trace at the Lagrangian level left of the original symmetry. It is hidden, but not gone. In fact, it is still perfectly there. 

To see this note first that the cubic coupling constant is not a free parameter of the theory, but it is uniquely determined by the other parameters. That is, as it should be, since by a mere field translation no new parameters should be introduced into the theory. Thus, though the symmetry is no longer manifest, it is still present: It ensures a relation between the couplings of various interaction vertices. In the quantum theory, this will resurface in the Ward identities. Thus, even though the shift \pref{sigmashift} seems to have manifestly broken the symmetry, this is not the case. If it would truly be broken, the couplings would all be independent. Thereby, the notion of a hidden symmetry arises \cite{O'Raifeartaigh:1978kv}.

In fact, the Lagrangian \pref{beforeid2} is still fully symmetric. This can be seen in the way the symmetry manifests. After \pref{sigmashift} the symmetry transformation \prefr{sxtrafo1}{sxtrafo2} reads
\bea
\sigma&\to&\sigma+\theta\chi\nn\\
\chi&\to&\chi-\theta(\sigma-v)\nn\\
v&\to&v+\theta\chi\nn,
\eea
\no i.\ e.\ $v$ has actually to be transformed. Therefore, it is useful to work with the Lagrangian \pref{beforeid2}, to see this. Because the identification $v=f$ has not been done yet it is manifest where changes occur. Note that especially the linear term and even the constant term, which do not participate in the dynamics, are still relevant to maintain the symmetry: Both change under this transformation, as they depend on $v$.

The other consequence is that in the Lagrangian \pref{beforeid2}, i.\ e.\ before the identification $v=f$ has been made, at tree-level both fields are still massive. However, these tree-level masses are not symmetry-invariant quantities, and there is no Goldstone boson.

All of this is always possible. It becomes particularly useful, if the symmetry is spontaneously broken. By introducing a source and aligning it with the vacuum expectation value in \pref{sigmashift} this particular direction freezes out. Any residual symmetry does no longer change $v$. Hence, $v$ can be treated as a parameter of the Lagrangian, and set to $f$, yielding \pref{lag:global}. Especially, there is now a massless Goldstone boson $\chi$, and symmetry transformations under any residual symmetry would only affect $\sigma$ and $\chi$, but not $v$. In the present case, this is only a discrete parity transformation, $\chi\to-\chi$.

\subsection{The Goldstone theorem}\label{ss:goldstone}

One\footnote{For selected steps of the standard treatment I follow here \cite{Bohm:2001yx}.} of the most remarkable consequences of spontaneous symmetry breaking is the masslessness of one of the degrees of freedom in \pref{lag:global}. This is a special case of Goldstone's theorem \cite{Goldstone:1961eq,Bohm:2001yx,Strocchi:2015uaa}. It is useful to formulate this theorem more generally, both at the classical and quantum level, to fully appreciate how the situation in the metastable case differs.

\subsubsection{Classical Goldstone theorem}\label{sss:cgoldstone}

The Goldstone theorem \cite{Goldstone:1961eq} states: If a symmetry group ${\cal G}$ of size $\mathrm{dim}{\cal G}$ is spontaneously broken, then there exist as many massless modes as there are generators. If the group is only partly broken then only as many massless modes appear as generators are broken. This theorem only applies if the state space is positive semi-definite, and therefore does not carry over to gauge theories, as will be discussed in chapter \ref{s:beh}.

Take as the symmetry group a (semi-)simple Lie-group ${\cal G}$ and write the fields in real components. Then the symmetry transformation of the associated real fields transforming under a real representation of the symmetry group are given by
\be
\delta\phi_i=iT^{ij}_a\phi_j\theta^a
\ee
\no with arbitrary infinitesimal parameters $\theta^a$ and the index $a$ counting from 1 to $\mathrm{dim}{\cal G}$. In a Lagrangian
\be
\La=\frac{1}{2}\pdm\phi_i\pd^\mu\phi^i-V(\phi)\nn
\ee
\no the kinetic term is trivially invariant. The extrema of the potential satisfy
\be
0=\frac{\pd V}{\pd\phi_i}\delta\phi^i=i\frac{\pd V}{\pd\phi_i} T_{ij}^a\phi^j\theta_a\nn.
\ee
\no Since the parameters are arbitrary, this can only be satisfied if
\be
\frac{\pd V}{\pd\phi_i}T_{ij}^a\phi^j=0\label{minpotcon}
\ee
\no holds. Assume that there are solutions to \pref{minpotcon} with non-zero $\phi_j$. These are the usual minima of spontaneous symmetry breaking. One of the minima, and thus its direction $v$, is now selected by the initial conditions.

The vacuum value $|v|$ of the field can then by identified with the parameters of the Lagrangian, as discussed at the end of section \ref{sss:hidden}. The symmetric matrix of second derivatives,
\be
\left.\frac{\pd^2 V}{\pd\phi_k\pd\phi_i}\right|_{\phi=v}=((M(v))^2)_{ki}\nn,
\ee
\no is positive semi-definite at a minimum, i.\ e., has only positive or zero eigenvalues. Shifting the fields by $\psi_i=\phi_i-v_i$ and expanding the potential around $v$ yields at quadratic order in the fields the Lagrangian
\be
\La=\frac{1}{2}\pd_\mu\psi_i\pd^\mu\psi^i-\frac{1}{2}(M^2)_{ki}\psi_k\psi_i+...\nn.
\ee
\no Because $M$ is positive semi-definite, all particles have then zero or positive mass at tree-level\footnote{This neglects the influence of terms beyond quadratic order, which can become relevant \cite{Strocchi:2005yk}.}.

In terms of this mass matrix the conditional equation for a classical minimum reads
\be
(M^2)^{ki} T^a_{ij} v^j=0\label{classicalgsm}.
\ee
\no The little group is the invariance group ${\cal H}$ of this minimum. As a consequence for generators $t_a$ of ${\cal H}$ the conditional equation reads
\be
t^a_{ij}v^j=0\nn.
\ee
\no Therefore, the value of the mass matrix is irrelevant for these directions, and there can be $\mathrm{dim}{\cal H}$ massive modes. However, for the coset space ${\cal G}/{\cal H}$ with generators $\tau^a$, the corresponding equations
\be
\tau^a_{ij}v^j\neq 0\nn.
\ee
\no are not fulfilled, and therefore the corresponding entries of the mass-matrix have to ensure \pref{classicalgsm}. Especially, if the mass matrix is diagonal they have to vanish. Since these represent $\mathrm{dim}({\cal G}/{\cal H})$ equations, there must be $\mathrm{dim}({\cal G}/{\cal H})$ massless modes, the Goldstone modes.

E.\ g.\ for the O($N$) linear $\sigma$-model, the generalization of \pref{o2model}, the only non-trivial minimum satisfies
\be
\phi^i\phi_i=\frac{v^iv_i}{\sqrt{2}}\stackrel{!}{=}\frac{f^2}{\sqrt{2}}>0\nn
\ee
\no provided $f^2$ is greater than zero. This minimum characterizes a vector of length $f^2/\sqrt{2}$ on the $N$-sphere, and is therefore invariant under the group O($N-1$), being the (only non-trivial) little group of the fundamental vector representation of the theory. Since O($N$) has $N(N-1)/2$ generators, there are $N-1$ generators spontaneously broken, and thus there exists $N-1$ massless modes.

\subsubsection{Quantized Goldstone theorem}\label{ss:qgoldstone}

To determine the consequence of hiding a symmetry at the quantum level, it is useful to investigate the normalized partition function
\be
\Omega[j]=\frac{Z[j]}{Z[0]}=\frac{1}{Z[0]}\int{\cal D}\phi\exp\left(i\int d^4x(\La+j_i\phi_i)\right)\nn,
\ee
\no with the same Lagrangian as before. The source $j$, inducing the spontaneous symmetry breaking, will be send to zero in the end, but it is useful to leave it non-zero for now.

Since the Lagrangian and the measure are invariant under a symmetry transformation, the variation of the partition function must vanish
\be
0=\delta\Omega[j]=\int {\cal D}\phi e^{iS+i\int d^4 xj_i\phi_i}\left(\frac{\pd\delta\phi_i}{\pd\phi_j}+\delta\left(iS+i \int d^4xj_i\phi_i\right)\right)\nn.
\ee
\no The first term is the deviation of the Jacobian from unity. As it is assumed that there are no anomalies, the measure is invariant, and the deviation vanishes. The second is the variation of the action, which also vanishes, as the action is invariant under the symmetry by construction. Only the third term can contribute. Since all variations are arbitrary, it thus follows
\be
\int d^4x j_i T_{ik}^a\frac{\delta \Omega[J]}{i\delta j_k}=0\nn,
\ee
\no where it has been used that $Z[0]$ is a constant, and the order of functional and ordinary integration has been exchanged. Furthermore, as all variations are independent this yields $\mathrm{dim}{\cal G}$ independent equations.

Since
\be
\delta \Omega\equiv\delta\left(e^{\Omega_c}\right)=e^{\Omega_c}\delta \Omega_c\nn,
\ee
\no this can be rewritten in terms of the generating functional $\Omega_c$ for connected correlation functions as
\bea
0&=&\int d^4x j_i T_{ik}^a\frac{\delta \Omega_c[j]}{i\delta j_k}\nn\\
\frac{\delta \Omega_c[j]}{i\delta j_i}&=&\la\phi_i\ra\nn.
\eea
\no This can be transformed into an equation for the vertex (i.\ e., connected and amputated correlation functions) generating functional $\Gamma$ by a Legendre transformation
\bea
i\Gamma[\phi]&=&-i\int d^4 x j_i\phi_i+\Omega_c[j]\nn\\
j_i&=&-\frac{\delta\Gamma[\phi]}{i\delta\phi_i}\label{legendre}.
\eea
\no This yields finally
\be
\int d^4x\frac{\delta\Gamma}{\delta\phi_i} T_{ik}^a\la\phi_k\ra=0\label{genfuncgreensym}.
\ee
\no To obtain the mass of a particle at the quantum level requires to study the poles of its propagators. The inverse propagator of the fields $\phi_i$ is given by
\be
\frac{i\delta^2\Gamma}{\delta\phi_i(x)\delta\phi_j(y)}=-((D(x-y))^{-1})_{ik}.
\ee
\no An expression for this object can be obtained by differentiating \pref{genfuncgreensym} with respect to the field once more yielding \cite{Bohm:2001yx}
\be
\int d^4x\left(\frac{\delta^2\Gamma}{\delta\phi_i(x)\delta\phi_j(y)}T_{ik}^a\la\phi_k\ra\right)=0\nn,
\ee
\no and all other terms vanish. This is just the sought-after Fourier-transform of the inverse propagator at zero momentum, yielding
\be
((G(p=0))^{-1})_{ij}T_{ik}^a\la\phi_k\ra=0\label{condgold}.
\ee
\no Thus, the inverse propagator needs to vanish at zero momentum exactly when the contraction with the vacuum expectation value does not vanish. Thus, the propagator has a pole at zero momentum in these tensor components, and there is one or more massless excitations. But this equation is nothing but the tree-level equation \pref{classicalgsm} where the mass matrix is replaced by the inverse propagator. Thus, there is the same number of massless particles at the quantum level as there is at tree-level, and in particular with the same quantum numbers.

\subsubsection{Metastable case}

It remains to understand what happens in the metastable case\footnote{I am indebted to Jeff Greensite, Christian B.\ Lang, and Erhard Seiler for discussions on this subject.}. Because the vacuum expectation value in equation \pref{condgold} explicitly vanishes, the equation is already fulfilled without a zero of the inverse propagator. Thus, the same argument for the presence of massless poles can no longer be made. This does not forbid the presence of massless poles, it just no longer implies their existence.

Furthermore, since there is no longer a preferred direction, the only invariant rank two tensor available is $\delta_{ij}$, and thus the propagator must be proportional to it, $G_{ij}=\delta_{ij}g(p^2)$, where $g(p^2)$ is a scalar function. The pole structure is then entirely carried by the function $g(p^2)$.

If the degeneracy of a state is counted by the algebraic multiplicity of the eigenvalues with poles of the corresponding propagator this implies that the multiplicities need to be compatible with the representations of the unbroken group. At fixed representation this would be in disagreement with the multiplicity for the case with breaking with an external source. Especially, if massless states exist, they may appear only in multiplets of the full symmetry, and thus in different representations than suggested by Goldstone's theorem. Hence, the metastable theory may exhibit a different multiplet structure than the spontaneously broken theory. It should be repeated that this does not imply that the number of massless poles is different, just that they need to be arranged differently into the different multiplets. E.\ g., all massless poles may occur in some group-singlet, with the appropriate degeneracy. If, and how, this occurs can no longer be derived in the same way as Goldstone's theorem. 

This has been a long-standing issue in lattice simulations \cite{Lang:pc}. Standard algorithms tend to get stuck, even with intact symmetry, in large volumes at a fixed value of the vacuum expectation value, thus mimicking an external source. This is either due to non-ergodicity of the algorithms or due to critical slowing down. As this is a systematic error, this is undesirable. If overcome, as discussed above, Goldstone's theorem is no longer applicable, and the symmetry is manifest on the level of expectation values \cite{Lang:pc,Neuberger:1987kt}. If it is desired to investigate the Goldstone dynamics, and thus the spontaneously broken theory, it is necessary to introduce an explicit symmetry breaking, typically an external source. This has been done in lattice simulations, especially for the O(4) case \cite{Luscher:1987ek,Hasenfratz:1988kr,Luscher:1988gc,Hasenfratz:1989ux,Hasenfratz:1990fu,Gockeler:1991sj,Gockeler:1994rx}. This corresponds to a numerical implementation of the prescription \pref{vevlimit}. Whether and where massless modes appear in the spectrum in the metastable case has unfortunately not yet been investigated in lattice simulations \cite{Lang:pc}.

\subsection{Classification of global symmetry breaking}\label{ss:classificationsb}

To summarize, the patterns of symmetry are
\begin{itemize}[leftmargin=0.1\textwidth]
 \item[Unbroken] The classical symmetry exists at the quantum level. Only invariant expectation values are non-zero. Individual measurements yield no preferred direction.
 \item[Hidden] Redefined fields no longer possess the original symmetry. A possibly involved and non-linearly realized version of the original symmetry persists, and the theory exhibits the symmetry as if unbroken.
 \item[Metastable] The classical symmetry exists at the quantum level. Only invariant expectation values are non-zero. Individual measurements yield a preferred direction. The theory is changing discontinuously when a source is applied.
 \item[Spon.\ broken] The classical symmetry is explicitly broken by a source. This breaking persist when the source is send to zero. In the limiting case the Goldstone theorem applies. Non-invariant expectation values are non-zero, and aligned with the source. Individual measurements are aligned with the source.
 \item[Expl.\ broken] A non-zero source or a term of the Lagrangian breaks the symmetry. The breaking can either vanish when taking the source or the offending term to zero or otherwise the symmetry is also spontaneously broken. Expectation values and individual measurements are aligned with the direction of the breaking.
 \item[Anomaly] A classical symmetry is explicitly broken by quantum effects. Usually this comes from the non-invariance of the path integral measure \cite{Bertlmann:1996xk}.
\end{itemize}
While the terms above are colloquially often used (partly) interchangeably, this is not really appropriate. Every notion has its very particular, and different, meaning. This has consequences for both calculational purposes and results, e.\ g.\ expectation values. It has also important implications for the observable charge structure of states \cite{Strocchi:2005yk}: Superselection sectors arise in context of the intact symmetries of the theory. Thus, a metastable and a spontaneously broken theory have different charge superselection sectors. It should also be noted that there is a deep connection to the existence of long-range order and thus how cluster decomposition operates \cite{Weinberg:1996kr}. This could also be used to classify the various possibilities.

It should be noted that both the unbroken and the metastable case behave as is expected for a linear realization, or Wigner-Weyl mode, of the symmetry. The difference is how they react to external perturbations or measurements. Conversely, the spontaneous broken case is the Nambu-Goldstone mode. This is also sometimes called a non-linear realization of the symmetry, which emphasizes that the states which belong to the coset of the spontaneously broken group still transform in a non-linear way under it, i.\ e.\ their transformation is no longer mediated by linear operators, as these would necessarily mix elements of the coset and any unbroken subgroup. Because these names cannot distinguish the unbroken and the metastable case here these notions are not employed.

\section{Brout-Englert-Higgs physics}\label{s:beh}

The next step is to upgrade the symmetry to a local one. This will have far-reaching consequences, especially when it comes to how the symmetry manifests itself. Again, the dimensionality is crucial \cite{Kennedy:1985yn,Kennedy:1986ut}, and implicitly assumed to be sufficiently large, i.\ e.\ usually three or more.

\subsection{Formulating the theory}\label{s:local}

The starting point is a theory with scalar fields $\phi$ and Lagrangian
\be
\La=\pdm\phi^\dagger\pd^\mu\phi-V(\phi)\nn,
\ee
\no which is invariant under some group ${\cal G}$. The symmetry is assumed to act linearly on the scalar fields, which are in some representation ${\cal R}$ of ${\cal G}$.

The next step is to gauge the theory. Exploiting the linear action on the fields leads to minimal coupling. The gauged theory is then
\bea
{\La}&=&-\frac{1}{4}W_\mn^aW^\mn_a+(D_\mu\phi_\alpha)^\dagger D^\mu\phi_\alpha-V(\phi)\label{la:hs}\\
W_\mn^a&=&\pdm W_\nu^a-\pdn W_\mu^a+gf^{abc}W_\mu^b W_\nu^c\nn\\
D_\mu&=&\pdm+gT^R\nn,
\eea
\no with the gauge fields $W_\mu^a$ in the adjoint representation. The $T$ are suitably normalized generators for the representation of the scalar field and $g$ is the newly introduced gauge coupling.

It is possible to promote either the full global group ${\cal G}$ to the gauge group $G$ or only a subgroup, as long as the resulting Lagrangian is gauge-invariant, and thus $G\subseteq{\cal G}$. This is particularly easy if the group is a product group ${\cal G}=G\times{\cal C}$. Otherwise the remainder may be either just a coset ${\cal G}/G$, or may contain another group, if there are additional subgroups of the original group ${\cal G}$ besides $G$.

If the remainder is not just a coset, and hence there exists a group ${\cal C}\subseteq{\cal G}/G$, this global symmetry group will be called the custodial group ${\cal C}$. If the potential $V$ in \pref{la:hs} is separately invariant under both groups, $G$ and ${\cal C}$, this will be an additional global symmetry group of the theory. Otherwise, the potential breaks the custodial group explicitly, either to a subgroup or completely. Despite its name, this symmetry is not different from, e.\ g., a flavor symmetry. It merely states that the scalar field has more degrees of freedom than are minimally necessary to write down the theory. Especially, the custodial group can also be a discrete group.

The most prominent example is the standard model Higgs sector. In this case the scalar is in the fundamental representation of ${\cal G}=O(4)$. Then\footnote{Actually, in the standard model $G=$SU($2$)/Z$_2$, due to the requirement of single-valuedness for the fermion fields \cite{O'Raifeartaigh:1986vq}. This will not play a significant role in the remainder of this review.} $G=\text{SU}(2)$ and ${\cal C}=\text{SU}(2)$, since loosely\footnote{More precisely SO(4)$\sim$(SU(2)$\times$SU(2))/Z$_2$.} O(4)$\sim$SU(2)$\times$SU(2). The custodial symmetry will be broken explicitly once either QED or fermions are introduced. This may happen also in general if there is an additional sector $\La_r$ which breaks ${\cal C}$ explicitly, even if $V$ does not. This will be discussed in sections \ref{ss:qed} and \ref{ss:flavor}.

The local and global symmetry structure in the standard model case can be made particularly well explicit by switching to a matrix representation, rather than a vector representation. For this write \cite{Shifman:2012zz}
\be
X=\bpm \phi_1 & -\phi_2^\dagger\\ \phi_2 & \phi_1^\dagger \epm\label{higgsx}
\ee
\no where $\phi_i$ are the usual components of the complex vector representation. In this representation the field transforms by a left-multiplication under a gauge transformation, $X\to g(x)X$ with the matrix $g(x)$ in the fundamental representation of SU(2). Under a custodial transformation $c$, with the matrix $c$ in the fundamental representation of SU(2), it transforms by a right multiplication, $X\to X c$. In terms of this representation, and for the standard model Higgs potential \cite{Bohm:2001yx}, the Lagrangian \pref{la:hs} reads \cite{Shifman:2012zz}
\be
\La=-\frac{1}{4}W_\mn^aW^\mn_a+\frac{1}{2}\tr\left((D_\mu X)^\dagger D^\mu X\right)-\lambda\left(\frac{1}{2}\tr\left(X^\dagger X\right)-f^2\right)^2\label{la:hsx}.
\ee
\no In this form the Lagrangian is manifestly invariant under both local and global transformations. Though such a manifest rewriting is not possible for general groups ${\cal G}\to G\times{\cal C}$, it will be very helpful to explicitly demonstrate general properties.

It is also useful to note that, if $\phi\neq 0$,
\be
X=\rho\alpha\label{lrdecomp},
\ee
\no where $\rho=(\det X)^\frac{1}{2}=(\phi^\dagger\phi)^\frac{1}{2}$ and $\alpha$ is a SU(2) matrix. If $\phi=0$, the transformation is ambiguous, as then $\rho=0$, but $\alpha$ is arbitrary. The reason is that this transformation corresponds to a mapping from $\cn^2$ to $\rn^1\times S^3$, where $S^3\sim SU(2)$ is the four-sphere. Both spaces are not topologically equivalent. Therefore, in general a rewriting like \pref{lrdecomp} modifies the target space of the theory, and both theories are no longer necessarily equivalent. This is irrelevant in perturbation theory, but may become relevant beyond perturbation theory. Note that this problem does not arise for the representation \pref{higgsx}.

\subsection{At the classical level}\label{ss:morse}

The first important observation is that the Lagrangian \pref{la:hs} is manifestly gauge-invariant. This is particularly obvious in the standard model case \pref{la:hsx}, or if the potential is a function of gauge-invariant combinations of the $\phi$ like $\phi^\dagger\phi$. For simplicity, only this polynomial case will be considered.

Since the potential is, by construction, a function of gauge-invariant quantities only, so are the minima given by gauge-invariant conditions. I.\ e.,
\be
\phi^\dagger\phi=f^2\label{minhpot}
\ee
\no is the usual condition. These minima are for a polynomial potential necessarily translationally invariant, and therefore the fields satisfying \pref{minhpot} need to be constant. The minima are also invariant under the custodial symmetry, as the potential is. It is essentially the existence of this minimum in a gauge theory which is at the root of the BEH effect, and everything else follows from it. The existence of this minimum, both classically or in the quantum effective action, will therefore be taken as being the BEH effect in the following.

In the standard model case \pref{la:hsx}, this condition takes the form
\be
X=v\alpha\nn,
\ee
\no with $v=f$ and $\alpha$ an arbitrary SU(2) matrix, since $\alpha^\dagger\alpha=1$. This also shows that the minimum is manifestly invariant under both, gauge transformations and custodial transformations.

In general, the potential can have multiple minima, $\phi^\dagger\phi=f_i$ where the $i$ now enumerates the minima, which may be global or local. These belong necessarily to different orbits. This will be considered in section \ref{ss:strata}. The general enumeration of these potentials and their minima is an exercise in group theory. The potential is build from invariants of the full group, ${\cal G}$. The enumeration of these invariants corresponds to the classification of the invariants of the group and the given representations. The identification of the possible minima is then given by Morse theory \cite{O'Raifeartaigh:1986vq,Sartori:1992ib}. Note, however, that the number and position of minima will be determined also by the parameters. Morse theory will only serve to enumerate all possible minima a potential can have at a given order of group invariants. The subset which are actually realized depends entirely on the potential. In the standard-model case the Higgs field is in the fundamental representation of SU(2), and the only group-invariant from which an invariant potential could be built is $\phi^\dagger\phi$. Therefore, in this case the minima necessarily have so far the same symmetry as the original potential, both locally and globally.

As a consequence, all possible minima belong to some little group of $G\times{\cal C}$, that is, they are characterized by a little group. Classically, it is possible to select a particular orbit as an initial condition, just as in the global case, e.\ g.\
\be
\phi_i^c=v_jn_i^j\nn
\ee
\no where no summation is implied and the $n^i_j$ is a suitable representative of an orbit. In the standard model case \pref{la:hsx}, a common choice is
\be
X^c=v1\label{xsmcase}.
\ee
\no If the little group is not just the trivial group consisting only of unity, there is a residual group,
\be
G\times{\cal C}\to G'\times{\cal C'}={\cal H}'\nn,
\ee
\no which may still have both a local part $G'$ and a global part ${\cal C}'$. It is not necessarily true that $G'\subset G$ and ${\cal C}'\subset{\cal C}$, but only\footnote{In other contexts, the same concept is also known as color-flavor locking \cite{Buballa:2003qv,Maas:2012ct}. Also there similar considerations apply as in the following when it comes to a manifest gauge-invariant description \cite{Maas:2012ct,Schafer:1998ef,Alford:1999pa}.} ${\cal H}'\subset G \times{\cal C}$. For any $\bar{h}'\in{\cal H}'$
\be
g(\bar{h}')\phi_i^c=\phi_i^c\nn,
\ee
\no showing that a whole orbit is actually selected as a minimum. Note that the index contains both, local and global indices.

As an example, in the standard model case \pref{xsmcase} this applies for any SU(2) matrix $A$ such that
\be
A^\dagger X^c A=v A^\dagger 1 A=v 1\label{colfmix}.
\ee
\no Thus, the remaining group ${\cal H}'$ requires to choose the same matrix $A$ from both the local and the global SU(2), but only from the global part of the gauge group. Thus, the remainder group is in this case indeed not a subgroup of either group, but rather a diagonal subgroup of both groups\footnote{Sometimes it is actually this diagonal subgroup which is referred to as the custodial symmetry. This makes sense in a purely perturbative setting, but would be problematic in section \ref{s:spectrum}, and will therefore not be done here.}. Colloquial \cite{Bohm:2001yx}, it is said that the condensate $\phi^c$ breaks the group $G\times{\cal C}$ to ${\cal H}'$. But it is once more the choice of an orbit by the initial conditions which performed the breaking, and not the dynamics. It is thus a case of spontaneous symmetry breaking.

\subsection{The quantum level, gauge-fixing and perturbation theory}\label{ss:qulev}

\subsubsection{Gauge-fixing}\label{sss:gf}

Going to the quantum level, the situation is best discussed in a path-integral language,
\be
Z=\pint\phi_i\cd W_\mu^a e^{i\int d^d x\La}\nn.
\ee
\no The path integral is over all field configurations and the Lagrangian and measure are invariant under the full group $G\times{\cal C}$. Therefore all orbits are treated equally.

However, as it stands, the path integral is ill-defined, because of the diverging integral over the gauge group \cite{Bohm:2001yx}. It is necessary to fix this. This can be done by gauge-fixing or by making the group integral finite by a lattice formulation. The latter will be done in section \ref{ss:lattice}. But to study the paradox alluded to in the introduction requires to fix the gauge.

To do so, the standard Faddeev-Popov procedure will be performed \cite{Bohm:2001yx}. There are a few subtleties, which will become important later on, and therefore it is useful to outline the most important steps. Start by selecting a local gauge condition $C^a(W_\mu^a,x,\omega)=0$, and assume that there is exactly one gauge copy on every orbit which satisfies this condition. The field $\omega$ here collects all other fields in the theory. I.\ e., for a set of gauge-fields related by gauge transformations $g(x)=\exp(i T^a\theta^a(x)/2)$ as
\bea
W_\mu&\to& gW_\mu g^{-1}-i(\pdm g) g^{-1}\approx T^a(W_\mu^a+D_\mu^{ab}\theta^b)+{\cal O}(\theta^{2})\label{gauget}\\
D_\mu^{ab}&=&\delta^{ab}\pdm+gf^{abc}W_\mu^c\nn
\eea
there is one, and only one, which satisfies the condition $C^a$. Herein $D_\mu^{ab}$ is the usual covariant derivative in the adjoint representation, a specialization of \pref{covderiv}. An example of such a condition is, e.\ g., the Landau gauge 
\be
C^a=\pd^\mu W^a_\mu\label{landaug}.
\ee
\no In perturbation theory, such a local condition is sufficient. The situation beyond perturbation theory is different and will be discussed in section \ref{ss:gribov}.

The next step is to factor off the gauge-equivalent field configurations, and remain with the one representative of each gauge orbit satisfying the gauge condition. To do so, introduce
\be
\Delta^{-1}[W_\mu^a]=\int {\cal D}g \delta(C^a[W_\mu^{ag}])\label{fpdelta},
\ee
\no which is an integration over the gauge group $G$ for a fixed physical field configuration $W_\mu^a$, but by the $\delta$-function only the weight of the one configuration satisfying the gauge condition is selected. The measure is taken to be the invariant Haar-measure, making the integral invariant under gauge transformations and therefore $\Delta$ is gauge-invariant by construction. Inverting $\Delta$ yields
\be
1=\Delta[W_\mu^a]\int{\cal D}g\delta(C^a[W_\mu^{ag}])\label{invertdelta}.
\ee
\no Inserting this into the functional integral yields
\bea
Z&=&\int{\cal D}W_\mu^a \Delta[W_\mu^a]\int{\cal D}g\delta(C^a[W_\mu^{ag}])\exp(iS[W_\mu^a])\nn\\
&=&\left(\int{\cal D}g\right)\int{\cal D}W_\mu^{a}\Delta[W_\mu^{a}]\delta(C^a[W_\mu^{a}])\exp(iS[W_\mu^{a}])\label{gffpexp}.
\eea
\no Using that the measure ${\cal D}W_\mu^a$ is gauge-invariant this isolates the gauge group integration, which can be absorbed in the normalization. Note that if the action would be replaced by any gauge-invariant functional, in particular expressions involving some gauge-invariant observable $f$ in the form $f[W_\mu^a]\exp(iS[W_\mu^a])$, the result would remain the same. Thus, this does not affect gauge-invariant observables. Due to the $\delta$-function now only one gauge-inequivalent field configurations contribute, making the functional integral well-defined. However, the appearance of the $\delta$ function becomes part of the measure, which now reads ${\cal D}W_\mu^a\Delta[W_\mu^{a}]\delta(C^a[W_\mu^{a}])$, and this expression is no longer gauge-invariant, and therefore gauge symmetry is explicitly broken by gauge fixing.

The integral \pref{fpdelta} can be explicitly evaluated with the rules for functional $\delta$-functions, yielding
\be
\Delta[W_\mu^a]=\left(\det\frac{\delta C^a(x)}{\delta\theta^b(y)}\right)_{C^a=0}=\det M^{ab}(x,y)\label{fpdet}.
\ee
\no $M$ is called the Faddeev-Popov operator, and the determinant the Faddeev-Popov determinant.

To get a more explicit expression the chain rule can be used
\be
M^{ab}(x,y)=\frac{\delta C^a(x)}{\delta\theta^b(y)}=\int d^4z\sum_{ij}\frac{\delta C^a(x)}{\delta \omega^i_j(z)}\frac{\delta \omega^i_j(z)}{\delta\theta^b(y)}\label{fpsumfield}.
\ee
In this case $i$ counts the field-type, while $j$ is a multi-index, encompassing gauge indices, Lorentz indices etc.. In the special case that $C$ only involves the gauge field, this reduces to
\be
M^{ab}(x,y)=\frac{\delta C^a(x)}{\delta W_\mu^c(y)}D_\mu^{cb}(y).
\ee
\no It is possible to recast the determinant also in the form of an exponential by introducing ghost fields,
\be
\det M\sim\int{\cal D}c^a{\cal D}{\bar c}^a\exp\left(-i\int d^4x d^4y \bar{c}^a(x) M^{ab}(x,y)c^b(y)\right)\nn.
\ee
\no The Faddeev-Popov ghost fields $c$ and $\bar c$ are Grassmann-valued scalar fields. The ghost fields are in general gauges not related, but may be so in particular gauges \cite{Alkofer:2000wg}. If the condition $C^a$ is local in the fields, the Faddeev-Popov operator will be proportional to $\delta(x-y)$, and this ghost term will become local. Note that if the gauge conditions involve fields other than the gauge fields this may introduce explicit interactions of these fields with the ghosts. This will happen below in the Lagrangian \pref{thla} for the gauge condition \pref{thooftg}.

\subsubsection{Standard gauge choices}\label{sss:stdgauge}

It is now useful to discuss a few of the standard gauge conditions employed, as they will be used throughout the rest of this review.

The first is the Landau gauge condition $C^a=\pd^\mu W^a_\mu$. In this case
\be
M^{ab}(x,y)=\pd^\mu D_\mu^{ab}\delta(x-y)\nn,
\ee
\no and the gauge is always well-defined. However, the gauge condition has an important incompleteness: It is invariant under global gauge transformations. Since global transformations have a finite volume, this does not impede the purpose of the gauge-fixing. But this implies that no direction is selected. Therefore, the global gauge transformations act like a global symmetry of section \ref{s:global}. Thus, any space-time-independent quantity having an open gauge index will still vanish \cite{Maas:2012ct}. In particular, it is impossible to have a non-zero Higgs expectation value in this gauge \cite{Maas:2012ct,Frohlich:1980gj,Lee:1974zg}.

A second useful choice is the class of linear covariant gauges. It is a different class of gauges as introduced before. They do not select a single gauge copy. Rather, they average over the complete orbit with a uniquely defined weight, which makes the integral over the orbit finite. However, this is implemented by averaging over gauge choices selecting a single gauge copy. Therefore, they generalize the above construction.

To do so, start from the choice $C^a=\pd^\mu W_\mu^a+\Lambda^a$ with $\Lambda^a$ some arbitrary function. This selects again a single gauge copy for each orbit. Afterwards, the gauge-fixed path integral is averaged with a Gaussian weight over the functions $\Lambda^a$, and thus over different gauge-fixings. Performing the Gaussian integral explicitly yields \cite{Bohm:2001yx}
\be
Z=\int{\cal D}W_\mu^a{\cal D}c^a{\cal D}{\bar c}^a\exp\left(iS-\frac{i}{2\xi}\int d^4 x\left(\pdm W^\mu_a\right)^2-i\int d^4x \bar{c}^a\pd^\mu D_\mu^{ab} c^b\right)\nn.
\ee
\no where $\xi$ is related to the width of the original Gaussian. Thus, ultimately, this gauge is equivalent to a Gaussian average over the gauge orbit. The limit $\xi\to 0$ recovers formally the Landau gauge condition. Note that also this gauge-fixed Lagrangian remains invariant under arbitrary global gauge transformations, and all expectation values not invariant under global gauge transformations still vanish\footnote{It should be noted that just leaving the global gauge degree of freedom untouched is by no means the only way of choosing a gauge with vanishing vacuum expectation values of non-invariant quantities \cite{Strocchi:1985cf}. Certain fully fixed gauges also achieve this by introducing local averages. There is also a connection to the Gribov-Singer ambiguity to be discussed in section \ref{ss:gribov}.}.

Thus, while both gauges are perfectly valid gauges they do have one serious problem, if perturbative calculations should be performed \cite{Lee:1974zg}: A perturbative expansion is in its nature a small-field amplitude expansion \cite{Rivers:1987hi}. Consider a situation with a Higgs potential\footnote{Of course, the potential needs to be convex at the quantum level, and is thus deformed, but this does not necessarily leads to a change in the size of the fluctuations around the average value of the fields.} of type \pref{complexpot}. If quantum effects are small this implies that the field amplitude is close to the minimum, and thus large. This is a serious problem, because the expansion point of standard perturbation theory is zero amplitude \cite{Bohm:2001yx}. Thus, in such a gauge the theory looks perturbatively like scalar QCD. This implies that the masses of the gauge bosons are zero to all orders in perturbation theory \cite{Bohm:2001yx}.

This can be remedied by a suitable choice of gauge, which makes a perturbative expansion around a non-zero Higgs field\footnote{This is also known as a mean-field expansion. At zeroth order this keeps only the vacuum expectation value of the Higgs field and neglects the fluctuation field. Also alternative expansion schemes than in the couplings are possible.} possible \cite{Lee:1972fj,Lee:1974zg,Lee:1972yfa,Bohm:2001yx}. Suitable for this purpose are 't Hooft gauges \cite{Bohm:2001yx}
\be
C^a=\pd^\mu W_\mu^a+ig\zeta\phi_iT^a_{ij} v_j+\Lambda^a\label{thooftg},
\ee
\no where the Higgs field $\phi$ and a constant vector $v$ of length $|v|=f$ in the representation of the Higgs appear. This implies that only the components of $\phi$ contribute to the gauge condition which yield a non-vanishing product with $T^av$, linking this term to section \ref{ss:qgoldstone}. They will become later the (would-be-)Goldstone bosons. Note that the choice of the direction of $v$ is purely part of the gauge choice\footnote{In fact, it would actually be a valid gauge by setting also the length of $v$ to an arbitrary value, but then the same problems with perturbative calculation reappear, especially for large discrepancies between $v^2$ and $f^2$.}. In particular, $v$ is not a vacuum expectation value, but a fixed (vector) parameter of the gauge choice. The corresponding gauge-fixed path integral becomes
\bea
Z&=&\int{\cal D}W_\mu^a{\cal D}c^a{\cal D}{\bar c}^a\exp\left(iS-\frac{i}{2\xi}\int d^4 x\left|\pd^\mu W_\mu^a+ig\zeta\phi_iT^a_{ij} v_j\right|^2\right.\nn\\
&&\left.-i\int d^4x \bar{c}^a(\pd^\mu D_\mu^{ab}-\zeta g^2 v_iT^a_{ij}T^b_{jk}\phi_k) c^b\right)\label{thla}.
\eea
\no There are a number of remarks to be made.

The {\bf first} is that the gauge parameters $\xi$ and $\zeta$ are independent. However, if $\xi\neq\zeta$ a mixing term of the gauge field and the Higgs field, or more precisely the Goldstone fields, would appear at tree-level. Therefore, usually both are set equal. But they receive different quantum corrections, and to avoid mixing at loop level requires to enforce the equality by making it part of the renormalization scheme \cite{Bohm:2001yx}. This will always be done in the following.

The {\bf second} remark is that the gauge condition reduces to the previous cases if $v$ is chosen to be zero.

The {\bf third} is that the gauge now explicitly prefers a particular direction, as $v$ is a constant vector. It therefore restores the situation of the classical level, but now by a gauge condition, rather than by an initial condition. This implies that a non-vanishing vacuum expectation value of the Higgs, $\la \phi\ra\neq 0$, is now possible, and will actually have the value $v$ at tree-level. In particular, just as in section \ref{sss:hidden}, $v^2$ can be identified with the minimum of the potential \pref{minhpot} determined by the parameters of the Lagrangian, $v^2=f^2$. But it needs to be kept in mind that again the vector $v$ is part of the gauge choice, while the number $f$ is a parameter of the Lagrangian. This is not relevant if, after fixing to this gauge, the gauge will no longer be changed, which is the usual approach in perturbation theory \cite{Bohm:2001yx}.

The {\bf fourth} is that if $v$ is identified with the vacuum expectation value $\la\phi\ra$ and if the BEH effect is not active, the gauge condition \pref{thooftg} reduces to the linear covariant gauges, as then $\la\phi\ra$ vanishes. Conversely, if $v$ is set to an arbitrary vector without identification, and if there is no BEH effect active, the condition $\la\phi\ra=v$ cannot be satisfied. Thus, the gauge condition cannot be fulfilled, and therefore this gauge does not create a well-defined quantum theory. It is therefore very important to distinguish how the gauge condition is defined, and what the meaning of $v$ is. In particular, if $v$ is a constant vector, it is not affected when performing transformations, while if it is identified with the vacuum expectation value it does change under transformations, e.\ g.\ custodial transformations. However, in the latter case it cannot be safely assumed that $v$ is non-zero, since this is a dynamical result of the theory. Thus, this gauge condition depends on the parameters of theory.

The {\bf fifth} is that the choice of this direction is not necessarily invariant under the global custodial symmetry. This happens, if $v$ is not invariant under custodial transformations. This is exactly the same situation as in the classical case, see \pref{colfmix} for an explicit example. To maintain the gauge condition therefore requires every custodial transformation to be accompanied by a compensating (global) gauge transformation. The remainder group is thus partly the (global) gauge symmetry, and therefore not itself observable. Instead, any gauge-invariant quantity constructed in this gauge will combine degrees of freedom such that the diagonal subgroup will be mapped to the custodial symmetry. This will be seen in explicit examples in section \ref{s:spectrum}.

{\bf Finally}, the interaction between the ghost fields and the scalar arise from the sum over all fields in \pref{fpsumfield}. It is therefore a direct consequence of the gauge condition.

\subsubsection{The gauge-fixed Lagrangian and the mass spectrum at tree-level}

Once this gauge is fixed and $f=|v|$ is set, it is common to perform in this Lagrangian the split \cite{Lee:1974zg,Bohm:2001yx}
\be
\phi(x)=v+\varphi(x)\label{qsplit}.
\ee
\no By construction, $\la\varphi\ra=0$, and thus the vacuum expectation value introduced by gauge-fixing is entirely identified\footnote{In fact, and to be pedantic, formally, the split is $\phi=w+\varphi(x)$. The constant background-field $w$ transforms both under gauge transformations and custodial transformations, and is only constant in the gauge where it is defined. In a second step, this background field is then identified with the constant vector $v$ of the gauge choice, with the usual consequences of identifying part of the field with a non-field quantity.} with $v$. However, fixing the gauge and performing this split are two independent operations and are performed separately, no matter that they are not very useful without each other.

It is important to note that the gauge and global transformation properties of $\varphi$ are influenced, just as in section \ref{sss:hidden}, by the fact that $v$ is not invariant under such transformations. The symmetry has been hidden. In fact, a gauge transformation away from \pref{thooftg} can shift $v$ even to zero. Thus, two of the mechanisms listed in section \ref{ss:classificationsb} are here at work simultaneously. The split \pref{qsplit} hides the symmetry. The gauge condition \pref{thooftg} explicitly breaks the gauge symmetry. The resulting theory therefore no longer has the symmetry, and even if it would have it, it would no longer be manifest.

Nonetheless, in this fixed class of gauges, everything is well-defined, with $\varphi$ being the dynamical degree of freedom. If the fluctuations of $\varphi$ are now small, i.\ e.\ $\la|\varphi|^2\ra\ll v^2$, then $\varphi$ describes small-amplitude fluctuations around zero. The field $\phi$ fluctuates then also only weakly around $v$. Thus, a perturbative description with $\varphi$ as dynamical variable seems possible \cite{Lee:1974zg,Damgaard:1985nb,Bohm:2001yx}.

This construction has a number of interesting consequences. It is useful for further reference to give as an explicit example the situation for the standard-model-like case. The starting Lagrangian is
\be
\La=-\frac{1}{4}W_\mn^a W_a^\mn+\frac{1}{2}\left(\left(\pdm\delta_{ij}-igW_\mu^aT^a_{ij}\right)\phi_j\right)^\dagger\left(\pdm\delta_{ik}-igW_\mu^bT^b_{ik}\right)\phi-\lambda(\phi^2-f^2)^2\label{higgsyml},
\ee
\no which has the classical minimum at $\phi^2=f^2$, and thus $v=fn$ will be set in the 't Hooft gauge condition \pref{thooftg}, keeping the actual direction $n$ arbitrary for the moment. Performing the Faddeev-Popov procedure and the split \pref{qsplit} yields.
\bea
\La&=&-\frac{1}{4}W_\mn^a W_a^\mn+\frac{1}{2}(D_\mu^{ij}\varphi_j)^\dagger(D_{ik}^\mu\varphi_k)\label{smkin}\\
&&+\frac{g^2f^2}{2}T^a_{ij}T^b_{ik}n_j^\dagger n_kW_\mu^a W_\mu^b-\lambda f^2|n\varphi|^2+\frac{ig^2\xi f^2}{2}\left|\varphi T^an\right|^2+i\xi g^2f^2 n_i^\dagger T^a_{ij}T^b_{jk}n_k\bar{c}^a c^b\label{smmass}\\
&&-2\lambda f((n\varphi)+(n\varphi)^\dagger)\varphi^2-\lambda(\varphi^2)^2+i\bar{c}^a\xi g^2 f(n_iT^a_{ij}T^b_{jk}\varphi_k^\dagger+n_i^\dagger T^a_{ij}T^b_{jk}\varphi_k) c^b\label{smint1}\\
&&+g^2 fT^a_{ij}T^b_{ik}(n^\dagger_j\varphi_k+n_k\varphi_j^\dagger)W_\mu^aW_\mu^b\label{smint2}\\
&&-\frac{i}{2\xi}(\pd^\mu W_\mu^a)^2-i\bar{c}^a\pd^\mu D_\mu^{ab}c^b\label{smgf}
\eea
\no Since the gauge will now remain fixed, cancellations of the constants $f^2$ and $v^2$, which have the same value, are permissible and have been performed. Also, constant and linear terms have been dropped. The cancellations already imply that the renormalization scheme must ensure that $f=|\la\phi\ra|=|v|$, as otherwise the cancellation would not be allowed at the quantum level, and additional terms would arise. Furthermore, by dropping terms the gauge is now fixed forever. These terms had contributions containing $v$, so they would transform under a gauge transformation.

This Lagrangian has a number of features. The terms \pref{smkin} are the ordinary kinetic terms. These two terms comprise together (massless) scalar QCD. The terms \pref{smgf} are the corresponding ordinary gauge-fixing terms for linear-covariant gauges. The terms, which are specific to the 't Hooft gauge are the remaining ones involving ghosts.

The terms \prefr{smint1}{smint2} are interactions. However, there are four interactions, which are determined by just three coupling constants, one of them the length of $v$. And one of them is the already appeared gauge coupling, and thus not independent. It is actually these relations, which are the manifest sign of a hidden gauge symmetry \cite{O'Raifeartaigh:1978kv}. In a theory where the symmetry would really be broken, these coupling constants are independent.

The terms \pref{smmass} are bilinears, and therefore are mass terms. It is quite instructive to study them term by term.

The first term in \pref{smmass} is the mass of the gauge bosons. It is matrix-valued. In the standard-model it is the remaining diagonal subgroup of the gauge and custodial symmetry which enforces a degeneracy of the eigenvalues of the mass matrix, giving all gauge bosons the same mass. This is, because the gauge bosons are in the adjoint representation of ${\cal H}'$. Again, as a hallmark of a hidden symmetry, these masses are not independent, but fixed by the coupling constants, and proportional to $gf$. The next term is a mass-term for the fluctuation field in the direction of $n$, as only the scalar-product with $n$ enters. This is an ordinary (correct sign) mass-term proportional to $\sqrt{\lambda}f$. This is what is usually called the Higgs mass \cite{pdg,Bohm:2001yx}, and the associated field component the Higgs. Note that the Higgs is, by construction, a singlet under the residual symmetry ${\cal H}'$. As can be seen, the value of the Higgs mass is completely independent of the masses of the gauge bosons, as the two independent couplings $g$ and $\lambda$ enter.

The next term contains the projection of the fluctuation field to $n$, but twisted by the representation matrices. This yields exactly all other degrees of freedom not contained in the Higgs part, what is usually called the (would-be-)Goldstones. Their mass is proportional to $\sqrt{\xi}gf=\sqrt{\xi}m_W$, and thus depends on the gauge parameter, signaling their unphysical nature. They also form a non-trivial representation of ${\cal H}'$. Similarly, the last term, involving the ghosts, yields also a gauge-dependent mass for the ghosts of the same size.

In addition, the quadratic term for the gauge bosons in \pref{smgf} also yields another gauge-dependent pole, which is best seen by writing down its tree-level propagator \cite{Bohm:2001yx}
\be
D_\mn^{ab}=\delta^{ab}\left(\left(\left(g_\mn-\frac{k_\mu k_\nu}{k^2}\right)\frac{-i}{k^2-m_W^2}\right)-\frac{i\xi k_\mu k_\nu}{k^2(k^2-\xi m_W^2)}\right)\label{wprop}.
\ee
\no The first contribution is the transverse part of the propagator, which carries the mass pole of line \pref{smmass}. The longitudinal part has the same type of gauge-dependent mass pole as the Goldstone bosons and the ghosts. The set of all of these gauge-dependent poles cancel in any perturbatively physical amplitude. This can be shown to all orders in perturbation theory, using a BRST construction \cite{Bohm:2001yx,DePalma:2013kua}. The same construction also shows that to all orders in perturbation theory only transverse gauge bosons and the Higgs appear as asymptotic states in physical amplitudes. This perturbative statement breaks down beyond perturbation theory, and this alters potentially the observable spectrum. This will be discussed in great detail in section \ref{s:spectrum}.

\subsubsection{Limits of the 't Hooft gauge}\label{ss:limitthooft}

The 't Hooft gauge has two particular limits. The first is the 't Hooft-Landau-gauge limit $\xi\to 0$. In this limit all perturbatively unphysical, i.\ e.\ non-BRST singlet, degrees of freedom become massless. However, the masses of the Higgs and the (transverse) gauge bosons remain non-zero.

It appears that in this limit the gauge condition \pref{thooftg} would degenerate to the Landau gauge condition \pref{landaug}. This would appear somewhat at odds with the situation discussed above for the ordinary Landau gauge. But this is not quite true. The difference arises from a subtle interchange of limits. In the 't-Hooft-Landau gauge first the global direction is fixed, and then the Landau gauge limit is taken. This corresponds to the sequence \pref{limssb}, while fixing the ordinary Landau gauge corresponds to the limit order \pref{limnssb}, yielding the differing results. In this case the explicit breaking by the presence of $v$ in \pref{thooftg} becomes a spontaneous breaking of the residual global gauge symmetry, in the sense of section \ref{ss:classificationsb}.

As a consequence of the difference between the spontaneously broken case and the case of genuine Landau gauge the masses of the (transverse) gauge bosons in both gauges remain different to all orders in perturbation theory. However, they remain in both gauges also gauge-parameter independent, once zero and once non-zero. But they are not gauge-independent. They can therefore not be physically observable quantities\footnote{Actually, the Oehme-Zimmermann superconvergence relation \cite{Oehme:1979ai} suggests already at the perturbative level that the gauge bosons may not be well-defined observables.}. This fact can be generalized to a larger class of gauges in form of the so-called Nielsen identities \cite{Nielsen:1975fs}, which are valid also non-perturbatively: The masses of the gauge bosons are gauge-parameter independent, but not (necessarily) gauge-independent, see section \ref{ss:masses}.

To distinguish both Landau gauge cases better, it is possible to call the standard gauge a non-aligned gauge and the 't Hooft-Landau gauge limit an aligned gauge, as they differ by the alignment of the (averaged) Higgs field \cite{Maas:2012ct}. 

The other interesting limit of the 't Hooft gauge is the opposite one, $\xi\to\infty$. This gauge sends the masses of the perturbatively unphysical degrees of freedom to infinity, and therefore effectively decouples them from the dynamics of the theory. It is therefore called the unitary gauge\footnote{In some special cases, like the standard model, the unitary gauge can also be explicitly constructed by performing a special gauge transformation \cite{Bohm:2001yx}. In the standard-model case, it reads $\exp(ig X/\sqrt{\det X})$, which explicitly cancels the Goldstone-degrees of freedom. This is only possible if the number of gauge bosons matches the number of Goldstone bosons, and will therefore not be used here. It creates unitarity by explicitly compensating the Goldstones and time-like gauge fields by a gauge transformation. Note that it also introduces (gauge) defects at points with $\det X=0$, see section \ref{ss:reformsm}. The latter do not arise in the limit prescription, indicating a possibility for a non-analyticity at $1/\xi=0$.}. While this procedure is evident at tree-level, this requires great care at loop level. If the masses are send to infinity before the regulator is removed, the theory is no longer manifestly renormalizable, but requires careful cancellations between newly arising counterterms at every order of perturbation theory \cite{Bohm:2001yx,Weinberg:1971fb,Lee:1973fw,Lee:1973xp,Appelquist:1972tn,Irges:2017ztc}, among other problems \cite{Bohm:2001yx,Bars:1972pe,Jackiw:1972jz}. This limit will therefore not be used here, as the emphasis is on non-tree-level physics\footnote{Formally, this gauge can be defined also in non-perturbative calculations, e.\ g.\ lattice calculations. But this still requires dealing with the renormalization issues, if not only renormalization-group-invariant quantities are studied.}.

This sets up the perturbative framework to be used in the following. It can be extended along standard lines to include fermions, and other interactions. As this is standard material, it will not be covered here, but can be found, e.\ g., in \cite{Bohm:2001yx,Barbieri:2007gi,O'Raifeartaigh:1986vq,Branco:2011iw,Langacker:1980js,Morrissey:2009tf}.

It should be noted that the perturbative framework has a number of intrinsic limits. This is even true for statements which have been proven to all orders in perturbation theory. The reason for these limits are manifold: First, perturbation theory yields qualitatively different results depending on the expansion point \cite{Lee:1974zg}, with the mass of the gauge bosons being the most notable example. The second is that perturbation theory is usually at best an asymptotic series \cite{Negele:1988vy,Rivers:1987hi}. The third is that even if the series seems to be convergent there are phenomena, which cannot even be qualitatively captured perturbatively \cite{Shifman:2012zz,Haag:1992hx,Rivers:1987hi}. Finally, the whole starting point of defining the series using gauge-fixing in section \ref{sss:gf} is not well-defined beyond perturbation theory, see section \ref{ss:gribov}. Most of the remainder of this review aims at a formulation without such restrictions. In addition, methods will be described which nonetheless allow to exploit the huge body \cite{pdg,Bohm:2001yx,Djouadi:2005gi,Dawson:2018dcd} of perturbative results. 

\subsection{The Gribov-Singer ambiguity}\label{ss:gribov}

While the previous discussion fully covered the gauge-fixed framework in the perturbative case, non-perturbatively a serious obstacle arises: The Gribov-Singer ambiguity \cite{Gribov:1977wm,Singer:1978dk,vanBaal:1997gu,vanBaal:1991zw,Dell'Antonio:1991xt,Maas:2011se,Vandersickel:2012tg,Sobreiro:2005ec}. The Gribov-Singer ambiguity is the statement that perturbative gauge-fixing conditions, like the covariant gauges and the 't Hooft gauges, are in general insufficient to define a gauge. The problem is that a condition like the Landau gauge, $\pd^\mu W_\mu^a=0$, has actually more than one solution. This problem does not appear in perturbation theory because these additional solutions are field-configurations which cannot be written as a power series in the gauge coupling \cite{Gribov:1977wm}. This implies that two gauge copies satisfying the same gauge condition cannot be connected by a sequence of infinitesimal gauge transformations, but only by a so-called large gauge transformation \cite{Gribov:1977wm}.

The origin of such gauge transformations can be tracked back to the geometrical properties of non-Abelian gauge theories \cite{Singer:1978dk}, showing also why it is absent from Abelian gauge theories. In terms of differential geometry the problem arises because the atlas of non-Abelian Lie groups contains at least two coordinate systems, and the transition functions between the coordinate systems need to be included in the construction of gauge conditions to uniquely single out gauge copies\footnote{The simplest example is SU(2), which is isomorphic to the surface of a three-sphere. Any single coordinate system leaves at least one point, usually a pole, ill-defined.}. Such transition functions are inherently global. Thus a local gauge condition is insufficient. In operational terms this is, e.\ g., seen by the fact that the Faddeev-Popov operator \pref{fpsumfield} develops zero modes \cite{Gribov:1977wm,Sobreiro:2005ec,Vandersickel:2012tg,Maas:2005qt}, and therefore an operation like \pref{invertdelta} is no longer admissible without dealing with the zero modes explicitly. Moreover, the Faddeev-Popov operator can also develop negative eigenvalues for Gribov copies. This implies potential cancellations in the path integral \pref{gffpexp}, which is at the core of turning perturbatively gauge-invariant into non-perturbatively gauge-variant statements. This is known as the Neuberger $0/0$ problem \cite{Neuberger:1986xz}.

The presence of Gribov copies has thus far-reaching implications. One is that the number of them could actually be infinite, and therefore the whole path integral could remain ill-defined. Another effect is that Gribov copies break the perturbative BRST symmetry \cite{Fujikawa:1982ss,Sorella:2009vt,Maas:2012ct}. The third is that potentially further gauge-fixing terms, introducing additional vertices in the Lagrangian, could be necessary to treat the Gribov-Singer ambiguity explicitly \cite{Maas:2011se,Vandersickel:2012tg,Serreau:2012cg}.

This problem is independent of the matter content of the gauge theory, and therefore also affects the Yang-Mills-Higgs systems. There is no argument known, why the Gribov-Singer ambiguity should be any less relevant \cite{Lenz:1994tb,Lenz:2000zt,Maas:2011se,Maas:2010nc,Capri:2012ah,Capri:2012cr,Capri:2013oja,Capri:2013gha,Capri:2016gut,Greensite:2004ke}, as long as the BEH effect is not operational, and this is also confirmed by lattice calculations \cite{Maas:2010nc}. In this case it appears possible to salvage a non-perturbative BRST symmetry \cite{Maas:2012ct,Sorella:2009vt,vonSmekal:2008ws,Dudal:2009xh,Kondo:2009ug,Dudal:2010fq,Kondo:2009gc,Boucaud:2009sd,Sorella:2010it,Neuberger:1986xz,vonSmekal:2007ns,vonSmekal:2008es,Fischer:2008uz} with essentially the same algebra, though at the expense of treating the Gribov-Singer ambiguity explicitly. Moreover, the construction then implies the absence of all gauge degrees of freedom, including transverse gauge bosons and all elementary scalars, from the physical spectrum \cite{Kugo:1979gm,Fischer:2008uz,Maas:2011se}. Thus, the physical spectrum then contains only manifestly gauge-invariant composite states, described by composite operators.

However, the situation appears different when the BEH effect is active \cite{Lenz:1994tb,Lenz:2000zt,Maas:2010nc,Ilderton:2010tf,Capri:2012ah,Capri:2012cr,Capri:2013oja,Capri:2013gha,Capri:2016gut,Greensite:2004ke}. While there is not yet a full understanding of why this is the case, the following reasoning can be given \cite{Lenz:2000zt}. Consider the Landau-gauge condition \pref{landaug} for a theory in which the gauge group is fully broken, i.\ e.\ the little group is trivial. The (Euclidean) Landau gauge condition can be reformulated using the functional \cite{Maas:2011se,Zwanziger:1993dh}
\be
F[W_\mu^a]=\int d^dx W_\mu^a W_\mu^a\label{fundmod}.
\ee
\no The extrema and inflection points of this functional satisfy the Landau gauge condition \cite{Zwanziger:1993dh}. In particular, the gauge orbit containing the trivial vacuum $W_\mu^a=0$ is an (absolute) minimum of \pref{fundmod}. In fact, every gauge orbit has, up to topological identifications, a unique absolute minimum \cite{Zwanziger:1993dh,vanBaal:1997gu,vanBaal:1991zw,Dell'Antonio:1991xt}, which is at the same time also the Gribov copy closest to the trivial vacuum. The set of these absolute minima forms the so-called fundamental modular region \cite{Zwanziger:1993dh,vanBaal:1991zw,Dell'Antonio:1991xt}, a convex and bounded region.

Considering now an aligned Landau gauge, the alignment does not change the functional \pref{fundmod}, as it is invariant under global gauge transformations. Since all gauge fields received a mass, their long-distance fluctuations are suppressed, up to irrelevant quantum noise. Take the field configuration closest to the vacuum. Taking the limit of infinite gauge boson masses, all gauge transformations
\be
W_\mu^a\to W^a_\mu+\pdm\omega^a+igf^{abc}W^b_\mu\omega^c\nn
\ee
\no become essentially Abelian, as the non-trivial part of the transformation are suppressed, as the gauge-fields fall off very fast due to the large mass. But Abelian transformations cannot create a Gribov copy. And all remaining Abelian gauge transformations are already taken care of by the perturbative gauge condition. Thus, all Gribov copies move, essentially, too infinity, and are no longer reachable. Hence, in the limit of infinite gauge boson mass, the Gribov-Singer ambiguity should become quantitatively irrelevant. Still, even in this case the field configurations of the path integral with zero or negative eigenvalues of the Faddeev-Popov operator remain, and thus the qualitative problem.

While this argument is more hand-waving, indeed in lattice simulations \cite{Maas:2010nc} no Gribov copies are anymore found when the BEH effect is active. However, these investigations have been rather exploratory, and the methods used to search for Gribov copies have been the same as when the BEH effect is not active. It may even be that this is therefore an algorithmic shortcoming. Still, this is the best available evidence in favor of the arguments of \cite{Lenz:2000zt} so far.

The non-aligned versions will not be different, as the gauge bosons are still found to obtain a non-zero mass \cite{Maas:2010nc}. There are not yet systematic investigations what happens when including the Higgs field in the gauge condition, i.\ e.\ at $\xi>0$, \pref{thooftg}, but there is no obvious argument how this should alter the situation\footnote{The special unitary gauge for the standard model case \cite{Bohm:2001yx} seems to avoid the Gribov problem. However, their role is taken over by the gauge defects, as they have also to be treated to define the gauge fully. In this way the ambiguity arises again, as it does in all purely perturbative gauges.}.

If the gauge condition does not fully fix the gauge, there remains a residual gauge group. If the residual gauge group is Abelian, the argument is essentially the same, as Abelian gauge theories have no Gribov-Singer ambiguity\footnote{Note that on a finite lattice even Abelian gauge theories have a Gribov-Singer ambiguity, but this is a lattice artifact vanishing in the thermodynamic limit \cite{deForcrand:1994mz,Giusti:2001xf}.} \cite{Lenz:1994tb,Lenz:2000zt}. However, if the residual gauge group is non-Abelian, it can be expected that some remnant of the Gribov-Singer ambiguity survives. Which form it then takes has not been investigated, but may be relevant, e.\ g., for grand-unified theories \cite{Georgi:1974sy,Georgi:1974yf,Langacker:1980js}.

All of the above gives only arguments why the Gribov-Singer ambiguity is quantitatively irrelevant. Even if this is the case, a local gauge condition can still not be formulated, because the structural reason is not resolved. Especially perturbative BRST symmetry is still broken. This has bearing on the spectrum of the theory, as will be discussed in section \ref{s:spectrum}.

\subsection{Lattice formulation}\label{ss:lattice}

Yang-Mills-Higgs theories suffer from the same problem as almost all interacting quantum field theories: There are no exact solutions. However, it is possible to obtain a number of exact statements using a lattice-regularized form of it. Also, many of the following results will be genuinely non-perturbative in nature, and thus required non-perturbative methods. So far, lattice has been the most used method for this purpose. Thus, here the lattice-regularized version of the theory will be introduced.

The simplest lattice discretization of the Lagrangian \pref{higgsyml} is on an Euclidean hypercubic\footnote{In practice often asymmetric lattices with an elongated time direction are used. This will be noted if relevant. Also, this introduces the aspect ratio as an additional systematic error source.} lattice with lattice spacing $a$ and size $N^d$. A suitable definition of the discretized theory is given by \cite{Montvay:1994cy}
\bea
S&=&\sum_x\left(\phi^\dagger(x)\phi(x)+\gamma(\phi^\dagger(x)\phi(x)-1)^2-\kappa\sum_{\pm\mu}\phi(x)^\dagger U^R_\mu(x)\phi(x+\mu)\right.\label{higgsymlat}\\
&&\left.+\frac{\beta}{d_F}\sum_{\mu<\nu}\Re\tr\left(1-U(x)_\mn\right)\right)\nn\\
U(x)_\mn&=&U_\mu(x)U_\nu(x+\mu)U_\mu(x+\nu)^\dagger U_\nu(x)^\dagger\label{plaq}\\
U_\mu(x)&=&\exp(iW_\mu^a\sigma_a)\label{link}\\
\beta&=&\frac{2C_F}{g^2}\label{beta}\\
-a^2(2\lambda v^2)&=&\frac{1-2\gamma}{\kappa}-2d\label{hopping}\\
\frac{1}{2\lambda}&=&\frac{\kappa^2}{2\gamma}\label{lambda}.
\eea
\no In comparison to \pref{higgsyml} some constant terms have been added. Note that also the fields are rescaled by $1/\sqrt{\beta}$ for the gauge field and $1/\sqrt{\kappa}$ for the Higgs field. The parameter $\kappa$ is known as the hopping parameter. $C_F$ is the fundamental Casimir of the gauge group, for SU($N$) $C_F=N$.  The $U_\mu^R$ are the group elements, in the representation $R$ of the Higgs fields. For the fundamental representation, $U^R=U$ and thus are just the ordinary links \pref{link}. In the adjoint representation, $U^A_{bc}=\tr\left(T^b U_\mu^\dagger T^c U_\mu\right)/2$ holds, where $T^a$ are the generators of the group in the fundamental representation, and so on. The notation $x+\mu$ denotes a lattice site shifted from $x$ by a vector in direction $\mu$ of length $a$.

Many differences are a straightforward consequence of the discrete lattice. But not all. The most important difference to the continuum formulation is that by \pref{link} the algebra-valued gauge fields $W_\mu^a$ are in the lattice formulation replaced by the group-valued link variables $U_\mu$ as dynamical variables. This is particularly important as there are different groups belonging to the same algebra, differing by discrete center groups \cite{O'Raifeartaigh:1986vq}. Thus, a choice must be made which of those different groups is selected.

This is even more relevant, as not every choice gives consistent, single-valued fields \cite{O'Raifeartaigh:1986vq}, depending on the field contents and representations in a theory. This issues becomes constraining for the full standard model, and many extensions of it, as there the choice of group is fixed because of the fermion charge structure \cite{O'Raifeartaigh:1986vq}.

For the particular example \pref{higgsymlat}, this is not a problem, and the usual choice for the Lie algebra su($N$) is the Lie group SU($N$). However, in the context of the standard model, it would need to be SU(2)/Z$_2$ \cite{O'Raifeartaigh:1986vq}. This subtlety is probably not a quantitatively important effect, as can be seen from the fact that QCD lattice simulations, despite using the group SU(3) rather than the correct group SU(3)/Z$_3$ \cite{O'Raifeartaigh:1986vq}, show a remarkable agreement with experiment \cite{Gattringer:2010zz,DeGrand:2006zz}. But this choice can be quite decisive for the efficiency of numerical simulations, which is, e.\ g., seen when comparing SU(2) to SU(2)/Z$_2\approx$SO(3) \cite{deForcrand:2002vs,Friedberg:1995cq,Burgio:2006xj,Burgio:2006dc}\footnote{For Yang-Mills theory, after early exploratory studies \cite{Greensite:1981hw,Halliday:1981te,Creutz:1982ga,Drouffe:1982fe}, recent results also support that the global structure of SU($N$) vs.\ SU($N$)/Z$_N$ is not relevant for the low-energy physics \cite{deForcrand:2002vx,Teper:2018aa}.}. Therefore, almost all lattice simulations of \pref{higgsymlat} used SU($N$) with mostly $N=2$ or $3$.

One of the fundamental advantages of a lattice formulation is that it is not necessary to fix a gauge to perform a calculation or numerical simulation \cite{Montvay:1994cy,Gattringer:2010zz,DeGrand:2006zz}. The reason is again the use of the group, rather than the algebra. The compact Lie groups, which appear in theories like \pref{la:gen}, have a finite volume. Integrals over compact groups are finite, in contrast to the continuum algebra integrals. Thus, the necessary group integral performed at every lattice site is finite. Then the original reason for fixing a gauge is no longer there, and gauge fixing becomes unnecessary. This does not mean it is impossible or irrelevant. If gauge-dependent quantities should be calculated fixing a gauge is still required. This can be done straightforwardly \cite{Maas:2011se}. But since the lattice formulation is non-perturbative, it is necessary to take the Gribov-Singer ambiguity of section \ref{ss:gribov} fully into account.

Of course, that finite integrals appear is due to the lattice regularization only. Once the continuum limit is taken, i.\ e.\ the lattice spacing $a$ is send to zero, the original divergence reappears, and needs to be treated in one way or the other. Therefore, the lattice can be considered to not only regularize ultraviolet divergences but also the divergence due to the gauge freedom by requiring only a finite number of finite group integrals.

Note that performing the limit of a lattice theory to a continuum theory is not only because of the reappearing divergences not straightforward \cite{Montvay:1994cy,Gattringer:2010zz,DeGrand:2006zz}. In particular, the triviality problem \cite{Callaway:1988ya} can substantially affect even the possibility to perform such a limit. This will be discussed in section \ref{ss:triv}.

Fortunately, just like in the continuum, even trivial theories are non-trivial at finite cutoff, here finite lattice spacing $a$. For the present class of theories, if the cutoff is large enough, this will not affect the low-energy physics \cite{Hasenfratz:1986za,Kuti:1987nr}. Thus, even in a full non-perturbative formulation the theory is as predictive as its continuum counterpart. In particular, it should be possible to provide reasonably reliable results for experimentally accessible energies. Note that this is not a problem exclusively applying to the lattice formulation, as for a trivial theory also continuum regulators cannot be removed \cite{Callaway:1988ya}. In the continuum, essentially the Appelquist-Carrazone theorem \cite{Bohm:2001yx} enables the use of the theory as a low-energy effective theory, in a very similar way as on the lattice. Thus, the triviality problem will not affect the low-energy physics to be discussed in this section and in sections \ref{s:spectrum} and \ref{s:scattering}, and can be safely ignored. For possible resolutions of it on a more fundamental level, see section \ref{ss:triv}.

In the following both exact calculations on the lattice and results of numerical Monte Carlo simulations will be discussed. There are many possibilities how to perform numerical simulations for theories like \pref{higgsymlat} \cite{Montvay:1994cy,DeGrand:2006zz}. This kind of technical details will not be dwelt on here, but rather the corresponding original literature will be referred to, except when some particular technicality is central to the point at hand. However, it should be noted that the computation time scales like some power of both the physical volume and the cutoff. To achieve the necessary scale separation to avoid lattice artifacts requires to have all relevant energies and masses sufficiently far away from either regulator, $1/L\ll E,m\ll 1/a$. Given current computer resources, mass hierarchies of about one order of magnitude can be covered. This implies that the standard model's mass hierarchy of at least 12 orders of magnitude between neutrino masses and the top mass will not be accessible in the foreseeable future in numerical simulations, and these have to contend with subsectors of the whole standard model. Especially, the inclusion of fermions also increases computing times substantially compared to purely bosonic theories even at fixed discretization and lattice volume, essentially because of the Pauli principle \cite{Gattringer:2010zz,DeGrand:2006zz}.

However, even if the computing time would not be an issue, there is an even more fundamental problem with regard to the weak interaction: Parity and charge-parity violation. A lattice regularization breaks chiral symmetry explicitly \cite{Montvay:1994cy,Gattringer:2010zz,DeGrand:2006zz}. As long as chiral symmetry is a global symmetry, this problem can be ameliorated by introducing a differing symmetry, which turns into chiral symmetry in the continuum limit \cite{Gattringer:2010zz,DeGrand:2006zz}. The weak interaction gauges chiral symmetry, and therefore the naive lattice regularization would introduce a gauge anomaly. It has been speculated that it should also be possible to find a replacement gauge symmetry, which becomes the weak interaction in the continuum limit \cite{Hasenfratz:2007dp}. And although there exists a number of proposals for such a replacement symmetry \cite{Grabowska:2015qpk,Gattringer:2008je,Cundy:2010pu,Igarashi:2009kj,Xue:2000du}, none of them has been unambiguously shown to be a suitable and sufficiently efficient formulation. Therefore, no lattice simulations with fermions and the weak interactions exist. This problem does not exist for pure Higgs-Yukawa theories, and there parity breaking can be and has been investigated on the lattice \cite{Fodor:2007fn,Gerhold:2007gx,Gerhold:2010bh,Gerhold:2011mx,Chu:2015nha,Bulava:2012rb}, but this is not subject of this review.

\subsection{The Higgs vacuum expectation value and Elitzur's theorem}\label{ss:vev}

Having now an accessible non-perturbative formulation of the theory at hand, it is possible to make many of the previous statements more precise. The first, and perhaps most central one, is the absence of a Higgs vacuum expectation value without gauge fixing, as argued for in section \ref{ss:qulev}. The problem posed by the lack of gauge invariance of the vacuum expectation value was recognized early on \cite{Lee:1974zg,Fischler:1974ue}, and shortly after formalized as Elitzur's theorem \cite{Elitzur:1975im}. However, in the original construction of Elitzur's theorem assumptions are made about the analyticity properties of expectation values in external sources, which especially for gauge sources are not necessarily justified \cite{Maas:2013sca}.

A reformulation of the same argument without sources can be found in \cite{Frohlich:1980gj}. Consider a gauge-invariant measure ${\cal D}\mu$, which includes the action, and a gauge-dependent quantity $f(W)\neq f(W)^g$ without any gauge-invariant contribution. Then \cite{Frohlich:1980gj}
\be
\la f(W)\ra=\int{\cal D}\mu f(W)=\int{\cal D}\mu^{g^{-1}}f(W)=\int{\cal D}\mu f(W^{g})=\int{\cal D}\mu f(W)^g=\la f(W)^g\ra\label{fmsel},
\ee
\no where $g$ is an arbitrary gauge transformation. Since, by assumption, $f$ is gauge-dependent, the equality can only be satisfied if $\la f(W)\ra=\la f(W)^g\ra=0$. The Higgs vacuum expectation value is just a particular case, and therefore needs to vanish. Note that gauge-fixing modifies the measure such that it is no longer gauge-invariant, as discussed in section \ref{ss:qulev}, and then this argument does no longer hold. On a finite (Euclidean) lattice, this is a well-defined expression, as the path integral can be decomposed into ordinary Riemann integrals, and the statement is exact. This has consequently been confirmed in numerical lattice calculations \cite{Caudy:2007sf,Maas:2012ct}. Even more, with the same argument it can be shown that any quantity which is not invariant under some subgroup or coset of the gauge group also vanishes identically \cite{Frohlich:1980gj}. This will be of particular importance in sections \ref{ss:strata} and \ref{sss:gut}.

Reversing the argument, if a gauge condition leaves some subgroup or coset of the gauge group untouched, the measure is still invariant under this subgroup or coset. Then, by the same argument, any quantity not invariant under the subgroup or coset still vanishes. An example are the non-aligned gauges in section \ref{ss:qulev}, where any quantity not invariant under global gauge transformations still vanishes. The latter is consequently also seen in lattice simulations \cite{Maas:2012ct}.

It is not incidental that the argument looks very similar to the one for the global metastable case of section \ref{s:hidssb}. Essentially, \pref{fmsel} is nothing but the statement that if all possible gauge transformations are included in the path integral, no particular direction reachable by a gauge transformation can survive. The difference to the global case is that introducing a source will not only prefer globally a direction, but locally, and thus break Poincar\'e invariance as well. Thus, spontaneous gauge symmetry breaking is incompatible with Poincar\'e symmetry, and hence forbidden.

There is one loophole in this argument. This loophole is whether the result has implications for the continuum theory, provided the latter exists in a mathematically strict sense at all. If the theory exists in the continuum limit, the argument requires that the path integral is well-defined even without gauge-fixing in the continuum\footnote{Alternatively, the continuum limit can be defined only for expectation values. Then their calculation can be performed in regularized form, and afterwards the limit is taken \cite{Seiler:1982pw}. In a sense, this is also what is done in perturbation theory.}. This is yet unknown, and even for non-interacting theories the individual integrals have to be extended from Riemann integrals to Ito integrals \cite{Rivers:1987hi}\footnote{Note that also the Wick rotation to Euclidean space-time can only be guaranteed for gauge-invariant quantities \cite{Seiler:1982pw}. For gauge-dependent quantities the indefiniteness of the state space has yet prevented a final answer.}. However, this question pertains to all lattice simulations, as they are performed without gauge-fixing \cite{Montvay:1994cy,Gattringer:2010zz,DeGrand:2006zz}. It is therefore assumed henceforth that such a construction is possible. Support for this assumption is provided by a limited number of investigations in which in continuum theories calculations without gauge fixing have been performed, yielding the same results as in the gauge-fixed setup \cite{Strocchi:1974xh,Arnone:2005fb,Arnone:2006ie,Rosten:2010vm}. While this does not constitute a proof, it is encouraging.

\subsection{The phase diagram and the Osterwalder-Seiler-Fradkin-Shenker construction}\label{ss:pd}

The previous section concluded that the Higgs vacuum expectation value is not a suitable order parameter for the Higgs effect, as its value can be zero or non-zero, depending on the choice of gauge. Any other gauge-dependent order parameter for the BEH effect will suffer from the same problem. In particular, for any order parameter constructed by utilizing some remnant (global) symmetry there will always be gauges with or without such a symmetry, and thus different behaviors. In fact, it was found that utilizing different remnant symmetries to detect the BEH effect yields differing points in the phase diagram where the BEH effect appears or vanishes \cite{Caudy:2007sf,Greensite:2008ss}.

On the other hand, there are regions in parameter space where the theory is expected to behave like a scalar version of QCD, which is also confirmed in explicit lattice calculations \cite{Knechtli:1998gf,Knechtli:1999qe,Maas:2010nc,Maas:2013aia,Maas:2014pba}. This also includes phenomena like confinement in the same sense as in QCD. In fact, there are multiple arguments both perturbative \cite{Bohm:2001yx,Kapusta:2006pm,Damgaard:1985nb} and non-perturbative \cite{Kugo:1979gm,Schaden:2013ffa,Alkofer:2000wg,Fischler:1974ue,Nielsen:1975fs,Greensite:2017ajx,Bricmont:1985sw,Bricmont:1985by,Bricmont:1987zh} that, in fixed gauges, both regimes are qualitatively different. Especially, gauge-dependent correlation functions are expected to change discontinuously between the regimes\footnote{Note that there are even counter examples to this: It appears that gauge-dependent correlation functions may change continuously across the silver-blaze transition in finite density (2-color) QCD \cite{Boz:2018crd}.}. Even for this lattice support exists \cite{Maas:2013aia}.

This naturally raises the question whether there is then a gauge-invariant order parameter for the BEH effect. But the answer is no. This hope is shattered conceptually by the Osterwalder-Seiler-Fradkin-Shenker (OSFS) construction \cite{Osterwalder:1977pc,Fradkin:1978dv,Seiler:2015rwa,Seiler:1982pw,Glimm:1987ng}.

To understand the OSFS construction\footnote{The central arguments are given in \cite{Osterwalder:1977pc}. The outline here follows \cite{Fradkin:1978dv}.}, consider the lattice model\footnote{The same construction can also be performed using the continuum action \pref{la:hsx} in continuum notation, but also using implicitly a suitable ultraviolet regulator. To emphasize this necessity here the lattice action is used.} \pref{higgsymlat} in the limit of $\gamma\to\infty$ for the standard-model case. The Higgs fields are then fixed to be unimodular, $\phi^\dagger\phi=1$, and there are no length fluctuations. Note that the value of $\gamma$ does never induce an explicit symmetry breaking, and therefore the arguments given in section \ref{s:global} when taking limits of coupling constants do not apply. Of course, there could be other kinds of non-analyticity, and therefore here explicitly the limit is taken, rather than selecting the unimodular theory from the outset.

Since the length is fixed to be non-zero, it is safe to use the decomposition \pref{lrdecomp}. Thus, the Higgs fields are now SU(2) group elements. In this situation it is possible to perform a gauge transformation
\be
g=\alpha^{-1}\nn
\ee
\no which implies $gX=1$, and thus this gauge transformation explicitly eliminates all Higgs fields. In vector notation, this is a transformation to a constant vector. This is always possible, and $\alpha=1$ can therefore be implemented as a gauge condition, the unitary gauge. Due to the fixed length, this is possible without gauge defects, and the gauge is also non-perturbatively, and on the lattice, well-defined.

In this way, the action becomes an expression only involving the links. In fact, the theory becomes gauge-fixed and operates on new links defined as $V_\mu=\alpha(x) U_\mu(x)\alpha(x+\mu)^{-1}$. This yields
\be
S=-\kappa H(V_\mu)-\beta S_\ym\nn,
\ee
\no where $H$ contains the Higgs part of the action in this gauge, and $S_\text{YM}$ is the ordinary Yang-Mills part of this action. In both cases the parameters are factored out to make them explicit in the following. The expectation value of a gauge-invariant operator takes the form
\be
\la\op\ra=\frac{\int\Pi_{x,\mu}dV_\mu(x)\op e^{-\kappa H(V_\mu)-\beta S_\ym}}{\int\Pi_{x,\mu}dV_\mu(x) e^{-\kappa H(V_\mu)-\beta S_\ym}}=\frac{\int\Pi_{x,\mu}dV_\mu(x)e^{-\kappa H(V_\mu)}\op e^{-\beta S_\ym}}{\int\Pi_{x,\mu}dV_\mu(x) e^{-\kappa H(V_\mu)}e^{-\beta S_\ym}}\nn.
\ee
\no Now, consider the Higgs part of the action. Since the Higgs fields are gone, the expression involving Higgs fields in \pref{higgsymlat} will give the smallest action for $V_\mu=1$. Thus, for $\kappa$ large, links close to unity will dominate the action. This incidentally also minimizes the Yang-Mills part of the action, and the expression $e^{-\beta S_\ym}$ will be close to one. A similar statement holds also true if $\beta$ is small. Thus, for now in the following assume that either $\beta$ is small or $\kappa$ is large.

Expanding
\be
e^{-\beta S_\ym}=\Pi_{x,\mn}(1+\rho_\mn)\nn,
\ee
\no where $\rho_\mn$ is the plaquette \pref{plaq}, but build from the gauge-invariant link variables $V_\mu$ instead of $U_\mu$. The products create from the elementary plaquettes larger Wilson loops, but build from the gauge-invariant plaquettes, just like the usual strong-coupling expansion \cite{Montvay:1994cy}. In addition, any gauge-invariant operator can necessarily also be written as some expression involving only the plaquettes \cite{Montvay:1994cy}. This yields \cite{Osterwalder:1977pc}
\be
\la\op\ra=\frac{\int\Pi_{x,\mu}dV_\mu(x)e^{-\kappa H(V_\mu)}\op \Pi_{x,\mn}(1+\rho_\mn)}{\int\Pi_{x,\mn}dV_\mu(x) e^{-\kappa H(V_\mu)}\Pi_{x,\mn}(1+\rho_\mn)}\nn
\ee
\no Up to the term involving $H$, the whole expression can therefore be considered as a sum of expressions involving only some arrangements ${\cal Q}$ of plaquettes, and afterwards summing over all such sets ${\cal Q}$. The crucial next step is that there is an upper bound \cite{Osterwalder:1977pc}
\be
\int\Pi_{{x,\mn}\in{\cal Q}}dV_\mu(x)e^{-\kappa H(V_\mu)}\op \Pi_{x,\mn}\rho_\mn< c_1c_2^n\nn,
\ee
\no where the $c_i$ are some finite constants and $n$ is the number of plaquettes in the set ${\cal Q}$. On a finite lattice, both $n$ and ${\cal Q}$ are always finite. In particular, the number of sets ${\cal Q}$ as well as the number of sets without this set grows less than exponentially fast. That follows from the geometry of the lattice.

The combination of these statements allows to bound the partition function by a geometric series, provided $\kappa$ is sufficiently large \cite{Osterwalder:1977pc,Fradkin:1978dv}. This implies uniform convergence of the series expansion of the partition function to the infinite-volume, and thus thermodynamic, limit. Therefore the free energy is an analytic function of both $\beta$ and $\kappa$ in the infinite-volume limit in the whole quantum phase diagram. However, if it is analytic, there is no phase transition separating the phase diagram in two (or more) parts. It is therefore possible to reach any point from any other point without crossing a phase boundary, and thus there is no qualitative distinction.

This does not imply that there is no phase transition at all in the phase diagram. Only that it can be circumvented, and thus it is either an isolated singularity or a cut with and endpoint, like the liquid-gas transition in nuclear matter or water. Therefore, there can exist no observable local\footnote{See \cite{Greensite:2017ajx,Greensite:2018mhh} for considerations on non-local order parameters.} order parameter to distinguish BEH effect-type physics and QCD-like physics. Both are physically qualitatively indistinguishable, and any difference can thus only be gauge-dependent and/or quantitative. This picture is confirmed in lattice investigations, as will be discussed in section \ref{ss:pdlat}. As stated in \cite{Seiler:2015rwa}, if driven to the extreme the qualitative distinction between QCD-like physics and BEH physics could be regarded as a pure gauge artifact.

While these insights concur with the picture developed in earlier sections, there are a number of caveats\footnote{There are very high bars to overcome for finding a physical distinction, or even definition, between confinement and a non-confining (BEH) physics in general \cite{Seiler:2015rwa}. But especially non-local and non-bulk observables still remain to be understood, see e.\ g.\ a recent proposal for a conjectured criterion between the Wilson criterion and the statement that confinement is essentially equivalent to gauge invariance \cite{Greensite:2017ajx,Greensite:2018mhh}.}. The phase with broken center symmetry manifests also features of the BEH effect \cite{Lang:1981qg,Drouffe:1984hb,Baier:1986ni,Capri:2012cr} to be made.

The {\bf first} is that the proof only works for theories in which the gauge group is fully broken by the BEH effect. In theories where this is not possible, the phase diagram could separate into distinct phases. An example is the case of SU(2) with an adjoint Higgs, where the phase diagrams decomposes into, at least, two phases \cite{Baier:1986ni}. These phases are distinguished  with respect to the global center symmetry, i.\ e.\ whether the Z$_2$ symmetry carried by the group-valued links can break spontaneously. The phases show similar behavior like the low-temperature phase and high-temperature phase of pure SU($N$) Yang-Mills theory, and can therefore be associated with different realizations of confinement in the Wilson sense. However, there is still no gauge-invariant Higgs vacuum expectation value, and the gauge-invariant physics is quite subtle \cite{Maas:2017xzh,Kondo:2016ywd}. Also, as soon as there are more complicated Higgs sectors, and thus more parameters, the argument also breaks down, even if the gauge group is fully broken by the BEH effect. The theory may then become quite involved with a rich phase structure \cite{Branco:2011iw,Ivanov:2017dad,Lewis:2010ps,Maas:2016qpu,Maas:2014nya}. This issue will be returned to in section \ref{ss:bsm}.

The {\bf second} is that in the construction the length of the Higgs field was frozen, by $\gamma\to\infty$. No similar proof is available at finite $\gamma$. It is therefore possible, at least in principle, that there exists a surface $\gamma(\beta,\kappa)$, which cuts through the whole three-dimensional phase diagram of the theory, separating two phases. This does not spoil the general statement that there is no difference between BEH-like physics and QCD-like physics, as both are present at $\gamma\to\infty$. But it allows for a more complicated structure, and additional phases\footnote{Note that the arguments that the gauge symmetry must remain unbroken still apply \cite{Elitzur:1975im,Haag:1992hx}, and thus only gauge-invariant quantities will have non-zero expectation values without gauge fixing.}.

A {\bf third} important issue is the following: A special role is played by the boundaries of the phase diagram. At $\beta=0$ the gauge degrees of freedom decouple, yielding the ungauged $\phi^4$ theory. At $\kappa=0$ the Higgs degrees of freedom decouple, yielding Yang-Mills theory. Only $\gamma=0$ does not decouple the two sectors. On a finite lattice, the limit to the first two cases are continuous\footnote{This has been explicitly seen, e.\ g.\ for the finite temperature phase transition, which can be continuously deformed on a finite lattice from the $\phi^4$ case to the BEH theory to pure Yang-Mills theory \cite{Wellegehausen:2011sc}.}. In a continuum theory, this cannot be expected, due to Haag's theorem \cite{Haag:1992hx}: The corresponding Hilbert spaces are not unitarily equivalent. Therefore, there can be non-analyticities on the boundaries in the thermodynamics limit. To avoid this unnecessary complication, in the present context only the interior of the phase diagram, including the limits of any of the coupling constants to zero, will be considered.

The {\bf fourth} is again that this construction was performed using the lattice as a regulator, and the usual caveats concerning the continuum limit apply.

\subsection{The phase diagram from the lattice}\label{ss:pdlat}

While section \ref{ss:pd} provided the qualitative structure of the phase diagram for some theories, it did neither for all theories, nor provided it quantitative insights. Consequently, there is a long history of investigations of the phase diagrams, especially with, but not limited to, lattice methods \cite{Lang:1981qg,Kikugawa:1982yh,Brower:1982yn,Kuhnelt:1983mw,Gerdt:1984ft,Evertz:1985fc,Langguth:1985eu,Langguth:1985dr,Gerdt:1985rb,Gerdt:1985xh,Olynyk:1985tr,Munehisa:1986jc,Evertz:1986vp,Hasenfratz:1986za,Bock:1989fk,Heller:1993yv,Campos:1997dc,Knechtli:1998gf,Knechtli:1999qe,Caudy:2007sf,Greensite:2008ss,Bonati:2009pf,Bonati:2009yi,Wurtz:2009gf,Lewis:2010ps,Maas:2014pba,Gongyo:2014jfa,Gies:2015lia,Biswal:2015rul,Maas:2016ngo,Gies:2016kkk,Bazavov:2019qih,Baier:1988sc}. These results will be reviewed in the following. Almost all of these investigations have been performed in four dimensions for the standard model case \cite{Kuhnelt:1983mw,Gerdt:1984ft,Evertz:1985fc,Langguth:1985eu,Langguth:1985dr,Gerdt:1985rb,Munehisa:1986jc,Evertz:1986vp,Hasenfratz:1986za,Bock:1989fk,Heller:1993yv,Campos:1997dc,Knechtli:1998gf,Knechtli:1999qe,Caudy:2007sf,Greensite:2008ss,Bonati:2009pf,Bonati:2009yi,Maas:2014pba,Gies:2015lia,Biswal:2015rul,Gies:2016kkk}, and only very few addressed other theories \cite{Kikugawa:1982yh,Brower:1982yn,Gupta:1983zv,Gerdt:1985xh,Olynyk:1985tr,Lee:1985yi,Shrock:1986fg,Shrock:1986bu,Gongyo:2014jfa,Maas:2016ngo,Lewis:2010ps,Wurtz:2009gf,Bazavov:2019qih,Baier:1988sc}.

The first question is how to define a phase modification, if there is no qualitative distinction. This problem can be addressed by using various quantities. Preferable are thermodynamic observables, most of all the free energy but also susceptibilities or generalized heat capacities are valuable. These can be unambiguously defined. But also less well-defined, probably even scheme-dependent, quantities are useful. In particular, the Polyakov loop has been employed. The latter is actually not a good observable, as it will not qualitatively differ anywhere in the phase diagram because of the explicitly broken center symmetry. However, the experience with QCD shows, that it is often reflecting phase modifications still quite well \cite{Gattringer:2010zz,Cossu:2007dk}.

The first results had been indeed quite similar \cite{Kuhnelt:1983mw,Gerdt:1984ft,Evertz:1985fc,Langguth:1985eu,Langguth:1985dr,Gerdt:1985rb,Evertz:1986vp} to the expectations of section \ref{ss:pd}\footnote{A similar picture is also found for the Abelian case \cite{Jansen:1985nh}.}. They showed a simply connected phase diagram. However, it was found that there exists still a phase transition of first order, starting at the $\beta=\infty$ axis, and extending into the phase diagram up to a critical end point, but extending through the whole phase diagram as a sheet of transitions. I.\ e.\ for every value of $\gamma$ existed a line $\beta(\kappa)$, starting from a critical point $\beta_c$ to $\beta=\infty$.

Later investigations at $\gamma=\infty$ showed that finite-volume effects were underestimated, but did not change the picture qualitatively \cite{Bonati:2009pf,Bonati:2009yi}. In these investigations a critical end-point was located at $\beta_c\approx2.727$ and $\kappa_c\approx0.7089$, indicating a continuum limit. The nature of this continuum limit, whether trivial or not, has not yet been studied in appreciable detail.

\begin{figure}[!htbp]
\begin{minipage}{0.7\linewidth}
 \includegraphics[width=\linewidth]{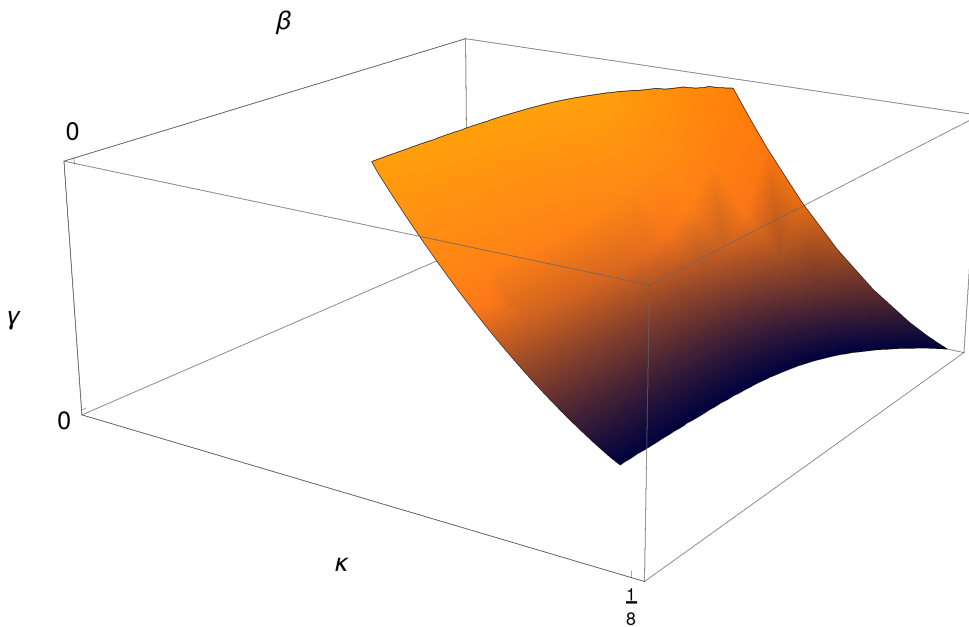}
\end{minipage}
\begin{minipage}{0.3\linewidth}
 \caption{\label{fig:pd}A sketch of the (conjectured) Higgs sector phase diagram. The sheet is a surface of first order transitions, which may terminate in the bulk by a line of second-order phase transitions for all values of $\gamma$. Based on the results of subsection \ref{sss:ground}, the separation surface shown in the top panel of figure \ref{fig:gdpdl} below should be the actual sheet.}
\end{minipage}
\end{figure}

While the value of this endpoint $\beta_c$ seems to be quite similar at all values of $\gamma$, the value for $\kappa$ seems to decrease with decreasing $\gamma$. However, this sheet of phase transitions has not yet been completely charted. A sketch of the phase diagram is shown in figure \ref{fig:pd}. Note that there are also arguments pointing to a possible second critical end-point, and thus second-order phase transition, at $\beta\to\infty$, $\kappa\to 0$, and $\gamma\to 0$ from functional methods \cite{Gies:2015lia,Gies:2016kkk}. There is not yet any systematic investigation of this using lattice simulation. This point will be taken up again in section \ref{ss:triv}. 

This phase diagram gives no direct insight into the distinction of regions of the phase diagram with a BEH-like physics in suitable gauges and regions where QCD-like physics is encountered in a fixed gauge, because this question depends on the choice of gauge \cite{Caudy:2007sf}. The origin of this comes from the fact that it is possible to define gauges with distinctively different residual global gauge symmetries. An example is given by the case of the non-aligned Landau gauge and the non-aligned Coulomb gauge. As noted in section \ref{ss:qulev}, there is never a Higgs vacuum expectation value in non-aligned gauges. But, it is still possible to define a criterion for metastability, as discussed in section \ref{s:hidssb}. In non-aligned Landau gauge this is e.\ g.\ \cite{Caudy:2007sf,Langfeld:2002ic}
\be
Q_L=\frac{1}{2}\la\left|\sum_x\phi(x)\right|^2\ra\label{olandau}
\ee
\no essentially the lattice version of \pref{vevd}. On a finite lattice, this quantity is always finite. If the theory in non-aligned Landau gauge is not metastable, this quantity vanishes as the volume goes to infinity, and in fact as $1/V$ \cite{Caudy:2007sf,Maas:2012ct}. In Coulomb gauge, it is possible to use \cite{Caudy:2007sf,Greensite:2003xf}
\be
Q_C=\frac{1}{2N_tV_{d-1}^2}\sum_t\la\tr\left(\sum_{\vec x,\vec y}U_0(t,\vec x)^\dagger U_0(t,\vec y)\right)\ra\label{ocoulomb}
\ee
\no where $U_0$ are the temporal link variables, which vanishes as $1/V_{d-1}$ in the infinite volume limit if there is no metastability. The important difference of both \pref{olandau} and \pref{ocoulomb} to $\la \phi\ra$ is that the are inherently non-local.

\begin{figure}[!htbp]
\includegraphics[width=\linewidth]{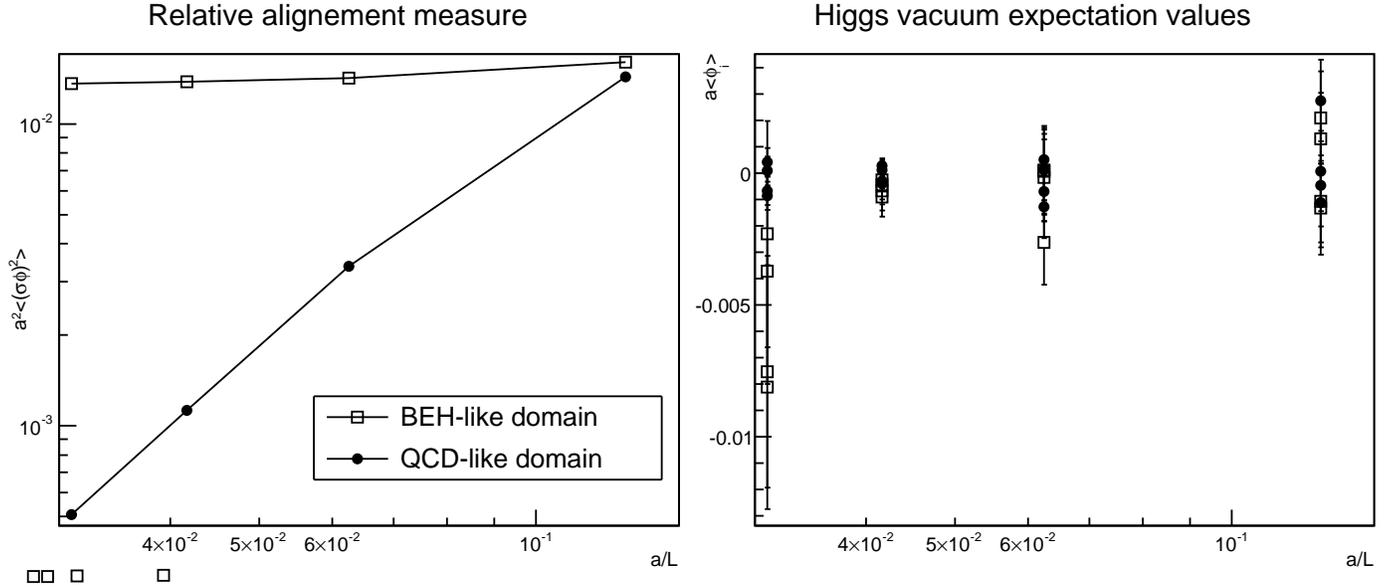}\\
\begin{minipage}{0.6\linewidth}
\includegraphics[width=\linewidth]{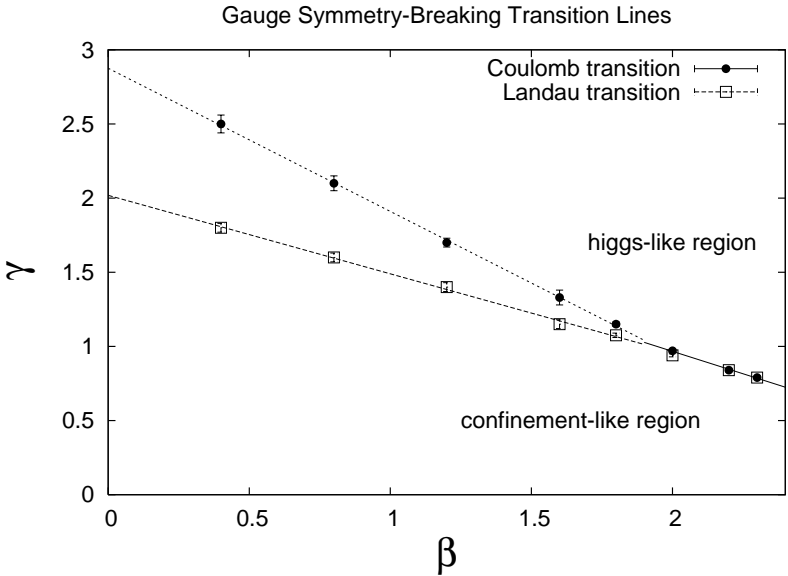}
\end{minipage}
\begin{minipage}{0.4\linewidth}
\caption{\label{fig:cg}The top left panel shows \pref{olandau} at a point in the QCD-like domain ($\beta=2.4728$, $\kappa=0.2939$, $\gamma=1.088$, closed symbols) and the BEH-like domain ($\beta=2.7984$, $\kappa=0.2954$, $\gamma=1.370$, open symbols) \cite{Maas:unpublished}. The right-hand side shows the expectation value of the Higgs components in a non-aligned Landau gauge \cite{Maas:2012ct,Maas:unpublished} at the same points. Note how indistinguishable the behavior in both cases is. Shown are individually the four real components. Errors are statistical errors only, and are smaller than the symbol size on the left-hand side. The lower panel shows the phase diagram seen from different gauges using \pref{olandau} and \pref{ocoulomb}, from \cite{Caudy:2007sf} (Reprinted with permission. Copyright (2007) by the American Physical Society.). Note that in this plot $\gamma=\kappa$ and the investigation is performed in the limit $\phi^\dagger\phi=1$, i.\ e.\ proper $\gamma\to\infty$.}
\end{minipage}
\end{figure}

This behavior is illustrated for the Landau gauge in figure \ref{fig:cg}. It is also shown that in a non-aligned Landau gauge the Higgs vacuum expectation value always vanishes. In the bottom panel of figure \ref{fig:cg} the result of sketching the phase diagram at $\gamma=\infty$ in nonaligned Landau gauge and Coulomb gauge, using \pref{olandau} and \pref{ocoulomb}, respectively, is shown \cite{Caudy:2007sf}. It is seen that at small $\beta$, substantially below the possible critical end-point, a phase transition is signaled in $Q_L$ and $Q_C$ but the location differs. Only if there is actually a physical phase transition at larger $\beta$ \cite{Bonati:2009pf,Bonati:2009yi}, the transition in both gauges start to coincide. This is not unexpected, as here all quantities, gauge-dependent and gauge-invariant, have non-analyticities. On the other hand, this implies that $Q_C$ and $Q_L$ show gauge-dependent pseudo-phase transitions even if there is no physical phase transition. Thus, fixed-gauge results on phase transitions do not necessarily coincide with a physical phase transition and/or with transitions in other gauges. Thus, phase transitions need to be seen always in gauge-invariant observables to justify the label of a phase transition.

This also implies that the standard methods in perturbation theory to detect a phase transition, i.\ e.\ using the Higgs vacuum expectation value \cite{Kapusta:2006pm}, cannot be expected to identify the phase transition. That this may still work in some cases and some gauges is due to a subtle effect to be discussed in section \ref{s:spectrum}. From figure \ref{fig:cg} it is, however, already clear that this can only work if there is an actual phase transition.

Finally, this emphasizes the results of section \ref{ss:pd} that there is no qualitative distinction between the two regions in the phase diagram. Therefore, these will not be denoted as phases, but rather as QCD-like domain and BEH-like domain. As is seen, the distinction cannot be sharp, and depends on the gauge at hand.

\begin{figure}[!htbp]
\begin{minipage}{0.7\linewidth}
\includegraphics[width=\linewidth]{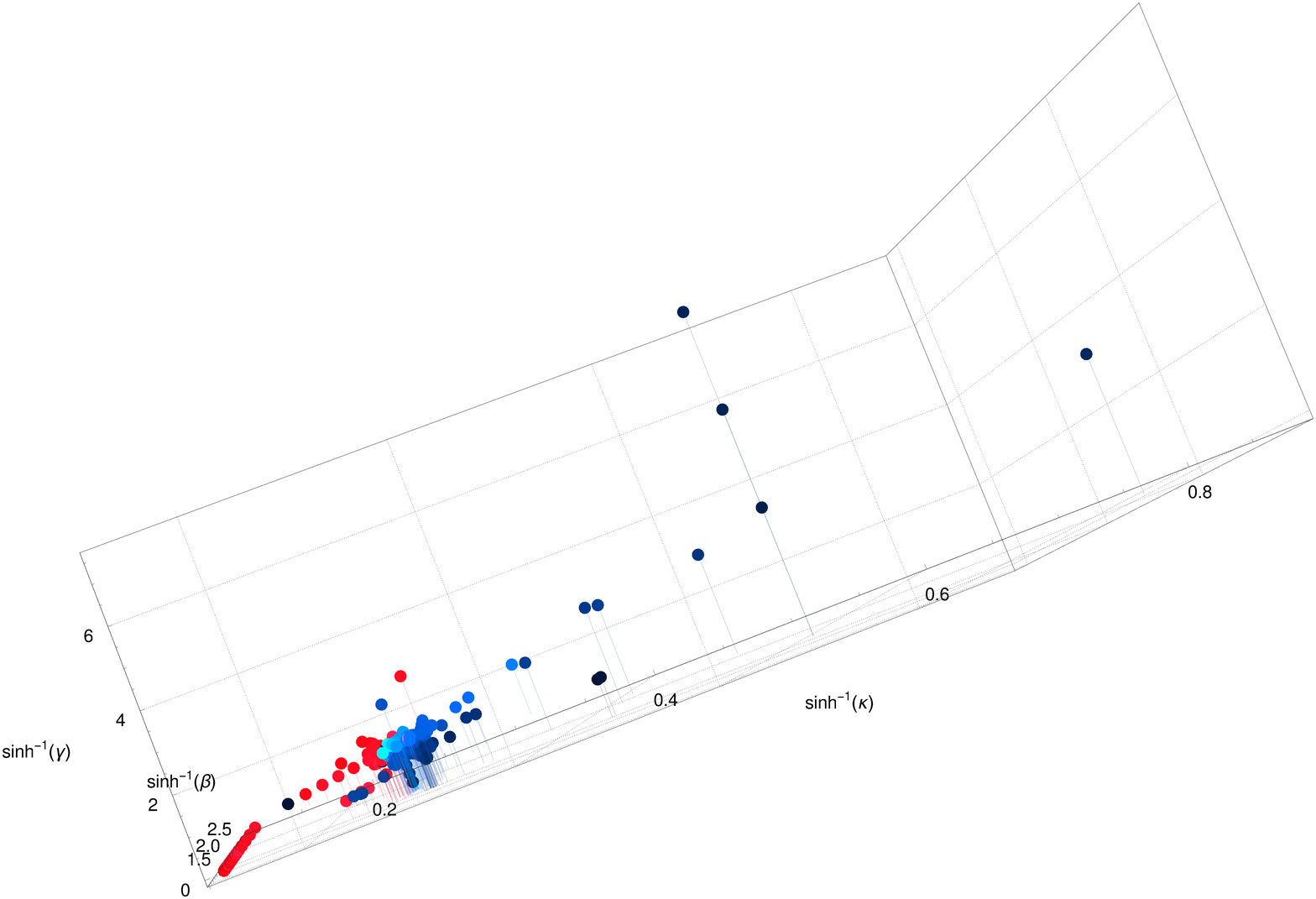}\\
\includegraphics[width=\linewidth]{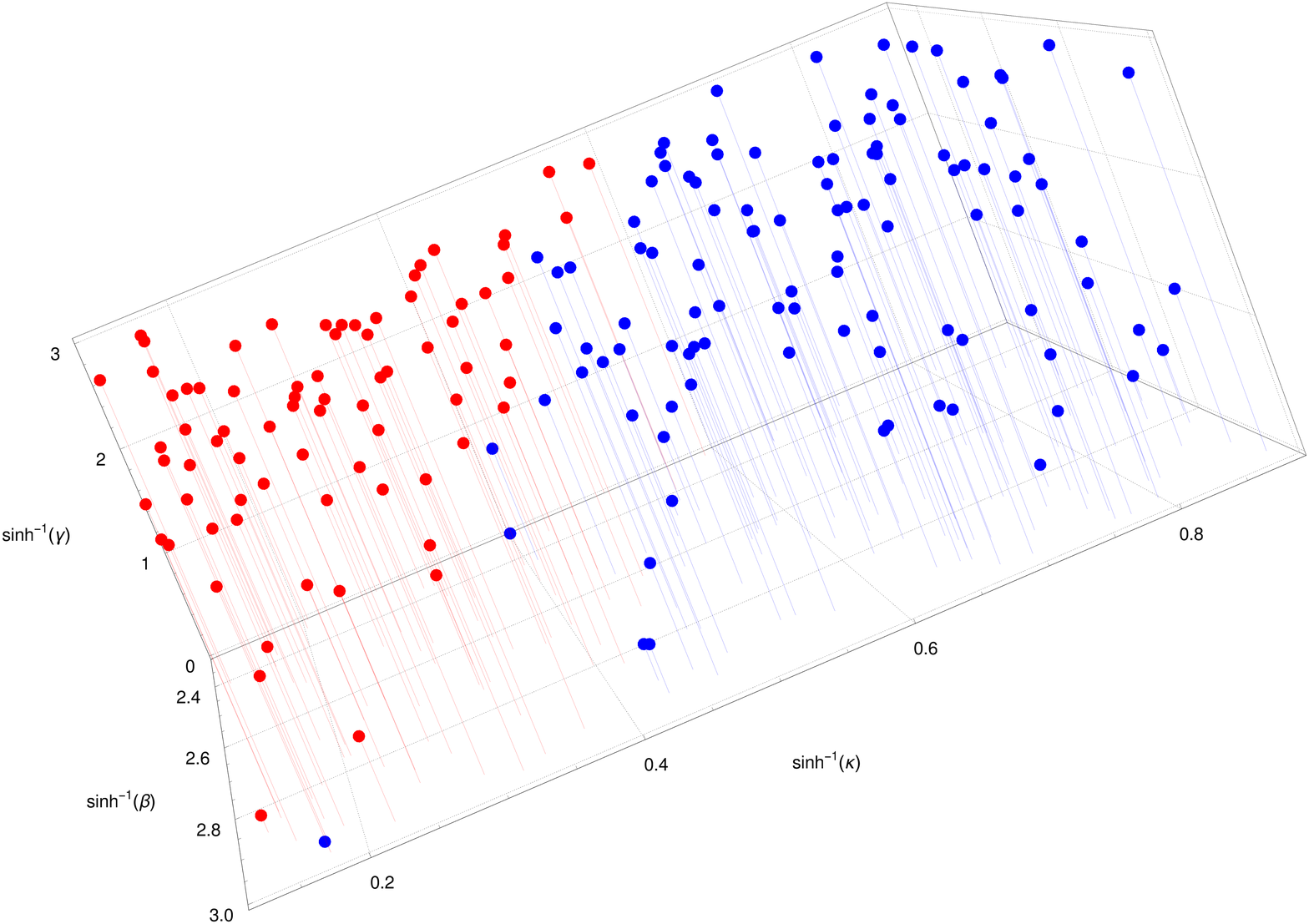}
\end{minipage}
\begin{minipage}{0.3\linewidth}
\caption{\label{fig:gdpdl}The gauge-fixed phase diagram according to \pref{olandau} in non-aligned Landau gauge for SU(2) (top panel) and SU(3) (bottom panel). Red points are in the BEH-like domain and blue points in the QCD-like domain \cite{Maas:2014pba,Maas:unpublished}.}
\end{minipage}
\end{figure}

\begin{figure}[!htbp]
\vspace{-0.5cm}
\begin{minipage}{0.7\linewidth}
\includegraphics[width=\linewidth]{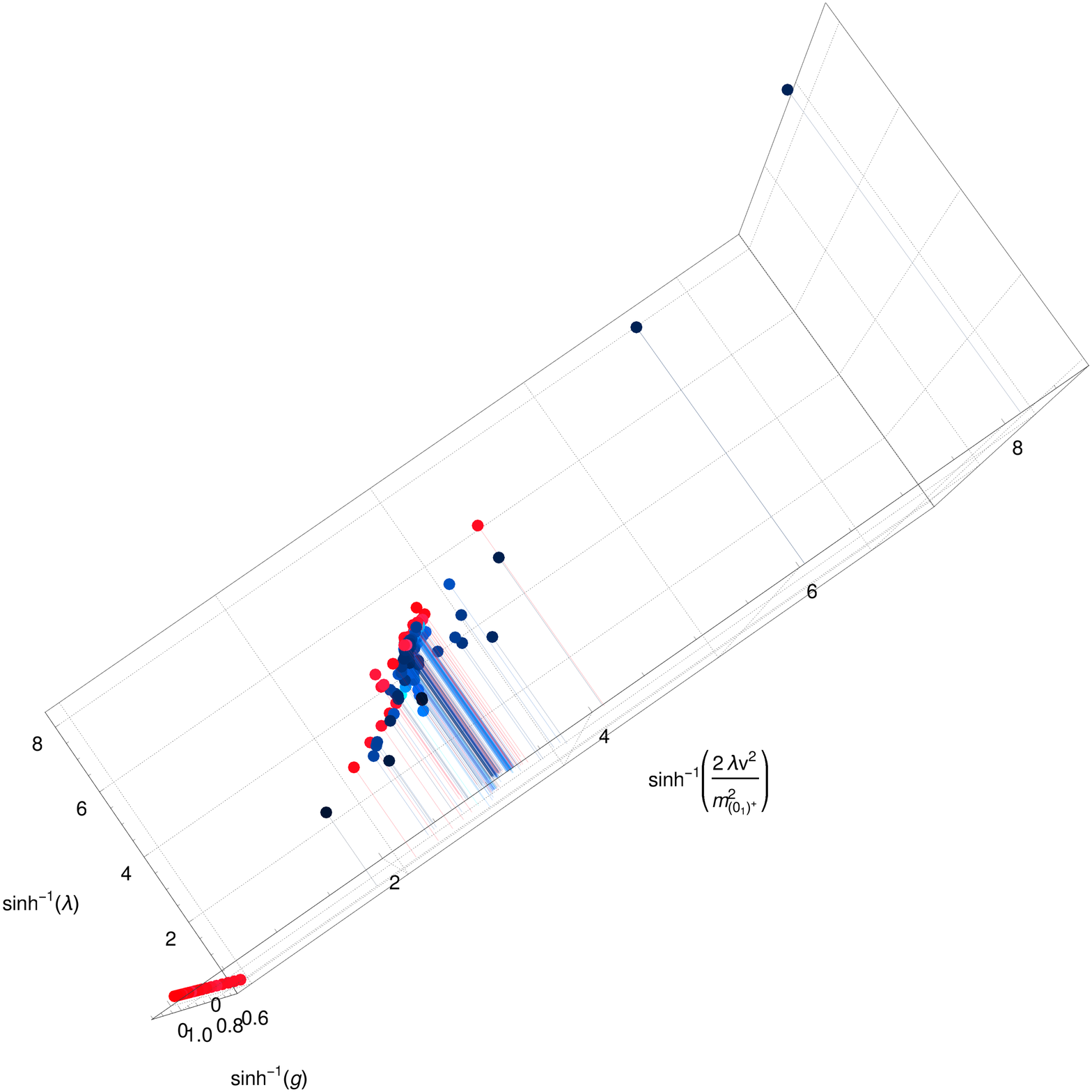}\\
\includegraphics[width=\linewidth]{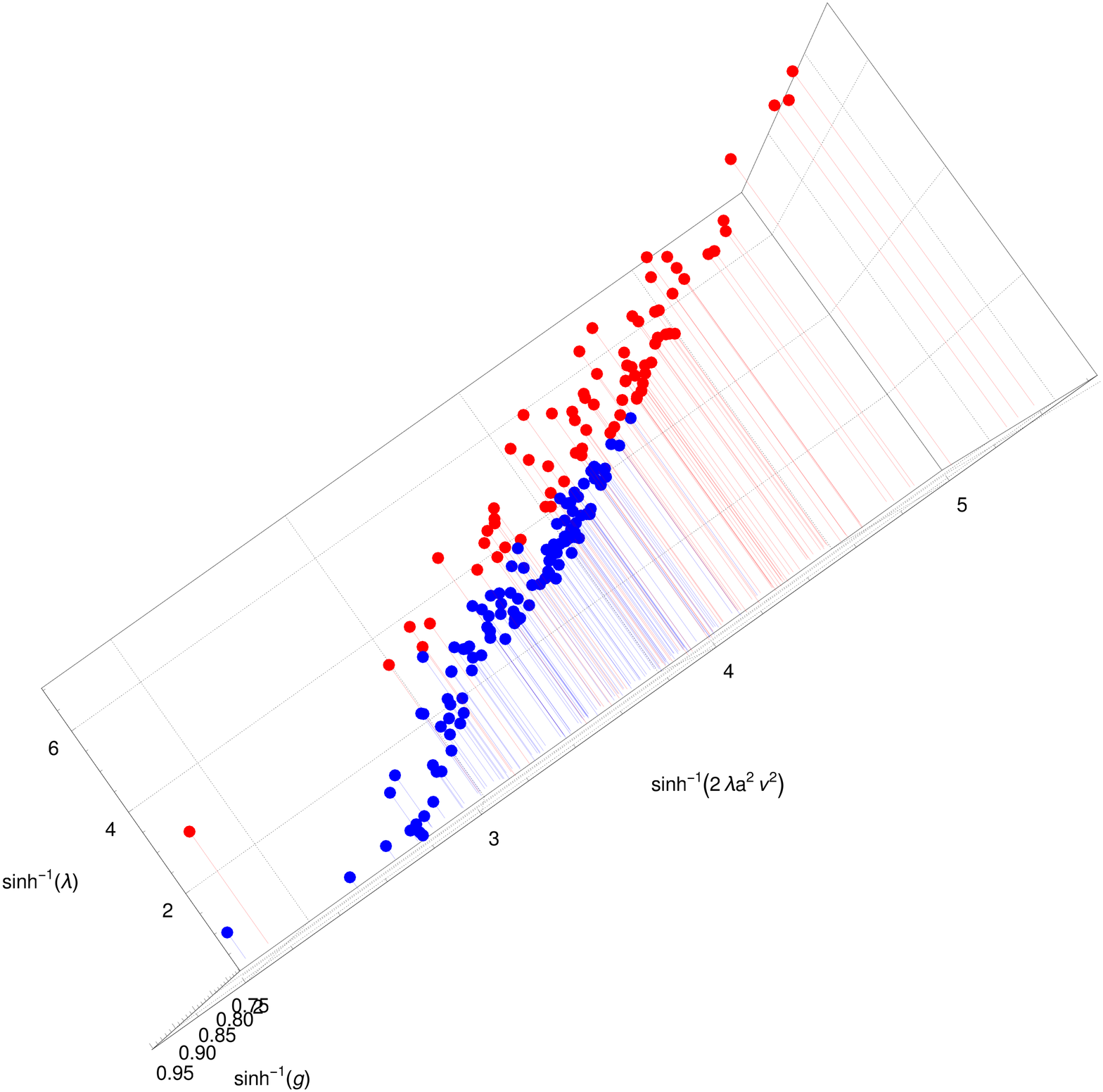}
\end{minipage}
\begin{minipage}{0.3\linewidth}
\caption{\label{fig:gdpdc}The same as figure \ref{fig:gdpdl}, but after translating the parameters to the continuum according to \prefr{beta}{lambda}. For SU(2), the dimensionful parameter is normalized  to the mass of the lightest excitation in the $0^+$ singlet channel. For SU(3), it is given in lattice units, as not in all cases spectroscopically data is available. For details on the spectroscopy see section \ref{s:spectrum}.}
\end{minipage}
\end{figure}

For theories with a single fundamental Higgs field for the gauge groups SU(2) \cite{Maas:2014pba} and SU(3) \cite{Maas:2016ngo} the phase diagram has been mapped using the order parameter \pref{olandau} on the lattice. The result is shown in figure \ref{fig:gdpdl} using the lattice parameters and in figure \ref{fig:gdpdc} using the continuum parameters. For the range of parameters plotted there is a clear distinction between both regions. Since there is not yet an equally good identification of physical transitions in these theories, or in other gauges, it is hard to predict where a physical separation is. However, this is actually far less of a problem for phenomenology as it may appear at first, as will be discussed in section \ref{s:spectrum}. Also, the results only cover a portion of the phase diagram, and a full map is, naturally, not available from lattice simulations. Interestingly, the situation in the snapshots for the phase diagram differs not that much for SU(2) and SU(3). There is a sheet-like separation structure, where the QCD-like domain starts for all $\gamma$ and $\beta$ at $\kappa=1/8$, i.\ e.\ the tree-level massless case. Then, at larger $\kappa$, the Higgs-like domain is present if $\gamma$ is smaller than a critical value, where the critical value depends on $\kappa$. Qualitatively, it looks as in the sketch depicted in figure \ref{fig:pd}. Thus, it is really the strength of the four-Higgs coupling, rather than the gauge-coupling or 'deepness' of the potential at tree-level, which determines the domain the theory is in. But, eventually, even in the limit $\gamma\to\infty$, there is a critical depth of the potential after which the theory is for any $\kappa$ in the BEH-like domain.

\begin{figure}[!htbp]
\includegraphics[width=\linewidth]{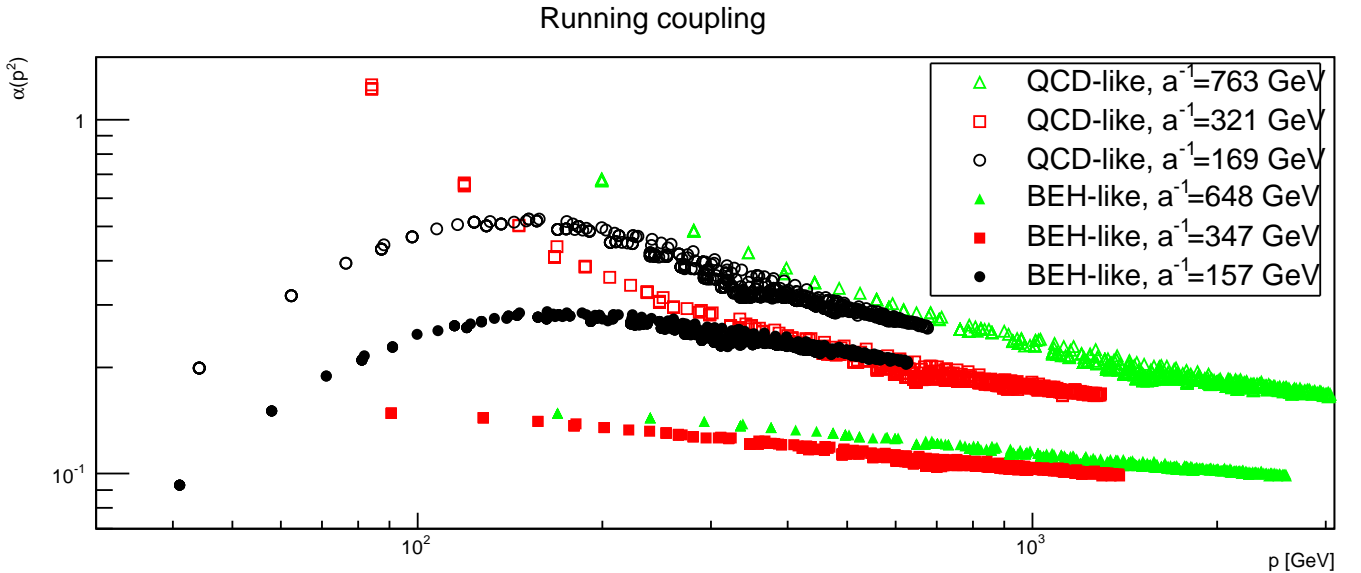}
\caption{\label{fig:alpha}The running gauge coupling in the miniMOM scheme \cite{vonSmekal:2009ae} for different physical situations and different lattice cutoffs \cite{Maas:2013aia,Maas:unpublished}. Note that the details of the physics in the QCD-like systems and BEH-like systems are not identical, as they do show, e.\ g., different masses in the spectrum \cite{Maas:2013aia}, explaining their different behaviors.}
\end{figure}

Note that while the results are from a gauge-dependent order parameter, this parameter shows to the best of our knowledge also always a transition if there is a true phase transition. This turning into a genuine phase transition will happen for a sufficiently large $\beta$, about $2.73$, corresponding to a sufficiently weak tree-level gauge coupling of $\alpha=g^2/(4\pi)\approx0.117$ for SU(2) \cite{Bonati:2009pf,Bonati:2009yi}, though the exact value depends also on the other parameters. However, it is important to note that this is a bare (lattice) coupling. The actual running gauge coupling in the miniMOM scheme \cite{vonSmekal:2009ae} is shown, for some sample systems, in figure \ref{fig:alpha}. It is noteworthy that, while it is generically larger for the QCD-like cases at low momenta, already at intermediate momenta at about a few hundred GeV there is not necessarily a large difference. In fact, the difference between different BEH-like or QCD-like points in the phase diagram can be larger than between QCD-like points and Higgs-like points. Nonetheless, it is found that the running coupling at sufficiently large momenta is in all cases entirely determined by one-loop (resummed) perturbation theory \cite{Maas:2013aia}. It is therefore important to keep in mind that the bare quantities of figures \ref{fig:gdpdl} and \ref{fig:gdpdc} do not allow to conclude the physical properties of a point.

\begin{figure}[!htbp]
\begin{minipage}{0.7\linewidth}
\includegraphics[width=\linewidth]{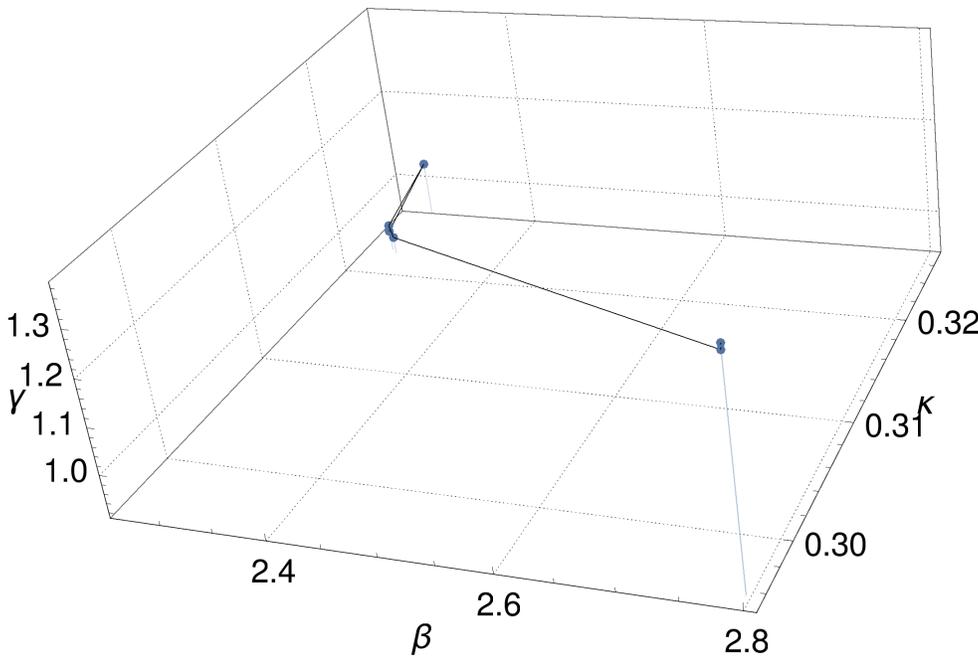}
\end{minipage}
\begin{minipage}{0.3\linewidth}
\caption{\label{fig:lcpex}A sample line of constant physics, fixed by the value of the lightest state in the scalar singlet and vector triplet channel (see section \ref{s:spectrum} for details) and the value of the running coupling in the miniMOM scheme at 200 GeV. The lattice spacing decreases from (131 GeV)$^{-1}$ to (440 GeV)$^{-1}$ towards larger $\beta$ values.}
\end{minipage}
\end{figure}

It should also be noted that, in principle, it is necessary to construct lines of constant physics \cite{Montvay:1994cy,Gattringer:2010zz,DeGrand:2006zz} to follow a fixed set of physical parameters to the continuum limit, or at least to a large cutoff. Unfortunately, the physical properties of the theory turn out to be very sensitive to the values of the bare parameters, especially to $\kappa$. This is a sign of the fine-tuning problem to be discussed in section \ref{ss:hierarchy}. It has therefore not yet been possible to follow any line of constant physics for any large number of points. An example of a short trajectory is shown in figure \ref{fig:lcpex}. It is noteworthy that in most cases identified short trajectories seem to proceed towards larger $\beta$, and $\kappa$ becoming closer to 1/8 for decreasing $a$. Such a tendency is actually predicted in studies using the functional renormalization group \cite{Gies:2015lia,Gies:2016kkk}, which also require $\gamma\to 0$. In fact, they predict an interacting, asymptotically free continuum theory in this limit. Whether this is indeed the case remains to be seen.

Another important lesson to be learned from the results comes from studying the parameter range where QCD-like physics is observed. The plots in figures \ref{fig:cg}, \ref{fig:gdpdl}, and \ref{fig:gdpdc} all cover only a region of the phase diagram where at the cutoff the classical potential has either a vanishing quadratic term, equivalent to $\kappa=1/8$, or where it has the characteristic wine-bottle-bottom shape. Thus, in a large region of the phase diagram quantum fluctuations drive the system out of the BEH-like regime and into a QCD-like regime, in contrast to the expectations from perturbation theory \cite{Bohm:2001yx,Kapusta:2006pm}. In particular, this occurs for essentially any value of the gauge coupling and the four-Higgs coupling. Thus this happens even at weak coupling, where naively perturbation theory is expected to work. Thus, quantum fluctuations are not negligible, and the decision where BEH-like physics prevails in a given gauge cannot always be answered by a tree-level calculation. Even if quantum corrections are taken into account perturbatively \cite{Coleman:1973jx,Kapusta:2006pm} this may still not be sufficient. In such cases non-perturbative methods are necessary. There is not yet any unambiguous prediction under which conditions the quantum fluctuations will alter the physics, and care has to be taken. But this effect was already observed in the earliest lattice calculations \cite{Evertz:1985fc,Langguth:1985eu,Langguth:1985dr}.

If more Higgs fields in the fundamental representation are added, the phase diagram becomes more complex. Especially, different physical phases arises, often differing in the realization of the custodial symmetry \cite{Branco:2011iw,Ivanov:2017dad}. This has been confirmed in lattice simulations \cite{Lewis:2010ps,Wurtz:2009gf}. However, these various phases may, or may not be connected to the presence of a BEH effect in any given gauge, a question essentially unexplored beyond perturbation theory.

Also when more gauge fields of different gauge groups are added, the structure of the phase diagram becomes quickly very involved. Except for some simulations including a U(1) sector \cite{Shrock:1985un,Zubkov:2010np,Zubkov:2011sk,Zubkov:2011ia} little exists so far in terms of lattice simulations. At infinite Higgs self-coupling they indicate a separation into two phases with one having the structure expected in the standard model \cite{Shrock:1985ur}.

The situation becomes even more complex when adding fermions. There are some investigations using functional methods of such theories, but focusing on a field content which makes the effective interaction strength weak \cite{Litim:2014uca,Litim:2015iea}. There are, of course, many results using perturbation theory \cite{Kapusta:2006pm,Branco:2011iw,Ivanov:2017dad}. But they are, for the reasons discussed above, not suitable as they are in fixed gauges and/or cannot guarantee the absence of a shift of the transition non-perturbatively. There are also many results for supersymmetric theories \cite{Weinberg:2000cr}, where such questions can be (partly) solved analytically. This is beyond the scope of this review. However, it is important to note that the arguments about a lack of gauge invariance also pertains to supersymmetric theories, if only vacuum expectation values are used.

\subsection{The phase diagram for multiple breaking patterns}\label{ss:strata}

If the Higgs is not in the fundamental representation then an additional problem arises: It is possible that even a potential renormalizable by power counting allows for multiple, different minima. This begs the question whether these multiple minima can be associated with different physical phases.

The, arguably simplest, case is a Higgs field in the adjoint representation with gauge group SU($N\ge 3)$. Take the potential to explicitly conserve the Z$_2$ symmetry $\phi\to-\phi$. Given the most general superficially renormalizable potential, a tree-level analysis shows that different Higgs vacuum expectation values are possible, each one leaving a different S(U($P$)$\times$U($N-P$)) subgroup of the gauge group invariant\footnote{The ranks of the little groups, and thus the size of the Cartans, play an important role in determining the number of different little groups \cite{O'Raifeartaigh:1986vq}.} \cite{Li:1973mq,Ruegg:1980gf,Murphy:1983rf,Maas:2017xzh}. Thus, there are $N/2$, rounded up, breaking patterns.

Naively applying the arguments of sections \ref{s:global} and \ref{ss:vev} yields that a particular of these breaking patterns, either by gauge-fixing or other means, would be selected. But it is more subtle than that. This subtlety arises in the following way. In the fundamental case, every gauge orbit is either only zero field, or belongs to an orbit with symmetry group SU($N-1$), as always a gauge transformation exists which rotates a given scalar field locally into a vector with only a single non-zero component. Thus, there are two strata, and the corresponding symmetry groups, SU($N$) for zero scalar field and SU($N-1$) otherwise, are the little groups \cite{O'Raifeartaigh:1986vq}. In this case there is only a single special orbit, the vacuum, and all others behave in the same way.

In the adjoint case, there are not only two strata and little groups, but more. E.\ g.\ for the SU(3) case with adjoint Higgs there are three: SU(3), SU(2)$\times$U(1), and U(1)$\times$U(1). Any value of the Higgs field can have only one of these little groups as invariance groups, and the set of all such orbits is again the corresponding stratum of the little group. Thus, there is no gauge transformation moving a value of the Higgs field from one stratum to another, and belonging to a stratum is a gauge-independent statement. Thus, corresponding gauge-invariant quantities to state this fact exist. These are merely the invariant polynomials of the group and the representation \cite{O'Raifeartaigh:1986vq}. E.\ g., 
\bea
&&\tr\left(\Sigma^{3}\right)\label{eq:sigma3}\\
&&\Sigma=T^a\phi^a\nn
\eea
\no yields to which stratum a field locally belongs. In fact, the values are discontinuously different if the field length is normalized.

There is now a twist to this group-theoretical observation in a field theory \cite{Maas:2017xzh}. The distinction is local. The scalar field is a field, and its value changes from point to point. Especially, a scalar field can belong to any stratum at different space-time points. Thus, the distinction is not meaningful in a global way. Still, because a gauge transformation acts on the Higgs field locally, this feature is again locally gauge-invariant. Thus, the function \eqref{eq:sigma3} locally characterizes the stratum of the scalar field.

Likewise, it is possible to characterize the space-time average of any scalar field configuration by the stratum to which it belongs. This will be independent under global gauge transformations. The question is now how this affects the phase diagram. Is this sufficient to distinguish phases?

First of all, this does not affect perturbation theory. Because perturbation theory is a small field expansion, perturbation theory will stay inside a given stratum, by definition, as the characterization in terms of invariants is discontinuous \cite{O'Raifeartaigh:1986vq}.

Secondly, any fixed field configuration will therefore belong to a given stratum. But the full average will again vanish, for the following reason. Inside a stratum, the average over the little group will erase any information which is not invariant with respect to the little group. In addition, it will be averaged over all possible orientations of the stratum inside the full group, yielding again zero for any naive vacuum expectation value. 

Can then something like \pref{eq:sigma3} be used? This is, unfortunately, not the case. Though \eqref{eq:sigma3} is gauge-invariant, its actual value is determined by weighting the value for every gauge orbit by the exponentiated action, and averaging over the orbits of the different strata. The path integral therefore decomposes into a sum of distinct parts. Each term contains the configurations for which the space-time average can be assigned to any of the strata. Consequently, the value of \pref{eq:sigma3} will be the sum of these terms. Therefore, the expectation value can, in principle, be continuous throughout the phase diagram of the theory\footnote{This is not necessarily so, but this requires either zero prefactors or cancellations. This is not impossible. After all, there is always the QCD-like phase, which technically belongs to the maximum little group, as no direction is preferred. However, the corresponding stratum has measure zero, it is only the vacuum, and it can thus not arise by any other means than cancellation.}. What actually happens, and what the actual value of \pref{eq:sigma3} is, depends therefore on the dynamics of the theory.

This can become a serious problem when attempting to fix a gauge like the 't Hooft gauge \pref{thooftg} beyond perturbation theory. If the vector $v$ in \pref{thooftg} belongs to a given stratum, and the Higgs field at a point $x$ belongs to a different stratum, then the term $\phi T v$ vanishes at this point. Thus, at this point the gauge condition degenerates to the covariant gauge condition.

Even more problematic is that the gauge was chosen such that the gauge condition rotates the space-time average of the Higgs field into the direction of $v$, which is part of a fixed stratum. This is impossible if the average of a gauge orbit belongs to a different stratum. Thus, the gauge condition cannot be fulfilled on this configuration and, in fact, on the whole gauge orbit. Instead, the orientation of the average remains within the stratum and is not affected. Thus, ultimately, an average over the directions inside the stratum is performed, yielding again a zero expectation value for averages on any gauge orbit belonging to this stratum. Hence, on gauge orbits having an average value belonging not to this stratum the gauge condition \eqref{thooftg} degenerates into the covariant gauge condition.

For the calculation of gauge-dependent expectation values this has the following consequence. Given the arguments above, the total vacuum expectation value of the Higgs has still the same direction\footnote{Provided there is no measure-zero problem or cancellations, which will be neglected.}, as all gauge orbits belonging to the corresponding stratum can still be fixed to satisfy the 't Hooft gauge condition. However, fields belonging to a different stratum do not have a small fluctuation around this vacuum expectation value. Thus, for some gauge-dependent operator $\op$ holds, symbolically,
\be
\langle\op\rangle=\int{\cal D}W\left(\int_\text{Selected stratum}{\cal D}\phi \op e^{iS}+\int_\text{Other strata}{\cal D}\phi \op e^{iS}\right)=\langle \op\rangle_s+\langle \op\rangle_o\label{mppi},
\ee
\noindent i.\ e.\ it decomposes into two expectation values, of which one is in the selected gauge, while the other is in the corresponding covariant gauge.

There are now two possibilities
\begin{itemize}
 \item The correlation function $\la\op\ra_o$ is negligible, in a suitable sense, with respect to $\la\op\ra_s$, and therefore the expectation value behaves as if \pref{thooftg} was fulfilled
 \item This is not the case, and the result is therefore a superposition of both results
\end{itemize}
What actually happens may depend on the theory and parameter values in question. This is yet unknown. E.\ g., in the adjoint case, if the global Z$_2$ custodial symmetry is explicitly broken, the absolute minimum always favors the maximal little group \cite{O'Raifeartaigh:1986vq}. Thus, it is entirely possible that if the potential is metastable with respect to the global symmetry the first case may be appropriate, and then only one of the strata contributes, and otherwise the second. This will require a full non-perturbative investigation to understand better. But as of now only very early exploratory investigations have been made using lattice simulations \cite{Gupta:1983zv,Lee:1985yi,Olynyk:1985tr,Kikugawa:1985ex,Azcoiti:1987ua}. Of course, though continuum notation has been used above, also these theories are affected by the possibility of triviality, and may require a regulator like the lattice.

It should be noted that this situation is quite prevalent in supersymmetric theories, especially in the presence of more than one supercharge. In this context the Higgs expectation values are also called moduli and the space of possible breaking patterns moduli spaces \cite{Weinberg:2000cr}. Nothing is yet known about the implications of the above considerations for this class of theories.

\section{Mass spectrum and mass gap}\label{s:spectrum}

As experiments are observing particles, the spectrum of the theories under investigation play a fundamental role. Gauge-invariance has important implications on which particles can actually be observed, as will be discussed in detail in subsection \ref{ss:physstates}. The remainder of this section will then be dedicated to how to obtain this spectrum in a manifest gauge-invariant, non-perturbatively well-defined, and practical way. For the sake of simplicity and to be able to use lattice results, most of this section will still focus on the Higgs sector \pref{la:hs} alone, except for sections \ref{ss:sm} and \ref{ss:bsm}.

\subsection{Physical particles}\label{ss:physstates}

The phase diagram has already shown that using gauge-dependent quantities is potentially misleading. This repeats itself, in a very subtle way, in the observable, physical spectrum and the dynamics of the theory. This requires a somewhat more extended discussion.

To start out, it is best to consider the standard approach. Usually \cite{Bohm:2001yx}, the gauge bosons and the Higgs particle, up to components belonging to BRST quartets, are considered as physical particles. Especially, it is assumed that these are the particles which are forming final states. Considering again the standard model case but only the Higgs sector alone this would imply that the Higgs and the $W$/$Z$ bosons, which are then stable, would be observable particles\footnote{Of course, neither the Higgs nor the $W$/$Z$ bosons are actually stable particles in the standard model. However, especially for the $W^\pm$, this is accidental due to the masses of the fermions in the standard model. Other parameters would allow the $W^\pm$ to be the lightest, electrically-charged particles, and thus stable. Also, with differing parameters, all of these states could be long-lived on experimental time scales. Finally, the question which particles can be observed needs to be answered also in a purely theoretical setting like \pref{la:hs} as a stand-alone theory.}.

However, these fields carry an open gauge index. In the QCD-like domain, they would act like partons, and would be expected to be confined. Indeed, the corresponding potential and string-tension in the QCD-like domain are quite similar to ones in true QCD \cite{Knechtli:1999qe,Knechtli:1998gf}. But if the phase diagram is continuously connected, the spectrum can only undergo quantitative, but not qualitative, changes. Thus, states which would be confined in the QCD-like domain should not be part of the physical spectrum anywhere in the phase diagram. Thus 'colored' states, i.\ e.\ states with an open gauge index, should not be observable \cite{'tHooft:1979bj,Banks:1979fi,Frohlich:1980gj}.

While this may be considered an esoteric problem of this particular theory, it is actually much more general. The reason why this problem becomes aggravated is that in a non-Abelian gauge theory it seems to be impossible to construct a gauge-invariant, (almost) local gauge charge \cite{Haag:1992hx,Ilderton:2007qy,Heinzl:2008bu,Strocchi:2013awa}, notwithstanding many attempts over the time \cite{Lavelle:1995ty,Polyakov:1978vu,Susskind:1979up,Gross:1980br,Marchesini:1981kt,Bagan:1999jf,Bagan:1999jk,Ilderton:2007qy,Lavelle:2011yc,Heinzl:2008bu}. Thus, objects with an open gauge index can generally not be observable, and corresponding correlation functions should therefore vanish without gauge fixing.

This is qualitatively different from the Abelian case, where it is possible to construct a physical, observable charge \cite{Frere:1974ia,Haag:1992hx,Heinzl:2008bu,Rubakov:2002fi}. There, a gauge-invariant charge can be constructed using a Dirac phase factor, which is almost local up to a Dirac string. Still, this means that even in an Abelian gauge theory the elementary fields do not describe physical observable particles, and in fact the corresponding propagators do not have a physical pole structure \cite{Maris:1996zg,Alkofer:2000wg}. Rather, it is necessary to dress them appropriately to obtain gauge-invariant states. Essentially, the elementary fields are dressed by a photon cloud. How this explicitly works in Abelian gauge theories can and has been investigated in lattice calculations and in continuum calculations, including in particular QED and the Abelian Higgs model \cite{Woloshyn:2017rhe,Evertz:1986ur,Fradkin:1978dv,Frere:1974ia,Haag:1992hx,Montvay:1994cy,Buchmuller:1995xm,Lavelle:2011yc,Lucini:2015hfa}. These confirm the necessity and correctness of this dressing.

That this appears to be not possible \cite{Haag:1992hx} in non-Abelian gauge theories can be traced back to the Gribov-Singer ambiguity \cite{Ilderton:2007qy,Lavelle:2011yc,Heinzl:2008bu}, discussed in section \ref{ss:gribov}. Thus, even though this ambiguity seems to be a quantitatively irrelevant problem, its qualitative implications are also felt even when the BEH effect is active.

Conversely, objects described by operators which have no open gauge index, and are thus gauge-invariant, should be observable. This means that observable particles need to be described by composite, gauge-invariant operators \cite{Banks:1979fi,Frohlich:1980gj,Frohlich:1981yi,Strocchi:2015uaa,Attard:2017sdn,Kondo:2018qus,DeWitt:2003pm,Karsch:1996aw,Philipsen:1996af,Philipsen:1997rq,Laine:1997nq,Maas:2012tj}\footnote{In this context also \cite{Lavelle:1994rh,Chernodub:2008rz,Faddeev:2008qc,Ilderton:2010tf} should be mentioned, which at least partly argue into this direction.}. Structurally, such operators are the same as bound state operators, e.\ g.\ hadron operators in QCD. In fact, on a mathematical level, they are exactly the same. Thus, states described by them should be regarded as bound states\footnote{Note that sometimes the argument is raised that the weak interactions cannot sustain bound states. However, it should be noted that the even weaker interacting QED not only sustains a plethora of bound states like atoms and molecules but also exhibits a multitude of other non-perturbative phenomena like phase transitions and even scale separations over many orders of magnitude in solid state systems. Also arguments based on quantum-mechanical approximations of the Higgs sector against the existence of bound states \cite{Grifols:1991gw} are not applying, as relativistic effects, like very large mass defects, turn out to play a major role.} of their constituents in the same sense as hadrons are considered as bound states of quarks and gluons \cite{Maas:2012tj}. This does not mean that they must have properties like those expected for QCD or QED bound states. As will be discussed below, this is indeed not so \cite{Frohlich:1980gj,Frohlich:1981yi,Maas:2012tj}. This problem of how physical states should be consistently defined appears to have been known at the time of the inception of the BEH effect \cite{Ivanov:pc,Strocchi:1977za,Englert:2004yk,Englert:2014zpa}, but was formalized only about fifteen years later \cite{Banks:1979fi,Frohlich:1980gj}.

These arguments show how, from a fundamental field-theoretical perspective, a treatment should occur. This is different from the usual argumentation in perturbation theory \cite{Bohm:2001yx}. As the perturbative treatment is proven to all order in perturbation theory, it is important to point out how it could still fail in a full treatment of the theory. There are two reasons, intrinsic to the perturbative approach. The first is that BRST symmetry, as discussed above, breaks down in the presence of Gribov copies \cite{Fujikawa:1982ss}, and therefore the usual quartet mechanism is not working. Even if this could be circumvented, the construction of the asymptotic state space in perturbation theory is based on a smooth limit of all coupling constants vanishing \cite{Bohm:2001yx}. This implies that at most the elementary fields in the Lagrangian can be asymptotic states. Thus, even a hydrogen atom could not be a stable asymptotic state in the perturbation theory of QED. This already shows that something is fundamentally awry. This comes about in the following way: If a smooth limit would be possible then asymptotically the gauge symmetry becomes a global symmetry. This is not a unitarily equivalent state space, as it carries a different group and orbit structure. Thus, the limit cannot be smooth. This absence of a unitary equivalence between the free theory and the interacting theory is actually only a particular example of the more general theorem of Haag, which also applies to non-gauge theories \cite{Haag:1992hx}. Thus, already in the very construction of the states around which perturbation theory should be made the root has been seeded why the full theory cannot be adequately captured\footnote{A different approach is to develop a perturbative expansion around interacting (gauge-invariant, bound) states. This is done e.\ g.\ in \cite{Hoyer:2014gna,Hoyer:2016orc,Attard:2017sdn}, and also \cite{Frohlich:1980gj,Frohlich:1981yi,Philipsen:1996af,Maas:2012tj} and section \ref{ss:gipt} is not too far away from this idea.}.

But now seemingly a paradox arises: Why is the description of experiments in the Higgs sector, and the whole standard model, using the Higgs and the $W$/$Z$ bosons as if they would be physical, observable particles then working so well \cite{Bohm:2001yx,pdg,Djouadi:2005gi,Dawson:2018dcd}? The answer to this is quite subtle, and deeply connected with the BEH effect \cite{Frohlich:1980gj,Frohlich:1981yi}. It will be discussed at length in section \ref{ss:gipt}. Before going to this general case, it is in the case of the standard-model Higgs sector actually possible to exemplify the previous discussion by a rewriting of the theory. This will be done in section \ref{ss:reformsm}.

Because of the subtlety of the situation, this warrants a special notation in the following. In this section, and the next section \ref{s:scattering}, the names and symbols of the elementary fields, i.\ e.\ those appearing in the Lagrangian and the path-integral measure, will be denoted by small (initial) letters, e.\ g.\ higgs, $w$-boson and $w_\mu^a$. The corresponding big (initial) letters will be reserved for gauge-invariant states, which will play the analogue of these elementary particles.

\subsection{A quasi-exact reformulation of the standard model Higgs sector}\label{ss:reformsm}

Consider the lattice version of the theory \pref{higgsymlat} in the standard model case. Define the composite, gauge-invariant operators
\bea
H(y)&=&\phi^\dagger(y)\phi(y)=\det x(y)\label{giv1}\\
V_\mu^c(y)&=&\frac{1}{\det x(y)}\tr\left(T^c x^\dagger(y) \exp(iT_bw^b_\mu(y))x(y+e_\mu)\right)\nn\\
&=&\frac{1}{H(y)}\tr\left(\tau^c x^\dagger(y) D_\mu x(y)\right)+{\cal O}(a^2)\label{giv2}\\
V_\mu(y)&=&T^cV_\mu^c(y)\nn
\eea
\no where $c$ is a custodial index, and $x$ is given in \pref{higgsx}. This defines a custodial singlet scalar and a custodial triplet vector. The action can now be rewritten as \cite{Frohlich:1980gj,Frohlich:1981yi,Langguth:1985eu,Philipsen:1996af,Masson:2010vx,Attard:2017sdn,Kondo:2018qus}
\bea
S&=&\beta\sum_{y\mu<\nu}\left(1-\frac{1}{2}\Re\tr V_{\mn}(y)\right)\nn\\
&&+\sum_y\left(H^2(y)-3\log H(y)+\lambda(H(y)-1)^2-\kappa\sum_{\mu>0}H(y+\mu)H(y)\tr V_\mu(y)\right)\label{giv3}
\eea
\no where $V_\mn$ is the plaquette formed from $V_\mu$. The $\log H$ term originates from the Jacobian when changing the variables to \prefr{giv1}{giv2} \cite{Langguth:1985eu}.

There are several observations. First, the theory has only a global custodial (SU(2)) symmetry, and no longer any gauge symmetry. Thus, the gauge degree of freedom has been completely absorbed in the field redefinitions. At the same time, the field $V_\mu$ is a massive vector field. This field does not suffer from unitarity problems as ordinary massive vector fields \cite{Bohm:2001yx}, because the term $\log H$ cancels the corresponding contributions. This term is also the second observation: This theory is no longer superficially renormalizable by power-counting, as the term $\log H$ corresponds to an infinite power series in $H$. However, since the underlying theory is both renormalizable and unitary, these effects must cancel each other making the theory well behaved.

Thus, the gauge theory corresponds to a non-gauge theory, with only physically observable states. These states correspond to composite operators in the original theory, manifesting the qualitative discussion above. In fact, for the parameters of the standard model, the field $H$ fluctuates weakly, and the $\log H$ term can be neglected in perturbative calculations to reasonable accuracy \cite{Philipsen:1996af}.

There is one caveat in this rewriting \cite{Maas:2013aia}. To obtain this expression requires the division by $H$ in \pref{giv2} and that $\log H$ is well-defined. This is only possible for $H\neq 0$. Thus, the theory is only equivalent if occurrences of $H=0$, so-called defects, are a measure-zero effect. If the BEH physics is an adequate description and the Higgs field only weakly fluctuates around a non-zero value, this seems to be a reasonable assumption. If not, such defects can play an important role, especially in the QCD-like domain, and possibly in the cross-over region. The origin of the problem is that the target space of theory is modified. The original theory has for the Higgs field $\cn^2$, while the reformulation maps this target space to $\rn\times S^3$, which is a topologically different manifold. There are arguments that this will not affect the theory qualitatively  \cite{Callaway:1988ya,Kenna:1993fp,Fernandez:1992jh}, but no full proof exists. Thus, when using this reformulation, this issue should be kept in mind\footnote{To the knowledge of the author, because in the BEH regime there is little influence by the topological defects, it is not yet excluded that the actual Higgs sector of the standard model could have either target space. Barring any final results, here the usual $\cn^2$ target space will be assumed.}. Still, this implies that there is a non-gauge theory which is equivalent to the original theory. The gauge symmetry only served to write the theory in a way which is renormalizable by power counting.

That it is possible to reformulate the theory in terms of gauge-invariant variables is actually not a big surprise\footnote{The so-called generalized Kretschmann objection states that for any theory a trivial reformulation as a gauge theory should be possible \cite{Francois:2017aa}. The statement here is, in principle, a reversed generalized Kretschmann objection that for every gauge theory there exists an equivalent non-gauge theory.}. This is already possible for theories like Yang-Mills theory \cite{Gambini:1996ik}. However, in this case the price to pay is that there are infinite number of variables, Wilson loops \cite{Gattringer:2010zz} of all sizes. The present theory is special, as this is possible with a finite number of variables, albeit at the cost of a superficially non-renormalizable action.

Unfortunately, in this form the transformation is no longer possible if either QED or fermions are added. Also, it was not yet possible to find such a rewriting for theories where the number of gauge bosons exceeds the number of higgs fields. Still, it is not impossible that eventually such a reformulation could be found in general \cite{Attard:2017sdn,Kondo:2018qus}. 

The definitions \prefr{giv1}{giv2} have an eerie resemblance to the unitary gauge of section \ref{ss:limitthooft}. This is because unitary gauge implements by the gauge condition a reduction to similar degrees of freedom. In addition, the gauge defects in unitary gauge arise at the same space-time points where the transformation \prefr{giv1}{giv2} is not well defined. But it lacks the relevant Jacobian in \pref{giv3} which makes the theory renormalizable. As a consequence, the strict order of limits between gauge-fixing and regularization is absent here. Finally, even at infinite mass, the Faddeev-Popov operator has negative and zero eigenvalues along the gauge orbit, and thus the Gribov-Singer problem remains.

\subsection{Gauge-invariant classification of states}\label{ss:classification}

As now discussed at length, the only way to classify observable states is in terms of observable quantum numbers. These quantum numbers need to be carried by gauge-invariant (composite) operators, and thus can be at best associated with global symmetries.

First and foremost, any operator can be characterized by it spin and parity, $J^P$, which is obtained from the corresponding spin and parity of the constituents, as well as possible relative momenta between them. Furthermore, states carrying no further quantum numbers can have a definite charge parity $C$, yielding $J^{PC}$. This works in the same way as in QCD or QED.

Besides this enter custodial quantum numbers. These are also obtained from the corresponding composite fields. This implies that an operator composed entirely of gauge fields cannot have any custodial quantum number. Since any integer spin, charge parity\footnote{Provided distinguishable anti-particle exists, i.\ e.\ the gauge group has complex representations.} and parity can be created from pure gauge field operators \cite{Chen:2005mg} always all custodial singlet states exist.

Concerning custodial charges, this is somewhat less trivial. Of course, to any pure gauge operator always a factor $\phi^\dagger\phi$ can be added to have a hidden custodial component. For open custodial charges, this depends on which gauge-invariant combinations of the higgs fields can be created. In the case of the standard-model Higgs sector this implies the absence of custodial doublets \cite{Wurtz:2013ova,Maas:2014pba}: A fundamental charge can only be screened by an anti-fundamental charge, but not by an adjoint charge. Thus, gauge bosons cannot screen the charge of the higgs, and only another (anti)higgs can. And since only the higgs carries custodial charge as a doublet it is impossible to construct a tensor product which is at the same time a custodial doublet and gauge-invariant. Only custodial singlets or multiplets with an integer representation, i.\ e.\ triplets etc., are possible in this theory. Of course, once more fundamental charges, e.\ g.\ the standard-model fermions, are introduced into the theory this will no longer be true.

Likewise, in an SU(3) gauge theory with a single fundamental higgs the custodial symmetry is a U(1). Since any gauge-invariant state, as in QCD, always contains either an higgs and an anti-higgs or three higgs particles or combinations thereof \cite{Iida:2007qp,Maas:2016ngo,Maas:2017xzh}, only states with zero or multiples of three times the higgs particles' custodial charge can be gauge invariant. If the theory has a more complicated custodial structure, the possibilities quickly proliferate \cite{Wurtz:2009gf,Lewis:2010ps,Maas:2016qpu,Maas:2017xzh}.

This has far-reaching consequences also for the remainder of the standard model, as will be discussed in sections \ref{ss:qed}, \ref{ss:flavor}, and \ref{ss:qcd}, but even more so beyond the standard model as will be discussed in section \ref{ss:bsm}.

\subsection{Masses and the Nielsen identities}\label{ss:masses}

Before really understanding how the masses of the bound states are related to the masses of the elementary particles, it is necessary to first understand how the masses of the elementary particles have to be understood. In perturbation theory, these are very well defined \cite{Bohm:2001yx}. However, as discussed in section \ref{ss:qulev}, this is insufficient, as already a different choice of gauge leads to a breakdown of perturbation theory. Of course, BRST non-singlets, i.\ e.\ perturbatively unphysical states, have already at tree-level a gauge-dependent mass \cite{Bohm:2001yx}. Thus, only perturbatively physical, i.\ e.\ BRST singlet states, will be discussed in the following.

It is possible to determine the masses using non-perturbative methods, and has been done so using lattice simulations \cite{Karsch:1996aw,Maas:2010nc,Maas:2012tj,Maas:2013aia,Maas:2016edk,Maas:2016ngo,Maas:2018sqz} and functional methods \cite{Benes:2008ir,Fister:2010yw}, in a fixed gauge. This is essentially done as in Yang-Mills theory or QCD alone \cite{Maas:2011se}. However, such calculations are always performed in a fixed gauge, and the question remains whether the results could be gauge-invariant, and thus define physical masses.

This is often argued using the so-called Nielsen identities \cite{Nielsen:1975fs}. They are essentially based on the effective potential in the class of 't Hooft and linear covariant gauges. But already in \cite{Nielsen:1975fs} it is pointed out that infrared divergences in linear covariant gauges are problematic, but they may be cured non-perturbatively.

The basic argument on which the identities are based is that any change in the gauge parameter can be compensated for by a non-trivial change in the higgs vacuum expectation value. This can already be guessed from the corresponding gauge condition \pref{thooftg}, where both enter as a product. Then, the actual effective gauge condition remains unchanged. Thus, the theory remains the same and the masses of the elementary particles need to be the same, and thus not depend on the gauge parameter. Thus, they are gauge-parameter independent. 

However, this is not the same as gauge-invariant. There are two reasons why this does not cover all gauges. First, as was pointed out already in the original work \cite{Nielsen:1975fs}, the same derivation cannot be done in gauges where the gauge condition does not only involve the would-be goldstone bosons but also the remainder higgs field in a non-trivial way. Such gauges are also well-defined, and have been considered, e.\ g.\ in \cite{Dolan:1974gu}. Second, and more importantly, the argument requires that there is a non-vanishing vacuum expectation value of the Higgs field, which can be used to offset the change. As has been discussed in section \ref{ss:qulev}, and as can be done explicitly in lattice calculations \cite{Maas:2012ct}, it is possible to construct gauges for which this is not true. Thus, the Nielsen identities do not hold for such gauges: The Nielsen identities guarantee gauge-parameter invariance in subclasses of gauges with a higgs vacuum expectation value. This is actually true even non-perturbatively. But they are not strong enough to ensure gauge invariance.

Note that these arguments only show that the Nielsen identities do not guarantee gauge-invariance, not that the masses are not gauge-invariant. However, there is an argument which also calls this into question. It comes from the analytic connection of the phase diagram. In the QCD-like region, all results so far strongly suggest that no mass pole exist for the gauge bosons, and thus no thing like a mass can be defined \cite{Maas:2010nc,Capri:2012cr,Maas:2013aia}. In fact, this may even be true for the higgs boson \cite{Maas:2016edk}. At the very least both have negative norm contributions, and thus a non-standard spectral representation. This changes when moving into the BEH-like domain, and at least a conventional mass pole arises, though some negative-norm contributions remain in the spectral functions \cite{Maas:2010nc,Maas:2013aia,Raubitzek:unpublished}. This is possible, because in a fixed gauge the gauge-dependent correlation functions can change discontinuously. But, since all gauge-invariant quantities remain analytic, a physical mass would need to continue to exist even in the QCD-like domain, which it does not. In addition, the appearance of a mass pole happens in different gauges at different points in the phase diagram \cite{Caudy:2007sf}, as discussed in section \ref{ss:pdlat}. Thus, also from this point of view a contradiction arises. These arguments make it also quite unlikely that some improved version of the Nielsen identities exist which could make the poles gauge-invariant. Or, at the very least, requires some baroque way the mass pole would need to be carried by different correlation functions in the different regions, depending on the gauge. This seems unlikely.

All of this does not diminish the technical usefulness of the Nielsen identities in fixed classes of gauges, especially in perturbation theory.

\subsection{Gauge-invariant perturbation theory}\label{ss:gipt}

\subsubsection{General strategy}\label{ss:rules}

The masses of the bound states are genuine\footnote{This is already visible heuristically when considering a stable bound state. In perturbation theory only elementary states are stable, asymptotic states. Therefore, a stable, and thus asymptotic, bound state can never exist in perturbation theory.} non-perturbative quantities\footnote{It should be noted that even in the perturbative picture of the weak interactions all states receive mass corrections from non-perturbative physics due to QCD. The non-perturbative chiral symmetry breaking acts like the BEH effect, and yields in the absence of the BEH effect masses for the $w$ and $z$ boson of order some tens of MeV \cite{Quigg:2009xr}. In fact, it is a realization of Technicolor \cite{Hill:2002ap,Morrissey:2009tf,Andersen:2011yj} within the standard model \cite{Quigg:2009xr}. However, if this occurs in addition to the BEH effect, both contributions are added in quadrature in a leading-order non-perturbative treatment \cite{Eichmann:pc}, and the hadronic mass contribution is diminished to about 5-6 keV, and therefore likely undetectable. Note that unitarity for such a correction can only be restored non-perturbatively.} \cite{Alkofer:2000wg,Hoyer:2014gna}. In principle, non-perturbative methods would therefore be needed to calculate them.

Fortunately, in presence of the BEH effect, there exists a possibility to calculate them analytically \cite{Frohlich:1980gj,Frohlich:1981yi}. This is done using so-called \cite{Seiler:2015rwa} gauge-invariant perturbation theory. As the name suggests, this is indeed a perturbative approach. This procedure can actually be extended beyond static properties \cite{Frohlich:1980gj,Frohlich:1981yi,Maas:2012ct,Egger:2017tkd}, as will be discussed in section \ref{s:scattering}.

At first, this seems a quite surprising statement, as it was just stated that the masses belong to non-perturbative bound states. However, in a very precise sense \cite{Frohlich:1980gj,Frohlich:1981yi} the BEH effect can be utilized to determine the masses nonetheless. The basic rules of this approach are similar to Feynman rules, and can be outlined in the following way \cite{Maas:2012ct,Egger:2017tkd,Torek:2016ede}:
\begin{itemize}
 \item[1)] Write down the gauge-invariant operators needed to describe the process in question, e.\ g.\ $\op$, and form the desired correlation functions, e.\ g.\ a propagator $\la\op^\dagger\op\ra$
 \item[2)] Fix a gauge with a non-vanishing vacuum expectation value, e.\ g.\ 't Hooft gauge \pref{thooftg}
 \item[3)] If there are higgs fields contained in the operators, split them in vacuum expectation value and fluctuation field. E.\ g., if $\op$ contains one higgs field this yields for the propagator
 \be
 \la\op^\dagger\op\ra=v^2n_in_j\la\op_i^{'\dagger}\op_j'\ra+vn_i\la\left(\op_i^{'\dagger}\op''+\op^{''\dagger}\op_i'\right)\ra+\la\op''\op''\ra\label{rule1}
 \ee
 \no where $\op'$ is the operator with the higgs field replaced by the direction of the vacuum expectation value $n$ and $\op''$ is the operator with the higgs field replaced by the fluctuation field $\eta$. This is the FMS mechanism \cite{Frohlich:1980gj,Frohlich:1981yi}, an expansion in $v$
 \item[4)] The three correlation functions on the right-hand side of \pref{rule1} can then be expanded in their usual perturbative series. This double expansion is then gauge-invariant perturbation theory
\end{itemize}
This recipe will now be applied to the standard-model case. Other cases will be treated in section \ref{ss:bsm}. In this course, how to interpret the results and all further applying subtleties of this procedure will be discussed. The validity of this double expansion will be tested using lattice methods in section \ref{ss:latgipt}. In section \ref{s:scattering} it will be applied to scattering processes.

\subsubsection{The scalar singlet}\label{sss:ss}

According to section \ref{ss:classification}, physical states are identified by their $J^{P(C)}$ and custodial quantum numbers.
Consider as the simplest case the scalar singlet in the standard model setup. The simplest operator carrying these quantum numbers is the composite operator \cite{'tHooft:1979bj,Frohlich:1980gj,Frohlich:1981yi}
\be
H(x)=(\phi^\dagger\phi)(x)\label{higgs},
\ee
\no already appearing in section \ref{ss:reformsm} in equation \pref{giv1}. Applying the rules of gauge-invariant perturbation theory of section \ref{ss:rules} up to step 3 yields\footnote{When passing to connected correlation functions the constant term will be no longer present.}
\bea
\la(\phi^\dagger\phi)(x)(\phi^\dagger\phi)(y)\ra = 
&&dv^4 + 
4v^2\la \Re\left[n_i^\dagger\eta_i\right]^\dagger(x)~\Re\left[n_j^\dagger\eta_j\right](y) \ra  \label{op0ppre}\\
+&&2v\left(\la (\eta_i^\dagger\eta_i)(x)~\Re\left[n_j^\dagger\eta_j\right](y) \ra + (x\leftrightarrow y) \right) + 
\la (\eta_i^\dagger\eta_i)(x)~(\eta_j^\dagger\eta_j)(y) \ra  \;,\nn
\eea
\no where terms with vanishing expectation values have been dropped and $d$ is a constant. So far, this is just an exact rewriting. However, the individual expectation values on the right-hand side are not separately non-perturbatively gauge-invariant, and only their sum is. Thus, it would still be necessary to evaluate them non-perturbatively to get an exact result.

Now apply step 4 of section \ref{ss:rules}. The correlation functions are expanded in the usual perturbative series. At lowest order, i.\ e.\ $g^0$ and $\lambda^0$, this yields
\bea
\la(\phi^\dagger\phi)(x)(\phi^\dagger\phi)(y)\ra& =& d'v^4+4v^2\la \Re\left[n_i^\dagger\eta_i\right]^\dagger(x)~\Re\left[n_j^\dagger\eta_j\right](y) \ra_\text{tl}\label{op0ppex} \\
&&+\la \Re\left[n_i^\dagger\eta_i\right]^\dagger(x)~\Re\left[n_j^\dagger\eta_j\right](y) \ra_\text{tl}^2+{\cal O}(g^2,\lambda)\nn\;,
\eea
\no where $d'$ is now another constant. The first non-constant term is the tree-level Higgs propagator, and the second term is the square of the Higgs propagator, with both propagators originating at the same space-time point and ending at the same space-time-point. Taking the connected and amputated part only, this amounts in Fourier space to\footnote{Note that terms, which vanish perturbatively because they are BRST non-singlets have been dropped, in accordance with performing perturbation theory}
\bea
D_{H}(P^2)&=&4v^2D_\eta(P^2)^\tl+2\int d^4q D_\eta^\tl((P-q)^2)D_\eta^\tl(q^2)=\frac{4v^2}{P^2-m_h^2+i\epsilon}+\Pi(P^2)\label{higgspoles}\\
&\approx&\frac{4v^2}{\left(P^2-m_h^2\right)\left(1+\frac{P^2-m_h^2}{4v^2}\Pi(P^2)\right)+i\epsilon}\nn\\
\Pi(p^2)&=&-2i\pi^2\left(\ln\frac{m_h^2}{\mu^2}+\frac{m_h^2}{p^2}\left(\frac{1}{r(p^2)}-r(p^2)\right)\ln r(p^2)\right)\nn\\
r(p^2)&=&\frac{2m_h^2-p^2-i\epsilon\pm\sqrt{(p^2-2m_h^2+i\epsilon)^2-4m_h^4}}{2m_h^2}\nn
\eea
\no where $D_H$ is the bound-state propagator to this order, $D_\eta^\tl$ denotes the tree-level higgs propagator, and $q$ could be considered to be the relative momenta of the higgs particles inside the scalar singlet. The second term is an elementary scalar one-loop integral, which is solved using standard techniques \cite{Bohm:2001yx}, and the expression is renormalized in the $\overline{MS}$ scheme. It is well visible, how corrections are suppressed close to the pole. This is the result to all orders in $v$, but to zeroth order in $g$ and $\lambda$. To zeroth order in all three couplings, only the first term would be present, and thus $D_H=4v^2D_\eta^\tl$.

The analytic structure of \pref{higgspoles} is a pole at $m_h$, the elementary higgs mass, from the first term and a cut starting at $P^2=4m_h^2$, describing a scattering state of twice the ground state mass. Because \pref{higgspoles} is the leading-order contribution in the double expansion \pref{op0ppex} for the singlet propagator, this predicts that the ground state in the scalar singlet channel has the mass of the elementary higgs. Also, it predicts that in this channel exists a scattering state of twice the ground state mass.

Thus, gauge-invariant perturbation theory predicts that the gauge-invariant singlet $0^+$ particle, i.\ e.\ the gauge-invariant Higgs $H$, has the same mass as the elementary higgs. Other than that, it predicts the (trivial) existence of a scattering state of two (gauge-invariant) Higgs particles $H$. And yes, this means twice the gauge-invariant Higgs, as two elementary higgs are only gauge-invariant when they form a composite operator, but not in the sense of two independent particles. Equation \pref{higgspoles} makes a statement about the pole structure, not the substructure of the state. After all, the right-hand side is now gauge-dependent, while the left-hand side is not. Still, at this level of approximation the Higgs propagator is given by \pref{higgspoles}. But this should be rather thought of in the same sense as to lowest order a hadron is made up out of quark propagators in the simplest constituent quark model.

It is the left-hand side of \pref{op0ppre} which describes an experimentally observable particle. Gauge-invariant perturbation theory was now used, through equation \pref{higgspoles}, to predict that the observed scalar particle should have the same mass as the elementary higgs particle. This is exactly what is observed experimentally \cite{pdg}. Thus, a field-theoretically more strict treatment than the usual perturbative treatment \cite{Bohm:2001yx} ends up with an equally good determination of the Higgs mass to this order. There are, of course, many further subtleties involved.

{\bf Firstly} with respect to gauge-invariance. At least, due to the Nielsen identities, which hold in such a class of gauges, the determined pole on the right-hand side is gauge-parameter-independent. However, it is not clear to which extent such a relation will actually hold in gauges like those discussed in \cite{Dolan:1974gu} or in non-covariant gauges with their intricate analytic structures and generically additional singularities \cite{Burnel:2008zz}. Like perturbation theory in general \cite{Lee:1974zg}, it should therefore be considered to be an approach working only in suitable gauges, in this case the 't Hooft gauges.

{\bf Secondly}, the result can be improved by going to higher orders in \pref{op0ppre}. This can be done in different ways. An alternative to \pref{higgspoles} would be to keep only the leading non-trivial order in $v$, yielding for the connected part
\be
\la(\phi^\dagger\phi)(x)(\phi^\dagger\phi)(y)\ra_c = 4v^2\la \Re\left[n_i^\dagger\eta_i\right]^\dagger(x)~\Re\left[n_j^\dagger\eta_j\right](y) \ra_c+\op(v)=4v^2D_\eta(x-y)+\op(v)\nn.
\ee
\no Then, to all orders in $g$ and $\lambda$, the propagator $D_H$ would equal the propagator $D_\eta$, and thus exactly the same results as in ordinary perturbation theory would be obtained for the Higgs' properties.

On the other hand, including higher orders in $v$, deviations to $D_\eta$ arise at every order in $g$ and $\lambda$, just as in \pref{higgspoles}. However, these additional contributions are likely quantitatively small, as a simple estimate shows: The Appelquist-Carrazone theorem suggests a suppression like $s/v^2$, where $s$ is the actual involved energy, not necessarily the collider energy. At LEP2, this would be at most a $2\times 10^{-3}$ correction\footnote{This shows for the first time that the standard model is particularly, and frustratingly, unsuited to detect this kind of field-theoretical effect. This insight will be made again and again in the following. In contrast, in many beyond-the-standard model theories differences already arise at the qualitative, rather than the quantitative, level. This will be discussed in section \ref{ss:bsm}.} to the (at LEP2) unobserved Higgs production. Naively, it would be expected that at the LHC, with substantial parton luminosity at 1 TeV or more, this would be different. But then, the next-to-leading contribution in \pref{op0ppex} is vanishing in leading-order in $g$ and $\lambda$, and thus will only contribute at loop order. The second contribution is already evaluated in \pref{higgspoles}, and thus will only arise for double-Higgs production, which is also not yet reached. An alternative estimate is the size of the quantum fluctuations with respect to the vacuum expectation value, $\la|\eta|\ra/v$, which in lattice simulations are found to be tiny \cite{Maas:2012ct}. Thus, these corrections are small, but may play a role eventually. Explicit evaluations start to probe this \cite{Maas:unpublished,Raubitzek:unpublished}. Whether and how they could be observed actually in experiments will be discussed further in section \ref{s:scattering}.

{\bf Thirdly}, another problem arises at loop order. The mass of the elementary higgs is scheme-dependent \cite{Bohm:2001yx,Einhorn:1992um}. Of course, the bound-state mass of the physical Higgs is both renormalization-group-invariant and renormalization-scheme-invariant. Thus, to make a useful association with the actual observed particle the pole scheme \cite{Bohm:2001yx,Einhorn:1992um} seems to be mandatory to keep the relation at loop level. However, this is an issue which has not yet been deeply explored \cite{Maas:2013aia}. It will be returned to in section \ref{ss:latgipt}. In particular, this implies that the apparent mass defect of $m_h$ of the scalar singlet becomes dependent on the renormalization. The mass defect is therefore no longer really a physical concept. This also ameliorates to some extent its enormous size, as the size of it becomes arbitrary at loop order. Still, it is a useful concept as also this is a genuine quantum effect. Therefore, the notion of mass defect will be kept in the following, but the issue of renormalization should always be kept in mind.

{\bf Finally}, there exists other operators with the same quantum numbers. These can be classified into three types.

One are operators which are made up of individually gauge-invariant operators. Physically, they correspond to scattering states of observable particles. In this case, rule 3 of section \ref{ss:rules} is applied to every gauge-invariant operator separately. The prediction is then that the scattering of the observable particles is mapped to scatterings of the corresponding elementary particles. The simplest such operator is
\be
{\cal O}_{2H}=H(y)H(y)\label{doublehiggs}.
\ee
\no It expands at leading order to the second term of \pref{higgspoles}, thus correctly predicting a scattering cut starting from twice the higgs mass, and thus of twice the ground state mass in the channel described by the operator $H$. Since the same scattering pole is already contained in the operator $H$ this shows that these operators actually do not form a tree-level mass basis.

Then, there are operators, which cannot be decomposed into other gauge-invariant operators, but still involve Higgs fields. Such operators can be obtained, e.\ g., by introducing Wilson lines in \pref{higgs}. They therefore can have a pole at the higgs mass. However, since the full operator has the same quantum numbers, it will again mix with both \pref{higgs} and \pref{doublehiggs}. Necessarily, such operators need to be included in the operator basis to find the mass eigenbasis in which only a single operator carries the higgs mass pole, while all other orthogonal combinations expand only to scattering states\footnote{Here the possibility of additional non-perturbative resonances on the left-hand-side is neglected. This will be considered in section \ref{ss:latgipt}.}. This one particular combination would be the ground-state in this channel. Such a basis would be the most suitable one. Fortunately, the few investigations of this question \cite{Maas:2013aia,Maas:2014pba,Wurtz:2013ova} all point towards \pref{higgs} dominating the ground state, simplifying the problem substantially.

The last type of operators are those which do not involve the Higgs field. Rule 3 of section \ref{ss:rules} therefore does not apply to them, and it is necessary to directly proceed to rule 4. In case of the $0^{+}$ singlet, this would be, e.\ g., the scalar $w$-ball, the equivalent of a glueball in Yang-Mills theory or QCD. A gauge-invariant operator of this type is, e.\ g.,
\be
\op_{w\mathrm{-ball}}=(w_\mn^a w^\mn_a)(y)\label{wball}.
\ee
\no Such operators would also be needed to form the mass eigenbasis. Of course, also operators like \pref{wball} can be generalized along the same way as the two previous classes. Applying rule 4 to \pref{wball} yields at leading order a scattering cut from two times $m_w$. With the same logic, this would imply that in the $0^+$ singlet channel a scattering state involving two physical particles with the same mass as the $w$s should exist\footnote{Note that the same reasoning in QCD would lead to a scattering state of two massless states, which is, due to the mass gap, not a good approximation.}.

\subsubsection{The vector triplet}

If the physical, gauge-invariant scalar singlet should have the scattering cut predicted by the expansion of \pref{wball}, it is necessary that a particle of mass $m_w$ exists in the spectrum. However, this cannot be the $w$-boson itself, as it is a gauge triplet, and therefore not observable. On the other hand, up to custodial breaking effects, experiments show the existence of a vector triplet \cite{pdg}. Thus, to accommodate both results, the experimental one and the need for the scattering state, requires that the spectrum has a triplet vector state.

The only possibility to obtain a triplet is to have a global symmetry with triplet representations. The custodial symmetry has exactly this structure. The simplest vector triplet operator is\footnote{In \cite{Frohlich:1980gj,Frohlich:1981yi} an operator based on the field-strength tensor is used, which, after contraction with a momentum, delivers the same operator.} \cite{'tHooft:1979bj,Evertz:1985fc,Maas:2012tj,Maas:2013aia}
\be
W(x)=\tr\left(T^ax^\dagger D_\mu x\right)(y)\label{op1t}.
\ee
\no where $x$ is defined in \pref{higgsx}. Applying the rules of section \ref{ss:rules} yields at leading order\footnote{Note that the Lorentz indices have to be contracted on both sides to have a non-vanishing correlation function. Thus, the correlation-function on the right-hand side is a combination of the transverse and the longitudinal dressing function of the $w$ propagator. Only in a Landau-'t Hooft gauge it will be entirely given by the transverse dressing function. However, the longitudinal part of a vector particle's propagator has never a pole, and thus this is irrelevant for the following \cite{Raubitzek:unpublished}.} in $v$
\be
\langle\tr(T^ax^\dagger D^\mu x)(z)\tr(T^bx^\dagger D_\mu x)(y)\rangle=
v^2c^{ab}_{kl}\langle w^{k\mu}(z) w^l_\mu(y)\rangle+{\cal O}(v)\label{wcor},
\ee
\no where  $c^{ab}_{kl}\sim\delta^{ab}\delta_{kl}$. Alternatively, $x$ can be replaced by $\alpha$ of \pref{lrdecomp}, without changing the quantum numbers \cite{Evertz:1985fc}. However, this would be just another operator in the same class as \pref{doublehiggs} for the triplet vector channel.

The result \pref{wcor} shows that the operator \pref{op1t} expands to the $w$ propagator. Therefore, by the same argument as before, the mass of the bound-state operator on the left-hand-side must be the same as the one of the elementary $w$. Since the mass of the $w$ is not affected by renormalization, this statement is even stronger than in the scalar singlet case. Thus, as required, there exists a vector triplet of mass $m_w$. As in the scalar singlet case, higher orders in the double expansion and other operators can be added.

Hence, gauge-invariant perturbation theory precisely predicts the same mass spectrum as perturbation theory: A single massive scalar, and a mass-degenerate vector triplet.

\subsubsection{Other channels}

To be fully compatible with experiment requires the absence of other (light) massive single-particle states. Performing similar calculations for other operators in the scalar singlet and vector triplet channels yield no additional states in gauge-invariant perturbation theory. While suggestive, this statement is not enough. There is still the possibility of genuine non-perturbative resonances in the left-hand side composite operators, i.\ e.\ internal excitations of these bound states. This requires an actual non-perturbative calculation. Such calculations have been performed using lattice methods. No such resonances have (yet) been found, as will be discussed in section \ref{ss:latgipt}.

A different question are states in other quantum number channels. For this, note that the matrix $c$ in \pref{wcor} cannot carry Lorentz indices, as otherwise Lorentz symmetry would not be manifest. Since only covariant gauges are considered, this should not happen. Thus, an expansion to a single-field operator can only occur for $J^{P(C)}$ quantum numbers which appear as elementary fields, i.\ e.\ scalars and vectors. In all other channels only scattering states involving multiple elementary particles can arise.

The only further alternative are scalars and vectors which have a different custodial structure, e.\ g.\ a scalar triplet or a (axial)vectorial singlet \cite{Wurtz:2013ova}. But the particular structure of the theory yields that it is not possible to construct higher tensors in custodial symmetry without at the same time constructing higher tensors in the gauge symmetry. The reason is that the elementary building block of the custodial symmetry, the higgs field, has the same representation and group for both the custodial symmetry and the gauge symmetry, the fundamental representation of SU(2). This is a very special situation. Hence, any higher representation of the custodial symmetry can only be build from gauge-invariant operators, and are already in ordinary perturbation theory scattering states \cite{Wurtz:2013ova}.

Thus, the prediction of the FMS mechanism and gauge-invariant perturbation theory is that there should only be at most two states in the physical spectrum, a scalar with a mass corresponding to the (tree-level) mass of the higgs, and a vector triplet with the mass of the $w$. Note that this prediction remains true for any mass of the Higgs, and is only limited by the same conditions as ordinary perturbation theory \cite{Bohm:2001yx}.

\subsection{Lattice spectra and tests of gauge-invariant perturbation theory}\label{ss:latgipt}

This is a non-trivial prediction. In particular, this implies a very different bound-state structure than in QCD and QED. Probably the most distinguished features are the large mass-defects, which is one higgs mass for the scalar and even two higgs masses for the vector. Thus, they are 50\% and about 76\% (at a Higgs mass of 125 GeV) of the sum of the masses of the constituents. Except for composite Goldstone bosons, like the pions, no other composite particles in QCD or QED do exhibit such large mass defects. This implies that any tests needs to be both non-perturbative and using quantum field theory. In addition, the bound-state poles of the stable states in the correlators \pref{op0ppre} and \pref{wcor} cannot appear as asymptotic states in perturbation theory. Simply because by definition only the elementary states appear as asymptotic states in perturbation theory.

So far, the only method with which the spectrum of these composite operators has been investigated have been lattice methods\footnote{This has also been done in three \cite{Karsch:1996aw,Philipsen:1996af,Philipsen:1997rq,Laine:1997nq,Capri:2012cr} and two dimensions \cite{Gongyo:2014jfa}, which will not be discussed in detail. Note, however, that especially three dimensions is an interesting model system, as the theory is non-trivial and has dynamical gauge degrees of freedom, and all of the arguments given still hold. It may therefore be promising to use this case to understand the continuum limit of the FMS mechanism. In fact, comments in \cite{Karsch:1996aw} indicate that the FMS mechanism also works in three dimensions.} \cite{Wurtz:2013ova,Maas:2013aia,Maas:2014pba,Maas:2012tj,Evertz:1986vp,Evertz:1985fc,Langguth:1985eu,Langguth:1985dr,Kuti:1987nr}\footnote{Note that in most lattice publications the gauge-invariant operators of section \ref{ss:gipt} are called already Higgs and $W$/$Z$, rather than the elementary degrees of freedom. The reason is that lattice methods do not (need to) fix a gauge. Therefore these are the only well-defined observables, and  hence have received theses name, without actually establishing their relation to the elementary degrees of freedom \cite{Lang:pc,Wittig:pc}. Thus in \cite{Wurtz:2013ova,Evertz:1986vp,Evertz:1985fc,Langguth:1985eu,Langguth:1985dr,Kuti:1987nr,Philipsen:1996af}, these names already refer to the Higgs and the $W$/$Z$ rather than to the higgs and $w$/$z$.}. In addition, only few calculations are available where simultaneously also the gauge-fixed propagators have been obtained \cite{Maas:2013aia,Maas:2012tj,Maas:2010nc} to check the FMS mechanism\footnote{Note that there are non-lattice non-perturbative calculations which address the masses of the elementary higgs boson and gauges bosons, see e.\ g.\ \cite{Benes:2008ir,Fister:2010yw,Capri:2013oja,Gies:2015lia} for recent examples, but they provide not at the same time results for the composite operators, and are therefore not relevant here.}.

The determination of the masses of the bound states in lattice calculation is straightforward \cite{Montvay:1994cy,Gattringer:2010zz}. Since lattice simulations are performed in Euclidean space-time, any gauge-invariant correlator in the present class of theories in position space has the form \cite{Seiler:1982pw,Frohlich:1981yi}
\bea
D(t)=\la\op(t)\op(0)\ra&=&\sum_n\left|\la n|\op\ra\right|^2e^{-E_n t}\stackrel{t\gg E_1^{-1}}{\sim}e^{-E_0 t}\label{lelevels}\\
\op(t)&=&\sum_{\vec x}\op(\vec x,t)\nn,
\eea
\no where the summation projects the states to zero momentum, and thus into the rest frame\footnote{Non-zero momentum is possible, but here an unnecessary complication, even though relevant in practice \cite{Gattringer:2010zz,Wurtz:2013ova}.}. On a finite, periodic lattice the exponentials have to be replaced by $\cosh$s. This will not affect the following qualitatively but is taken into account in actual calculations \cite{Montvay:1994cy,Gattringer:2010zz}.

$E_0$ is the mass of the lightest state in the corresponding quantum number channel. However, at times $t\lesssim E_1^{-1}$ the correlator will have contributions also from higher levels. Also, the overlap $\left|\la n|\op\ra\right|^2$ needs to be non-zero. All of these issues can \cite{Gattringer:2010zz} and are taken into account \cite{Wurtz:2013ova,Maas:2013aia,Maas:2014pba} for the following results. With this it is possible to isolate the rest masses of particles. Furthermore, various operators can mix, as discussed above. This can also be taken into account \cite{Gattringer:2010zz}. In practice, only an ansatz can be made which operators could have overlap with the ground state. Within this subset of possible operators then the lowest-lying state is isolated. This has been done for various operator bases in practice \cite{Wurtz:2013ova,Maas:2013aia,Maas:2014pba}, yielding coinciding results. However, the choice of operators is ad hoc, and therefore cannot be guaranteed to be systematically controlled. The choice is usually to include the operators with the lowest number of elementary fields, up to a certain maximum. This yields good results in QCD, and yields consistent results for the Higgs sector.

A useful concept is the effective mass, defined as \cite{Maas:2011se,Gattringer:2010zz,Montvay:1994cy}
\be
m(t)=-\ln\frac{D(t)}{D(t+a)}\label{effmass},
\ee
\no where $a$ is the lattice spacing, and $D$ a correlator like \pref{lelevels}. If a single exponential dominates \pref{lelevels}, this effective mass will be just the mass $E=m$ for this state, and there is no time-dependence. If additional contributions arise, this will give a deviation, which increase the effective mass at shorter times. Thus, for a physical particle the effective mass is necessarily a monotonously decreasing function of time \cite{Seiler:1982pw}.

Finally, because of the $\cosh$ behavior due to the finite volume, the effective mass will decrease even for a single exponential contributing at times close to the maximum temporal extent of a lattice. In principle, this can be corrected for. However, not doing so shows explicitly at which times finite-volume artifacts start to play a significant role. This will therefore not be done here. 

\subsubsection{The ground states of the gauge-invariant spectrum}\label{sss:ground}

Following the prescribed procedure, it is possible to obtain the masses of the ground states in both the scalar and vector triplet channels. This has been done using a variety of operator bases \cite{Wurtz:2013ova,Maas:2013aia,Maas:2014pba,Maas:2012tj,Langguth:1985eu,Langguth:1985dr,Evertz:1986vp,Evertz:1985fc}. Technical details can be found in the corresponding references. However, it should be remarked that, like glueballs in QCD, the operators in the Higgs sector are rather noisy, and thus large statistics and/or advanced noise reduction techniques are necessary to obtain statistically reasonably reliable results \cite{Philipsen:1996af,Wurtz:2013ova,Maas:2014pba}. Note that whenever units are given in the following, they are obtained by setting the mass of the lightest state in the spectrum to 80.375 GeV.

Generically the results \cite{Wurtz:2013ova,Maas:2013aia,Maas:2014pba,Maas:2012tj,Langguth:1985eu,Langguth:1985dr,Evertz:1986vp,Evertz:1985fc,Maas:unpublished} find that the lightest state is the scalar singlet in the QCD-like domain and the vector triplet in the Higgs-like domain. The respective other of these states can have any larger mass. However, the scalar singlet can decay into two vector triplets. For some parameter sets it is found to be above the decay threshold, but the resonance properties have not yet been determined\footnote{This requires a so-called L\"uscher-type analysis \cite{Gattringer:2010zz,Luscher:1990ux,Luscher:1991cf}, which has been applied successfully to ungauged higgs sectors \cite{Zimmermann:1991xx,Gerhold:2011mx}. This is numerically quite demanding, especially when trying to cross the inelastic threshold \cite{Briceno:2014oea,Briceno:2017tce}.}. The vector triplet seems to be always the lightest state with non-zero custodial charge, and thus always stable, allowing for a scale separation. There are a few more observations.

{\bf First}, it is actually a quite striking observation \cite{Langguth:1985eu,Langguth:1985dr,Evertz:1986vp,Evertz:1985fc,Maas:2014pba} that if the vector triplet is lighter than the singlet scalar, the physics is BEH-like, rather than QCD-like. I.\ e., the phase diagram shown in figure \ref{fig:gdpdc}, intended as a gauge-dependent statement, actually also says something about gauge-invariant physics. In fact, in the cross-over region the mass of both states seem to become degenerate \cite{Maas:2014pba}. In ordinary perturbation theory this ratio is free in presence of the BEH effect, while the gauge boson mass is strictly zero if not \cite{Bohm:2001yx}. Gauge-invariant perturbation theory only maps this perturbative statement on the gauge-invariant states, and thus inherits this feature. Of course, this is as so far only a numerical result obtained in a certain region of the phase diagram. Also, lattice calculations cannot span a too different sets of scales, as the computation time scales at least as the fourth power of the ratio between the most extreme scales. On a finite lattice, this is the (inverse) lattice extent and the lattice spacing. As corrections due to the volume vanish exponentially with the lattice extent for massive states \cite{Luscher:1985dn}, this will likely not be the relevant problem.

\begin{figure}[hbtp!]
\begin{minipage}{0.6\linewidth}
\includegraphics[width=\linewidth]{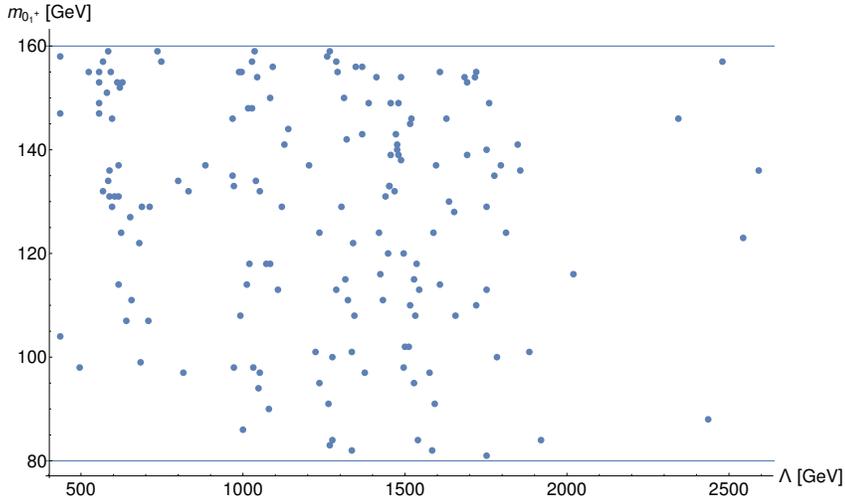}
\end{minipage}
\begin{minipage}{0.4\linewidth}
 \caption{\label{fig:hml}The mass of the singlet scalar, i.\ e.\ the Higgs, as a function of the lattice cutoff, defined to be the magnitude of the largest possible four-momentum. Each point corresponds to a fixed set of (lattice) parameters, i.\ e.\ to one point in the phase diagram in figure \ref{fig:gdpdl}. The lower line gives the mass of the vector triplet, which is used to set the scale. If the scalar mass is below this value, the theory becomes QCD-like. The upper line is the elastic threshold, above which no reliable data is available yet. To avoid clutter, lattice errors are not shown. Data is from \cite{Maas:unpublished,Maas:2014pba,Torek:2016ede}. Note that this corridor is narrower than the perturbative \cite{Altarelli:1994rb,Djouadi:2009nu} and non-perturbative \cite{Eichhorn:2015kea,Gies:2017ajd,Gies:2017zwf} expectations.}
\end{minipage}
\end{figure}

However, as the mass corridor for the possible Higgs masses depends on the cutoff \cite{Altarelli:1994rb,Callaway:1988ya,Bohm:2001yx,Djouadi:2009nu,Eichhorn:2015kea,Gies:2017ajd,Gies:2017zwf}, the comparatively low cutoff values reachable in current lattice simulations may be the reason for this. This is shown in figure \ref{fig:hml}. So far, the available data fill the corridor between the limit to pass into the QCD-like domain and the elastic threshold rather homogeneously. Of course, the selection of which points are included in the simulations is also biased, and represents the points in figure \ref{fig:gdpdl}.

Thus, this may change if the cutoff is send to very large values and/or other regions of the phase diagram are sampled. Still, it is quite intriguing, and so far lacks any explanation. Especially, as nothing was done in selecting the points in the parameter space to obtain such a behavior. Furthermore, as will be discussed in section \ref{sss:gut}, for gauge groups SU($N>2$) and a single fundamental higgs gauge-invariant perturbation theory may suggest an explanation for this lower mass limit.

{\bf Second}, a surprising result about the singlet scalar is that in large regions of parameter space it is light, i.\ e.\ with a mass below the elastic threshold \cite{Wurtz:2013ova,Maas:2013aia,Maas:2014pba,Maas:2012tj,Evertz:1986vp,Evertz:1985fc,Langguth:1985eu,Langguth:1985dr}. In figure \ref{fig:gdpdl} this applies to about two thirds of the points in the Higgs-like domain. Based on fine-tuning, naively it would have been expected that any random set of lattice parameters would not yield a light Higgs. However, this can again be a bias artifact and/or an effect due to the comparatively low cutoff. Still, it is more probable to find a heavier scalar than a lighter scalar, as can already be seen from the clustering of points at larger masses in figure \ref{fig:hml}.

This does not mean that the theory is not fine-tuned. In fact, the lattice spacing and the mass-ratio depend very sensitively on the hopping parameter $\kappa$ if $\kappa$ is close to the boundary of the domains, and even small changes of $\kappa$ can have substantial effects.

{\bf Third}, another interesting observation is the composition of the scalar. As far as it has been investigated \cite{Maas:2014pba,Wurtz:2013ova}, the ground state in the Higgs-like domain is dominated by the operator \pref{higgs}, while in the QCD-like domain it is the $w$-ball operator \pref{wball}. Together with the scalar singlet being the lightest state in the QCD-like domain, this points to a gauge-dynamics-dominated QCD-like domain, as is also the case for QCD itself.

\subsubsection{Testing gauge-invariant perturbation theory for the $W$/$Z$ bosons}\label{sss:w}

The most simple object to test the gauge-invariant perturbation theory is the $w$ and the custodial triplet vector \cite{Maas:2012tj}. The reason is that the mass of the $w$ boson does not depend on the renormalization scale and scheme. However, one additional problem arises in this context. Because the $w$ is not a gauge-invariant state, its propagator in position space does not need to have the form \pref{lelevels}. In particular, at sufficiently short distances and weak coupling it will have the same type of (perturbative) shape as a gluon propagator. The gluon propagator has  non-positive contributions to its spectral function \cite{Bohm:2001yx} already in perturbation theory. Therefore it cannot be of the form of a sum of exponentials with positive coefficients. Still, at long distances it should have the behavior of an ordinary massive particle. Rather than just attempting to fit the $w$ propagator with \pref{lelevels}, is necessary to address these contributions directly. An alternative is to extract the mass from its momentum-space propagator, i.\ e.\ the Fourier-transform of \pref{lelevels} \cite{Maas:2011se}. At tree-level, it is just \pref{wprop}, and at least perturbative higher-order contributions can be quantified and included in the fit as well.

\begin{figure}
\includegraphics[width=\linewidth]{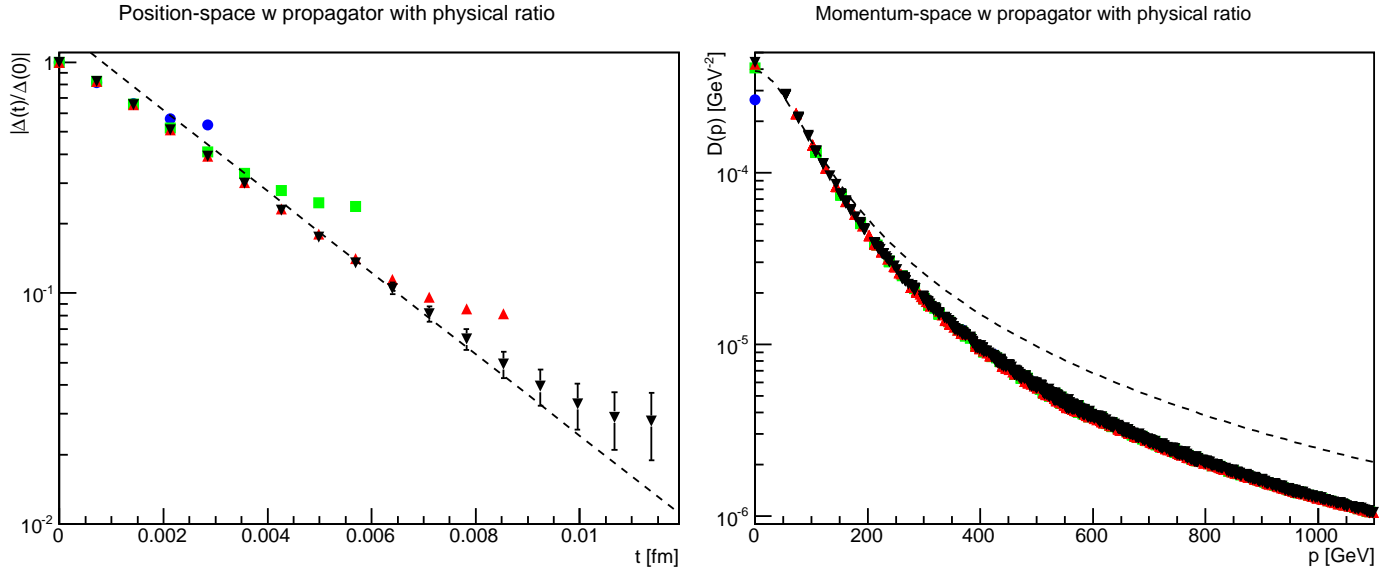}
\caption{\label{fig:w}The position-space propagator (left panel) and momentum-space propagator (right panel) for the $w$ for parameters where the mass ratio of the higgs and $w$ is about the physical one of $80.4/125\approx0.643$ ($\beta=2.4728$, $\kappa=0.2939$, $\gamma=1.036$). The dashed lines is the (renormalized) tree-level behavior \pref{wprop}. The different symbols correspond to different (lattice) volumes. From \cite{Maas:2013aia}.}
\end{figure}

Example results are shown in figure \ref{fig:w}. These are prototypical, and more can be found in  \cite{Maas:2013aia,Maas:unpublished,Maas:2012tj,Maas:2010nc}. First, it is visible that the $w$ propagator only weakly, and only at large momenta or short times, starts to deviate from the tree-level propagator. These are the expected perturbative corrections, which behave logarithmically with momentum \cite{Bohm:2001yx}. The deviations at long time in position space stem from the finite volume, deforming the exponential to a $\cosh$ behavior. As the position-space propagator approaches an exponential only asymptotically with volume, the effective mass \pref{effmass} will deviate from the actual mass. Because both perturbative and volume effects will reduce it, it will be underestimated, and needs to be extrapolated eventually to infinite volume. However, within statistical errors for lattices with reasonably large physical volumes the mass reaches the asymptotic limit up to a few GeV.

\begin{figure}
\includegraphics[width=\linewidth]{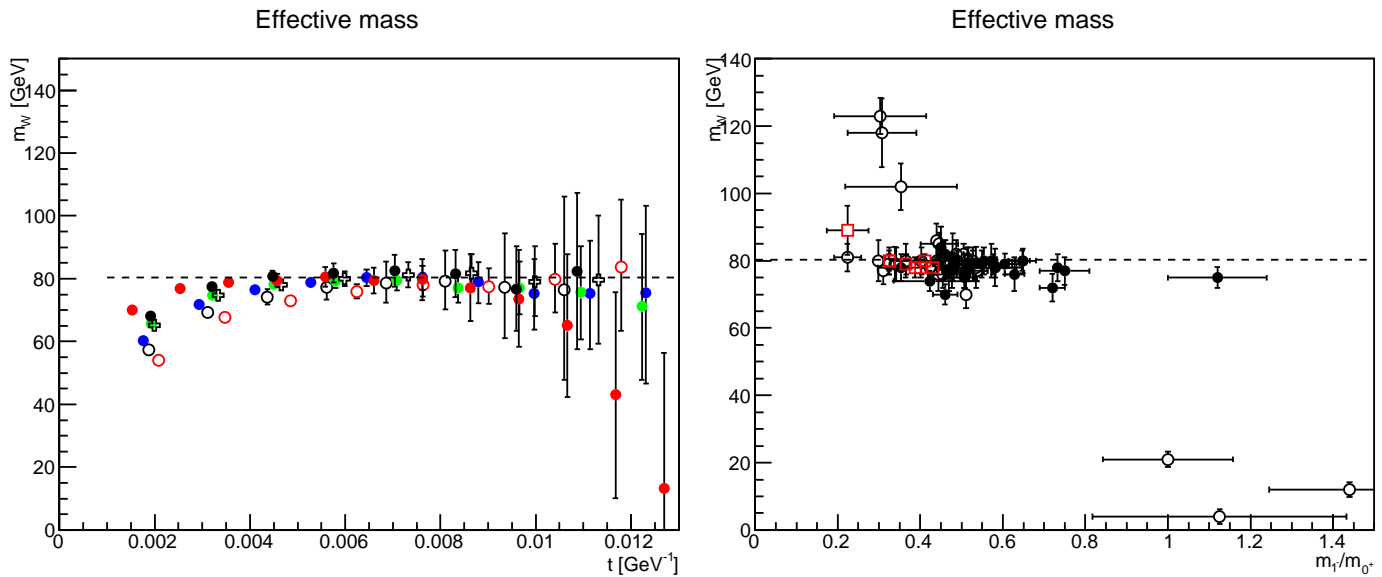}
\caption{\label{fig:wmass}The left panel shows the obtained effective masses \pref{effmass} of the $w$ for various ratios of the mass of the vector triplet to the mass of the scalar singlet. The right-hand side gives the mass obtained from the effective mass as a function of the mass ratio of the vector triplet to the scalar mass. Note that at 1/2 the elastic threshold is reached. Points below 1/2 have been obtained by taking the ratio with the first non-trivial level, but this position should be taken to be illustrative rather than actual. In the right-hand plot, the reliability of the data is indicated by the symbols, where filled symbols are most reliable, and open squares are least reliable. The reliability is assessed by the size of the discretization errors. Data from \cite{Maas:2013aia,Maas:unpublished}.}
\end{figure}

\begin{figure}
\includegraphics[width=\linewidth]{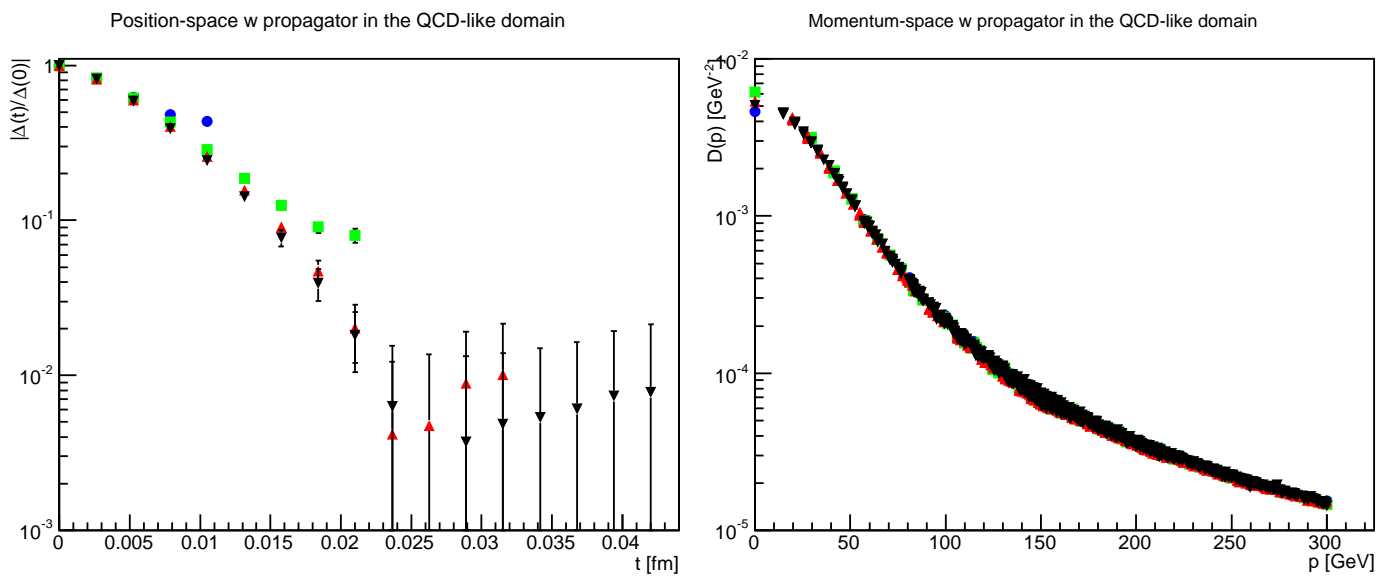}
\caption{\label{fig:w-qcd}The position-space propagator (left panel) and momentum-space propagator (right panel) for the $w$ for parameters in the QCD-like domain ($\beta=2.3724$, $\kappa=0.125$, $\gamma=0$). The different symbols correspond to different (lattice) volumes. Note that to reach sufficiently long times to show the characteristic zero-crossing of the position-space propagator a rather coarse lattice has been used, though this does not substantially affect this type of long-distance physics \cite{Maas:2011se}. Data from \cite{Maas:unpublished}.}
\end{figure}

The effective mass for various points in the Higgs-like domain are shown in figure \ref{fig:wmass}. It is seen that they always approach the mass of the vector triplet, up to the artifacts mentioned before. In this plot the extracted mass from the largest volume is shown as a function of the ratio of the vector triplet to the scalar mass. The points all cluster in the Higgs-like domain around the predicted mass of the vector triplet, even in those cases where the scalar is no longer stable. Only in the QCD-like domain, i.\ e.\ for ratios larger than one, the masses substantially deviate. However, in the latter case often the propagator of the $w$ exhibits a QCD-like behavior \cite{Maas:2011se} with no discernible exponential behavior at all, and no known association with any kind of mass pole. Such points are therefore not included. An example of this situation is shown in figure \ref{fig:w-qcd}.

The results are in pretty good agreement with the predictions of double-leading-order gauge-invariant perturbation theory. The few points, which are not, have rather large error bars, and therefore require further scrutiny. They are also all in regions where the mass of the scalar exceeds the elastic threshold, and furthermore their lattice spacing is comparatively coarse, giving rise to further systematic errors, which have not yet been fully accounted for. This supports strongly the FMS mechanism, and is a non-trivial field-theoretical test of it.

\subsubsection{Testing gauge-invariant perturbation theory for the Higgs boson}\label{sss:higgs}

Because the Higgs mass is not scheme-invariant \cite{Bohm:2001yx} it is not so simple to test the FMS mechanism non-perturbatively. As noted above, the best choice is probably to select a pole scheme, which needs to be defined in the complex momentum plane. On the lattice, this cannot be directly implemented, as this would require to analytically continue the propagator. Since in an actual simulation only a finite number of points is available, this is not possible unambiguously. 

Instead, here a prescription in the space-like domain will be employed \cite{Maas:2010nc,Maas:2016edk}. The renormalized propagator has the form \cite{Bohm:2001yx}
\be
D^{ij}(p^2)=\frac{\delta^{ij}}{Z(p^2+m_r^2)+\Pi(p^2)+\delta m^2}\nn,
\ee
\no where $m_r^2$ is the renormalized mass and $\Pi(p^2)$ is the unrenormalized self-energy. This self-energy is obtained from the unrenormalized propagator. The renormalization scheme is
\bea
D^{ij}(\mu^2)&=&\frac{\delta^{ij}}{\mu^2+m_r^2}\label{prc}\\
\frac{\pd D^{ij}}{\pd p}(\mu^2)&=&-\frac{2\mu\delta^{ij}}{(\mu^2+m_r^2)^2}\label{dprc},
\eea
\no which determines the renormalization constants $Z$ and $\delta m^2$ as
\bea
Z&=&\frac{2\mu-\frac{d\Pi(p^2)}{dp}(\mu^2)}{2\mu}\nn\\
\delta m^2&=&\frac{(\mu^2+m_r^2)\frac{d\Pi(p^2)}{dp}(\mu^2)-2\mu\Pi(\mu^2)}{2\mu}\nn.
\eea
\no In the following, always $\mu=m_r$ will be chosen. Thus, at space-like $p^2=\mu^2=m_r^2$ the propagator and its first derivative are forced to be the tree-level ones. If the analytic structure is such that on the circle $p^2=m_{\text{Pole}}^2$ a trivial rotation is possible, i.\ e.\ $m_\text{Pole}=m_r$, this is equivalent to the pole scheme.

This can be tested. Once the renormalization has been performed, it is possible to determine the renormalized position-space propagator by a Fourier transform. Then the effective mass \pref{effmass} can be determined, and thus the pole mass extracted. This has so far been only investigated in great detail in the quenched case, and thus for QCD-like physics \cite{Maas:2016edk}. There, it turned out that the pole scheme is not always equivalent to this scheme. In fact, it was found that the pole mass satisfies $m_{\text{Pole}}\ge m_\Lambda$, with $m_\Lambda$ some characteristic mass scale, even if $m_r\ll m_{\Lambda}$. However, if $m_r\gtrsim m_\Lambda$ it was found that $m_{\text{Pole}}\approx m_r$ \cite{Maas:2016edk}. Thus, only in the latter case both schemes seem to be equivalent when it comes to determining the pole mass.

Similar problems arise whenever an explicit solution in the complex plane is not possible. This also affects functional methods \cite{Benes:2008ir,Fister:2010yw,Gies:2015lia}, though here progress is made \cite{Strauss:2012as,Pawlowski:2015mia}. In semi-perturbative calculations \cite{Capri:2012cr,Capri:2017abz}, this is less of an issue, but here again access to the composite states is far less trivial. Thus, for now the scheme \prefr{prc}{dprc} will be used here.

\begin{figure}
\includegraphics[width=\linewidth]{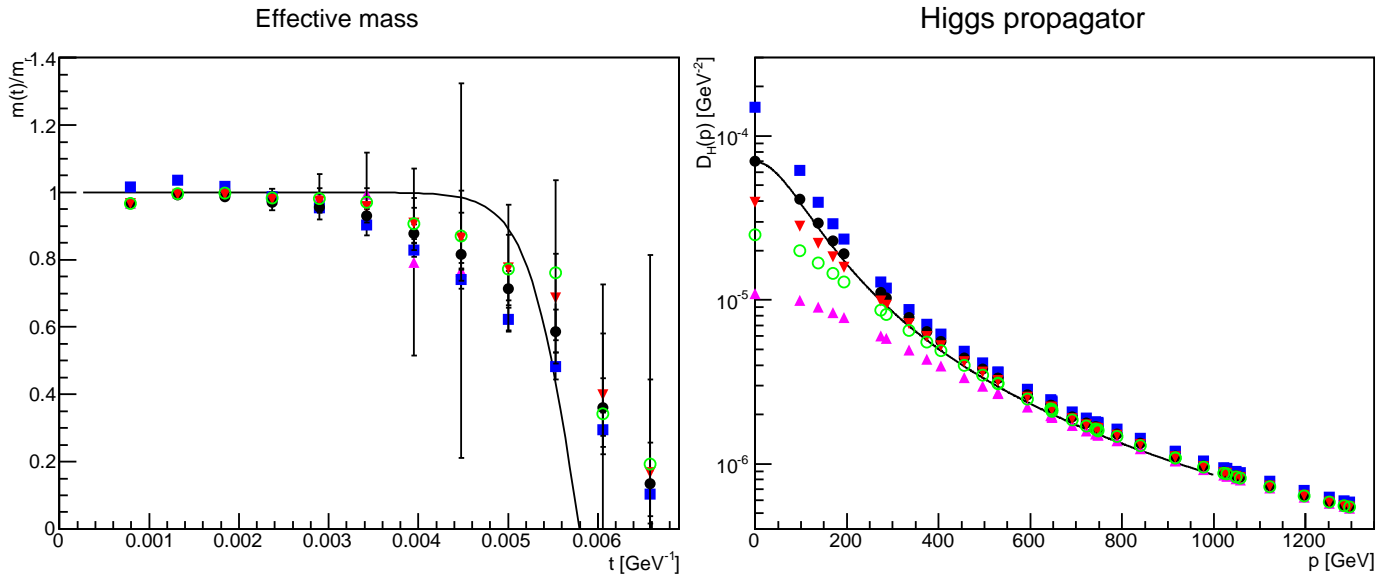}\\
\caption{\label{fig:h}An example of the effective mass in units of the renormalized mass (left panel) and propagator (right panel) for the higgs for various renormalized masses $m_r=80$, $120$, $160$, $200$, and $300$ GeV in the Higgs-like domain. The lattice parameters are the same as in figure \ref{fig:w}. Data from \cite{Maas:2013aia,Maas:unpublished}. The scalar singlet has a mass of 120 GeV. The solid line is the tree-level result for $m_r=120$ GeV, as are the solid black circles.}
\end{figure}

A result for various choices of $m_r$ is shown in figure \ref{fig:h} for the case that $m_{0^+}=120$ GeV. As is visible, only small deviations from the tree-level behavior are seen. More importantly, the ratio of the effective mass to the renormalized mass is essentially one, up to the expected lattice artifacts and perturbative short-range effects. While this needs much more scrutiny, this suggests that the pole scheme and this scheme are roughly equivalent. Thus, it is possible to find a scheme such that the poles agree.

Now it can be asked, whether this is suitable comparison. However, to what this actually has to be compared to is to the usual way of setting the renormalization scheme in perturbation theory. After all, what is tested here is gauge-invariant perturbation theory, which is a perturbative scheme. Thus, the question is, whether this leads to the same result as when using in perturbation theory the pole scheme. In that case, the pole mass is set by the experimentally observed peak in the scattering cross-section \cite{Bohm:2001yx,Einhorn:1992um}. Thus, the question is whether in a scattering process the same kind of peaks and poles arise. This will be discussed in section \ref{s:scattering}, as this is beyond a lattice question. The answer will be in the affirmative, thus justifying the FMS mechanism even in this case.

Conversely, choosing deliberately a different scheme will make the FMS mechanism fail. But this is actually not unexpected. After all, the FMS mechanism is a perturbative prescription. Just as a bad choice of scheme, and even of the renormalization point, will render perturbative predictions unreliable \cite{BeiglboCk:2006lfa,Baglio:2011wn,Baglio:2011hc}, this will certainly also apply to gauge-invariant perturbation theory. E.\ g., gauge-invariant perturbation theory will necessarily inherit the residual unphysical scheme-dependence at two loop order from ordinary perturbation theory.

Thus, the FMS mechanism can be made correct for the scalar and the higgs. As the situation in the quenched case discussed above shows, this is already a non-trivial statement.

An alternative test would be to determine the renormalization constants $Z$ and $\delta m^2$ using different quantities in the pole scheme, and then determine the higgs mass from the such renormalized higgs propagator. Then, the test would be also non-trivial. However, this will necessarily involve higher $n$-point functions, and appears to be out of reach of lattice simulations for computer time reasons within the foreseeable future.

\begin{figure}
\includegraphics[width=\linewidth]{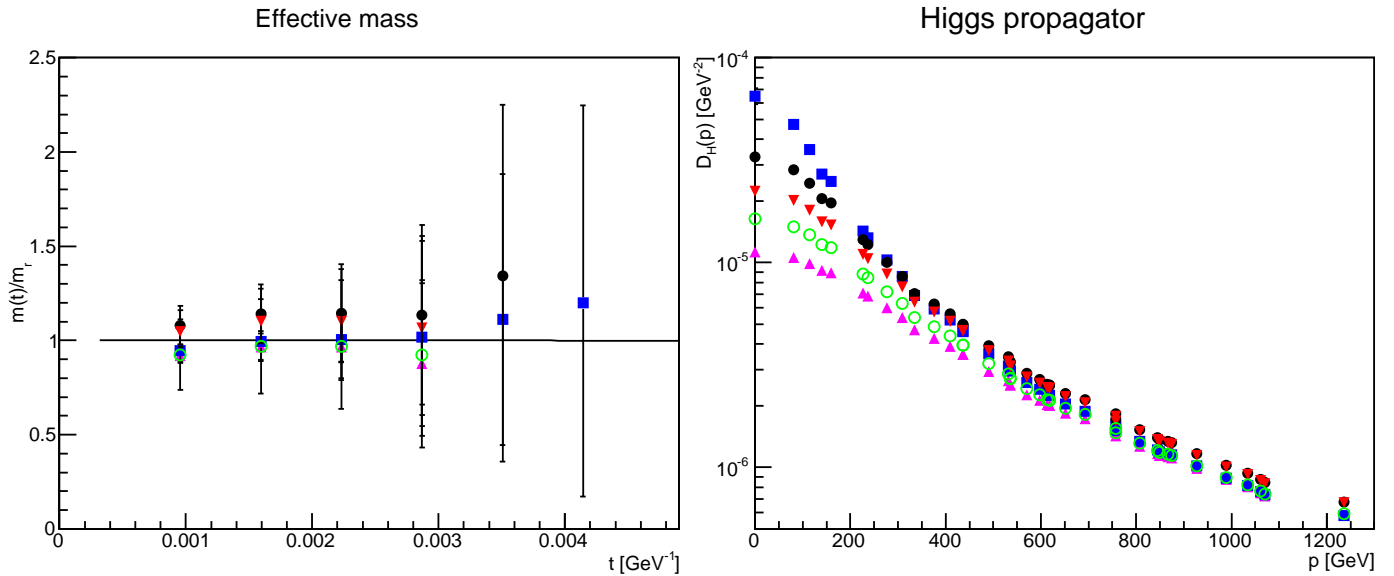}\\
\caption{\label{fig:h-qcd}An example of the effective mass in units of the renormalized mass (left panel) and propagator (right panel) for the Higgs for various renormalized masses $m_r=120$, $160$, $200$, $250$, and $300$ GeV in the QCD-like domain. The lattice parameters are $\beta=3.9282$, $\kappa=0.125$, $\gamma=0$. Data from \cite{Maas:2013aia,Maas:unpublished}.}
\end{figure}

The situation in the QCD-like domain is actually not too different from the Higgs-like domain \cite{Benes:2008ir,Fister:2010yw,Gies:2015lia,Maas:2013aia}, as is shown in figure \ref{fig:h-qcd}. In fact, the propagator shows a very similar behavior to the quenched case \cite{Maas:2016edk}. Given that a mass of the higgs arises always already perturbatively by loop effects, no matter the detailed physics \cite{Bohm:2001yx}, this is not too surprising. However, whether as in the quenched case also in the dynamical case an explicit minimum mass arises is not yet known.

\subsubsection{Other channels}\label{sss:other}

Besides the coincidence of the poles in the signal channels, the other prediction of gauge-invariant perturbation theory is that in the Higgs-like domain in all other channels only scattering states should appear. Such states are also expected in perturbation theory. Moreover, the prediction is that besides the poles corresponding to the higgs and the $w$ mass no additional poles in the scalar and triplet vector channel, respectively, should arise, except for scattering states. Or, more directly said, there are no non-trivial additional single-particle states.

\begin{figure}
\includegraphics[width=\linewidth]{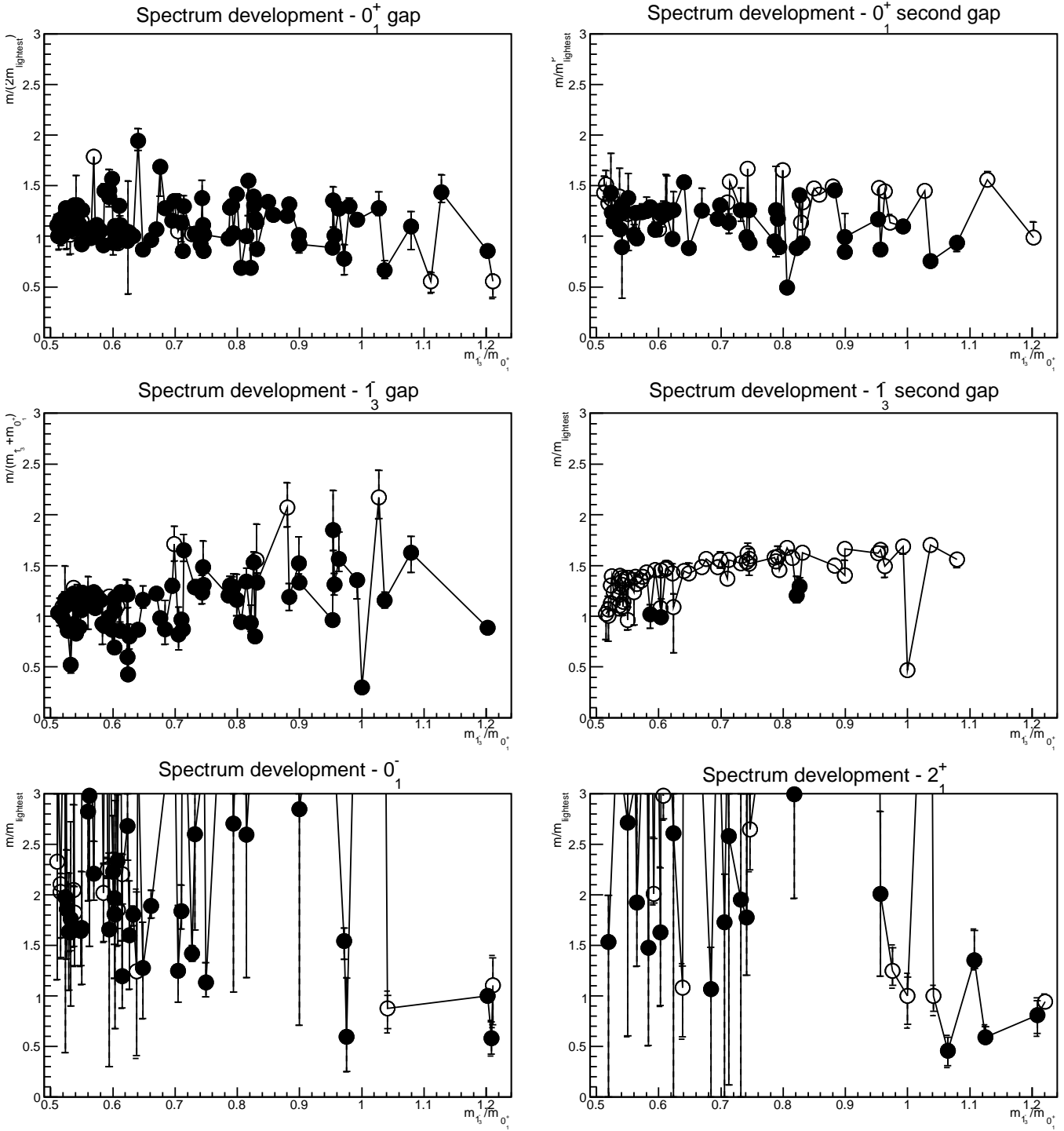}\\
\caption{\label{fig:spectrum}The development of various energy levels as a function of the mass ratio of the vector triplet mass to the scalar mass, divided by the expected value if the state is just a scattering state, or to the lightest mass in the spectrum, from \cite{Maas:2014pba,Maas:unpublished}. Note that at a ratio smaller than one-half the scalar ground state itself becomes a scattering state, and a ratio larger than one corresponds to being in the QCD-like domain. Open symbols are masses which in lattice units exceed one, and therefore are considered to be strongly affected by discretization artifacts. However, these errors have not yet been quantified, and the displayed errors are statistical only.}
\end{figure}

This question has been investigated in the Higgs-like domain in \cite{Maas:2014pba,Wurtz:2013ova}. Such investigations are quite expensive in terms of computing time. Therefore only at a few points in parameter space a very detailed investigation was possible \cite{Wurtz:2013ova}. Large-scale scans \cite{Maas:2014pba}, of which a sample is shown in figure \ref{fig:spectrum}, are thus not yet too reliable. Nonetheless, all results are consistent with the absence of low-lying resonances and/or bound states in other channels than the scalar singlet and vector triplet, in agreement with the predictions. Especially, the trivial scattering states, as e.\ g.\ \pref{higgspoles}, are explicitly seen. Thus, it is not a problem of seeing additional states. There are just no surplus states which are candidates for additional resonances or bound states (yet?).

The situation seems to change in the QCD-like domain, as is also shown in figure \ref{fig:spectrum} \cite{Maas:2014pba}. Here indications of additional stable bound states, especially in the singlet pseudoscalar channel, have been seen. In fact, in some cases the pseudoscalar singlet becomes almost degenerate with the scalar. At the same time no very light states in the singlet tensor channel $2^+$ have been observed. Thus, the spectrum seems to be qualitatively different from the pure Yang-Mills case, where the tensor is lighter than the pseudoscalar \cite{Mathieu:2008me}. Again, the reliability of this statement is not yet as good as desired.

In particular, in the QCD-like domain the only state investigated carrying a custodial quantum number, the vector triplet, becomes quite heavy. It would be very interesting to see whether this is true for all non-singlet states, since then a non-Yang-Mills spectrum arises but without any light open custodial contribution. This is especially interesting as the lightest non-singlet state is necessarily stable. Thus, this would correspond to a large scale separation between stable states.

\subsection{The standard model}\label{ss:sm}

The previous sections show that the pure Higgs sector can be well described using gauge-invariant perturbation theory. In the following it will be shown that this also works for the remainder of the standard model, starting with QED in section \ref{ss:qed}, and including flavor in section \ref{ss:flavor} and finally QCD in section \ref{ss:qcd}. However, for these sectors not yet any non-perturbative tests have been performed\footnote{Though note attempts like \cite{Aoki:1989xe}.}. Lattice simulations are possible for the Higgs sector and QED together \cite{Zubkov:2010np,Zubkov:2011sk,Zubkov:2011ia,Lucini:2015hfa,Shrock:1985un}, but have not yet been done for the purpose of testing the FMS mechanism. Including flavor faces the aforementioned problems of parity violation \cite{Hasenfratz:2007dp,Grabowska:2015qpk,Gattringer:2008je,Cundy:2010pu,Igarashi:2009kj}. Not to mention the enormous computing costs given the scale separations involved, which make already the bottom-top mass splitting difficult to cover even without the gauge interactions \cite{Gerhold:2007gx,Gerhold:2011mx}. Thus, it is likely that other methods will be necessary for a genuine non-perturbative test of larger fractions of the standard model, e.\ g.\ functional methods \cite{Maas:2011se,Alkofer:2000wg,Gies:2006wv,Kopietz:2010zz,Binosi:2009qm,Boucaud:2011ug,Roberts:2015lja}.

Thus, the best test to date of gauge-invariant perturbation theory is the fact that to leading order in the vacuum expectation value it agrees with standard perturbation theory, and therefore with experimental results \cite{pdg}. But this assumes subleading contributions in the vacuum expectation value are small, which needs to be tested. This is, e.\ g., possible against lattice data for the Higgs sector alone \cite{Raubitzek:unpublished,Maas:unpublished} or by evaluating them and testing against experiment \cite{Egger:2017tkd,Fernbach:unpublished}. This would be simultaneously a test of the underlying quantum field theory of the standard model, because exact gauge-invariance in the standard model is the only driving force behind all of the presented material. Thus a failure to comply with predictions, provided they are theoretically sufficiently under control, could likely hint at a problem with the field-theoretical description of the standard model.

\subsubsection{QED and mixing}\label{ss:qed}

Adding in hypercharge, and thus ultimately QED, makes the situation somewhat more involved. One of the reasons is that hypercharge, as an Abelian gauge theory, yields an observable charge \cite{Haag:1992hx}. The observable charged states are obtained by dressing the gauge-dependent states with a Dirac phase. This structure is recovered for the bound states by the FMS mechanism \cite{Frohlich:1980gj,Frohlich:1981yi,Shrock:1985ur,Shrock:1986av}.

This works because of the direct product structure of the weak and hypercharge interaction, and because of the structure of the standard model. Consider how custodial symmetry and hypercharge are linked. The kinetic term of \pref{la:hsx},
\be
\frac{1}{2}\tr\left((D_\mu x)^\dagger D^\mu x\right)\nn,
\ee
\no is manifestly invariant under custodial symmetry. The Higgs field transforms like $\phi\to g\phi=\exp(i\alpha(y))\phi$ under a hypercharge gauge transformation $g(y)$. However, $x$ does not transform linearly under a hypercharge transformation, but by $x\exp(i\alpha(y)\tau^3)$, explicitly demonstrating that hypercharge is nothing but gauging the U(1) subgroup of the SU(2) custodial symmetry. Since SU(2)/U(1) is merely a coset, but not a group, there is no remainder global symmetry group left. Only because an Abelian gauge symmetry allows for the existence of an (almost) global U(1) group, the electric charge remains to classify states.

Consider the observable particles. Neither for the physical $W^\pm$ nor the Higgs anything changes, as they are either just linear combinations or unchanged compared to the previous case. Their operators are always mapped to the elementary $w^\pm$ and higgs. Thus, gauge-invariant perturbation theory yields the same results for them as before. In fact, despite their name, the $W^\pm$ bosons are not even charged under hypercharge, and thus need no dressing factors. Just as in the perturbative treatment of the $w^\pm$ \cite{Bohm:2001yx} their apparent electric charge comes from the interaction with the $w^3$ component. This arises from the mixing of the physical $W^3$ with the physical hypercharge boson to yield the physical $Z$ and the physical photon $\Gamma$.

For this mixing, the situation is more involved, but follows the sames rules as before. Gauge-invariant equivalent states can be constructed as
\bea
\op_Z&=&\sin\theta_Wd\op_{1^-_3}^3+\cos\theta_W Db\approx \sin\theta_W w^3+\cos\theta_W b+\op(v^{-1})=z+{\cal O}(v^{-1})\label{opz}\\
\op_\Gamma&=&\sin\theta_Wd\op_{1^-_3}^3-\cos\theta_W Db\approx \sin\theta_W w^3-\cos\theta_W b+\op(v^{-1})=\gamma+{\cal O}(v^{-1})\label{opg}
\eea
\no where $D$ and $d$ are the necessary Dirac projector and phase \cite{Lavelle:1995ty}, respectively, $b$ is the hypercharge gauge field, and $\theta_W$ is the usual Weinberg angle \cite{Bohm:2001yx}, and a factor $v^2$ has been absorbed in the relative normalization of $w^3$ and $b$. It should be noted that both have the same quantum numbers, i.\ e.\ gauge-invariant vectors with no conserved quantum numbers, because $d\op_{1^-_3}^3$ was the custodial charge multiplet member with third component zero. The rules of section \ref{ss:rules} map in gauge-invariant perturbation theory these operators to the usual elementary ones, the elementary $z$ boson and the elementary photon $\gamma$, with the usual approximations for the Dirac projector and phase \cite{Lavelle:1995ty}. Thus, the corresponding propagators also have the correct pole structure. As the $Z$ and $\Gamma$ can mix, this is the first example where operator mixing and two singe-particle states appear in the same quantum number channels, as was anticipated in section \ref{sss:ss}. The basis \prefr{opz}{opg} represents the mass eigenbasis of this channel.

Moreover, consider the three-point function for the charged $W$ bosons and Photons. Expanding it in according to the rules of section \ref{ss:rules} yields in gauge-invariant perturbation theory
\bea
\la\left(\op_{1_3^-}^+\right)_\mu\left(\op_{1_3^-}^-\right)_\nu\left(\op_\Gamma\right)_\rho\ra&\approx& v^4\la w_\mu^+w_\nu^-\gamma_\rho\ra+\op(v^3)\label{electric}\\
&=&v^4\left(\sin\theta_W\la w_\mu^+w_\nu^-w^3_\rho\ra+\cos\theta_W\la w_\mu^+w_\nu^-b_\rho\ra\right)+\op(v^3)\nn\\
&\stackrel{\mathrm{tree-level}}{\approx}&v^4\sin\theta_W\la w_\mu^+w_\nu^-w^3_\rho\ra_{\mathrm{tl}}\nn.
\eea
\no The left-hand-side is a combination of three separately gauge-invariant quantities. Following the rules discussed in section \ref{ss:gipt}, each of them has been expanded separately. This used that the operator \pref{op1t} in the standard choice \pref{colfmix} expands like
\be
\tr\tau^a x^\dagger D_\mu x=v^2w_\mu^b\tr\tau^a\tau^b+\op(v)=w_\mu^a+\op(v)\nn.
\ee
\no Hence, to leading order, the physical $W^+W^-\Gamma$ vertex is given by the perturbative one $w^+w^-\gamma$. As in perturbation theory \cite{Bohm:2001yx}, this vertex comes at tree-level actually from the SU(2) vertex due to the $w^+w^-w^3$ interaction only, combined with a factor of $\sin\theta_W$, thus reproducing the usual electric charge as coupling constant\footnote{The prefactor $v^4$ seems to deny this possibility at first. However, at leading order in $v$ always the same combinations of $v$ will appear, and thus a common factor can be absorbed in the renormalization process. This will only start to play a role when subleading terms in $v$ are included, and relative weights become important.}. This explicitly shows that the would-be electric charges of the physical $W^\pm$ are only a mapping of custodial indices to gauge interactions.

In the same manner also the other vertices arise. Thus, the electromagnetic interaction is, up to the level of higgs fluctuations, not modified when using the observable particles instead of the elementary degrees of freedom. Thus, in the same sense as at the perturbative level, the $W^\pm$ are the electrically charged vector operators, and the two operators \pref{opz} and \pref{opg} are electrically uncharged vector singlets. The spectrum and interactions observed in experiment is recovered.

The usual singlet scalar, that is the physical Higgs, remains a singlet with respect to electric charge. It would now be possible to construct a non-singlet, gauge-invariant scalar operator, e.\ g.\ by a combination of a $Z$ and a $W^\pm$. However, these states expand to scattering states, and thus yield no additional states in the physical spectrum. In particular, they do not expand to, e.\ g., charged Goldstone bosons, as the contribution of Goldstones vanish in the expansion because they are BRST non-singlets.

This picture has so far only be exploratory investigated on the lattice, and only with respect to the spectrum. But these investigations indeed indicate the emergence of massless photons in the prescribed way \cite{Shrock:1985ur,Shrock:1985un,Lee:1985yi}, supporting the FMS mechanism for the electroweak sector.

In other theories, this does not need to be true, as will be discussed in section \ref{ss:bsm}. Moreover, the subleading corrections in \pref{electric}, and in other vertices, imply that at some point deviations should become relevant and modify the perturbative results. This would show up eventually in the experimental searches for anomalous gauge couplings \cite{Gounaris:1996rz,Baak:2013fwa}, as long as the reference gauge couplings are the perturbative ones. Of course, they are not really anomalous and, e.\ g., will not be a sign of new physics. Especially, any unitarity violation at the perturbative level due to these corrections \cite{Bohm:2001yx} would again be non-perturbatively compensated. Indeed, first exploratory investigations on the lattice support the existence of such subleading contributions \cite{Raubitzek:unpublished}.

\subsubsection{Flavor symmetries}\label{ss:flavor}

The situation becomes substantially more involved when introducing fermions. For simplicity, ignore for now the Yukawa couplings as well as hypercharge. Also, consider only a single generation of left-handed leptons. The rest will be added later.

In the standard model, these leptons are doublets under the weak interactions \cite{Bohm:2001yx}. Therefore, these fermions can again not be observable particles, as they depend on the gauge \cite{Frohlich:1980gj,Frohlich:1981yi}. Gauge-invariant states can be constructed for the fermion spinor $\psi$ describing these leptons by a dressing with a Higgs field \cite{Frohlich:1980gj,Frohlich:1981yi,Egger:2017tkd},
\be
\mathcal{O}_\Psi(y)= (x\epsilon)^{\dagger}(y)\psi(y)\label{opferm},
\ee
\no where $\epsilon$ is the antisymmetric tensor of rank two\footnote{Alternatively either the assignment of the hypercharge values of the fermions can be reversed, or the convention of \cite{Egger:2017tkd} can be used to avoid $\epsilon$.}. This is by construction gauge-invariant, and remains a (left-handed) fermion. It is important to note that every component of the operator $\op_\Psi$ is a left-handed spinor. Because $x^\dagger$ transforms by a left-multiplication under the custodial symmetry, the state is a custodial doublet. Thus, the weak gauge charge, which is perturbatively associated with the up/down or lepton/neutrino distinction \cite{Bohm:2001yx} is replaced by a custodial charge for the physical states. In fact, the same charge which distinguishes the physical $W^\pm$ and $Z$ now distinguishes also the components of a physical fermion. These two custodial charge states are the physically observable fermions, e.\ g.\ the Electron $E$ and the Electron-Neutrino $N$.

To identify their connection to the elementary fermions, consider the electron-neutrino case. Denoting the two fundamental weak states $\psi^1$ and $\psi^2$ by $e$ and $\nu$, the rules of section \ref{ss:rules} yield \cite{Frohlich:1980gj,Frohlich:1981yi,Egger:2017tkd}
\be
\mathcal{O}_{NE}=\bpm N\cr E\epm=(x\epsilon)^\dagger \bpm \nu \\ e \epm = \bpm \phi_2 \nu  -  \phi_1 e  \\  \phi_1^* \nu + \phi_2^* e  \epm\approx v\bpm \nu\cr e\epm+{\cal O}(v^0)\label{opfermexp}
\ee
\no Thus, the usual electron and neutrino appear to leading order as the custodial eigenstates. Likewise, forming the propagator from $\op_{NE}$ this implies that the two custodial states expand to the tree-level propagators of the electron and the neutrino, and therefore have the same mass as the perturbative electron and neutrino. For this identification it is absolutely crucial that the Higgs is a scalar particle, as in any other case this would alter the spin-parity quantum numbers of the fermions. Thus, the perceived flavor symmetry of left-handed fermions needs to be identified with the custodial symmetry SU(2)$_c$. In addition, there is a global U(1)$_l$ fermion number symmetry, which only affects the fermions. Thus, at this point there is a global U(2)$_L\sim$SU(2)$_c\times$U(1)$_l$ symmetry.

An extension to three generations\footnote{The following could possibly also have quite interesting implications for the generation structure of the standard model. Since this is highly speculative, this will not be reviewed here, but can be found in \cite{Egger:2017tkd}.} is straightforward, as a generation index would be just pushed through to the bound state, and pushed back by the expansion. This gives an SU(3)$_g$ additional global symmetry. Note that the total global symmetry is then SU(3)$_g\times$U(2)$_L$.

Return to the one-generation case. So far, there is no way to generate a tree-level mass for the fermions. This requires the Yukawa interaction. Introduce two right-handed additional fermions, a right-handed electron and a right-handed neutrino\footnote{Here for simplicity, as then the quark and lepton sector work in the same way, the existence of a right-handed neutrino will be assumed. If the neutrino sector is different from this minimal version, this would require corresponding amendments to the construction.}, not interacting weakly. Without any further distinguishing feature, this implies an additional U(2)$_R\sim$SU(2)$_\text{flavor}\times$U(1)$_\text{number}$ flavor and right-handed counting symmetry. Thus at this point the global symmetry of the theory is SU(2)$_c\times$U(1)$_l\times$U(2)$_R$=U(2)$_L\times$U(2)$_R$.

Due to the gauge symmetry at the Lagrangian level no conventional mass-term is allowed for the fermions \cite{Bohm:2001yx}. By assumption, there are right-handed neutrinos, and thus Majorana masses are also not an option. Combining the two right-handed leptons in a flavor spinor $\Psi_R$, a possibility is a Yukawa interaction
\be
\La_Y=g_Y\left(\bar{\op}_{NE}\Psi_R-\bar{\Psi}_R\op_{NE}\right)\nn,
\ee
\no with $g_Y$ the Yukawa coupling. Because this locks the symmetries, this breaks the custodial and flavor symmetries down to a diagonal subgroup U(2)$_d$. This diagonal subgroup can be identified with the  conventional, physical Flavor group. This group can be further broken down to U(1)$_E\times$U(1)$_N$, separate counting symmetries of both electrons and neutrinos, by introducing a matrix $\Omega$, which is not invariant under an SU(2) transformation, as
\be
\La_Y=g_Y\left(\bar{\op}_{NE}\Omega\Psi_R-\bar{\Psi}_R\Omega^\dagger\op_{NE}\right)\label{yukawa}.
\ee
\no If the matrix $\Omega$ is diagonal $\diag=(g_\nu/g_Y,g_e/g_Y)$, this gives two independent interaction terms
\be
\La_Y=g_\nu(\phi_2\bar{\nu}_L-\phi_1 \bar{e}_L)N_R+g_e(\phi_1^*\bar{\nu}_L+\phi_2^*e_L)E_R+\text{h.c.}\nn,
\ee
\no which for $x\approx v1$ are nothing but the usual mass terms for the neutrino and the electron of the standard model \cite{Bohm:2001yx}. Then, the tree-level propagators obtain their usual mass \cite{Bohm:2001yx}, and by virtue of the FMS mechanism so does the propagator of the bound state \pref{opferm}. If $\Omega$ is not diagonal, a change of the basis in the fermion fields can make it so, at the expense of introducing a matrix in the weak interactions, in the same way as the CKM/PMNS matrices introduce such an effect at the level of generations \cite{Bohm:2001yx}. Furthermore, in the presence of multiple generations, intrageneration mixing can be absorbed as usual also in the CKM/PMNS matrices.

Adding further generations proceeds almost as before. However, so far this upgrades the right-handed flavor symmetry to an (S)U(6) symmetry, which is then explicitly broken by the Yukawa terms by selecting individual flavors for the interactions with the three left-handed generation, in contrast to the generation structure before. The global generation structure for left-handed fermions and right-handed fermions can therefore still differ. The same global generation structure for the left-handed and the right-handed symmetry will be enforced by the hypercharge.

Before proceeding, return to the case of a single left-handed doublet. In section \ref{ss:qed} it was pointed out that the hypercharge is actually gauging the custodial symmetry. Thus the composite state \pref{opferm} would apparently already transform under a hypercharge transformation. Now, in addition the fermions become coupled to the hypercharge, in principle with an arbitrary value. But actually, this has to happen with the same hypercharge for both members of the elementary fermion doublet, as it would otherwise break the weak gauge symmetry. Thus, $\psi$ transforms as $\psi\to g\psi$. Because the hypercharge gauge transformation acts like $\diag(g,g^\dagger)$ on $x$ this yields a total transformation like $\diag(1,g^2)$ for the state \pref{opferm}. With the usual hypercharge assignment for the elementary fields, $g^2$ is equal to a transformation with twice the hypercharge value, as it is an Abelian group. This shows explicitly that the physical Neutrino state is uncharged under the hypercharge, while the physical Electron is charged, with twice the hypercharge of the elementary electron.

The actual electric charge is then, as in \pref{electric}, obtained by considering the expansion of $\la\bar{\op}_\Psi\op_\Psi\op_\Gamma\ra$. This expands for the various combinations of custodial indices to the usual tree-level interactions, yielding the correct assignment of the electromagnetic charges to the various states \cite{Bohm:2001yx}.

The only thing left is the assignment of hypercharge to the right-handed fermions. To make \pref{yukawa} gauge-invariant enforces the standard assignment \cite{Bohm:2001yx}, thereby breaking the right-handed flavor symmetry already down to U(1)$^3$, or SU(3)$\times$U(1)$^2$ for three generations, as the right-handed neutrino and electron cannot have the same hypercharge value. Therefore the breaking pattern for left-handed and right-handed flavor due to the presence of hypercharge is different, as in one case the custodial symmetry is actually gauged, while in the other case the assignment of the hypercharges breaks the generation symmetry explicitly. Still, it leads to the same remaining global symmetry in the left-handed and the right-handed sector\index{This is probably the most baroque structure required to have a certain effect the authors has ever witnessed.}. Note that differing hypercharge assignments to the differing generations would be compatible with all results. They would merely introduce a different breaking pattern of the generations, though would prevent the introduction of intergeneration transitions except for very special relative values of the hypercharges, just as in perturbation theory.

The quark sector follows exactly the same structure. From the point of view of the electroweak and Higgs interaction, it is just a copy of the lepton sector, and operates, but for different values of the couplings, just like three more generations. But it is distinguished from it by color. Color has no bearing on the FMS mechanism or any of the present results. However, reversely the change of the meaning of flavor and the global symmetry structure of the theory have implications for QCD, as will be discussed in the next section \ref{ss:qcd}.

Before this, it should be noted that also the anomaly cancellation is not altered at all. After all, the anomaly arises from the path integral measure \cite{Bohm:2001yx,Bertlmann:1996xk} which involves the elementary fields only. The observable, physical states, and the FMS expansion acting on it, are merely expectation values. They therefore never interfere with the mechanism for the intrageneration anomaly cancellation.

There exists exploratory lattice simulations of this theory with vectorial fermions \cite{Lee:1987zu,Aoki:1988aw}, though yet without detailed spectroscopy. In this case it can be expected that the FMS mechanism works in the same way, though with operators representing this vectorial symmetry. Note also that if the BEH effect would be switched off, this is the situation encountered in the Abbott-Farhi model \cite{Abbott:1981re}, except that the fermions would be vectorially coupled instead of chirally. This situation is also relevant to so-called partial compositness models \cite{Kaplan:1991dc,Sannino:2016sfx}. In this case, the FMS mechanism is not applicable, and the theory can manifest also other phases. Such theories have also been subject to exploratory lattice simulations \cite{Lee:1987zu,Aoki:1988aw,Aoki:1988fg,Hansen:2017mrt}, and may host many interesting phenomena \cite{Lee:1988ut,Hsu:1993zc}.

\subsubsection{QCD and hadrons}\label{ss:qcd}

As noted before, with respect to the electroweak interactions the same applies to quarks as to leptons. Considering only low-energy QCD, it is possible, because of \pref{opfermexp}, to first apply the FMS mechanism, to obtain the standard quarks, and then reduce to QCD alone. This leads to the usual QCD, and nothing changes. Thus, QCD in isolation emerges as a low-energy effective theory in the same way as in a purely conventional treatment of the standard model. This is therefore not particularly interesting in the present context.

However, this will only be correct as long as it is good to keep just the leading term of \pref{opfermexp}. As discussed repeatedly, this may only be good for energies below which Higgs production is not appreciable and/or virtual Higgs corrections are negligible. Since modern experiments probe well into this region, it is worthwhile to have a closer look at how QCD operates within the context of the full standard model.

Still, even in the full standard model confinement is operative\footnote{It is already in QCD alone not a trivial issue to even phrase the question what confinement is precisely, much less how it operates \cite{Greensite:2011zz,Greensite:2017ajx,Greensite:2018mhh}.}, as has been discussed at the beginning of this section. In the present subsection, confinement will be used in the operative sense that quarks and gluons are necessarily bound inside hadrons in such a way that the hadron is gauge-invariant with respect to color. There are thus three type of gauge-invariant operators describing hadrons, which need to be considered in the following.

One are those involving only gluons, the glueballs. Since no other charges are involved, they work in precisely the same way as in pure QCD. Most importantly, they are always 'flavor' singlets, and do not carry electric charge.

The second type are operators involving quarks, but with the same quantum numbers as glueballs. Therefore, they do not carry flavor and/or electric charge. With respect to the electroweak sector they are therefore uncharged, and thus completely gauge-invariant. Also for these nothing changes. An example of this is the $\sigma$ meson, which carries the quantum numbers of the vacuum,
\be
\op_\sigma=\bar{u}_L\sigma u_L+\bar{d}_L\sigma d_L+\bar{u}_R\sigma u_R+\bar{d}_R\sigma d_R\label{sigma}
\ee
\no where the matrix $\sigma$ transforms the Weyl spinors from left-handed to right-handed, and thus acts only in Dirac space to create Lorentz scalars \cite{Aitchison:2007fn}. Thus, this operator is a singlet for all global symmetries. Another example is the $\omega$ meson, which is a vector singlet. It should be noted that such operators mix with all singlet operators, like the (physical) Higgs. Fortunately, this mixing seems to be small experimentally \cite{pdg}, so this subtlety can be ignored here.

A change happens for the third type of operators, which is any operator not belonging to either of the two previous classes. Thus, these are all flavor-non-singlet operators from the perspective of pure QCD \cite{Egger:2017tkd}. Consider for the moment the one-generation case only, and ignore both hypercharge and the Yukawa interactions. Then an eight-component spinor can be constructed as
\be
\Psi=\bpm (x\epsilon)^\dagger\psi_L\cr U_R \cr D_R\epm\label{qcdquark},
\ee
\no where the Weyl spinors are not detailed further. Note that these states are denoted by capital letters as they are gauge-invariant with respect to the weak interaction, but not yet gauge-invariant with respect to the strong interaction.

QCD constructed from theses spinors is invariant under a global symmetry of SU(2) $\times $SU(2)$\times $U(1), dropping another U(1) due the axial anomaly \cite{Bohm:2001yx}. The first SU(2) acts only on the first two components, the custodial symmetry, and the second one only on the right-handed flavor symmetry. Since the custodial and right-handed flavor symmetry are not explicitly broken, the left-handed and right-handed part do not mix under the symmetry transformation. Thus, this symmetry replaces the usual chiral and flavor symmetry of QCD. By a change of basis it is possible to reestablish this symmetry, but this has also impact on the weak gauge boson physical particles, which are charged under the custodial symmetry.

Consider now a state like a pion, which has the structure
\be
\pi^a=\bar{\Psi}\tau^a\gamma_5\Psi\nn,
\ee
\no where the Dirac representation is chosen for the $\gamma$ matrices. This operator mixes the various states, resulting in
\bea
\pi^{+}&=&\bar{D}_R ((x\epsilon)^\dagger\psi_L)^u+\overline{((x\epsilon)^\dagger\psi_L)}^dU_R\label{pi1}\\
\pi^{-}&=&\bar{U}_R ((x\epsilon)^\dagger\psi_L)^d+\overline{((x\epsilon)^\dagger\psi_L)}^uD_R\\
\pi^0&=&\bar{D}_R ((x\epsilon)^\dagger\psi_L)^d+\overline{((x\epsilon)^\dagger\psi_L)}^dU_R-\bar{U}_R ((x\epsilon)^\dagger\psi_L)^u-\overline{((x\epsilon)^\dagger\psi_L)}^uU_R\label{pi3},
\eea
\no where the indices $u$ and $d$ identify the components of the spinors $(x\epsilon)^\dagger\psi_L$. The first remarkable insight is that, in contrast to the $\sigma$ meson \pref{sigma}, such operators always involve three valence particles, the two quarks and a higgs in form of $x$. This is necessary as otherwise the mixing of left-handed and right-handed particles to create a pseudo-scalar would not be possible gauge-invariantly.

Of course, applying the rules of gauge-invariant perturbation theory of section \ref{ss:rules} will reduce \prefr{pi1}{pi3} immediately to the usual operators of QCD to leading-order in $v$, reestablishing the pions as quark-anti-quark bound states, but multiplied with the Higgs vacuum expectation value. But this is a statement in a fixed gauge. The genuine gauge-invariant expression inside the standard model has a valence higgs besides the valence quarks. This is remarkable for multiple reasons. One is that the mass defect is actually large. But this also implies that to excite the higgs substructure will require a substantial amount of energy, and the corrections at the sub-leading order are generically small. Whether this could be probed in actual experiments will be discussed in more detail in section \ref{s:scattering}.

Concerning global symmetries, the pions therefore mix already the symmetries, and the operators \prefr{pi1}{pi3} are neither custodial nor right-handed flavor eigenstates. If both would be separately conserved, any pion current would be needed to be decomposed into the two component currents. But this becomes irrelevant as soon as the Yukawa couplings are turned on, as both symmetries are then explicitly broken, and can mix freely. This explicit breaking also translates into the usual explicit breaking of chiral and flavor symmetry in the stand-alone QCD and/or the FMS expansion to QCD.

The introduction of hypercharge emphasizes this even more. Because the left-handed quarks and the right-handed quarks have a different hypercharge just building a Dirac spinor from them is not possible in a gauge-covariant way. However, by the combination with the higgs field in \pref{qcdquark} having also hypercharge, \pref{qcdquark} transforms in a gauge-covariant fashion, and in particular allows the operators \prefr{pi1}{pi3} to have a well-defined electric charge. Note that any other relative assignments of hypercharges to the higgs and the quarks would not create suitably electrically charged hadrons. That this involves also the higgs and not just the quarks in stand-alone QCD makes the particular assignments of hypercharges to the elementary particles of the standard model even more mysterious than it already is.

All of this becomes much more dramatic for baryons \cite{Egger:2017tkd}. Because any baryon has an odd number of quarks, it is impossible to couple left-handed fermions together such that they form at the same time a gauge-invariant state under the strong and the weak interaction. Therefore, any baryon involving a left-handed component necessarily involves a valence higgs. Moreover, as hadrons are essentially parity eigenstates \cite{pdg,Gattringer:2010zz,DeGrand:2006zz}, any baryon contains such a left-handed component.

Consider for a moment just a system of left-handed fermions. Then \cite{Egger:2017tkd}
\bea
N_L&=&\epsilon^{IJK} c_{ijkl} q_i^I q_j^J q_k^K (x\epsilon)^{\dagger}_{\tilde{i}l}\label{protonl}\\
\Delta_L&=&\epsilon^{IJK} q_i^I q_j^J q_k^K (x\epsilon)^{\dagger}_{\tilde{i}i} (x\epsilon)^{\dagger}_{\tilde{j}j} (x\epsilon)^{\dagger}_{\tilde{k}k}\label{deltal}
\eea
\no form two possible three-quark operators, which are gauge-invariant under both the strong interactions, denoted by capital indices, and the weak interaction. The operator \pref{protonl} corresponds to cases which have one open custodial index, like the nucleons, while the operator \pref{deltal} corresponds to cases with three open custodial indices, like the $\Delta^{++}$ and the $\Delta^{-}$. The matrix $c$ depends on the other properties of the baryon in question. E.\ g.\ for a proton it becomes \cite{Egger:2017tkd}
\be
c_{ijkl}^p=a_1\epsilon_{ij}\delta_{kl} + a_2\epsilon_{ik}\delta_{jl} + a_3\epsilon_{jk}\delta_{il}\nn
\ee
\no while for the $\Delta^{++}$ it suffices to set in \pref{deltal} $\tilde{i}=\tilde{j}=\tilde{k}=1$.

Returning to the full theory with both left-handed and right-handed quarks, a nucleon operator is, e.\ g., constructed as \cite{Gattringer:2010zz}
\be
N^{rst}=\epsilon^{IJK}F^{rst}_{uvw}\Psi_{Iu}\left(\Psi^T_{Jv}C\gamma_5\Psi_{Kw}\right)\label{nucleon}
\ee
\no where $F$ acts in custodial/right-handed flavor space and the charge-conjugation matrix is $C=i\gamma_2\gamma_0$. The expression in parentheses is a scalar diquark, and as such mixes left-handed and right-handed components. The fermionic nature is carried entirely by the first quark field. To obtain definite parity states requires projection with a parity operator \cite{Gattringer:2010zz}, essentially $(1\pm\gamma_0)/2$. Since parity transforms left-handed to right-handed \cite{Itzykson:1980rh}, a parity eigenstate necessarily contains both left-handed and right-handed components. Therefore, in the full standard model there is also a higgs valence contribution in the nucleon. Actually, because the scalar diquark combines products of left-handed and right-handed components of the $\Psi$ field, the proton, being a positive parity state, is a mixture of a state with one and three valence higgs fields.

It is, of course, possible to write down operators with definite valence higgs contributions. However, these are no longer mass eigenstates, and therefore are only of limited use when considering the low-energy limit of stand-alone QCD.

The discussion here used only a particular set of operators for the hadrons. Of course, all operators of the same quantum numbers contribute, and therefore also operators with more fields. However, because of gauge-invariance and the particular quantum numbers they will always involve some higgs component. Thus, for all hadrons, which cannot mix with the glueballs, there is necessarily a valence higgs contribution\footnote{Especially, a valence higgs, in addition to also present sea higgs \cite{Bauer:2017isx,Bauer:2018arx,Bauer:2018xag}. This also implies that valence particles always form full multiplets of the weak interaction.}. The FMS expansion shows that these states have, to leading order in $v$, just their usual masses. However, at some point, the valence higgs contribution should show. Why this has (not yet) any experimental consequences will be discussed in section \ref{s:scattering}, and whether it could eventually be observed at the LHC or future hadron colliders is currently under investigation \cite{Fernbach:unpublished}.

And again, only because the higgs is a scalar particle its presence does not alter the quantum numbers of the hadrons. If it would be, e.\ g., a pseudoscalar, stand-alone QCD as the low-energy limit of the standard model would not describe hadron physics correctly. This is consistent with the experimental status that the higgs is indeed most likely a scalar particle \cite{Aad:2015mxa,Aad:2013xqa,pdg}.

\begin{table}[hbtp!]
\begin{tabular}{|l|c|c|c|r|}
\hline
Name & Spin and Parity & Custodial representation & Example operator & Equation \cr
\hline
Higgs & $0^+$ & Singlet & $\phi^\dagger\phi$ & \pref{higgs} \cr
\hline
$W$ & $1^-$ & Triplet $\pm 1$ & $W^a=\tr\left(T^a_{ij}x_j^\dagger D_\mu x_i\right)$ & \pref{op1t} \cr
$Z$ & $1^-$ & Mix triplet 0 and singlet & $\sin\theta_Wd\op_{1^-_3}^3+\cos\theta_W Db$ & \pref{opz} \cr
$\Gamma$ & $1^-$ & Mix triplet 0 and singlet & $\sin\theta_Wd\op_{1^-_3}^3-\cos\theta_W Db$ & \pref{opg} \cr
\hline
$N_R$ & $\frac{1}{2}^+$ & Singlet & $\nu_R$ & \cr
$E_R$ & $\frac{1}{2}^+$ & Singlet & $de_R$ & \cr
\hline
$E_L/N_L$ & $\frac{1}{2}^+$ & Doublet & $d(x\epsilon)^\dagger \bpm \nu_L \\ e_L \epm$ & \pref{opferm} \cr
\hline
Mesons & Integer$^\pm$ & Any integer & & \pref{sigma},\cr
 & & & & \prefr{pi1}{pi3} \cr
\hline
Glueballs & Integer$^\pm$  & Singlet & & \cr
\hline
Baryons & Half-integer$^\pm$ & Any half-integer & $d\epsilon^{IJK}F^{rst}_{uvw}\Psi_{Iu}\left(\Psi^T_{Jv}C\gamma_5\Psi_{Kw}\right)$ & \pref{protonl},\pref{deltal}, \cr
& & & & \pref{nucleon}\cr
\hline
\end{tabular}
\caption{\label{tab:sm}The gauge-invariant, physical spectrum of the standard model. The custodial symmetry is explicitly broken, but can still serve to order states in multiplets. Note that only the first generation is shown explicitly.}
\end{table}

To summarize, the observable standard model spectrum, together with the corresponding operators, is listed in table \ref{tab:sm}.

\subsection{Beyond the standard model}\label{ss:bsm}

Consider now theories with a different structure than the standard model, i.\ e.\ beyond-the-standard model scenarios \cite{Maas:2015gma}. The results will then depend on the relative properties of the gauge and custodial symmetries, as well as the involved representations of the Higgs fields. In addition, it is possible to have a custodial symmetry involving also fermions in supersymmetric theories \cite{Morrissey:2009tf,Aitchison:2007fn}. This is a possibility not yet explored, and which will therefore not be discussed here. And, of course, there is the possibility of the absence of a BEH effect, e.\ g.\ in composite models \cite{Hill:2002ap,Andersen:2011yj,Sannino:2009za,Sannino:2008ha,Morrissey:2009tf,DeGrand:2015zxa}. The latter will be taken up in section \ref{ss:tc}. There is also the possibility of more than four dimensions in form of gauge-Higgs unification \cite{Morrissey:2009tf}, which has been studied also in lattice simulations, see e.\ g.\ \cite{Lang:1986kq,Knechtli:2016pph,Alberti:2015pha,Irges:2013rya,Irges:2012mp}. While this scenario certainly increases the technical complexity, because of the anisotropy in space-time, it is not conceptually different with respect to gauge-invariance. However, it has not yet been treated using gauge-invariant perturbation theory to understand the validity of perturbation theory, though the lattice results suggest \cite{Alberti:2015pha,Knechtli:2016pph} that this should be possible.

In the following, only the situation for the gauge sector and the higgs fields will be analyzed. For an enlarged custodial symmetry the remainder of the standard model seems to be possible to include as in the standard model case \cite{Maas:2016qpu}. This is not even perturbatively true if the gauge group is enlarged beyond a direct-product structure \cite{Bohm:2001yx,Langacker:1980js}. Then, it is first necessary to somehow decompose the gauge degrees of freedom to arrive at the spectrum. This has not yet been studied in detail.

\subsubsection{Enlarged custodial symmetry}\label{sss:nhdm}

The simplest case of an enlarged custodial symmetry arises in two-Higgs doublet models (2HDMs) \cite{Morrissey:2009tf,Branco:2011iw,Ivanov:2017dad}. In such a theory an additional complex doublet of higgs fields is added. This can enlarge the custodial symmetry up to\footnote{The maximum SU(4) symmetry is not compatible with the standard model.} SU(2)$\times$SU(2), depending on the explicit breaking in the Higgs potential. In addition, there may be further discrete Z$_N$ symmetries \cite{Branco:2011iw}, which will be ignored for the moment.

Because the higgs fields in 2HDMs are all in the fundamental representation of the weak gauge group, it is possible to perform field redefinitions by linear field transformations, essentially changes of basis in the higgs fields. Two bases are of particular importance \cite{Branco:2011iw,Ivanov:2017dad}. One is the Higgs basis, in which in perturbation theory all of the vacuum expectation value is pushed into a single higgs doublet. The other is the mass eigenbasis, where the mass matrix in the Higgs sector has no off-diagonal elements.

In the Higgs basis it is possible to copy verbatim for each of the custodial groups the results of section \ref{ss:rules}, as long as the entity is a singlet with respect to the other custodial group \cite{Maas:2016qpu}. Thus, for the custodial group involving the higgs condensate this creates again a scalar and a triplet of vector bosons, with the mass of the corresponding higgs and the $w$ bosons. At the same time, the operators in the second custodial group have no expansion to tree-level propagators, and therefore expand to scattering states.

It becomes more interesting when considering operators which are tensor products, i.\ e.\ carrying quantum numbers of both custodial groups. Then, there arises one more operator which expands to a tree-level one \cite{Maas:2016qpu}
\be
\op_{0^+}^{ab}=\tr\left(x\tau^a\tau^by\right)\label{hdmtensor}
\ee
\no where $y$ is formed analogously to \pref{higgsx} for the second higgs doublet. This is a bifundamental tensor and therefore has four different states. They expand precisely to the four additional higgs propagators of perturbation theory at tree-level following the rules of section \ref{ss:rules}. Therefore, the additional Higgs particles appear again in the spectrum with the same masses \cite{Maas:2016qpu} as in perturbation theory \cite{Branco:2011iw,Ivanov:2017dad}. Furthermore, if the second custodial symmetry should be broken\footnote{The first one cannot be broken by the higgs potential without contradicting standard-model phenomenology}, the breaking pattern manifest in the masses carries over to these states \cite{Maas:2016qpu}. Though not explicitly shown, this was already suspected in \cite{Frohlich:1981yi}. Thus, the prediction of gauge-invariant perturbation theory is that, as for the standard model, perturbative results give the correct leading behavior in $v$.

Lattice simulations of 2HDMs are possible and have been done \cite{Wurtz:2009gf,Lewis:2010ps,Maas:2014nya}. But the predictions of gauge-invariant perturbation theory have not yet been checked, as in none of these simultaneously the physical and elementary spectrum has been determined. Other results have been established, nonetheless. Especially, it has been observed that the custodial symmetry can break spontaneously \cite{Wurtz:2009gf,Lewis:2010ps}. This implies a much richer phase diagram that in the standard-model case. How this affects the physical spectrum has not been investigated in detail, though the FMS mechanism predicts that even in this case the physical spectrum could be mapped to the perturbative spectrum \cite{Maas:2016qpu}.

Likewise, so far no predictions exist what will happen with more higgs doublets, so-called $n$-Higgs-doublet models \cite{Ivanov:2017dad}. However, the suspicion is that in the Higgs basis the FMS mechanism will again yield agreement with perturbation theory to leading order in $v$.

It should be noted that it was crucial to use the Higgs basis for this calculation. Using a basis where the vacuum expectation value is spread between different Higgs fields will naively create additional massive particles \cite{Maas:2015gma}. However, in such a case these are not in the correct multiplet structure for the symmetry, and must be unmixed first, similar to the case of weak gauge and hypercharge.

\subsubsection{Enlarged gauge symmetry}\label{sss:gut}

Consider first SU($N$). This case is relevant, e.\ g., to grand-unified theories\footnote{Note that non-perturbative investigations yield arguments why GUTs may be problematic due to their large field content \cite{Llanes-Estrada:2017zws,Shaposhnikov:2009pv,Dona:2013qba,Eichhorn:2016esv,Litim:2011qf}. These arguments are orthogonal to the discussion here, and are not necessarily connected to BEH physics.} (GUTs) \cite{Bohm:2001yx,Langacker:1980js}.

Note first that, because of its pseudo-reality, SU(2) is qualitatively different from SU($N>2$). This implies in particular that if the gauge group is SU($N>2$), it is possible to have a single higgs in the fundamental representation. Thus, the minimal custodial symmetry is reduced to a U(1) in contrast to the SU(2) of the standard-model case.

Consider the simplest case of SU(3). This theory is expected to have a QCD-like and a BEH-like region in the phase diagram, which has also been seen on the lattice \cite{Maas:2016ngo}, and is shown in figure \ref{fig:gdpdl}. Perturbatively \cite{Bohm:2001yx,Maas:2016ngo}, this theory is broken by the gauge condition down to SU(2). This leads to a single non-Goldstone scalar degree of freedom, identified as the higgs particle. The unbroken SU(2) shows up in a triplet of massless gauge bosons, making up the adjoint representation of the SU(2). There is also a quadruplet of heavy gauge bosons, forming a fundamental and an anti-fundamental representation of SU(2), which by pseudoreality once more coincide. Finally, there is a singlet with respect to the SU(2), with a mass which is at tree-level $\sqrt{4/3}$ larger than that of the quadruplet. Only the higgs carries a custodial U(1) charge, which will be set to 1/3, by convention.

\begin{figure}
\includegraphics[width=0.5\linewidth]{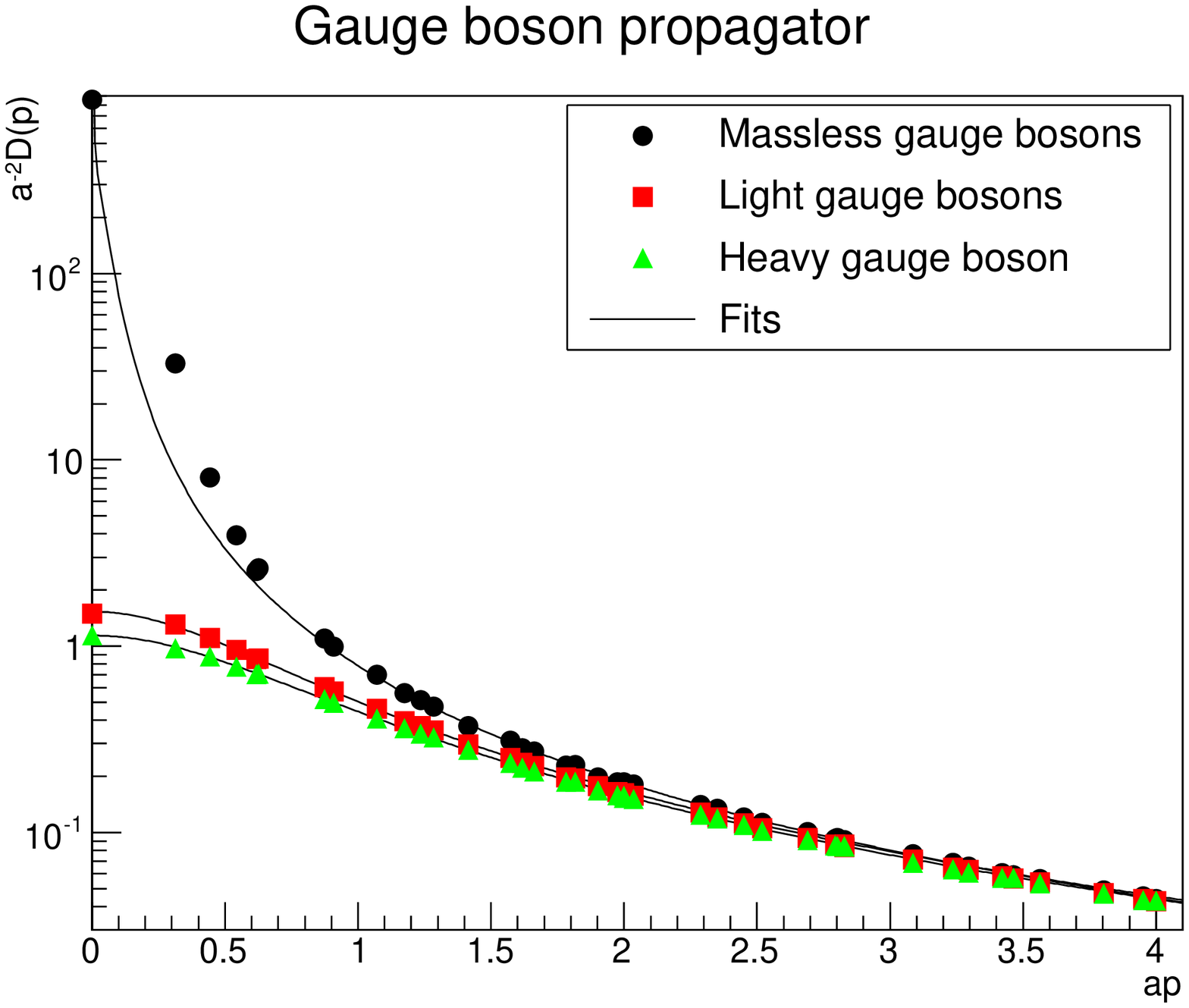}\includegraphics[width=0.5\linewidth]{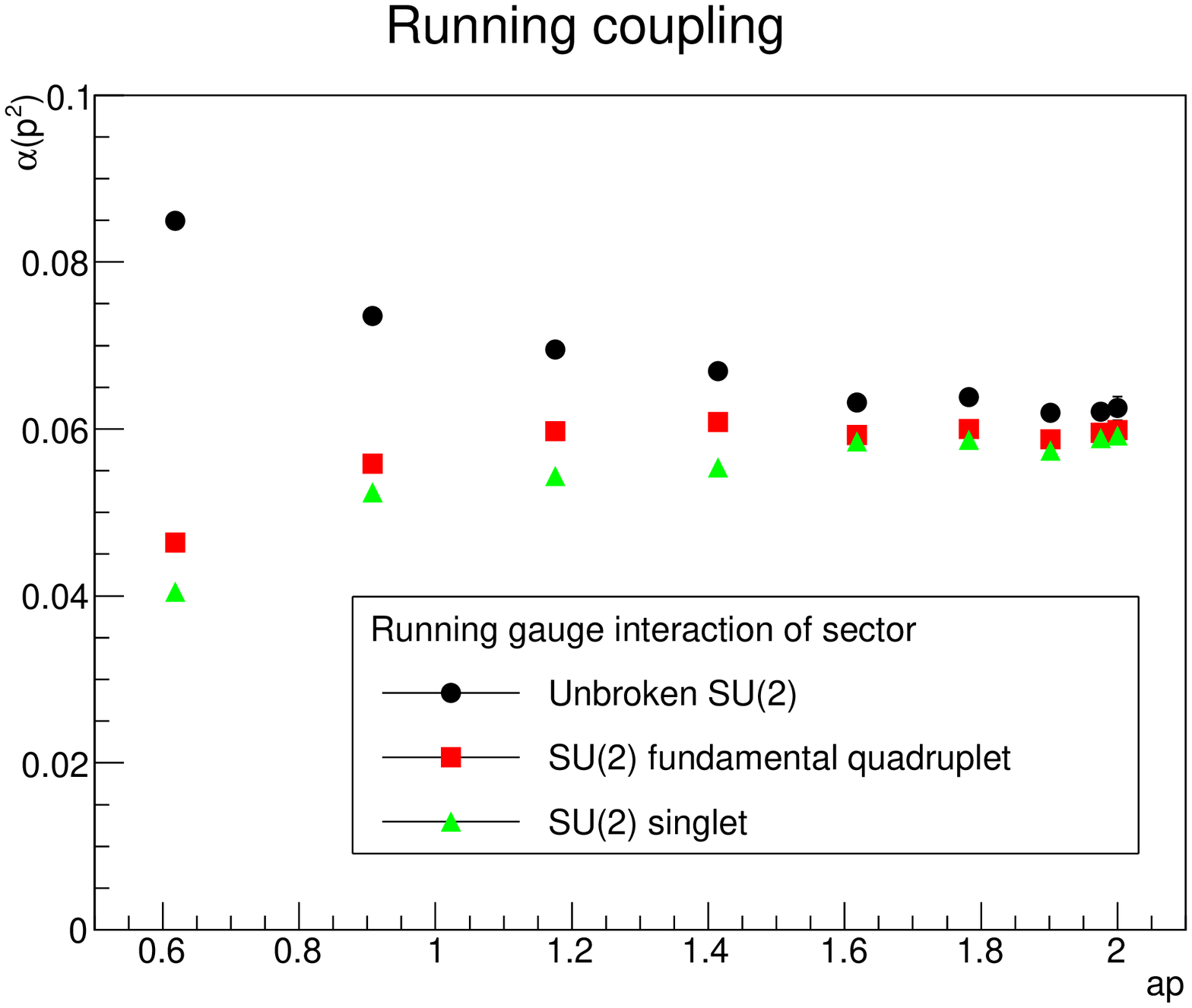}
\caption{\label{fig:gut}The left panel shows the gauge boson propagators and the corresponding fits \cite{Maas:2016ngo}. The right-hand side shows the running gauge coupling in the miniMOM scheme \cite{Maas:unpublishedtoerek}.}
\end{figure}

Lattice simulations in 't Hooft gauge in the BEH-like region confirm this behavior of the elementary higgs and gauge bosons \cite{Maas:2016ngo,Maas:unpublishedtoerek}. This is shown in the left-hand plot of figure \ref{fig:gut}. The results can be fitted well using a one-loop ansatz for the massive modes. Even the so obtained mass ratio between the SU(2)-quadruplet and the SU(2)-singlet are very close to the tree-level ratio. While stronger affected by finite-volume artifacts, also the massless modes behave as expected. Moreover, even in the unbroken sector the gauge interaction remains weak, as is also shown in figure \ref{fig:gut}. Thus no interference from strong interactions of the unbroken subgroup will play a role at the typical scale of the BEH effect. Constructing gauge-invariant states follows the recipe of section \ref{ss:classification}. The gauge-invariant states can either be singlets or charged under the U(1) symmetry \cite{Maas:2017xzh}.

The singlet scalar is essentially the same as in the standard model, described by an operator like \pref{higgs}. Gauge-invariant perturbation theory yields that the only particle in this channel has the same mass as the higgs particle of perturbation theory \cite{Torek:2015ssa,Maas:2016ngo,Maas:2017xzh}, though the scheme issues are naturally the same as in the standard model. Lattice simulations support this results, though at a much less systematic level than in the standard model \cite{Maas:2016ngo,Maas:unpublishedtoerek}.

\begin{figure}
\begin{minipage}{0.7\linewidth}
\includegraphics[width=\linewidth]{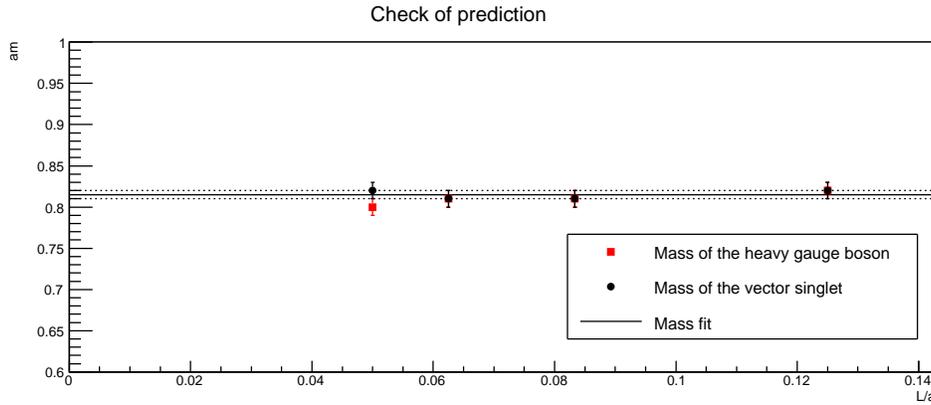}
\end{minipage}
\begin{minipage}{0.3\linewidth}
\caption{\label{fig:gut2}Comparison of the $1^-$ singlet mass compared to the mass of the heaviest gauge boson as a function of volume, from \cite{Maas:2016ngo}. The two masses are predicted to be the same in gauge-invariant perturbation theory. There is no significant volume-dependence, indicating that both states are massive and stable \cite{Luscher:1985dn}.}
\end{minipage}
\end{figure}

The situation becomes more interesting for the singlet vector. The simplest operator in the singlet channel and its FMS expansion is \cite{Torek:2015ssa,Maas:2016ngo}
\bea
\langle \op_\mu^{1^-_0}(x) \op^{\dagger1^-_0}_\mu(y)\rangle&=& \frac{v^4g^2}{4}\langle w_\mu^8(x)w_\mu^8(y)\rangle + \mathcal{O}(\eta)\label{su3w}\\
\op_\mu^{1^-_0}(x) &=& i(\phi^\dagger D_\mu \phi)(x)\nn
\eea
\no where $w^8$ is the field describing the heaviest gauge boson. Thus, the mass of the singlet vector should be equal to the one of the singlet gauge boson. This is indeed found generically on the lattice \cite{Maas:2016ngo,Maas:unpublishedtoerek}, and shown in figure \ref{fig:gut2}. That none of the other states show up is actually not surprising from the point of view of gauge-invariance: After all, there is no global symmetry which could create the necessary degeneracy pattern\footnote{Not withstanding the possibility of an emergent symmetry, but this is not observed.}. Thus, in this theory the physical spectrum does not coincide with the elementary spectrum. Still, gauge-invariant perturbation theory seems to be able to predict it correctly.

There are, of course, a multitude of other singlet channels. Some of them have been investigated, but none of them have shown any signal of a state below the elastic decay threshold in lattice simulations, as predicted by gauge-invariant perturbation theory \cite{Maas:unpublishedtoerek}.

\begin{figure}
\includegraphics[width=0.5\linewidth]{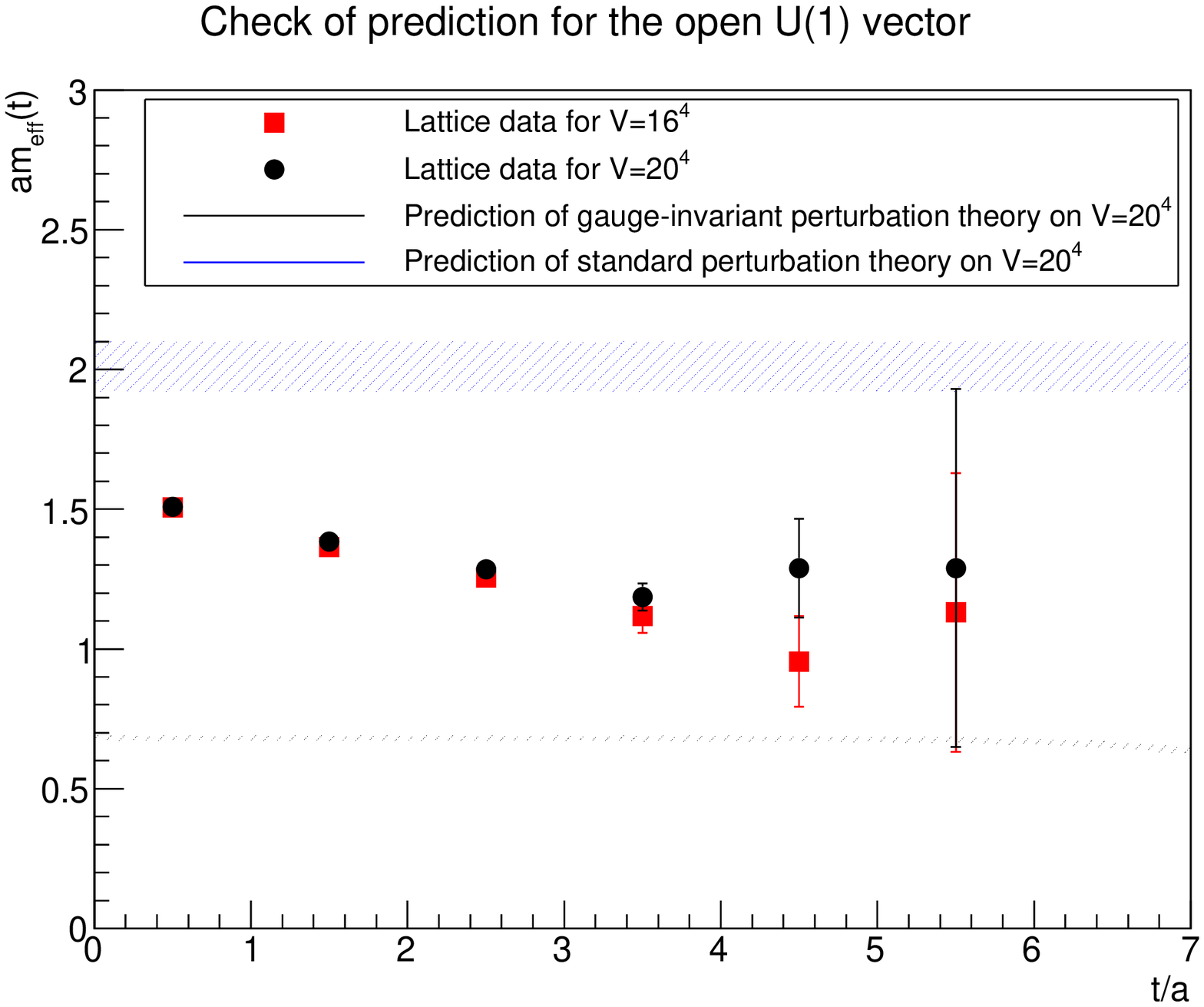}\includegraphics[width=0.5\linewidth]{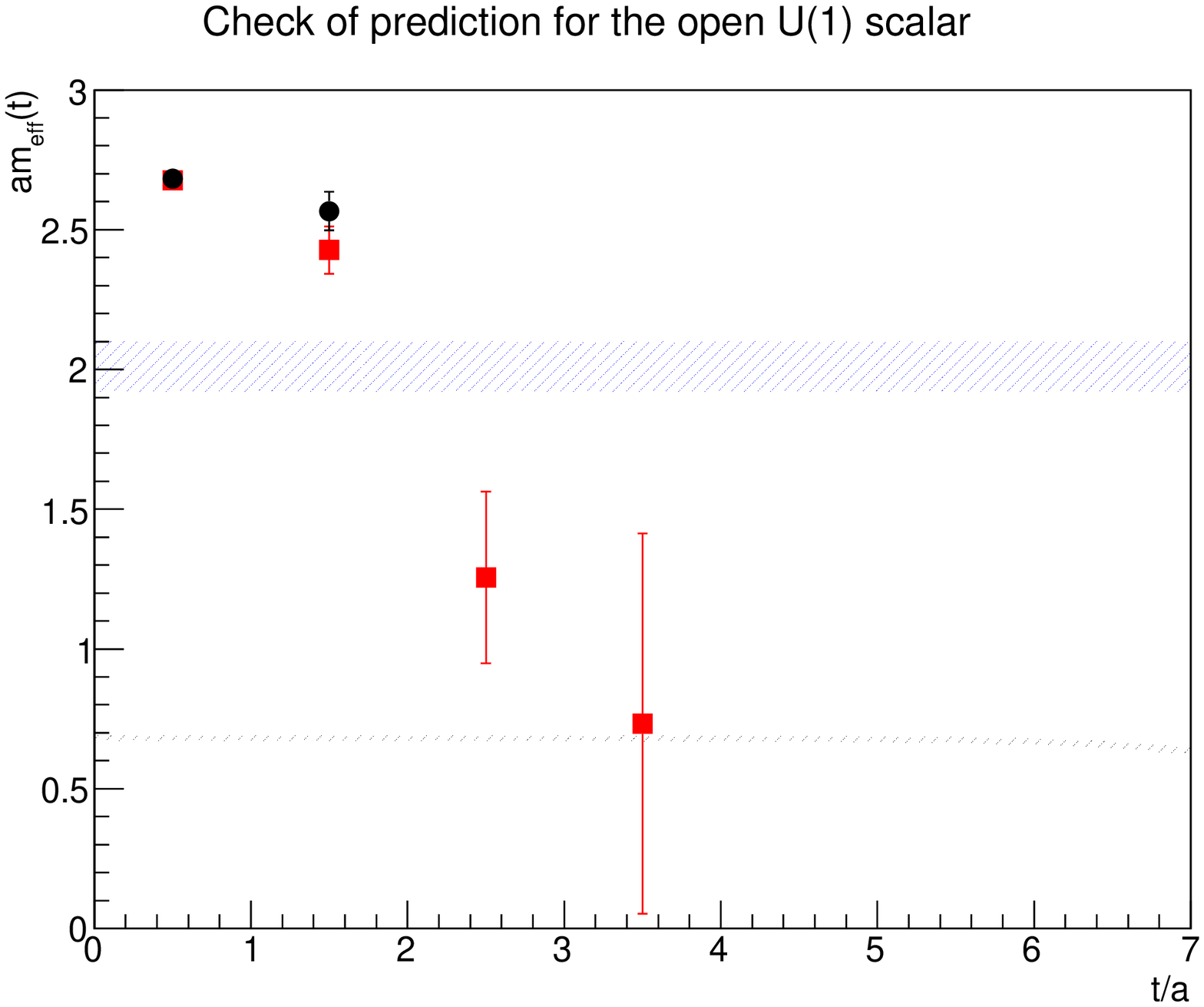}
\caption{\label{fig:gutu1}The left panel shows the effective mass \pref{effmass} of the open U(1) vector channel and the right panel of the open U(1) scalar channel \cite{Maas:2017pcw,Maas:unpublishedtoerek}. In the two predictions no contamination of heavier (scattering) states has been taken into account, yielding the smallest possible effective mass. Allowing for such contributions leads to acceptable agreement \cite{Maas:unpublishedtoerek}. Note that the results are for a different set of lattice parameters, and thus different masses, than in figure \ref{fig:gut2}.}
\end{figure}

\begin{figure}
\includegraphics[width=\linewidth]{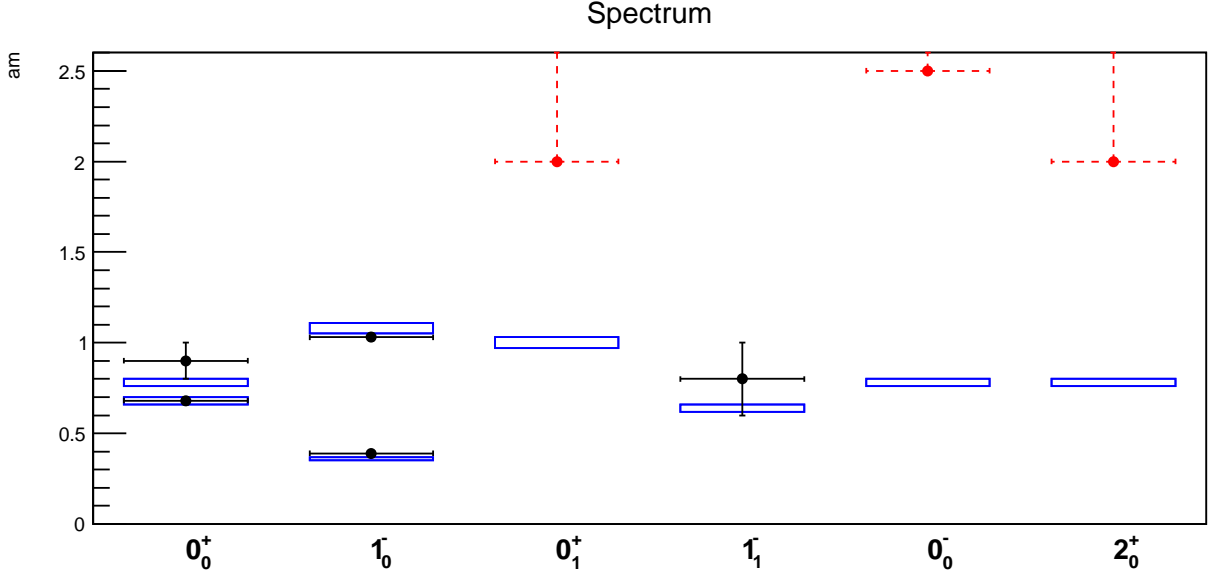}
\caption{\label{fig:gutspec}The level spectrum at fixed volume of the toy theory in U(1) singlet (lower index 0) and non-singlet (lower index 1) channels, for different $J^P$ values. The blue boxes are the predictions from gauge-invariant perturbation theory, the black data points are the lattice data from \cite{Maas:unpublishedtoerek}, and the red lines are upper limits from the same lattice calculation.}
\end{figure}

The situation is quite different for the non-singlet channels. Because the global U(1) is not explicitly broken, there must be a lightest stable state carrying this quantum number. Such additional stable states are usually dark matter candidates \cite{Morrissey:2009tf}, emphasizing the importance of understanding such states. Because of gauge-invariance these states will require at least three valence higgs fields, and have at least three times the U(1) charge of the elementary higgs field, in analogy to baryons in QCD in section \ref{ss:qcd}. The simplest scalar and vector operators of this type are\footnote{They become much simpler if there are two higgs fields \cite{Iida:2007qp}. Note that for the special case of SU(3) there exists also simpler operators \cite{Maas:2017xzh}.} \cite{Maas:2017xzh}
\bea
\op^{0^+_1}&=&\epsilon_{abc}\phi_a D_\mu\phi_b D_\mu D_\nu D_\nu\phi_c\label{dm0}\\
\op^{1^-_1}_\mu&=&\epsilon_{abc}\phi_a D_\nu\phi_b D_\mu D_\nu\phi_c\label{dm1}.
\eea
\no These states do not expand to a single elementary particle state in gauge-invariant perturbation theory, but to scattering states. Still, the lightest state is stable, and thus necessarily bound. To estimate its mass, a possibility is to use the approximation of gauge-invariant perturbation theory of it by a scattering state in terms of the elementary states \cite{Maas:2017xzh}. This can be done in a similar way as determining the mass of a baryon in QCD using the quark model \cite{BeiglboCk:2006lfa}: There it is to leading order three times the mass of the non-interacting quarks. Here, the higgs fields do not contribute to the scattering state, and it is their vacuum expectation value that carries the global U(1) charge - after all the 't Hooft gauge breaks this symmetry to a diagonal U(1) subgroup of the SU(3) gauge group and the U(1) custodial group. In fact, performing the quite cumbersome expansions explicitly yields that both states should be to leading order mass degenerate, and having the mass of twice the light non-zero $w$ mass \cite{Maas:2017xzh}. Thus, they should be about 1.7 times heavier than the singlet vector. Lattice simulations, though expensive and plagued by discretization artifacts and statistical uncertainties for such a heavy state, find indeed for the vector a mass in this range \cite{Maas:unpublishedtoerek}, being $1.0(3)m_\text{predicted}$. It should be noted that a naive counting of masses, like in the quark model, would yield a mass of $3m_h$ in both \pref{dm0} and \pref{dm1}. In the corresponding lattice simulations, this provides a lower bound of $3.5(8)m_\text{predicted}$, substantially above the FMS prediction. A measurement for the non-singlet scalar has not yet been possible. The corresponding results for the effective mass \pref{effmass} from \cite{Maas:2017pcw,Maas:unpublishedtoerek} are shown in figure \ref{fig:gutu1}. All in all, the comparison between the predicted and observed spectrum shown in figure \ref{fig:gutspec} is reasonably good, especially when comparing to the effort invested in the prediction.

Hence, the FMS mechanism and gauge-invariant perturbation theory seem to provide not only for the channels which expand to elementary fields an adequate description, but also for more complicated states. In addition, this is a non-trivial test of the underlying field-theoretical concepts, which are based on the necessity of gauge-invariance. Especially, triply charged stable states do not exist in perturbation theory, and such dark matter candidates would therefore be missed in a perturbative analysis, while incorrectly having a singly charged higgs. The physical spectrum is quite different.

Another remarkable result is the following \cite{Maas:2017xzh}. Expanding the operator \pref{su3w} to the next order in the fluctuation field yields
\be
\op_\mu^{1^{-}_0}=-\frac{v^{2}g}{2}w_\mu^8+\frac{v}{\sqrt{2}}\pdm\eta-\sqrt{2}gv w_{\mu}^8\eta+\mathcal{O}(\eta^{2}).\label{subleadpole}
\ee
\no The third term involves two particles, and will therefore only contribute in scattering states. But the second term involves a single higgs field. Such a term does not arise in the SU(2) case because of the differing custodial structure. The remarkable implication is that there should be a second pole, besides the one at the gauge boson's heaviest mass, at the mass of the higgs. This second pole is therefore degenerate with the scalar. In addition, this implies that the lowest mass in this channel is the minimum of the two masses. This implies that the scalar singlet can never be lighter than the singlet vector.

There is a subtlety to be added here. The higgs field in \pref{subleadpole} has an additional derivative. When forming expectation values, the Lorentz indices will be contracted, to obtain a non-vanishing Lorentz scalar\footnote{A quantity like a propagator $\la\op_\mu\op_\nu\ra$ is always identical zero, for exactly the same reason as any non-invariant quantity is zero if there is a corresponding intact global symmetry, as discussed in section \ref{s:global}. Poincar\'e symmetry is no exception. The Wigner-Eckart theorem implies that the operator has the structure $\op_\mu\op_\nu=T_\mn A+L_\mn B$, where $T$ and $L$ are transverse and longitudinal Lorentz tensors, respectively, with dressing functions $A$ and $B$. The latter can be calculated by $\la T_\mn\op_\mu\op_\nu\ra\sim A$ and $\la L_\mn\op_\mu\op_\nu\ra\sim B$ while $\la\op_\mu\op_\nu\ra=0$.}. It is only from the Wigner-Eckart theorem and the fact that a longitudinal tensor structure does not carry poles from which it would be reconstructed that such a pole could be located in the transverse part of the correlator. However, the additional factor of $p^2$ arising may alter this behavior. This is not yet fully understood. This additional pole has not yet been confirmed on the lattice \cite{Maas:unpublishedtoerek}, but finding such a second, heavier state is not trivial either. Intriguingly, though the FMS mechanism provides no explanation for this, this hierarchy is also observed in lattice simulations for the SU(2) case, as discussed in section \ref{sss:w}.

The situation with multiple poles identifiable with elementary degrees of freedom is actually not exceptional \cite{Maas:2017xzh}. Already the physical $Z$ and $\Gamma$ in section \ref{ss:qed} gave an example of this in \prefr{opz}{opg}. This is therefore a fourth possible outcome of the FMS mechanism besides the identification of a single pole with an elementary field, a scattering state, and a non-trivial bound state. Unfortunately, so far no situation has been found in lattice calculations where the second pole was statistically and systematically reliably identifiable, and therefore no positive confirmation has yet been possible. If the pole would be above the elastic threshold, this will require a L\"uscher-type analysis \cite{Gattringer:2010zz,Luscher:1990ux,Luscher:1991cf}, but this demands substantially more statistics than currently available.

Enlarging the gauge group to arbitrary $N>3$ yields qualitatively the same results \cite{Maas:2017xzh}. The perturbative spectrum remains essentially the same, up to changes in the degeneracies of multiplets and values of the mass ratios \cite{Bohm:2001yx,Maas:2017xzh}. On the other hand, the physical spectrum keeps its degeneracies, because the global custodial symmetry remains U(1). In addition, the predicted mass spectrum is mapped to the same quantities, and thus qualitatively the same physical spectrum arises independently of $N$. Unfortunately no lattice test of this prediction has yet been made. It will be necessary to perform these investigations to see whether this is generically true. But the SU(3) case is already a highly non-trivial test of the mechanisms.

Interestingly, when increasing in the $N=3$ case the number of higgs fields to two, increasing the global symmetry to a SU(2)$\times$U(1) symmetry, gauge-invariant perturbation theory again predicts that more of the perturbative states reappear in the spectrum \cite{Torek:2015ssa,Maas:2017xzh}. Whether it is completely recovered requires further investigations, especially for generalizations of the tensor operator \pref{hdmtensor}. From a group-theoretical point of view, this can be understood. After all, the additional elementary states build SU(2) representations, which are then mapped on the global SU(2) representations, provided the latter is not explicitly broken. Thus, this is very similar to what occurs in the 2HDM of section \ref{sss:nhdm}. Whether this actually works in this way has unfortunately not been tested on the lattice yet. If the gauge group becomes again larger but the number of higgs fields is not increased then again the representations do not match, and problems with the degeneracy patterns would be expected. This has not yet been studied in detail.

What happens in case of gauge groups different than SU($N$) has not yet been explored.

\subsubsection{Changing the representation of the Higgs}\label{sss:rep}

A third possibility to modify the basic setup is to change the representation of one, or more, higgs fields. This arises regularly in grand-unified theories \cite{Bohm:2001yx,Langacker:1980js}.

The simplest change is to use the standard-model SU(2) gauge group, but to put the Higgs in the adjoint representation\footnote{This theory, also known as the Georgi-Glashow model, has been variously used as a low-energy effective theory for the gauge-dependent degrees of freedom of pure Yang-Mills theory in monopole confinement/Abelian dominance scenarios \cite{Greensite:2011zz,Ripka:2003vv,Kondo:2016ywd,Nishino:2018mwi}. This will not be discussed here. The three-dimensional version is also appearing as the infinite-temperature limit of pure Yang-Mills theory, where then only the QCD-like region is of interest \cite{Kajantie:1995dw,Maas:2011se,Hart:1999dj,Cucchieri:2001tw,Maas:2004se}.}. In this case the global symmetry is only a Z$_2$, which changes $\phi\to-\phi$. In addition in lattice realizations a second Z$_2$ connected to the link variables appears, which is also present in the pure Yang-Mills case \cite{Gattringer:2010zz}.

Perturbatively, this theory breaks to a U(1) gauge theory, and in fact, it is not possible to break the gauge symmetry completely \cite{O'Raifeartaigh:1986vq}. Thus, the arguments of section \ref{ss:pd} do not apply. Consequently, non-perturbative studies found a separation of the phase diagram into (at least) two phases. They are distinguished by the spontaneous breaking of the global symmetry \cite{Lang:1981qg,Drouffe:1984hb,Baier:1986ni,Capri:2012cr,Kondo:2016ywd}. The phase of broken symmetry exhibits also features like the BEH-like region in the fundamental case, while the other one is QCD-like. This fits with the idea of the Wilson confinement criterion, where an unbroken Z$_2$ symmetry of the links is connected to a linear rising string tension \cite{Greensite:2011zz,Gattringer:2010zz}. Hence, in this theory appears to be a phase separation between both types of physics.

As far as has been investigated, the QCD-like phase is not qualitatively different from the fundamental case when it comes to spectral properties and other physical features. Therefore, nothing new arises in this case. Thus, the interesting phase is the one in which the BEH effect operates. 

As noted, perturbatively there should remain an unbroken U(1) subgroup. The hallmark of this would be a massless gauge boson. This is especially interesting, as a U(1) gauge theory allows for a physically observable charge \cite{Haag:1992hx}. This would imply that there should be an observable consequence of this breaking. This is even more important as that the hypercharge group of the standard model is expected to be obtained from a GUT in exactly this way \cite{Bohm:2001yx,Langacker:1980js}.

However, from a gauge-invariant perspective, this is not possible. Since the gauge-symmetry is intact, there cannot be a subgroup, which is in any way different from the remainder, for the very same reasons as the unbroken SU($N-1$) groups in the fundamental case could not be identified. Otherwise, a gauge transformation would need to be able to distinguish between this particular subgroup, and any other possible U(1) subgroup. Especially, the gauge symmetry remains non-Abelian, and thus no simple gauge-invariant charge is possible.

That said, it is not impossible that the physical spectrum could resemble the existence of a U(1) group, without actually having one. Remarkably, this is what seems to happen \cite{Maas:2017xzh}.

This requires two things. First, there must be an observable, massless vector particle. The second is that this massless vector particle interacts with matter particles with an interaction of the kind expected of an Abelian gauge theory. In this way, an effective U(1) arises. Of course, even in the perturbative setting the later emerges also as an effective interaction as being 'just' the ordinary non-Abelian interaction evaluated for a particular combination of elements of the representations \cite{Bohm:2001yx}, rather than a genuine independent interaction. This is not unlike how electromagnetism in the standard model arises as a combination of the non-Abelian weak isospin and the Abelian hypercharge interaction.

This can be investigated using gauge-invariant perturbation theory \cite{Kondo:2016ywd,Maas:2017xzh}. The first step is to classify states by their global quantum numbers. They can therefore be singlets or non-singlets under the global symmetry. Furthermore, they can be classified according to their spin.

First of all, a singlet scalar arises as in \pref{higgs}, with the mass of the higgs, using essentially the same operator. Everything which has applied to this operator also applies here. A non-singlet scalar operator cannot be constructed in a similar way such that it expands to a single-field operator, because of the pseudoreality of SU(2). This will change for SU($N>2$) below. Reversely, also no simple singlet vector operator can be constructed. Since, however, a non-singlet vector can be constructed, non-singlet scalar and singlet vector operators can be constructed from gauge-invariant operators with non-trivial momentum configurations, like in the case of \pref{doublehiggs}. However, these operators do only expand to scattering states in terms of the elementary states.

So, the only non-trivial further operator is a non-singlet vector operator. A generalization of \pref{op1t} does not yield an operator which expands to a scattering state. However, a generalization of the non-local operator used in \cite{Frohlich:1981yi} to construct a gauge-invariant vector in the fundamental case expands non-trivially \cite{Maas:2017xzh},
\be
\op^\Gamma_\mu=\frac{\pd^\nu}{\pd^2}\tr\left(\phi^a T^aw_\mn\right)\approx-v\tr\left(T^3\left(\delta_\mn-\frac{\pdm\pdn}{\pd^2}\right)w^\nu\right)+\op(\eta)\label{masslessadjvec}.
\ee
\no In the second step the vacuum expectation value in the 't Hooft gauge was put into the Cartan direction $T^3$. This Cartan choice is always possible by a global gauge transformation. Thus, this operator expands to the transverse component of the $w$ boson in the Cartan direction, which is perturbatively massless \cite{Maas:2017xzh,O'Raifeartaigh:1986vq}. This predicts a massless non-singlet vector, notwithstanding that it is a composite, bound state. It thus acts like a physical Photon. Note that this is not a Goldstone boson, because even if the global Z$_2$ symmetry would be broken, it is a discrete symmetry, and thus does not fall under the purview of Goldstone's theorem\footnote{Note that the masslessness of the photon in massless QED is associated with it being a Goldstone boson of the broken dilatation symmetry \cite{Alkofer:2000wg,Lenz:1994tc}. It has not been investigated whether a similar argument could be made here.}. Furthermore, as in the case of \pref{subleadpole}, higher orders in the quantum fluctuations yield additional poles at the masses of the higgs and twice the mass of the massive gauge boson \cite{Maas:2017xzh}.

Note that these results imply that none of the other states can actually be stable, as two non-singlet massless vector bosons can always be coupled to a singlet scalar, and from there all other states become reachable. Unfortunately, this implies that any lattice investigation will be rather complicated, as this implies polynomial volume corrections \cite{Gattringer:2010zz,Luscher:1985dn}. Consequently, so far only exploratory investigations for the SU(2) case exists \cite{Lee:1985yi}, but they support a massless pole in the correlator \pref{masslessadjvec}. This is a dramatic confirmation of the sketched scenario.

However, the masslessness of the vector is not sufficient to mimic an Abelian gauge theory at low energies. The low-energy limit of the interactions needs to be also that of an Abelian gauge theory. Concerning a possible interaction vertex between this vector particle and any of the massive particles, in analogy to QED in section \ref{ss:qed}, leads unfortunately not to the desired result. Expanding e.\ g.\ the scalar channels always reduces to expectation values of fields with the same gauge-indices, which vanish because of the anti-symmetry of the structure constants at tree-level \cite{Sondenheimer:pc}. Thus, the interaction structure needs to arise either at loop-level or in more involved operators. However, if it would arise at loop level this would provide an interesting hint how such an interaction could be weaker than the original gauge interaction. This is still subject of future research.

As noted above, the importance of this observation is that only such a mechanism can actually rescue the idea of GUTs that the hypercharge is part of an overarching single gauge interaction to explain the relative hypercharges of the particles in the standard model \cite{Langacker:1980js}. The mechanism outlined above therefore deserves further studies. This is especially true as the GUT scenarios requires larger gauge groups with more involved Higgs sectors \cite{Langacker:1980js}.

However, when considering large groups with a higgs in the adjoint representation, another problem arises. As was discussed in section \ref{ss:strata} multiple breaking patterns become possible. Especially, a gauge condition involving an explicit vacuum expectation value can no longer be satisfied for every configuration, if the vacuum expectation value in the gauge condition and the configuration belong to different strata. This also affects gauge-invariant perturbation theory. In analogy to \pref{mppi}, the following situation arises \cite{Maas:2017xzh}
\bea
\langle \op^\dagger(x) \op(y)\rangle&=&\int{\cal D}w\left(~\int_\text{Selected stratum}{\cal D}\phi \op^\dagger(x)\op(y)e^{iS}+\int_\text{Other strata}{\cal D}\phi\op^\dagger(x)\op(y)e^{iS}\right)\nonumber\\
&=&\langle \op^\dagger(x)\op(y)\rangle_e+\langle \op^\dagger(x) \op(y)\rangle_n\nn.
\eea
\no Thus, any correlator decomposes into a part ($e$) which is evaluated on those configurations where the gauge condition can be fulfilled, and a part ($n$) in which it is evaluated on configurations for which the gauge condition degenerates to a non-aligned covariant gauge. On the first part gauge-invariant perturbation theory can be applied. This is not true for the second part.

As discussed in section \ref{ss:strata}, it now depends on the actual behavior of the second part what the physics is. For the mass spectrum this boils down to the question what the pole structure of the $n$-part is. If there are no poles in it, then the spectrum is determined by the pole structure of the $e$-part. However, this implies that for a given set of parameters there is only one gauge condition associated with a particular stratum which can be used for the purpose of gauge-invariant perturbation theory. This seems odd at first. However, when recalling that only aligned gauges can be used in the standard model case for the same end, it does not seem odd that there is a 'preferred' gauge. However, it is important to note that this does not physically prefers a gauge, but only technically. After all, gauge-invariant perturbation theory is not necessary to evaluate the left-hand side, it is just convenient.

The alternative is that the $n$-part exhibits a pole structure determined by some or all other breaking patterns. Then the physical spectrum would be determined by the union of all poles. This union can be determined by applying gauge-invariant perturbation theory to each stratum to determine the set of all poles.

As discussed in section \ref{ss:strata} the actual correct result will likely depend on the dynamics of the theory. It may remain a non-perturbative question which strata contribute to the union of poles determining the spectrum. This awaits non-perturbative study. Note, however, that the existing quantum number channels do not depend on this. They are only determined by gauge-invariant physics. Thus, even a continuous deformation between different sets contributing to the union could be possible. Therefore, in the following just a few remarks on the situation for SU($N>2$) will be made. Especially, as the number of breaking patterns quickly proliferates with the size of the gauge group \cite{Li:1973mq,Ruegg:1980gf,Murphy:1983rf,O'Raifeartaigh:1986vq,Kojima:2016fvv,Maas:2017xzh}.

When proceeding to $N>2$, the group SU(3) is also special. This has to do with the potential structure for the adjoint higgs \cite{O'Raifeartaigh:1986vq,Maas:2017xzh}, under the condition of power-counting renormalizability. Only for $N\ge 4$ the most generic case for the adjoint representation has been reached.

In the SU(3) case two breaking patterns\footnote{If the Z$_2$ symmetry is explicitly broken, only SU(2)$\times$U(1) remains \cite{O'Raifeartaigh:1986vq}.} are possible, SU(2)$\times$U(1) and U(1)$\times$U(1). Perturbatively, this gives in the first case rise to a three-fold hierarchy of levels for the gauge bosons, and the second case to a two-fold hierarchy. In both cases also not all components of the higgs field play the role of would-be goldstone bosons, and additional massless higgs bosons remain.

The situation is, however, qualitatively similar in both strata for the gauge-invariant spectrum \cite{Maas:2017xzh}. First of all, both Z$_2$ channels harbor now non-trivially expanding operators. Remarkably, the states in both channels are found to be degenerate. In the vector channel the massless mode remains, and thus there are two massless vector bosons in the spectrum. Also, the additional poles show up in both cases. Even more interesting is the scalar case. Here, the massless modes are also pushed through to the physical spectrum, and thus two massless scalars, one in each Z$_2$ channel, exist. In addition, poles at the higgs mass exist as well. Thus, such a theory yields additional massless degrees of freedom compared to what is needed for a GUT. However, once the potential breaks the Z$_2$ symmetry explicitly, these vanish again from the spectrum \cite{Maas:2017xzh}. These quite spectacular, results have unfortunately not yet been checked in lattice investigations, and only exploratory simulations of this theory exist so far \cite{Gupta:1983zv,Kikugawa:1985ex}.

In the SU($N>3$) case the massless scalar states again vanish from the physical spectrum and, depending on the stratum, also the degeneracy between the Z$_2$ even and odd states are lifted \cite{Maas:2017xzh}. This shows that $N=2$ and $N=3$ are special in the adjoint representation. This also gives again hope that a low-energy effective version of the GUT idea can still be realized by a deliberate choice of fields and representations.

Ultimately, the need to have a U(1) seems to suggest the necessity for an adjoint Higgs field for a GUT setup. Such an adjoint Higgs field also appears in perturbative treatments of most grand-unified theories \cite{Bohm:2001yx,Langacker:1980js}. However, an explicit test of the standard SU(5) GUT setup of one higgs in the adjoint representation and one higgs in the fundamental representation \cite{Bohm:2001yx,Langacker:1980js} showed again that a discrepancy in the spectrum of the vector bosons could arise \cite{Maas:2017xzh,Pedro:2016hnd}. The reason is that two, instead of one, massless vectors arise as physical Photons, while too few massive degenerate gauge bosons are present which could play the role of the physical $W$ and $Z$ bosons. If the Z$_2$ symmetry is broken explicitly, even the mass hierarchy in the vector boson sector is no longer faithfully reproduced. This seems to indicate that more higgs fields in either representations would be necessary to allow for an adequate degeneracy pattern in the global symmetries.

Theories with multiple higgs in different representations have been subject of exploratory studies on the lattice \cite{Greensite:2008ss}, but once again so far not including spectroscopy. It therefore remains to be seen whether these theories can sustain a phenomenological relevant spectrum.

\subsubsection{Model-building 2.0}\label{sss:model}

Usually, building a new model is based upon the definition of the elementary degrees of freedom and their interactions based on a particular (set of) mechanism(s) \cite{Morrissey:2009tf}. Then, their experimental implications are derived, and checked against experiment. 

Taking the results of sections \ref{sss:nhdm}, \ref{sss:gut}, and \ref{sss:rep} at face value suggests a somewhat different approach, based upon the available experimental data. Therefore, the first step is defining the physical, observable spectrum. Ideally, the model could then be build led by this physical spectrum of particles. This would suggest the following procedure:
\begin{itemize}
 \item[1)] Choose the desired physical, gauge-invariant particle spectrum in absence of explicit symmetry breaking, and determine a custodial group to sustain the multiplet structure
 \item[2)] Add explicit breaking to split multiplets
 \item[3)] Find a group which has a little group sustaining the same multiplet structure \cite{O'Raifeartaigh:1986vq}
 \item[4)] Arrange Higgs fields and the Higgs potential such that it has the appropriate custodial group, including multiplet structure \cite{O'Raifeartaigh:1986vq}
 \item[5)] Find a gauge with a vacuum-expectation-value which breaks the gauge group to the required little group
 \item[6)] Use gauge-invariant perturbation theory to verify that the spectrum is correct
 \item[Optional)] Use non-perturbative methods to confirm that in the relevant parameter ranges gauge-invariant perturbation theory is applicable, and, if multiple strata are present, the correct union of pole sets contributes
\end{itemize}
However, this leads to physical mechanisms being a mere consequence of the desired particle spectrum. It therefore does not allow to base a model upon a new mechanism. If this is not desired, it is therefore not a useful procedure. Then, still, it is necessary to first define the model based on the need to realize the new mechanism. But it seems then to be prudent to ensure various properties of the constructed model as a new intermediate step between defining the model and determining experimental predictions:
\begin{itemize}
 \item There must be a global symmetry which allows a triplet structure to create the pattern of physical $W$ and $Z$ bosons, or some mechanism creating a similar pattern
 \item Likewise, an explicit breaking of the global symmetry, or some replacement, to create the splitting in the $W$ and $Z$ sector
 \item A global symmetry to provide left-handed flavor, including an explicit breaking mechanism
 \item If the hypercharge should be realized as a part of an underlying theory a mechanism to obtain a global analogue
 \item Ensure that the physical states under gauge-invariant perturbation theory expand to the desired masses
 \item Ensure that no additional light (stable) physical particles exist in other channels, except as dark matter candidates, axions or otherwise acceptable candidates
 \item[Optional] Use non-perturbative methods to confirm the applicability of gauge-invariant perturbation theory and, if necessary, that only the desired strata contribute to the union of poles
\end{itemize}
Of course, the FMS mechanism only works for BEH physics. For a model without BEH effect, the situation becomes quite different, as will be discussed in section \ref{ss:tc} \cite{Maas:2015gma}. While such programs have yet only been partially followed through \cite{Maas:2015gma,Maas:2016qpu,Maas:2017xzh}, these guidelines should be useful to develop model building using gauge-invariant perturbation theory.

\section{Scattering processes}\label{s:scattering}

So far only static properties have been considered. Most of our knowledge stems, however, from dynamic processes, especially scattering experiments. It is therefore necessary to develop a description\footnote{Interestingly, an approach based on a confining rather than BEH-type physics in the standard model leads to formally quite similar results \cite{Calmet:2000th,Calmet:2001rp,Calmet:2001yd,Calmet:2002mf}, though for completely different physics reasons. Similar lines of arguments \cite{Dosch:1983hr,Dosch:1984ec} also follow in the Abbott-Farhi model \cite{Abbott:1981re}, but again for different physics reasons.} of scattering processes based on gauge-invariance, which at the same time also explains quantitatively why conventional perturbation theory works so well for the standard model \cite{pdg}.

Of course, this is not a new problem. After all, bound-state-bound-state scattering is the standard situation in QCD \cite{BeiglboCk:2006lfa,Bohm:2001yx,Dissertori:2003pj}. However, the different type of physics leads to quite different results. Necessarily, it must be explained why in experiments so far no detectable sign of the substructure has been seen \cite{pdg}. It is here important to remember that the bound-state structure is not speculative new physics, but rather a consequence of the field-theoretical underpinnings of the standard model. In fact, it is a consequence of enforcing strict gauge invariance. Thus, the two decisive questions are: Why has nothing been seen so far? What needs to be done to see something? Both questions, and whatever partial answers are yet available, will be discussed in the following.

\subsection{Asymptotic states}\label{ss:asymp}

Before doing so, it is worthwhile to discuss both how initial and final states operate in the present context. The gauge-invariant operators themselves are just bound-state operators. Thus, scattering processes which involve these states as initial and final states can be described by the usual LSZ construction for bound states in terms of the corresponding matrix elements, having the bound states as in and out states \cite{Alkofer:2000wg,Bohm:2001yx}. These can be either calculated entirely non-perturbatively, or using gauge-invariant perturbation theory. The latter will be done in sections \ref{ss:gscattering}, \ref{ss:pheno}, and \ref{ss:qscattering} to study how quantitative and qualitative similarities and differences to the usual results emerge.

The more interesting question regards the elementary states, and especially, why they are not suitable in states and out states. Technically, this is not an issue, as the example of QCD shows, where quarks and gluons are never considered as in and out states. But there this is linked to confinement. Thus, this leaves only the question of why this should not occur, or even why the composite states should not decay into the elementary states. This latter question is probably the best one to discuss the problem.

Hence, consider a matrix element
\be
\la \op(x)^\dagger\varphi(y)\varphi(z)\ra\label{decay},
\ee
\no where $\op$ is a gauge-invariant composite state and the fields $\varphi$ are some of the elementary, gauge-dependent states. To define suitable asymptotic states, it will be necessary to take eventually the limit that $x$, $y$, and $z$ are sufficiently far separated.

Of course, the final state is not gauge-invariant. But the usual argument is that at asymptotically large distances the out states become non-interacting, and thus only global transformations apply anymore. While this is true perturbatively, this is not correct non-perturbatively. Haag's theorem states that the free theory is not unitarily equivalent to the interacting one \cite{Haag:1992hx}, and thus no unitary time evolution can yield non-interacting states. The interacting states, however, cannot be made gauge-invariant, as was discussed at length in section \ref{ss:physstates}, where isolated particles were addressed. Thus, the final state is not gauge-invariant, and is unphysical. This implies especially that cluster decomposition fails for the elementary states non-perturbatively.

This does not mean that \pref{decay} is an irrelevant expression. Consider The situation, where the state described by $\op$ can decay into gauge-invariant composite states $\op'=\phi\varphi$, which in leading order in the FMS expansion behave as $\op'\sim v\varphi$. Then
\be
\la\op(x)\op'(y)\op'(z)\ra=v^2\la\op(x)\varphi(y)\varphi(z)\ra+\op(v)\nn.
\ee
\no Thus, to leading order in $w$, the decay can be approximated by \pref{decay}. If the residual, asymptotic gauge-dependence is sufficiently small compared to the matrix element itself, this will be a very good approximation. This is the case in the standard model, as will be explored in sections \ref{ss:gscattering} and \ref{ss:pheno}. But, of course, this will only work if the correction in $v$ are small. Otherwise, this may even be qualitatively wrong. This will explored in section \ref{ss:qscattering}.

In this sense, there also exists an overlap between gauge-dependent and gauge-invariant operators. Even though $\varphi\varphi$ does not belong to the same superselection sector as $\op$, $v^2\varphi\varphi$ does. Still, this overlap is gauge-dependent, but this gauge-dependence is of order $v$ compared to the leading-order $v^2$, and may therefore be negligible in actual calculations, if a suitable gauge is chosen which makes the prefactors small.

It is, of course, possible to expand even \pref{decay} further, by also expanding $\op$ according to the FMS mechanism. If such a full expansion is performed, this is just the gauge-invariant perturbation theory of section \ref{ss:gipt}, applied to a decay process. This is the strategy to actually calculate cross sections and decays using gauge-invariant perturbation theory, as will be exemplified in the sections \ref{ss:gscattering} and \ref{ss:qscattering}. However, the gauge-invariant matrix element is then approximated by a matrix element entirely from gauge-dependent fields. Such a gauge-dependent matrix element needs not to follow the usual rules of physical matrix elements \cite{Seiler:1982pw}. This is already visible for the $w$-boson-propagator in figure \ref{fig:w}, from which it can be read off that it does not have a positive spectral functions, and does not support a K\"allen-Lehmann representation \cite{Maas:2011se}. Other examples can be found in \cite{Raubitzek:unpublished}.

\subsection{Scattering processes in the standard model}\label{ss:gscattering}

The simplest example will be a lepton collider, avoiding the complications due to the QCD background in hadron colliders. The simplest possible process is probably\footnote{In this section capital letters denote gauge-invariant, composite states, and small letters denote elementary states.} $E^+E^-\to{\bar F}F$, which is at the same time also experimentally reasonably easy to access with high sensitivity. And in fact, this process has been studied intensively up to LEP(2) \cite{pdg} and this will be continued (hopefully) at either ILC, CEPC, CLIC, and/or FCC-ee in the future.

\begin{figure}[!htbp]
\begin{center}
\includegraphics[width=0.8\linewidth]{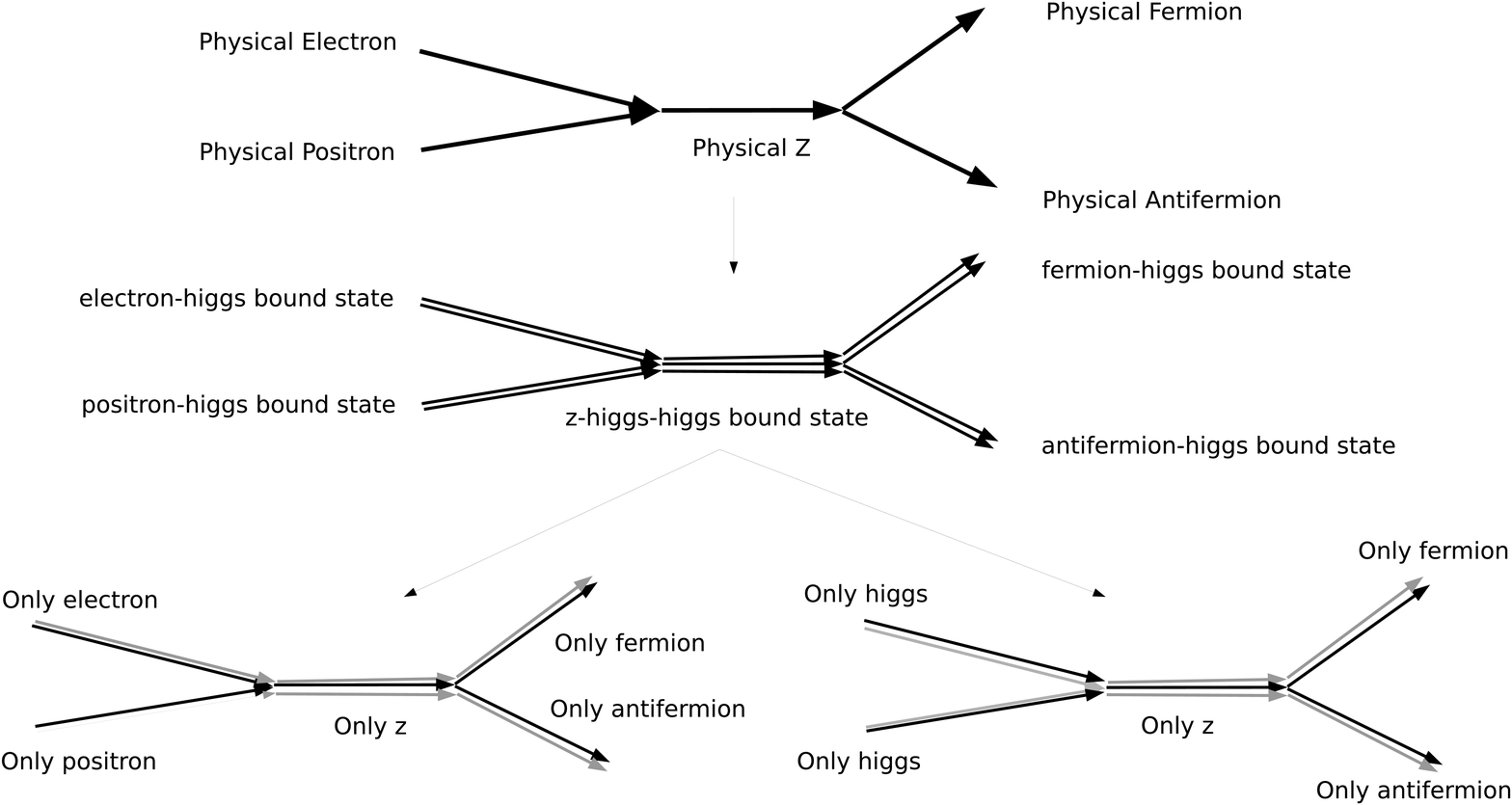}
\end{center}
\caption{\label{fig:scattering}A cartoon of a scattering process in leading order in gauge-invariant perturbation theory for $E^+E^-\to{\bar F}F$ with an exchange of a $Z$ in the $s$-channel \cite{Maas:2017swq}.}
\end{figure}

Given the discussion of section \ref{ss:flavor}, the incoming Electron and Positron need to be considered as higgs-electron/positron bound states. A hand-waving argument why no effect has been seen so far is that the very massive valence higgs acts entirely as a spectator particle, like the non-participating quarks and gluons in a hadron collision \cite{BeiglboCk:2006lfa}. Then a picture emerges like it is sketched on the left of  figure \ref{fig:scattering} \cite{Maas:2012tj,Egger:2017tkd}. The valence higgs simple does not participate in the interaction, and only dresses the initial, the intermediate, and the finial state. The actual interaction therefore does not differ to this order from the conventional perturbative picture, and thus should be described to leading order just by ordinary perturbation theory. Only if the higgs does no longer act as a spectator, like it is sketched on the right-hand side, this will change.

Since off-shell contributions are suppressed at leading order, the requirement will be to get the higgs on-shell. Though due to the renormalization-scheme dependence this is a non-trivial statement, this will essentially always be some kind of typical electroweak scale, and in the pole scheme the singlet scalar mass. Since the higgs-electron-Yukawa coupling is tiny in comparison to the electroweak coupling or the three-higgs coupling \cite{pdg} the most relevant process will require an interaction involving both higgs from both bound states to participate\footnote{Or an experiment sensitive on the level of the higgs-electron-Yukawa coupling.}. This suggests a scale of $\sim2m_h$. Hence, even at LEP2 this process will be strongly suppressed. The Applequist-Carrazone theorem \cite{BeiglboCk:2006lfa} suggests a suppression of at least $s/(4m_h^2)$, which is at the working point of LEP2, the $Z$ mass, at least a suppression by a factor ten. This seems at first sight not a sufficiently strong suppression to avoid detection there.

Is is therefore necessary to investigate the process further to see why it is actually even more suppressed. To do so, it is instructive to follow the principles of gauge-invariant perturbation theory of section \ref{ss:rules}. The relevant matrix element is then given by \cite{Egger:2017tkd}
\be
{\cal M}=\la {\cal O}^{NE}_2(p_1){\bar{\cal O}}^{NE}_2(p_2){\cal O}^F_i(q_1){\bar{\cal O}}^F_i(q_2)\ra\label{eeff}
\ee
\no where only the lower custodial component, i.\ e.\ the electron component in the sense of \pref{opfermexp}, is considered in the initial state. The final-state is defined to be of two fermions of the same type, an exclusive measurement, and therefore the custodial index is not summed over.

At leading order in $v$ this yields\footnote{Note that this result is perturbatively gauge-invariant, though non-perturbatively it is not.}
\be
{\cal M}\sim v^4\la e^+e^-\bar{f}f\ra+\op(v^2)\label{eeffv0}
\ee
\no and thus corroborates the deduction above. In this order the matrix element is just the ordinary, full matrix element. Expanding this matrix element in perturbation theory further reproduces the conventional result \cite{Bohm:2001yx} to all orders in the couplings. Thus, there is no difference at leading order in $v$.

Including further terms yields \cite{Egger:2017tkd}
\be
{\cal M}\approx v^4\la e^+e^-\bar{f}f\ra+v^2\la\eta^\dagger\eta e^+e^-\bar{f}f\ra+\la\eta^\dagger\eta\eta^\dagger\eta e^+e^-\bar{f}f\ra + \text{rest}\label{sexp}.
\ee
\no Terms with an odd number of fields will have additional particles in the initial or final state, and will therefore not contribute if an exclusive measurement is done. While, for the sake of brevity, not all terms are written down and arguments are suppressed, all relevant structures appear. It is now seen that the further terms are suppressed by two or four powers in $v$. Thus, besides the relative suppression of matrix elements from the Appelquist-Carrazone theorem a suppression of order $s/v^2$ arises, which is at the $z$ mass of order ten as well. Both are multiplied, given a total factor 100 of suppression in the matrix element alone on dimensional grounds. This is now much more in-line with the order of sensitivity with which the process has been probed.

However, investigating the present terms further yields even more suppression. The second term in \pref{sexp} adds a pair of Higgs fluctuation fields. Depending on the arguments, they can form an interaction in the initial or final state, or can add a propagating Higgs. The last term includes all possibilities where the Higgs can contribute in both, initial and final state. However, it is even more suppressed, and will therefore be neglected.

To understand what the processes of the second term in \pref{sexp} involve, it is useful to expand the second term further, keeping only leading interactions \cite{Egger:2017tkd}
\bea
\la\eta^\dagger\eta e^+e^-\bar{f}f\ra\approx\la\eta^\dagger\eta\ra\la e^+e^-\bar{f}f\ra  +  \la e^+e^-\ra\la\eta^\dagger\eta\bar{f}f\ra+  \la\bar{f}f\ra\la e^+e^-\eta^\dagger\eta\ra\label{sexp2}.
\eea
\no Thus, there appear three types of corrections to the leading process of \pref{eeffv0}. The first term adds a correction to the perturbative leading term, which is suppressed by a factor $1/v^2$ and a non-interacting spectator higgs. Such a contribution will therefore only affect the process like the presence of a spectator in a hadron collision, and thus is not relevant, except for the formation of the initial and final state, and/or as in QCD by an escape in the beampipe.

The second and third term in \pref{sexp2} correspond to an interaction of the initial state or final state higgs with the corresponding fermions, and the other fermions acting as spectators. The latter two processes therefore correspond to a reaction of the second constituent of the fermionic bound-state in the initial or final state. At leading order perturbative corrections can then be calculated just by expanding all appearing correlation functions to the corresponding order in perturbation theory. Since double-higgs production has not been observed at LEP2, the last matrix element will not be relevant, as it is already negligible as a leading effect. So only the one with a higgs in the initial state could contribute. However, at leading-order this matrix-element is in the $s$-channel proportional to the product of the three-higgs coupling and the higgs-fermion Yukawa coupling and in $t$ and $u$ channel to the higgs-fermion Yukawa coupling squared. Thus, it is negligible compared to the leading term for anything but the the top. Hence, it will again be irrelevant at LEP2.

These considerations show why no contribution at LEP2 of the bound-state structure should be expected. While this was a particular investigation, similar considerations will apply to any other process. And most other processes have not been measured as well at these energies. Thus, this answers the first question from above, why so far nothing has been seen.

Thus, the question remains what is necessary to see something. Performing similar considerations for hadrons at the LHC is highly non-trivial, as the effect of the constituent higgs needs to be isolated somehow from the QCD background. While available parton energies and luminosities do cancel the argument of being off-shell, the suppression by factors of $v$ remain for sub-leading effects, as does the question which processes have a large coupling to the higgs in the initial state. Processes involving tops are natural candidates. This is certainly a challenging, but important topic, which is under investigation \cite{Fernbach:unpublished}. In fact, this needs to be understood. If there would be a measurable effect, this would constitute additional, not yet accounted for, standard-model background. Hence, slight deviations at the LHC with respect to a purely perturbative treatment could be not due to new physics but rather due to corrections from these subtle field-theoretical issues.

The situation is a bit better for future lepton colliders. All presently discussed machines will work at, or substantially above, $\sqrt{s}=2m_h$. Therefore, no off-shell suppression will arise. Also, double higgs and/or double top production is reachable with reasonable sensitivity. Hence, they are the perfect machines to look for these processes or at least give a reasonable bound on them.

Of course, it would be good to have a theoretical estimate. One possibility is to evaluate expressions like \pref{sexp2}, or even \pref{sexp}, perturbatively. This will give a perturbative estimate of the size of the effect. While straightforward, this has not yet been completed \cite{Maas:unpublished} given that the number of involved diagrams, especially when it comes to six-point functions like in \pref{sexp}, is not small even at tree-level. Not to mention loop-level. This approach misses, of course, part of the bound-state effects in the initial and final states.

\subsection{A phenomenological study}\label{ss:pheno}

An alternative is to take a cue from the situation in QCD, where these kind of problems routinely arise \cite{BeiglboCk:2006lfa,Bohm:2001yx,Dissertori:2003pj,Brodsky:2010an}. There, also bound states collide. The only difference in the treatment would be that no alternative path using gauge-invariant perturbation theory is available. In QCD, these effects are directly taken into account starting from an expression like \pref{eeff}, and the bound-state effects are taken care of by parton-distribution functions and fragmentation functions \cite{BeiglboCk:2006lfa,Bohm:2001yx,Dissertori:2003pj}.

Such an approach has been explored in \cite{Egger:2017tkd}. This yields formally
\bea
\sigma_{E^+E^-\to\bar{F}F}(s)&=&\sum_i\int_0^1 dx\int_0^1 dy f_i(x) f_i(y)\sigma_{\bar{i}i\to\bar{f}f}(xp_1,yp_2)\label{pdf}\\
&=&\theta(4m_h^2-s)\sigma_{e^+e^-\to\bar{f}f}\nn\\
&&+\theta(s-4m_h^2)\times\sum_i\int_0^1 dx\int_0^1 dy f_i(x) f_i(y)\sigma_{\bar{i}i\to\bar{f}f}(xp_1,yp_2)\nn,
\eea
\no where the $f$ are the parton distribution functions, and fragmentation was assumed to be to 100\% in the bound-states corresponding to the intermediate elementary fermion $f$. In the second step the approximation was made that the elementary particles need to be on-shell in the initial state. Thereby automatically no changes arise if $s\le 4m_h^2$, implementing in a hard way the discussion above that below this threshold bound-state effects should be negligible. This assumption could be relaxed using generalized parton distribution functions \cite{Lorce:2013pza,Diehl:2015uka}.

Of course, using factorization assumes that $s\gg\max(4m_h^2,4m_f^2)$ to be strictly valid, and thus this requires at least the highest energies of the future linear colliders. Otherwise corrections of order $s/\max(4m_h^2,4m_f^2)$ are expected to arise, due to the Appelquist-Carrazone theorem. Furthermore, the higgs interactions are not asymptotically free. While this appears as a subleading problem also in the usual QCD factorization, these interactions contribute here at leading order. But since the scale where this becomes a non-negligible effect is many orders of magnitude larger than currently reachable, this should not yet be a problem. In addition, violations of factorization due to strong interactions, as happen in QCD \cite{Brodsky:2010an} will likely not be a severe problem, though other problems may be inherited as well \cite{Baumgart:2018ntv}.

This still requires to determine the PDFs. The standard approach using experimental input \cite{Gao:2017yyd} is not yet possible, as so far no effects have been observed. They could therefore only constraint the PDFs. In principle, non-perturbative methods \cite{Nguyen:2011jy,Lin:2014zya,Chen:2016utp,Lin:2017snn} and/or a combination of them with experimental data \cite{Nocera:2017war,Lin:2017snn} would be possible. But even for QCD these techniques are only developing.

\begin{figure}
\includegraphics[width=0.5\linewidth]{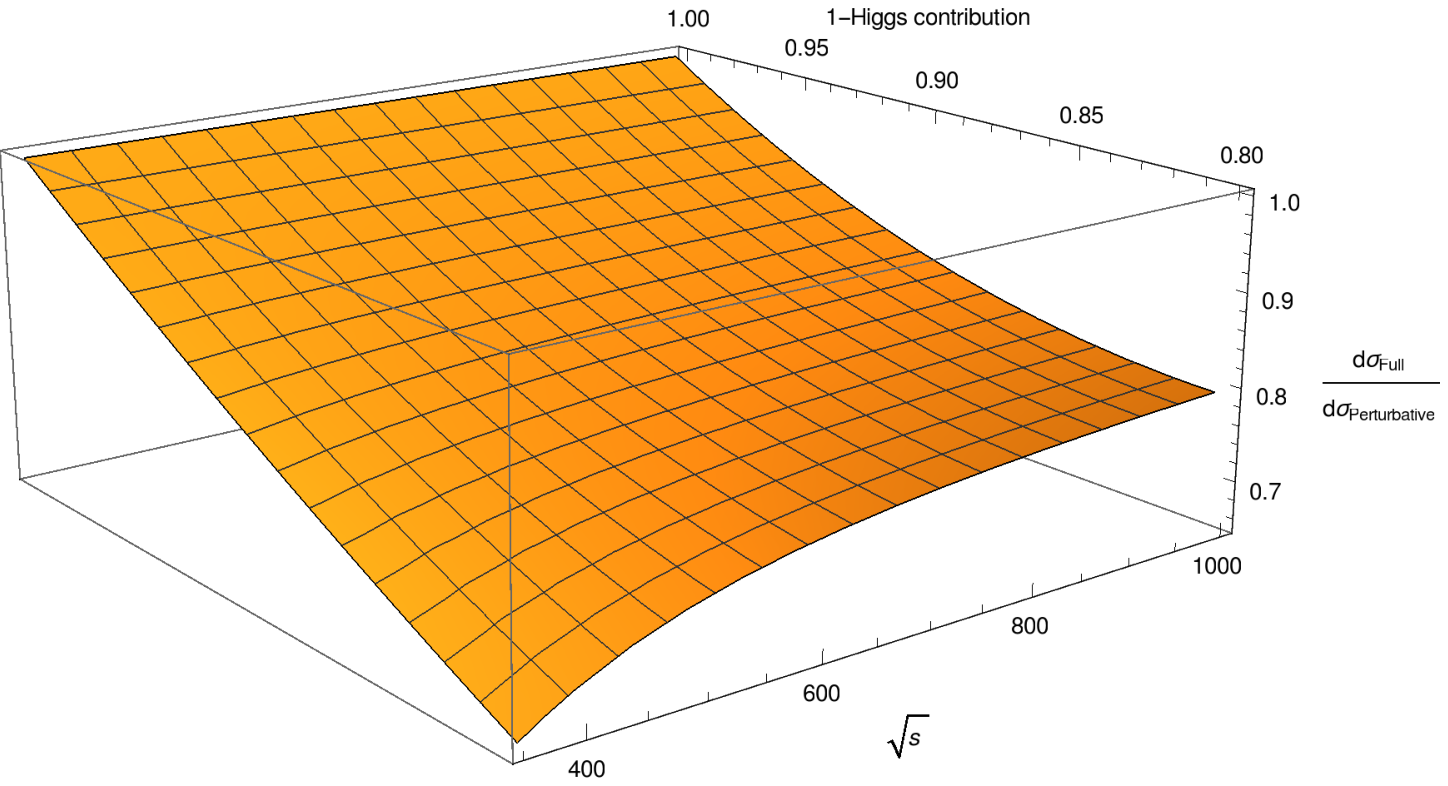}\includegraphics[width=0.5\linewidth]{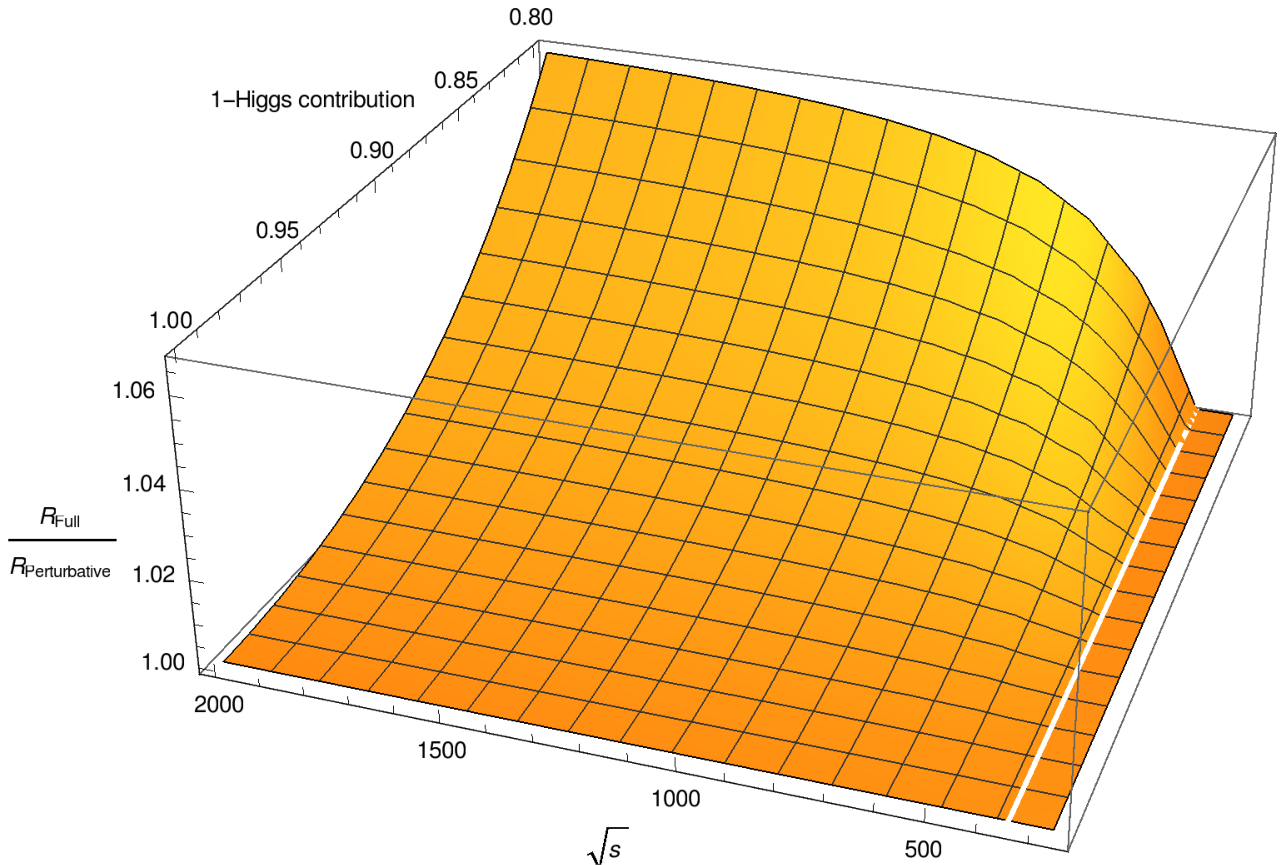}
\caption{\label{fig:pdf}Deviations of the full differential cross section \pref{pdf} to the perturbative one at rapidity zero for the process $\sigma_{E^+E^-\to\bar{T}T}$ (left panel) and for the $R$ ratio (right panel) as a function of the center-of-mass energy and the higgs contribution in \pref{pdfansatz} to the bound state structure, using the approximations and assumptions in \cite{Egger:2017tkd,Maas:2017swq}.}
\end{figure}

Thus, exploratory, in \cite{Egger:2017tkd} the ansatz\footnote{This is the valence contribution to the higgs PDF. Besides the $Q^2$ evolution also higgs sea contributions \cite{Bauer:2017isx,Bauer:2018arx,Bauer:2018xag} are not included in this ansatz.}
\be
f_i(x)=a_i\delta(x)+c_i\delta(x-1)\label{pdfansatz}
\ee
\no was made, where $c_e\gg c_\eta$ was used and the $a_i$ were then fixed by sum rules, implying a quite large $a_\eta$. This realizes to some extent the perturbative picture: The electron carries essentially all of the momentum, and acts almost as a single particle. The higgs carries almost no momentum, and thus fills out the bound state almost constantly, just like the vacuum expectation value would. The results for the cross-section $\sigma_{E^+E^-\to\bar{T}T}$ and for the $R$ ratio $\sigma_{E^+E^-\to\text{hadrons}}/\sigma_{E^+E^-\to M^+M^-}$ is shown in figure \ref{fig:pdf} \cite{Egger:2017tkd,Maas:2017swq}. While the $R$-ratio is essentially not affected, the production cross-section shows a drop, which is well within experimental reach if $c_\eta$ is not too small.

The study can only be considered exploratory at best and probably with assumptions only valid at higher energies then reachable at the currently planned colliders. However, it shows that effects could be limited to only some observables, but it hints where such effects could arise and how to search for them.

Reducing the large amount of approximations needed, especially by getting more information on the PDFs, using off-shell PDFs, and comparing the purely perturbative contribution in \pref{sexp} and \pref{sexp2} are logical next steps. Applying the same rationale to the LHC is certainly more demanding, but ultimately necessary as this is the currently available machine able to reach into kinematically interesting regions. Such investigations are underway \cite{Maas:unpublished,Fernbach:unpublished}.

\subsection{Beyond quantitative effects}\label{ss:qscattering}

An obvious question in the context of searches for new physics is, if the qualitative changes in the spectra discussed in section \ref{ss:bsm} imprint themselves on observable scattering processes. Especially, the physical vector bosons and Higgs particles are generically not stable, but will ultimately decay into standard model particles. A full investigation would require a corresponding extension of the standard model, which is not yet available. However, a toy theory can be used to assess the impact \cite{Torek:2018qet}.

Consider the SU(3) theory with a single higgs in the fundamental representation of section \ref{sss:gut}. There, only two massive vector bosons appear in the spectrum, rather than eight as in perturbation theory. Now, couple in addition a fermion $\psi$ in the fundamental representation to this theory. For the following, it will not matter whether the fermion is chirally or vectorially coupled, nor will it be relevant if it is interacting with the higgs through a Yukawa interaction.

Following the rules of section \ref{ss:flavor}, a physical fermion-number one state is obtained by the fermionic bound state $\Psi=\phi^\dagger\psi$, with $\phi$ the higgs field. This state carries, besides the fermion number, the U(1) custodial charge, which takes over the role of the would-be flavor. Hence, there is only a single physical fermion state. To leading order, its propagator expands like
\be
\big\langle(\phi^\dagger\psi)^\dagger(x)\gamma_0(\phi^\dagger\psi)(y)\big\rangle =v^2\big\langle\bar{\psi}_3(x)\psi_3(y)\big\rangle_\mathrm{tl}+\cdots\;\nn.
\ee
\no Thus, the ground state in this channel has the mass of the perturbative fermion state in the broken subsector. While the appearance of the state is consistent with perturbation theory, it should be noted that no analogue state of the other two fermions $\psi_{1,2}$ appear, in contrast to perturbation theory. They only appear as quantum corrections. Thus, there is again a qualitatively difference in the fermion-number one sector\footnote{There are also three-fermion states, similar to nucleons. They carry a three-times larger fermion number and thus belong to a different charge sector, and will therefore be ignored here.}.

This toy theory is also well suited to apply the full machinery of gauge-invariant perturbation theory of section \ref{ss:rules} to a scattering process. Describing bound-state scattering using the machinery of the LSZ formalism of section \ref{ss:asymp} has been fully developed \cite{Bohm:2001yx}. The cross-section for the elastic scattering process $\bar{\Psi}(p_1)\Psi(p_2)\to\bar{\Psi}(q_1)\Psi(q_2)$ is given by
\be
d\sigma=\frac{(2\pi)^4}{4\sqrt{(p_1p_2)^2-m^4}}|M_{\bar{\Psi}\Psi\to\bar{\Psi}\Psi}|^2\delta(p_1+p_2-q_1-q_2)\frac{d^3\vec q_1}{(2\pi)^32E_{q_1}}\frac{d^3\vec q_2}{(2\pi)^32E_{q_2}}\nn,
\ee
\no where $m$ is the mass of the bound state $\Psi$. The matrix element can then be expanded using the rules of section \ref{ss:rules} to leading order in $v$ and at tree-level in the couplings as
\be
M_{\bar{\Psi}\Psi\to\bar{\Psi}\Psi}=\big\langle\bar{\Psi}(p_1)\bar{\Psi}(q_1)\Psi(p_2)\Psi(q_2)\big\rangle =v^4\big\langle\bar{\psi}_3(p_1)\bar{\psi}_3(q_1)\psi_3(p_2)\psi_3(q_2)\big\rangle_\mathrm{tl}+\cdots\;\nn
\ee
\no For a fundamentally coupled fermion the tree-level vertex for the gauge-boson-fermion vertex is proportional to the Gell-Mann matrices $\lambda^a$. To leading order the matrix elements on the right-hand side consists out of diagrams connected by the expression $\lambda_{33}^aD_{\mu\nu}^{ab}\lambda_{33}^b$, where $D_{\mu\nu}^{ab}$ is the gauge boson propagator \cite{Bohm:2001yx}. But only $\lambda^8$ has a non-vanishing $33$ component, and thus only the propagator $D^{88}$ contributes, which has just the pole at $M_A$, the mass of the heaviest gauge boson. Neglecting interactions with the higgs and the possible masses of the fermions, the result for the spin-averaged matrix element is
\be
|M_{\bar{\Psi}\Psi\to\bar{\Psi}\Psi}|^2=\frac{32g^4}{9}\left(\frac{s^2+u^2}{(t-M_A^2)^2}+\frac{t^2+u^2}{(s-M_A^2)^2}+\frac{2u^2}{(s-M_A^2)(t-M_A^2)}\right)\nn
\ee
\no where $g$ is the gauge coupling and $s$, $t$, and $u$ the usual Mandelstam variables, and the odd prefactor in comparison to usual Bhabha scattering comes from the involved Gell-Mann matrix.

Hence, in the scattering cross-section only poles at $M_A$ arise. But this is exactly the mass of the physical vector state! Thus, even in the cross-section measured in an experiment at leading order only the physical state shows up as a resonance. This supports that only the gauge-invariant states are observable. Of course, investigations beyond leading order and in other processes are needed to confirm this. Still, this shows that differences in the physical spectrum, no matter the open decay channels, can potentially carry over to scattering cross-sections. This would deliberately alter the signature of a theory qualitatively compared to the expectations in standard perturbation theory.

\section{Further aspects}\label{s:further}

There are a number of further aspects of the BEH effect, which are neither directly affecting the manifestation of the BEH effect, nor the physical spectrum, but are still very relevant to phenomenology. These will be discussed in the following. For most of these not yet a conclusive final statement can be made beyond perturbation theory, and therefore are rather evolving subjects, in contrast to the main part of this review where the qualitative features are already quite well traced out.

\subsection{The triviality problem}\label{ss:triv}

One problem allured to in section \ref{s:beh} is the question whether the theory itself exists as an interacting quantum-field theory in the continuum, i.\ e.\ whether it is not trivial \cite{Callaway:1988ya}. In the remainder of this text this was taken as an ultraviolet problem, and it was assumed not to be relevant to the low-energy physics mostly treated.

However, this is only true to some extent. If the theory is trivial, it has only interactions as long as an ultraviolet cutoff is in place. Thus, even in the sense of an effective theory, the results will depend on the regulator, e.\ g.\ the lattice spacing.

In a perturbative setting, the theory is considered trivial due to the presence of Landau poles \cite{Callaway:1988ya}. Fully non-perturbative, this is far less clear. Lattice calculation have found indications of a continuum limit at infinite Higgs self-coupling \cite{Bonati:2009pf}. However, it is not yet clear whether this is an interacting theory \cite{Dashen:1983ts,Fernandez:1992jh}, and no such strong statements as in case of the $\phi^4$ theory \cite{Luscher:1987ay,Luscher:1987ek,Hasenfratz:1988kr,Luscher:1988gc,Luscher:1988uq,Zimmermann:1991xx,Heller:1993yv} are actually available. In fact, arguments using the functional renormalization group have been given that there should exist an interacting continuum limit at weak Higgs self-coupling \cite{Gies:2015lia,Gies:2016kkk,Gies:2019nij}, which is not excluded by studies of these regions on the lattice \cite{Wurtz:2013ova,Maas:2014pba}. Furthermore, fermions can potentially influence the qualitative behavior already at the perturbative level \cite{Callaway:1988ya,Litim:2014uca,Litim:2015iea,Bezrukov:2012sa}, but also non-perturbatively \cite{Gies:2013pma}. This does not even mention the influence of more extreme settings, like extended Higgs potentials\footnote{This possibility has also been studied for the ungauged theory \cite{Gies:2014xha,Chu:2015nha,Borchardt:2016xju}, to essentially the same effect.} \cite{Gies:2017zwf} or the influence of quantum gravity \cite{Dona:2013qba,Eichhorn:2016esv,Eichhorn:2017ylw,Gonzalez-Martin:2017bvw}.

In the end, even if the theory is trivial, this does not prevent its usefulness as a low-energy effective theory. In this case the ultraviolet cutoff necessarily acts as an additional parameter of the theory. This does not affect the low-energy behavior of the theory \cite{Hasenfratz:1986za}, except possibly for the range of values observable quantities can take \cite{Eichhorn:2015kea}. Thus, even if the class of theories treated here should be trivial, it will not matter for the bulk of this review\footnote{The only exception would be, if the gauge symmetry would turn into a non-gauge symmetry in the ultraviolet-complete theory.}, as long as the cutoff is sufficiently high compared to the required energy range.

Moreover, if such a deviation would be observed in an experiment, this will be a hint for the ultraviolet-complete theories of which the theories treated here are the low-energy effective ones. This is very similar to the case of chiral perturbation theory \cite{BeiglboCk:2006lfa} or to the use of low-energy effective theories to parameterize perturbatively new physics in electroweak physics \cite{deFlorian:2016spz}. In fact, the problem can be turned around. Under the assumption that the theory is trivial, and a firm control over how this manifests, predictions can be made for possibly new physics to achieve non-triviality \cite{Bezrukov:2012sa,Eichhorn:2017ylw,Christiansen:2017cxa}.

There is also a quite subtle issue. Already for the $\phi^4$ theory Haag's theorem \cite{Haag:1992hx} forbids unitary equivalence between the interacting theory and the non-interacting theory. Thus, the dependence of the regulated theory on the regulator needs to be highly non-trivial, though not necessarily non-analytic \cite{Glimm:1987ng}. This problem is amplified in the gauged theory. There, the physical and unphysical state space need to change discontinuously in the case of triviality when the regulator is removed: States which were beforehand gauge-dependent, as the Higgs, are no longer so afterwards. Whether this qualitative difference between the ungauged and gauged case is relevant is not yet clear.

\subsection{The hierarchy problem and the Higgs mass}\label{ss:hierarchy}

One issue which has played an important role in phenomenology is the hierarchy problem \cite{Feng:2013pwa,Morrissey:2009tf}. In its simplest form it is often stated that the value of the Higgs mass is arbitrary and for random values of the coupling constants it should be of order of any cutoff, i.\ e.\ the scale of new physics. At loop level, this problem becomes more intricate, as the Higgs mass becomes scheme-dependent, and a definition of a physical Higgs mass is much less clear.

However, there are two physical ways of posing the same question:
\begin{itemize}
 \item Why is the scale of the standard model physics so small compared to the Planck scale?
 \item Is the theory very sensitive\footnote{Phrasing mathematically what 'very sensitive' means is always a question of taste. Here, it will be used in the sense of what most people consider to be very sensitive at the time of writing. This amounts to a dependence on the parameters of the theory which for dimensionless quantities is stronger than logarithmically.} to small changes of the parameters, and if yes, why? If yes, this requires fine-tuning of the parameters of the theory for particular quantitative features.
\end{itemize}
The first question is clearly beyond the scope of this review, and is likely requiring a full quantum theory of gravity to answer \cite{Shaposhnikov:2009pv,Dona:2013qba,Eichhorn:2016esv,Litim:2011qf}.

The second question is, however, directly linked to the present subject. At the perturbative level, it is indeed so \cite{Feng:2013pwa,Callaway:1988ya,Morrissey:2009tf} that the theory is sensitive, especially to the Higgs mass parameter, i.\ e.\ the coefficient of the quadratic term in the Higgs potential in the Lagrangian \pref{la:hs}. This seems to be also true non-perturbatively \cite{Maas:2014pba,Gies:2015lia,Gies:2016kkk}. Especially, realizing large values of the cutoff requires a fine-tuning of the parameters of the theory. Thus, from this point of view, also non-perturbatively this theory is fine-tuned. This does not imply that any overarching theory needs to be not fine-tuned. The theory itself in isolation is just so.

But the low-energy physics of the theory seems to be quite independent of this fine-tuning issue. E.\ g.\ finding a light scalar seems not to be connected to finding a small or large cutoff, at least for the standard model case \cite{Maas:2014pba}. This shows that the masses of the observed particles are actually not a good argument itself for the existence of fine-tuning. Whether this is true also in more general settings is unclear. This therefore points to a genuine BSM resolution of the fine-tuning problem, if it is not a feature of nature.

\subsection{Vacuum stability}\label{ss:vs}

When regarding the BEH effect from the point of view of spontaneous symmetry breaking, it is at the perturbative level straightforward to find that multiple minima can emerge \cite{Sher:1988mj,EliasMiro:2011aa,Alekhin:2012py}. Consequently, if minima of different depths are found, it is possible that metastability can occur. Strictly perturbatively, the standard model seems to be on the edge of such a scenario \cite{Alekhin:2012py,Buttazzo:2013uya,Gabrielli:2013hma,Bednyakov:2015sca,Iacobellis:2016eof}. If this would be naively true, then, like in a superheated gas, the system could undergo a spontaneous phase transition.

However, a purely perturbative setup suffers from a number of conceptual problems \cite{Holland:2003jr,Branchina:2005tu,Branchina:2008pc,Bulava:2012rb,Branchina:2014rva,Eichhorn:2015kea,Borchardt:2016xju}. It is therefore necessary to perform full non-perturbative studies. These strongly suggest that such a metastability cannot arise \cite{Bulava:2012rb,Eichhorn:2015kea,Borchardt:2016xju} in a generic theory with a Higgs potential having at most a $\phi^4$ term. Only with higher-dimensional operators at tree-level an instability can be introduced either in the Higgs sector \cite{Chu:2015nha,Eichhorn:2015kea,Sondenheimer:2017jin} or the Yukawa sector \cite{Gies:2017zwf}. However, these are so far mostly hints from the ungauged theory or involve some level of other approximations. Nonetheless, this underlines the importance of non-perturbative effects even at weak coupling.

From the point of view of gauge invariance this problem appears even more unclear. The perturbatively distinct minima arise because of different values of the vacuum expectation value of the Higgs. They do still have the same breaking pattern. The Higgs vacuum expectation value is, however, gauge-dependent, and may differ between different gauges, not only in direction, but also in magnitude. The discussed zero and non-zero values have only been two particular cases.

This does not prevent per se a phase structure with multiple quantum phases coexisting in the vacuum. The phase structure is then just necessarily driven by a different physical mechanism than the BEH effect. Because of the arguments in section \ref{ss:pd}, this is excluded for the Higgs sector alone. Since, however, the problem is driven by the interaction with the top sector \cite{EliasMiro:2011aa,Alekhin:2012py}, this is not a sufficient argument as long as the argument of section \ref{ss:pd} cannot be extended to the Higgs-top sector. This seems so far not possible.

Thus, if such a metastability exists, it is not characterized by different values of the Higgs vacuum expectation value, as without gauge fixing it is always zero. Also, as it was discussed at length in section \ref{ss:pd}, gauge-fixed signals do not necessarily imply physical effects \cite{Caudy:2007sf}. There are, of course, other possible quantities to characterize phases and phase transitions, in particular thermodynamic bulk quantities like the free energy. It thus remains ultimately to determine, e.\ g., the free energy as a function of the parameters of the theory to decide this question.

\subsection{The finite-temperature transition}\label{ss:ft}

The vacuum stability problem and the whole phase diagram of sections \ref{ss:pd} and \ref{ss:pdlat} address the quantum aspects of BEH physics. However, also the thermodynamic aspects\footnote{Similarly, theories of the type considered here also appear in other context as effective theories to describe phase structures \cite{Yamamoto:2018}.} are of great importance to cosmology \cite{Kapusta:2006pm,Dolgov:2006xi,Morrissey:2009tf}. In particular, the question whether there is a (strong) first-order phase transition at finite temperature could be decisive for the baryon abundance problem and baryogenesis in general. If there is none, this would be a strong indication that at this scale new physics needs to be active.

This subject, especially when considering its cosmological implications, is vast, and certainly deserves a review in its own right. But there are also very direct consequences arising from the perspective on BEH physics studied here.

First of all, as has been discussed at length in sections \ref{ss:vev} and \ref{ss:pd}, it is insufficient to use the Higgs vacuum expectation value as the only indicator of the presence of BEH physics, as is done in the simplest possible perturbative treatments \cite{Kapusta:2006pm}. Otherwise, also at finite temperature it is possible that a phase transition is only present in some gauges. Modern treatments use therefore instead physical observables, like the free energy or other gauge-invariant quantities \cite{DOnofrio:2015gop,Laine:2015kra,Wellegehausen:2011sc,Reichert:2017puo}.

In fact, investigations \cite{Kajantie:1996mn,Csikor:1998ge} showed that no phase transition sufficient to explain the baryon asymmetry is observed for a Higgs mass as large as 125 GeV. Interestingly, one is seen only at or below a Higgs mass around 80 GeV, and thus in a regime which seems to be dominated by QCD-like physics, as discussed in section \ref{ss:pdlat}. Furthermore, it appears that a Higgs-sector with an extended potential seems to be able to provide a more favorable thermodynamic behavior \cite{Grojean:2004xa,Reichert:2017puo}, as is also indicated by non-gauge theories \cite{Chu:2017vmc,Hegde:2013mks,Akerlund:2015fya}. This may indicate a possible gateway to beyond-the-standard model physics or at least to a non-standard realization of BEH physics \cite{Reichert:2017puo}.

However, a few of the standard arguments in this respect will possibly need a reappraisal, when taking the results reviewed here at face value.

One is the expectation that above the phase transition the BEH effect is gone, and thus, up to thermal masses, all particles become essentially massless \cite{Kapusta:2006pm}. But the theory at zero temperature is gaped throughout the phase diagram, as discussed in sections \ref{ss:pd} and \ref{ss:pdlat}. Especially, the physical vector states are always massive, while the elementary gauge bosons are also either massive or do not exhibit a mass-pole at all, see section \ref{sss:w}. Of course, this will only affect the situation around the phase transition, as at even higher temperatures any masses become irrelevant compared to thermal effects anyways. But it is precisely this region which is of most interest to baryogenesis. Note that the implications for fermions stand distinct from this, as here too little is yet known, though even there a gaping may still always occur \cite{Iida:2007qp}. Also, the fact that there appears to be an endpoint in the Higgs-mass-parameter and temperature plane \cite{Kapusta:2006pm,Kajantie:1996mn,Csikor:1998ge} seems to indicate that there is an analytic connection between the low-temperature phase and the high-temperature phase, and thus no qualitative differences can arise.

The other issue is that the gauge symmetry is restored \cite{Kapusta:2006pm}. As discussed at length in section \ref{s:beh}, it is never really broken in the first place\footnote{Although it is observed that certain features seem to suggest it. E.\ g.\ when increasing the Higgs mass the phase transition temperature shifts on the lattice between the two different values of the theories with different gauge groups \cite{Brower:1982yn,Kikugawa:1985ex,Wellegehausen:2011sc}. However, this is observed when the bare Higgs mass is pushed above the lattice cutoff, and thus the ordering of limits becomes incorrect and the lattice theory is probed, rather than an approximation of the continuum theory.}. What may happen, however, is that a gauge condition involving explicitly a non-zero Higgs vacuum expectation value can no longer be satisfied at finite temperature, as discussed in section \ref{ss:qulev}. This indeed leads to non-analyticities in the gauge-dependent correlation functions, just as is the case in the quantum phase diagram. But as in case of the quantum phase diagram in section \ref{ss:pd}, this does not necessarily imply a physically observable phase transition. Also, because in this case the FMS expansion fails gauge-invariant perturbation theory can no longer be applied. Thus, the physical spectrum may no longer be mapped in a one-to-one way to the elementary degrees of freedom, as is the case also for, e.\ g., Yang-Mills theory and QCD. In particular, the elementary states may no longer exhibit physical features \cite{Maas:2011se}.

Thus, while the quantitative determination of the phase transition temperature from bulk thermodynamics will be accurate, a microscopic description of the dynamics above the phase transition may be much more involved than usually anticipated. Especially, the idea of (almost) non-interacting, massless degrees of freedom above the phase transition may well be as inaccurate as it is in QCD \cite{Maas:2011se}, not withstanding the essentially trivial thermodynamic properties. Thus, full non-perturbative calculations of the gauge degrees of freedom and the fermions would be desirable to ultimately understand the details of the electroweak phase transition and its implications for baryogenesis\footnote{This is especially true if the generation structure, as speculated e.\ g.\ in \cite{Egger:2017tkd}, would be dynamically generated by the gauge-Higgs dynamics.}.

\subsection{Theories without an elementary Higgs}\label{ss:tc}

It may not be  directly obvious, but the considerations here have also substantial implications for theories without an elementary Higgs, and thus without a BEH effect \cite{Maas:2015gma}. Such theories are especially theories like Technicolor or composite Higgs models \cite{Hill:2002ap,Morrissey:2009tf,Andersen:2011yj,Sannino:2009za}.

The reason is that even in such theories the weak gauge interaction remains. Thus, the necessity to have a gauge-invariant replacement to create an observable, gauge-invariant equivalent to the $W$ and $Z$ bosons remains. It is only no longer the case for the Higgs, or, more appropriately, the observed scalar particle. As it is a composite state of additional, new particles in such theories, it can be just a normal, gauge-invariant scalar state, and no issue of gauge-dependence arises for it. So it is actually the Higgs which is least problematic from this point of view.

But the $W$ and the $Z$ bosons are a different story entirely. Due to the absence of the Higgs there is no custodial symmetry. And thus no global symmetry to provide the multiplet structure. Still, observation dictates to have a (nearly) degenerate triplet of vector states. But it is not only the degeneracy pattern that is a challenge. Considering only such a new extended (strongly-interacting) sector and the weak interaction in isolation, these vector states need\footnote{Of course, if somehow this is strongly influenced by the standard model fermions this is not necessary. But the author is not aware of any such scenario and would be grateful to hear about it.} to be the lightest states\footnote{Barring the existence of axions or light dark matter particles.}. Thus, a symmetry with a triplet representation must exist.

In a Technicolor-like scenario, the degeneracy pattern can be provided by a suitable techniflavor symmetry. Then, the observed vector bosons, mimicking the gauge-dependent $W$ and $Z$ bosons, will form a techniflavor triplet. However, this still requires to have as lightest, but still massive, particles composite vector states. In theories with (strongly-interacting) fermions so far only either scalars or pseudoscalars have been observed as the lightest particles \cite{Gattringer:2010zz,Andersen:2011yj,Kuti:2014epa}. However, in all full non-perturbative studies so far  only one gauge interaction was considered, and thus the possibility remains that this is a genuine feature of two-gauge-interaction theories. Of course, it is also possible that this will arise in single-gauge interaction theories with parameters and/or particle content not yet studied.

In addition, this implies new problems with the standard model fermions. As discussed in section \ref{ss:flavor} the weak isospin needs to be exchanged for a global symmetry to create gauge-invariant states. This cannot be achieved by creating bound states either with the additional gauge bosons nor with single additional fermions. The prior are not suitably charged and the latter would alter the quantum numbers. Thus, they need to form at least more complicated bound-states, including multiple of the new fermions, being effectively mixed baryons with the new states.

Effectively, at low energies, such a construction would still look like the usual standard-model, as long as the substructure of the new bound states cannot be resolved. But then still the substructure of the gauge-invariant standard-model fermions and the scalar would differ compared to the standard model case alone. This may be detectable by precision experiments, as was already argued for in the pure standard-model case in section \ref{s:scattering}.

\section{Summary, conclusion, and future challenges}\label{s:sum}

Appreciating fully the fact that the weak interactions are a non-Abelian gauge theory leads to a very intricate structure of the standard model as a whole. In particular, it requires to reevaluate our view of what is the structure of the observed particles on a very fundamental level, with the possible exception of right-handed neutrinos. The standard-model then tells a story of intricate cancellations, but a simple effective theory. The latter leads to the same phenomenology established using perturbation theory, and is thus in beautiful agreement with existing experimental results.

While the agreement with current experiments is reassuring, it can only be the first step. Such an intricate interplay on a fundamental level needs to be established in detail theoretically, and eventually experimentally. All lattice simulations of restricted sectors of the standard model are in agreement with the underlying field theory \cite{Shrock:1985un,Shrock:1985ur,Lee:1985yi,Maas:2012tj,Maas:2013aia,Raubitzek:unpublished}. But as any non-perturbative methods they are yet far from exact. Also, they are limited yet to bosonic sectors. A full non-perturbative analysis of the full standard model, or at least including part of the Yukawa sector, at the level of gauge-invariant and observable quantities remains desirable. But given the inherent technical challenges this will remain a future perspective for some time to come.

A quite different challenge is the fact that such an intricate structure needs to induce eventually deviations from a phenomenology based entirely on perturbation theory. The parameters of the standard model, and the very good agreement to experiment so far, strongly suggest that these will be small, or even tiny, deviations. Thus, true precision measurements will be needed to uncover them. For theoretical calculations it remains a challenge to quantify them, and make a prediction what kind of experimental effort would be needed to detect them. Provided that fermionic corrections are not on a very large quantitative, or even qualitative, level this appears easier and more within reach than a full inclusion of fermions. Methods to calculate the size of bound-states \cite{Koponen:2015tkr,Gockeler:2003ay} and determine or infer (quasi) distribution functions \cite{Nguyen:2011jy,Lin:2014zya,Chen:2016utp,Nocera:2017war,Lin:2017snn,Gao:2017yyd} are available or are being developed, and could be and are applied in a straight-forward way to the observable scalars and vectors \cite{Egger:2017tkd,Raubitzek:unpublished,Fernbach:unpublished}. This should provide a first estimate of the size of the effects.

In total, while quantitatively (yet) a tiny effect, gauge invariance invites us to reevaluate the qualitative way of how we think about the elementary particles we know and we hunt for. For the standard model, these insights only change the way how we perceive particles. It does not necessarily change how we treat them. And the identification of the elementary particles with the observed particles, while not literally correct, will remain certainly a valid, and pretty good, approximation.

But this is not the case for beyond-the-standard model scenarios. As has been seen in section \ref{ss:bsm}, and has been supported by all available lattice simulations to date \cite{Maas:2016ngo,Lee:1985yi,Maas:unpublishedtoerek}, there could be distinct, qualitative differences between the observable and elementary spectrum of such theories. This problem still needs full classification \cite{Maas:2017xzh}. But it seems likely that a correct prediction of the physical spectra needs to be based on gauge invariance. Gauge-invariant perturbation theory of section \ref{ss:gipt} appears to be a suitable tool to do so. The implications moreover also pertain to theories without elementary Higgs, as discussed in section \ref{ss:tc}. Fortunately, the technical changes appear to be comparatively small \cite{Maas:2017xzh}, and much what has been done can be reused.

In the end, what remains is that particle physics is on a fundamental level more complex than expected. Recognizing the observed particles for what they are - relatively involved composite objects - may actually be an inspiring principle. Especially, as sections \ref{ss:qed}-\ref{ss:qcd} showed, in the standard model a relatively baroque structure of many wheels interact in a quite intricate pattern. But then, whenever it was understood in particle physics that something has a composite, intricate structure, this gave us a new handle to understand the systematics of what lies below.

\section*{Acknowledgments}

Working through half a century of research largely not widely known is always begging for oversights. I apologize to everyone I have missed out, and will be happy to add anything brought to my attention regularly to the arXiv version of this review. The same goes for any corrections.\\

I am very grateful to all the people I had the opportunity to discuss with about this topic, and learn from them. After all, the development of these questions has taken now more than fifty years. With much of it being hidden in the literature, as it was not directly relevant to experiment due to the coincidental structure of the standard model, much of preparing this review was as much scientific archaeology as it was research.

I am especially grateful to Christian B.\ Lang for many important hints, and many insights about lattice gauge theory. His knowledge was indispensable over the last eight years in which I worked on this topic.

Particular thanks go also to my collaborators with whom I worked on this topic, Larissa Egger, Simon Fernbach, Tajdar Mufti, Dominik Nitz, Leonardo Pedro, Simon Pl\"atzer, Sebastian Raubitzek, Robert Sch\"ofbeck, and especially Ren\'e Sondenheimer and Pascal T\"orek.

I am grateful to all those, with whom I could discuss this and related topics, and which helped me to understand it, or which helped me understand aspects which were necessary before understanding this topic. Special thanks should be given here to Reinhard Alkofer, Luigi Del Debbio, Dennis Dietrich, Philippe de Forcrand, Mathias Garny, Christof Gattringer, Holger Gies, Leonid Glozmann, Jeff Greensite, J\"org J\"ackel, Sebastian J\"ager, Frank Krauss, Jan Pawlowski, Claudio Pica, Tilman Plehn, Michael M\"uller-Preussker, Francesco Sannino, Erhard Seiler, Luca Vecchi, Andreas Wipf, Hartmut Wittig, Luca Zambelli, Daniel Zwanziger, and Roman Zwicky.

I am grateful to Philippe de Forcrand, Jeff Greensite, Christian Lang, Leonardo Pedro, Erhard Seiler, Ren\'e Sondenheimer, Franco Strocchi, and Milan Vujinovic for a critical reading of the manuscript and helpful comments, corrections, and improvements. I am indebted to Robert Shrock for pointing me to a wealth of additional lattice references.

This work was supported by the DFG under grant numbers MA 3935/5-1, MA-3935/8-1 (Heisenberg program) and the FWF under grant numbers M1099-N16 and W1203-N16. Some of the results were obtained from simulations on the HPC clusters at the Universities of Jena, Graz, TU Graz, and the Vienna Scientific Cluster (VSC). I am grateful for the teams of the HPC centers for the very good support. Partly the ROOT framework was used \cite{Brun:1997pa}.\\

Finally, but foremost, I dedicate this work to my beloved wife Renate Knobloch-Maas. Thank you for everything.

\appendix

\section{Frequently-asked questions}\label{s:faq}

As discussed in the introduction, most of the results presented here are away from the main stream of both perturbation theory and lattice gauge theory. Thus, most of it is widely unknown. This leads to a set of questions, which very often arise upon first encounter with this topic. To facilitate such first encounters and improve the accessibility of the topic here the questions my collaborators and myself have encountered most often over the years are assembled and briefly answered, together with pointers to the corresponding sections of this review and the most pertinent original literature, if appropriate.

\begin{enumerate}
 
 \item {\bf Is all of this necessary?}\\
 Yes, for several reasons. One is that it gives additional contributions in scattering cross sections, although it is yet unclear how large, which could fake new physics, see section \ref{s:scattering} and \cite{Egger:2017tkd,Raubitzek:unpublished}. The other is that the consistency of the theory demands it, see section \ref{ss:physstates} and \cite{Frohlich:1980gj,Frohlich:1981yi}, and if not confirmed experimentally the theoretical description of the standard model is actually questionable.
 
 \item {\bf Does section \ref{s:global} imply that current treatments of (spontaneous) global symmetry breaking are incorrect?}\\
 No, not all. The purpose of section \ref{s:global} is to clearly define what the various terms mean, and to make often implicit left assumptions or limits explicit. Because of these implicit assumptions and limits, all the standard treatments are completely adequate. However, to fully appreciate the subtleties in the following sections on local symmetries it is useful to make them explicit in the case of global symmetries first. It therefore also does not contain any new claims on global symmetries at all. Also, it is very important to distinguish expectation values and individual measurements, as discussed in section \ref{sss:metastable}.
 
 \item\label{q2} {\bf Is perturbative BRST symmetry not sufficient to ensure gauge invariance?}\\
 No, because the Gribov-Singer ambiguity breaks perturbative BRST symmetry, and this applies also when a BEH effect is active. Thus, the Kugo-Ojima construction fails in the usual form. See sections and \ref{ss:gribov} and \ref{ss:physstates} and \cite{Fujikawa:1982ss,Lenz:2000zt}.
 
 \item {\bf What is the status of proofs to all orders in perturbation theory?}\\
 Any quantity $A$ can be split as
 \be
 A=A_p+A_n\nn
 \ee
 \no where $A_p$ can be expanded in a perturbative series, and $A_n$ is any remainder. There are general arguments why $A_n$ can in general not be zero \cite{Haag:1992hx}. Proofs to all orders in perturbation theory are statements on the contribution $A_p$, and are valid. However, depending on $A_n$ taking only $A_p$ can be an almost perfect, good, or qualitative wrong approximation of $A$. This cannot be determined from $A_p$ alone. In addition, in sections \ref{s:spectrum} and \ref{s:scattering} it is shown that standard perturbation theory only captures a part of $A_p$, $A_p=A_s+A_g$, where $A_s$ is the standard perturbative result based on elementary degrees of freedom as asymptotic states and $A_g$ are additional contributions from the FMS expansion. Here, statements to all orders in perturbation theory only apply to $A_s$.
 
 \item {\bf What is about the perturbative proof of gauge invariance of perturbation theory?}\\
 As discussed in \cite{Lee:1974zg} perturbation theory in the theories treated here makes only sense in a certain subclasses of gauges, preventing any notion of perturbative gauge invariance. As an example, see the discussion in section \ref{ss:masses} on the mass of the vector bosons. See also question \ref{q2}.
 
 \item {\bf How does the violation of gauge invariance in a fixed gauge in perturbation theory explicitly arises?}\\
 This happens due to appearance of zero and negative eigenvalues of the Faddeev-Popov operator, and thus by cancellations in the path integral. These zero or negative eigenvalues are genuine non-perturbative, and do not occur in perturbation theory. Such cancellations can remove perturbatively gauge-invariant states non-perturbatively from the spectrum. These additional eigenvalues arise due to the Gribov-Singer ambiguity discussed in section \ref{ss:gribov}, and is known as the Neuberger $0/0$ problem \cite{Neuberger:1986xz}. See also \cite{Maas:2011se} for details on the subject of the Gribov-Singer ambiguity.
 
 \item {\bf What is about the Nielsen identities, saying that the masses are gauge-invariant?}\\
 The Nielsen identities only claim gauge-parameter invariance, which is a weaker statement, see section \ref{ss:masses} and the original article \cite{Nielsen:1975fs}. This argument is often made in connection with perturbation theory, but, as discussed in section \ref{sss:stdgauge} and indicated in \cite{Lee:1974zg}, the masses of, e.\ g., the elementary gauge bosons differ perturbatively in gauges, which cannot be perturbatively deformed into each other.
 
 \item {\bf Is the Higgs vacuum expectation not a measurable, and thus physical, quantity?}\\
 No. What is measured are quantities like decay widths, cross-sections, and masses. In a fixed gauge with a non-vanishing vacuum expectation value, they depend, often at tree-level, in a simple way on the vacuum expectation value. A well-known example is the Fermi constant governing the $\beta$-decay of the neutron \cite{Bohm:2001yx}. However, there are gauges without vacuum expectation value \cite{Maas:2012ct,Lee:1974zg}, and in fact it cannot be a physical quantity \cite{Osterwalder:1977pc,Seiler:2015rwa,Fradkin:1978dv}. In such gauges, however, perturbation theory usually does not work, and therefore non-perturbative methods would be needed to calculate the observable quantities, see section \ref{sss:stdgauge} and \cite{Lee:1974zg}.
 
 \item {\bf Does not the triviality problem needs to be solved first?}\\
 No. Even if the theory is trivial, taking it as a low-energy effective theory generates the same problems. Also, there is no conceptual problem in applying the reviewed construction to an effective theory, see section \ref{ss:triv} and \cite{Hasenfratz:1986za,ZinnJustin:2002ru}.
 
 \item {\bf If the theory is trivial, does this not preclude any comparisons of the lattice calculations and continuum calculations?}\\
 Trivial theories are non-trivial once a cutoff is introduced. They then depend on the cutoff, the regularization, and the renormalization scheme. Thus, it is not a question of lattice against continuum, but between different such schemes, e.\ g.\ also MOM and MS schemes in the continuum. However, if the dependence on the regulator and renormalization is suppressed by a large scale, e.\ g.\ the cutoff itself, results will agree up to corrections of this order. While there is no exact, non-perturbatively valid proof available for the theories under investigation, this is supported by \cite{Hasenfratz:1986za} and the Appelquist-Carrazone theorem \cite{BeiglboCk:2006lfa}.
 
 \item {\bf Could not fermions alter this, as they are not included in the lattice calculations?}\\
 Fermions fit straightforwardly in the picture, see \cite{Frohlich:1981yi} and sections \ref{ss:flavor} and \ref{ss:qcd}, and do not alter anything at the level of gauge-invariant perturbation theory. Testing them on the lattice is currently impossible for chirally coupled fermions, see section \ref{ss:lattice}, but it would be possible for vectorial fermions. The arguments for vectorial fermions are identical to the ones for chiral fermions, and this would allow to test the qualitative mechanism, if not the quantitative one. However, so far no such calculations exist, as they would be still quite expensive, even if no standard-model-like mass hierarchies would be used.
 
 \item {\bf How large are the quantitative effects in the standard model?}\\
 Qualitatively small, as discussed in sections \ref{ss:gscattering} and \ref{ss:pheno}, but reliable numbers are only in progress \cite{Egger:2017tkd,Maas:unpublished,Raubitzek:unpublished,Fernbach:unpublished}.
 
 \item {\bf How do the physical bound states (e.\ g.\ the scalar singlet) differ from the elementary ones?}\\
 If the composite state expands under the FMS mechanism to an elementary state, both will be identical on-shell, see section \ref{ss:gipt}. However, off-shell their properties are different, due to the additional terms in equations like \pref{op0ppre}. Quantitative statements are currently under development \cite{Raubitzek:unpublished,Maas:unpublished}. In a situation where the composite state is under the FMS expansion not in a one-to-one mapping to an elementary state, like in the example of section \ref{sss:gut}, this question is actually not well-defined.
 
 \item {\bf Could the composite states described in section \ref{s:spectrum} decay into their constituents, e.\ g.\ the composite Top into an elementary bottom and higgs and elementary leptons?}\\
 No, as the constituents are not physical particles, since they are gauge-dependent. This is quite similar to QCD, where a hadron also cannot decay into quarks. And in fact, confinement is not conceptually different from the BEH effect, see section \ref{ss:pd} and \cite{Osterwalder:1977pc,Seiler:2015rwa,Fradkin:1978dv}. Of course, just like hadrons can decay into lighter hadrons, they still can decay in kinematically allowed other composite states, giving e.\ g.\ the decay of a physical, composite Muon into a composite, physical Electron and suitable, physical Neutrinos, thus providing the experimentally well-known decay patterns \cite{pdg}. See also section \ref{ss:asymp}.
 
 \item\label{q13} {\bf What is about operator mixing?}\\
 This is indeed relevant, as discussed in section \ref{ss:gipt}. In principle, it would be necessary to find suitable operators for asymptotic states to describe the actually observed particles also in gauge-invariant perturbation theory. In lattice calculations, this is done, see section \ref{ss:latgipt}, and usually the simplest operator approximates the ground states well \cite{Wurtz:2013ova,Maas:2014pba,Maas:unpublishedtoerek} but not perfectly \cite{Raubitzek:unpublished}.
 
 \item {\bf Is it guaranteed that any operator can be used?}\\
 No, there is always in principle the problem that an operator has no overlap with a given state \cite{Gattringer:2010zz,DeGrand:2006zz}. There is no known method, aside from exact solutions, to figuring this out. But, see question \ref{q13}, often the simplest one seems to have.
 
 \item {\bf Should the elementary field operators not have overlap with the gauge-invariant operators with the same spin?}\\
 No, because they belong to different superselection sectors due to the gauge charge they carry, see section \ref{ss:classification} and \cite{Frohlich:1980gj,Frohlich:1981yi}. Note that the spontaneous breaking of gauge invariance, as it is called, is only a figure of speech, as this merely refers to gauge-fixing. This does not change the superselection structure, see section \ref{s:beh} and \cite{Banks:1979fi,'tHooft:1979bj,Frohlich:1980gj,Seiler:2015rwa}.
 
 \item{\bf Is this still true in a fixed gauge?}\\
 This is a subtle issue. Consider the scalar channel with operator $\la(\phi^\dagger\phi)(x)(\phi^\dagger\phi)(y)\ra$, but apply the FMS expansion only to one of the operators. This yields
 \bea
 \la(\phi^\dagger\phi)(x)(\phi^\dagger\phi)(y)\ra&=&v^\dagger v\la(\phi^\dagger\phi)(x)\ra+v^{a\dagger}\la(\phi^\dagger\phi)(x)\eta^a(y)\ra+v^a\la(\phi^\dagger\phi)(x)\eta^{a^\dagger}(y)\ra\nn\\
 &&+\la(\phi^\dagger\phi)(x)(\eta^\dagger\eta)(y)\ra\nn.
 \eea
 \no The first term is just a constant. However, the second and third term provide in a fixed gauge a non-zero overlap of the elementary states and the gauge-invariant states. So, in a fixed gauge, this seems to exist. But the last term is needed to maintain gauge invariance beyond leading order, and will contain in next-to-leading-order a contribution at the pole of the Higgs. Thus, beyond tree-level, even on-shell the second and third terms alone are not non-perturbatively gauge-invariant alone, and thus neglecting the fourth, e.\ g.\ in the LSZ construction, would make the result non-perturbatively gauge-dependent even on-shell. Thus, there is a gauge-dependent overlap in any fixed gauge.
 
 Even when neglecting this gauge-dependence this would not be sufficient to recover unphysical poles, as was discussed e.\ g.\ in section \ref{sss:gut}. Here, the physical operators carry projection matrices, which projects the overlaps to the physical states only. Thus, only physical states remain visible, even in the LSZ formulation. See also section \ref{ss:asymp}.
 
 \item {\bf Does a different particle content not affect anomaly cancellations?}\\
 No, because the different particle content is only with respect to the observable states. The path integral measure and Lagrangian, which are the crucial ingredients for anomaly cancellations \cite{Bohm:2001yx}, are not affected, see section \ref{ss:flavor}. The only changes necessitated by gauge invariance are on the level of matrix elements and expectation values.
 
 \item {\bf Could the 'missing' gauge-dependent states, e.\ g.\ in the SU(3) example of section \ref{ss:bsm}, not hide non-perturbatively in some operators?}\\
 Without an exact solution nothing can ever be excluded. However, this appears unlikely for several reasons. First, none of these operators carries a multiplicity structure which could be mapped to the elementary fields. Thus, degenerate states need to arise without being in a multiplet. Second, if they arise non-perturbatively, then there is no reason not to suspect even more resonances without analogue. Third, none of the lattice investigations described in section \ref{ss:latgipt} and \ref{ss:bsm} see any hint of additional light states, especially not massless states. Fourth, there is no obvious reason why the FMS mechanism should work only for some states, but not all.
 
 \item {\bf Can the gauge-dependent degrees of freedom not be merely a technical tool?}\\
 It is, of course, possible to consider the gauge-fixed, perturbative construction as an effective description, and defining the standard model to be just this part, and not a full field theory, but merely a technical implement. However, this cannot be true for the QCD subsector, where a full, non-perturbative theory is needed. To avoid getting conflicts this then requires a full treatment, leading to the topics of this review. Note that also from a philosophy of science perspective this is the adequate treatment \cite{Lyre:2008af,Francois:2017aa}.
 
 \item {\bf Who did it first?}\\
 The first explicit discussion of the necessity to formulate the observable states in a manifestly gauge-invariant way was in \cite{Banks:1979fi,'tHooft:1979bj,Frohlich:1980gj}, but the issue seems to have been known earlier \cite{Ivanov:pc,Strocchi:1977za,Osterwalder:1977pc,Englert:2004yk,Englert:2014zpa}. The formal solution was given by Fr\"ohlich, Morchio, and Strocchi in \cite{Frohlich:1980gj,Frohlich:1981yi}. During the (early) 1980ies, this was known in some part of the lattice community, e.\ g.\ \cite{Lang:pc,Shrock:1985ur,Shrock:1986av,Lee:1985yi,Olynyk:1985tr}. In the late 1980ies, early 1990ies, this seems to have been essentially forgotten, as the citation history of \cite{Frohlich:1980gj,Frohlich:1981yi} shows. In the 1990ies, BEH physics became again interesting in the lattice community, especially due to its thermodynamics, as described in section \ref{ss:ft}. In these works it appears to have been independently rediscovered that a gauge-invariant formulation of states is needed, see e.\ g.\ \cite{Karsch:1996aw,Philipsen:1996af,Philipsen:1997rq,Laine:1997nq}, but no explicit discussion of how this relates to standard perturbation theory has been made. The reason has been that for the purposes of these works, this was irrelevant \cite{Wittig:pc}. After this, the topic lay again dormant, before in 2012 in \cite{Maas:2012tj} the first explicit test using lattice methods of  \cite{Frohlich:1980gj,Frohlich:1981yi} was performed, and after that implications for beyond-the-standard model physics were being started to be studied. This is also visible in the citation history of  \cite{Frohlich:1980gj,Frohlich:1981yi}, which started to steadily rise since then.
\end{enumerate}

\itemsep -2pt
\bibliographystyle{bibstyle}
\bibliography{bib}

\end{document}